\pdfoutput=1

\documentclass[cernpreprint,texlive=2011,txfonts,UKenglish]{latex/atlasdoc} 

\usepackage{latex/atlaspackage}
\usepackage{siunitx}
\usepackage{latex/atlasbiblatex}

\usepackage{latex/atlascontribute}

\usepackage{latex/atlasphysics}

\graphicspath{{logos/}{figures/}}

\usepackage{main-defs}
\usepackage{subfig} 
\usepackage{dcolumn}
\usepackage{longtable}

\hypersetup{pdftitle={ATLAS draft},pdfauthor={The ATLAS Collaboration}}

\AtlasTitle{Search for production of vector-like quark pairs and of four top quarks in the lepton-plus-jets final state 
in $pp$ collisions at $\sqrt{s}=8\tev$ with the ATLAS detector}

\author{The ATLAS Collaboration}

\AtlasRefCode{EXOT-2013-18}

\PreprintIdNumber{CERN-PH-EP-2015-095}

\AtlasJournal{JHEP}

\AtlasAbstract{%
A search for pair production of vector-like quarks, both up-type ($T$) and down-type ($B$), as well as for four-top-quark production, is presented. 
The search is based on $pp$ collisions at $\sqrt{s}=8\tev$ recorded in 2012 with the ATLAS detector at the CERN Large Hadron Collider and
corresponding to an integrated luminosity of 20.3 fb$^{-1}$. Data are analysed in the lepton-plus-jets final state, characterised by an isolated 
electron or muon with high transverse momentum, large missing transverse momentum and multiple jets. Dedicated analyses are performed 
targeting three cases:  a $T$ quark with significant branching ratio to a $W$ boson and a $b$-quark ($T\bar{T} \to Wb$+X), and both a $T$ quark 
and a $B$ quark with significant branching ratio to a Higgs boson and a third-generation quark ($T\bar{T} \to Ht$+X and $B\bar{B} \to Hb$+X 
respectively). No significant excess of events above the Standard Model expectation is observed, and 95\% CL lower limits are derived on the 
masses of the vector-like $T$ and $B$ quarks under several branching ratio hypotheses assuming contributions from $T \to Wb$, $Zt$, $Ht$ and 
$B \to Wt$, $Zb$, $Hb$ decays. The 95\% CL observed lower limits on the $T$ quark mass range between $715\gev$ and $950\gev$ for all possible
values of the branching ratios into the three decay modes, and are the most stringent constraints to date. Additionally, the most restrictive upper bounds 
on four-top-quark production are set in a number of new physics scenarios.
}

\AtlasCoverSupportingNote{Search for $T\bar{T}\to Ht$+X, $B\bar{B}\to Hb$+X and 4-top}{https://cds.cern.ch/record/1696772}
\AtlasCoverSupportingNote{Search for $T\bar{T}\to Wb$+X}{https://cds.cern.ch/record/1952998}
\AtlasCoverCommentsDeadline{Comments deadline}

\AtlasCoverAnalysisTeam{Henri Bachacou~(*), Clement Helsens, Aurelio Juste~(*), Eve Le Menedeu, Javier Montejo, Francesco Rubbo, Antonella Sucurro}

\AtlasCoverEdBoardMember{Chiara Debenedetti, Zach Marshall, Elizaveta Shabalina~(*), Oliver Stelzer-Chilton}

\AtlasCoverEgroupEditors{atlas-EXOT-2013-18-editors@cern.ch}

\AtlasCoverEgroupEdBoard{atlas-EXOT-2013-18-editorial-board@cern.ch}

\begin{document}

\tableofcontents


\section{Introduction}
\label{sec:intro}

The discovery of a new particle consistent with the Standard Model (SM) Higgs boson by the ATLAS~\cite{Aad:2012tfa} and CMS~\cite{Chatrchyan:2012ufa} collaborations
is a major milestone in high-energy physics. However, the underlying nature of electroweak symmetry breaking remains unknown.
Naturalness arguments~\cite{Susskind:1978ms} require that quadratic divergences that arise from radiative corrections to the Higgs boson mass must be cancelled by some 
new mechanism in order to avoid fine-tuning. To that effect, several explanations have been proposed in theories beyond the SM (BSM).
In supersymmetry, the cancellation comes from assigning superpartners to the SM bosons and fermions.
Alternatively, Little Higgs~\cite{ArkaniHamed:2002qy,Schmaltz:2005ky} and Composite Higgs~\cite{Kaplan:1983sm,Agashe:2004rs} 
models introduce a spontaneously broken global symmetry, with the Higgs boson
emerging as a pseudo--Nambu--Goldstone boson~\cite{Hill:2002ap}. Such models predict the existence of vector-like quarks, 
defined as colour-triplet spin-1/2 fermions whose left- and right-handed chiral components have the same transformation properties under
the weak-isospin SU(2) gauge group~\cite{delAguila:1982fs,AguilarSaavedra:2009es}. In these models vector-like quarks are expected 
to couple preferentially to third-generation quarks~\cite{delAguila:1982fs,Aguilar-Saavedra:2013wba} and they can have flavour-changing 
neutral current decays, in addition to the charged-current decays characteristic of chiral quarks. 
As a result, an up-type quark $T$ with charge $+2/3$ can decay not only to a $W$ boson and a $b$-quark, but also to a Higgs or $Z$ boson and a
top quark ($T \to Wb$, $Zt$, and $Ht$). Similarly, a down-type quark $B$ with charge $-1/3$ can decay to a Higgs or $Z$ boson and a $b$-quark,
in addition to decaying to a $W$ boson and a top quark ($B \to Wt$, $Zb$, and $Hb$). 
In order to be consistent with the results from the precision electroweak measurements, a small mass splitting between vector-like quarks belonging to the
same SU(2) multiplet is required~\cite{Aguilar-Saavedra:2013qpa}, which forbids cascade decays such as $T \to WB$ and leaves 
direct decays into SM particles as the only possibility.
Couplings between the vector-like quarks and the first and second quark generations, although not favoured, are not excluded~\cite{Atre:2008iu,Atre:2011ae}. 
This leads to a rich phenomenology at the LHC, which the experiments are investigating.

Early searches for the pair production of exotic heavy quarks published by the ATLAS and CMS collaborations focused on  
exclusive decay modes assuming a 100\% branching ratio.
These include searches for $T\bar{T} \to W^+bW^-\bar{b}$~\cite{Aad:2012xc,Aad:2012bt,Chatrchyan:2012vu,CMS:2012ab}, 
$B\bar{B} \to ZbZ\bar{b}$~\cite{Aad:2012pga,Chatrchyan:2012af,Chatrchyan:2011ay}, 
and $B\bar{B} \to W^+tW^-\bar{t}$~\cite{ATLAS:2012aw,Chatrchyan:2012af,Chatrchyan:2012yea}.
The limits derived from these searches cannot easily be applied to other branching ratio values, due to the potentially large
expected signal contamination from mixed decay modes. A more general search strategy should consider simultaneously 
all three decay modes, providing a more extensive coverage of possible signal contributions. In absence of an excess, 
quasi-model-independent limits would be set in the plane defined by the branching ratios to two of the decay 
modes\footnote{The branching ratio to the third decay mode is fully determined by the requirement that the sum of branching 
ratios equals unity.} as a function of the heavy-quark mass. 
The first search that considered all three decay modes in the interpretation of results, performed by the ATLAS  
Collaboration using $pp$ collisions at $\sqrt{s}=7\tev$, primarily targeted the $T\bar{T} \to W^+bW^-\bar{b}$ process~\cite{ATLAS:2012qe}.
Using the full dataset collected at $\sqrt{s}=8\tev$, the ATLAS Collaboration has recently published searches for heavy quarks 
decaying to a $Z$ boson and a third-generation quark~\cite{Aad:2014efa}, and searches for heavy quarks decaying predominantly to $Wt$ 
in events with one lepton and jets~\cite{Aad:2015mba} and in events with two leptons of the same charge or three leptons~\cite{Aad:2015gdg}.
In the context of vector-like quarks, these searches are used to probe $T\bar{T}$ and $B\bar{B}$ production, and the three decay modes 
are considered in the interpretation of the results. 
The CMS Collaboration has published an inclusive search for $T\bar{T}$ production~\cite{Chatrchyan:2013uxa} resulting from the combination of 
several analyses in lepton-plus-jets and multilepton final states at $\sqrt{s}=8\tev$. This search set 95\% confidence level (CL) lower limits 
on the $T$ quark mass ranging between $690\gev$ and $780\gev$ for all possible values of the branching ratios into the three decay modes.

The results presented in this paper complete the program of searches for pair production of vector-like quarks decaying 
into third-generation quarks by the ATLAS Collaboration using the $pp$ dataset collected at $\sqrt{s}=8\tev$.
Three separate searches are presented, all of them focused on the pair production of vector-like quarks in final states involving 
one isolated electron or muon, high missing transverse momentum from the undetected neutrino and multiple jets.
The first search, referred to as $T\bar{T} \to Wb$+X, is optimised for $T\bar{T}$ production with at least one $T \rightarrow Wb$ decay, 
where the resulting $W$ boson acquires a high momentum from the large $T$ quark mass.
The second search, referred to as $T\bar{T} \to Ht$+X,  targets $T\bar{T}$ production with at least one $T \rightarrow Ht$ decay, with $H \to \bbbar$, resulting 
in events with high jet multiplicity and a large number of jets tagged as originating from $b$-quarks. 
The third search, referred to as $B\bar{B} \to Hb$+X, is instead focused on $B\bar{B}$ production with at least one $B \rightarrow Hb$ 
decay and $H \to \bbbar$, in events with the same final-state signature probed by the $T\bar{T} \to Ht$+X search. 
In all three searches the isolated lepton and the high missing transverse momentum are provided by the leptonic decay of a $W$ boson 
originating in the decay of a vector-like quark, a top quark, or a Higgs boson.

The large mass of the top quark makes it a prime candidate to help uncover the dynamics behind electroweak symmetry breaking and/or new physics at the
electroweak scale. In many new physics models the top quark plays a prominent role, often participating in new interactions 
related to electroweak symmetry breaking, or preferentially coupling to new degrees of freedom. 
Such BSM scenarios usually predict an enhanced rate of events containing four top quarks ($\fourtop$) in the final state, 
compared to the SM production via the strong interaction. Examples include top quark compositeness~\cite{Pomarol:2008bh,Lillie:2007hd,Kumar:2009vs}, 
Randall--Sundrum extra dimensions~\cite{Guchait:2007jd}, models with coloured scalars~\cite{Plehn:2008ae,Choi:2008ub,Kilic:2009mi,Kilic:2008pm,Burdman:2006gy,Calvet:2012rk},  
or universal extra dimensions~\cite{Cacciapaglia:2009pa,Cacciapaglia:2011kz,Arbey:2012ke}.
The CMS Collaboration has performed a search for SM $\fourtop$ production at $\sqrt{s}=8\tev$ in the lepton-plus-jets final state~\cite{Khachatryan:2014sca},
setting an observed (expected) 95\% CL upper limit on the production cross section of 32~fb (32 fb). 
Using multilepton final states, the ATLAS Collaboration has also searched for SM  $\fourtop$ production at $\sqrt{s}=8\tev$, setting an observed (expected) 
95\% CL upper limit of 70 fb (27 fb)~\cite{Aad:2015gdg}. The observed limit is higher than the expected one owing to an excess of data above the background
expectation with a significance of 2.5 standard deviations. In addition, the ATLAS multilepton search sensitively probes several of the above BSM scenarios 
giving rise to large enhancements in $\fourtop$ production. Given its sensitivity to a wide range of models, the $T\bar{T} \to Ht$+X search presented
in this paper is also used to search for a $\fourtop$ signal, within the SM as well as in the same BSM scenarios as the ATLAS multilepton search,
with comparable sensitivity.

\section{ATLAS detector}
\label{sec:detector}

The ATLAS detector~\cite{atlas-detector} consists of the following main subsystems: an inner tracking system, 
electromagnetic and hadronic calorimeters, and a muon spectrometer.
The inner detector provides tracking information from silicon pixel and microstrip detectors in the pseudorapidity\footnote{ATLAS 
uses a right-handed coordinate system with its origin at the nominal interaction point (IP) in the 
centre of the detector and the $z$-axis coinciding with the axis of the beam pipe.  The $x$-axis points from
the IP to the centre of the LHC ring, and the $y$-axis points upward. Cylindrical coordinates ($r$,$\phi$) are used 
in the transverse plane, $\phi$ being the azimuthal angle around the beam pipe. The pseudorapidity is defined in 
terms of the polar angle $\theta$ as $\eta = - \ln \tan(\theta/2)$.  For the purpose of the fiducial selection, this is calculated 
relative to the geometric centre of the detector; otherwise, it is relative to the reconstructed primary vertex of each event.} 
range $|\eta|<2.5$ and from a straw-tube transition radiation tracker covering $|\eta|<2.0$, all immersed in a 2 T axial magnetic field provided 
by a superconducting solenoid.  The electromagnetic (EM) sampling calorimeter uses lead as the absorber material 
and liquid-argon (LAr) as the active medium, and is divided into barrel ($|\eta|<1.475$) and end-cap ($1.375<|\eta|<3.2$) regions.  
Hadron calorimetry is also based on the sampling technique, with either scintillator tiles or LAr as the active medium, and with 
steel, copper, or tungsten as the absorber material. The calorimeters cover $|\eta|<4.9$. The muon spectrometer measures the deflection 
of muons with $|\eta|<2.7$ using multiple layers of high-precision tracking chambers located in a toroidal field of 
approximately 0.5~T and 1~T in the central and end-cap regions of ATLAS, respectively. The muon spectrometer is also 
instrumented with separate trigger chambers covering $|\eta|<2.4$.
A three-level trigger system~\cite{atlas-trigger-2010} is used to select interesting events.
The first-level trigger is implemented in custom electronics and uses a subset of detector information to reduce the event rate to at most 75~kHz.
This is followed by two software-based trigger levels exploiting the full detector information and yielding a 
typical recorded event rate of 400~Hz during 2012.

\section{Object reconstruction}
\label{sec:object_reco}

The main reconstructed objects considered in this search are electrons, muons, jets, $b$-jets and missing transverse momentum. 

Electron candidates~\cite{Aad:2014fxa} are reconstructed from energy deposits (clusters) in the EM
calorimeter that are matched to reconstructed tracks in the inner
detector.  The candidates are required to have a transverse energy\footnote{The electron transverse energy is defined as $\et=E_{\rm cluster}/\cosh{\eta_{\rm track}}$, where
$E_{\rm cluster}$ is the energy of the cluster in the calorimeter and $\eta_{\rm track}$ is the pseudorapidity of its associated track.} $\et$ greater than $25\gev$ 
and $|\eta_{\rm cluster}| < 2.47$, where $|\eta_{\rm cluster}|$ is the
pseudorapidity of the cluster associated with the electron
candidate.  Candidates in the EM calorimeter transition region $1.37 < |\eta_{\rm cluster}| < 1.52$ are excluded.  
Electrons are required to satisfy ``tight'' quality requirements~\cite{Aad:2014fxa}, which include stringent selection 
requirements on calorimeter,  tracking and combined variables that provide good separation between prompt electrons and jets.
The longitudinal impact parameter of the electron track with respect to the event's primary
vertex (see section~\ref{sec:data_presel}), $z_{0}$, is required to be less than 2 mm.
To reduce the background from non-prompt electrons resulting from semileptonic decays of $b$- or $c$-hadrons, and 
from jets with a high fraction of their energy deposited in the EM calorimeter, 
electron candidates must also satisfy calorimeter- and track-based isolation requirements.
The calorimeter isolation variable is based on the energy sum of cells  within a cone of radius 
$\Delta R = \sqrt{(\Delta\phi)^2 + (\Delta\eta)^2} = 0.2$ around the direction of each electron candidate, 
and an $\eta$-dependent requirement is made, giving an average efficiency of 90\% for prompt electrons from $Z$ boson decays. 
This energy sum excludes cells associated with the electron cluster and is corrected for
leakage from the electron cluster itself and for energy deposits from additional $pp$ 
interactions within the same bunch crossing (``pileup'').
A further 90\%-efficient isolation requirement is made on the track transverse momentum ($\pt$) sum around the
electron in a cone of radius $\Delta R = 0.3$.

Muon candidates~\cite{Aad:2014zya,Aad:2014rra} are reconstructed from track segments in the various layers of the muon spectrometer 
and matched with tracks found in the inner detector.  The final candidates are refitted using the complete
track information from both detector systems and are required to satisfy $\pt > 25\gev$ and $|\eta|<2.5$. 
Muons are required to have a hit pattern in the inner detector consistent with a well-reconstructed track 
to ensure good $\pt$ resolution.
The longitudinal impact parameter of the muon track with respect to the primary vertex, $z_{0}$, is required to be less than 2 mm.
Muons are required to satisfy a $\pt$-dependent track-based isolation
requirement: the scalar sum of the $\pt$ of the tracks within a cone of 
variable radius $\Delta R=10\gev/\pt^\mu$ around the muon (excluding the muon track itself) must be less than 5\% of the muon $\pt$ ($\pt^\mu$). 
This requirement has good signal efficiency and background rejection even under high-pileup conditions, 
as well as in boosted configurations where the muon is close to a jet.  For muons from $W$ decays in simulated $\ttbar$ events
the average efficiency of the isolation requirement is about 95\%.

Jets are reconstructed with the anti-$k_t$
algorithm~\cite{Cacciari:2008gp,Cacciari:2005hq,Cacciari:2011ma} with a
radius parameter $R=0.4$ from calibrated topological
clusters~\cite{Cojocaru:2004jk,topoclusters} built from energy deposits in the
calorimeters.  Prior to jet finding, a local cluster calibration scheme~\cite{Aad:2011he}
is applied to correct the topological cluster energies for the effects of non-compensating response of the calorimeter,
dead material and out-of-cluster leakage. The corrections are obtained
from simulations of charged and neutral particles. 
After energy calibration~\cite{Aad:2014bia}, jets are required to have 
$\pt > 25\gev$ and $|\eta| < 2.5$.  
To reduce the contamination due to jets originating from pileup interactions,
a requirement that the so-called ``jet vertex fraction" (JVF) be above 0.5 is applied to jets with $\pt<50\gev$ and $|\eta|<2.4$.
This requirement ensures that at least 50\% of the scalar sum of the $\pt$ of the tracks matched to
the jet comes from tracks originating from the primary vertex.
During jet reconstruction, no distinction is made between identified
electrons and jet energy deposits.  Therefore, if any of the jets lie 
within $\Delta R=0.2$ of a selected electron, the closest jet
is discarded in order to avoid double-counting of electrons as jets.
Finally, any electron or muon within $\Delta R=0.4$ of a selected jet is discarded.

Jets are identified as originating from the hadronisation of a $b$-quark ($b$-tagged) 
via an algorithm~\cite{CLTaggingEfficiency} that uses 
multivariate techniques to combine information from the impact
parameters of displaced tracks as well as topological properties of
secondary and tertiary decay vertices reconstructed within the jet.
For each jet, a value for the multivariate $b$-tagging discriminant is calculated,
and is considered $b$-tagged if this value is above a given threshold.
The threshold used in this search corresponds to 70\% efficiency to tag
a $b$-quark jet, with a light-jet rejection factor\footnote{The rejection factor is defined as the reciprocal of the selection efficiency.} 
of $\sim$130 and a charm-jet rejection factor of 5, as determined for jets with
$\pt >20\gev$ and $|\eta|<2.5$ in simulated $\ttbar$ events.

The missing transverse momentum ($\met$) is constructed~\cite{Aad:2012re} from the vector sum of all calorimeter energy 
deposits\footnote{Each cluster in the calorimeter is considered a massless object and
is assigned the four-momentum $(E_{\rm cluster}, \vec{p}_{\rm cluster})$, where $E_{\rm cluster}$ is the measured energy and $\vec{p}_{\rm cluster}$ is a vector
of magnitude $E_{\rm cluster}$ directed from $(x,y,z)=(0,0,0)$  to the centre
of the cluster.} contained in topological clusters. All topological cluster energies are corrected
using the local cluster calibration scheme discussed above.
Those topological clusters associated with a high-$\pt$ object (e.g. jet or electron)
are further calibrated using their respective energy corrections. 
In addition, contributions from the $\pt$ of selected muons are included in the calculation of $\met$.

\section{Data sample and event preselection}
\label{sec:data_presel}

This search is based on $pp$ collision data at $\sqrt{s}=8\tev$ collected by the ATLAS experiment between April and December 2012. 
Only events recorded with a single-electron or single-muon trigger under stable beam conditions and for which 
all detector subsystems were operational are considered.  The corresponding integrated luminosity is $20.3\pm 0.6$~$\ifb$~\cite{Aad:2013ucp}.
Single-lepton triggers with different $\pt$ thresholds are combined in a logical OR in order to increase the overall efficiency.
The $\pt$ thresholds are 24 or 60~\gev\ for the electron triggers and 24 or 36~\gev\ for the muon triggers.  The triggers with the lower $\pt$
threshold include isolation requirements on the candidate lepton, resulting in inefficiencies at high $\pt$ 
that are recovered by the triggers with higher $\pt$ threshold. Events satisfying the trigger selection are required to have at least one 
reconstructed vertex with at  least five associated tracks with $\pt>400\mev$, consistent with originating from the beam collision region in the $x$--$y$
plane. The average number of $pp$ interactions per bunch crossing is approximately 20, 
resulting in several vertices reconstructed per event. If more than one vertex is found, the hard-scatter primary vertex is taken
to be the one which has the largest sum of the squared transverse momenta of its associated tracks. For the event 
topologies considered in this paper, this requirement leads to a probability to reconstruct and select the correct hard-scatter 
primary vertex larger than 99\%.

Events are required to have exactly one reconstructed electron or muon and at least four jets satisfying the quality and 
kinematic criteria discussed in section~\ref{sec:object_reco}. 
The selected lepton is required to match, with $\Delta R < 0.15$, the lepton reconstructed by the trigger.
The background from multijet production is suppressed by a requirement on 
$\met$ as well as on the transverse mass of the lepton and $\met$ ($\mtw$).\footnote{$\mtw = \sqrt{2 p^\ell_{\mathrm T} \met (1-\cos\Delta\phi)}$, where
$p^\ell_{\mathrm T}$  is the transverse momentum (energy) of the muon (electron) and $\Delta\phi$ is the
azimuthal angle separation between the lepton and the direction of
the missing transverse momentum.}
For both lepton selections the requirements are  $\met >20\gev$ and $\met +\mtw>60\gev$.
Further suppression of the background not including $b$-quark jets is achieved by requiring at least one $b$-tagged jet in the
$T\bar{T} \to Wb$+X search, and at least two $b$-tagged jets in the $T\bar{T} \to Ht$+X and $B\bar{B} \to Hb$+X searches.
In the following, events satisfying either the electron or muon selections are combined and treated as a single analysis channel.

\section{Signal modelling}
\label{sec:signal_model}

This section describes the different signal scenarios considered in the interpretation of the results, together with details of how
they are modelled in the analysis.

\subsection{Vector-like quark pair production}
\label{sec:vlq_model}

Vector-like quarks with mass below approximately $1\tev$ are mostly produced in pairs via the strong interaction in $pp$ collisions at $\sqrt{s}=8\tev$.
For higher masses, single production mediated by the electroweak interaction can potentially dominate, depending on the strength of the interaction 
between the new quarks and the weak gauge bosons.
The predicted pair-production cross section ranges from 5.3~pb for a quark mass of $350\gev$ to 3.3~fb for a quark mass of $1000\gev$,
with an uncertainty that increases from 8\% to 14\% over this mass range.  This cross section is independent of the electroweak quantum numbers of the
new heavy quark and just depends on its mass. It was computed using {\sc Top++} v2.0~\cite{Czakon:2011xx} at next-to-next-to-leading 
order (NNLO) in QCD, including resummation of next-to-next-to-leading logarithmic (NNLL) soft gluon 
terms~\cite{Cacciari:2011hy,Baernreuther:2012ws,Czakon:2012zr,Czakon:2012pz,Czakon:2013goa}, 
and using the MSTW 2008 NNLO~\cite{Martin:2009iq,Martin:2009bu} set of parton distribution functions (PDF). 
Theoretical uncertainties result from variations on the factorisation and renormalisation scales, as well as from uncertainties on the
PDF and $\alpha_{\rm S}$. The latter two represent the largest contribution to the overall theoretical uncertainty on the cross section 
and were calculated using the PDF4LHC prescription~\cite{Botje:2011sn} 
with the MSTW 2008 68\% CL NNLO, CT10 NNLO~\cite{Lai:2010vv,Gao:2013xoa} and NNPDF2.3 5f FFN~\cite{Ball:2012cx} PDF sets.

As discussed in section~\ref{sec:intro}, vector-like quarks can couple preferentially to third-generation quarks, as the mixing between weak eigenstates of the same electric charge 
is proportional to the mass of the SM quark~\cite{delAguila:1982fs,Aguilar-Saavedra:2013wba}, and thus present a rich phenomenology.
In particular, a vector-like quark has neutral-current tree-level decays to a $Z$ or $H$ boson plus a SM quark, in
addition to the charged-current decay mode to a $W$ boson and a SM quark, which is the only decay mode chiral quarks can have. 
Figure~\ref{fig:VLQ_FD} depicts representative Feynman diagrams for the signals probed by the searches discussed in this paper. The branching ratios to each of
these decay modes vary as a function of the heavy-quark mass and depend on its weak-isospin (SU(2)) quantum numbers~\cite{AguilarSaavedra:2009es}.
Figure~\ref{fig:VLQ_BR}(a) shows the branching ratios as a function of $T$ quark mass in the SU(2) singlet and doublet hypotheses.\footnote{The branching 
ratios in figure~\ref{fig:VLQ_BR} are valid for small mixing between the new heavy quark and the third-generation quark~\cite{AguilarSaavedra:2009es,Aguilar-Saavedra:2013wba,Aguilar-Saavedra:2013qpa}.}  
In the case of a singlet, all three decay modes have sizeable branching ratios, while the charged-current decay mode $T \to Wb$ is absent in the doublet cases.
The doublet prediction is valid for an $(X,T)$ doublet, where the charge of the $X$ quark is $+5/3$, as well as a $(T,B)$ doublet when a mixing assumption 
of $|V_{Tb}| \ll |V_{tB}|$ is made, where $V_{ij}$ are the elements of a generalised Cabibbo--Kobayashi--Maskawa matrix~\cite{AguilarSaavedra:2009es}.
Since the $T$ quark branching ratios are identical in both doublets, in the following no distinction between them is made when referring to 
the $T$ quark doublet hypothesis.
Similarly, figure~\ref{fig:VLQ_BR}(b) shows the branching ratios as a function of $B$ quark mass 
in the singlet and doublet hypotheses.  In the case of a $(T,B)$ doublet with the mixing assumption $|V_{Tb}| \ll |V_{tB}|$, $\BR(B \to Wt)=1$, 
while such a decay mode is absent for the $(B,Y)$ doublet case,
where the charge of the $Y$ quark is $-4/3$. The $Y$ quark is equivalent to a chiral quark since it only has charged-current decays, $Y \to W^- b$.

Simulated samples of $T\bar{T}$ and $B\bar{B}$ are generated with the leading-order (LO) generator {\sc Protos}~v2.2~\cite{protos} using the MSTW 2008 LO PDF set 
and passed to {\sc Pythia} 6.426~\cite{Sjostrand:2006za} for parton showering and fragmentation. The AUET2B~\cite{ATLASUETune1} 
set of optimised parameters for the underlying event (UE) description, referred to as the ``UE tune'', is used.
The vector-like quarks are forced to decay with a branching ratio of $1/3$ to each of the three 
modes ($W,Z,H$).  Arbitrary sets of branching ratios consistent with the three decay modes summing to unity are obtained by reweighting the samples using particle-level 
information. Samples are generated assuming singlet couplings and for heavy-quark masses between $350\gev$ and $1100\gev$ in steps of $50\gev$.  
Additional samples are produced at two mass points ($350\gev$ and $600\gev$) assuming doublet couplings in order to confirm that kinematic differences 
arising from the different chirality of singlet and doublet couplings are negligible in this analysis. In all simulated samples (both signal and background) used in
this search, the top quark and SM Higgs boson masses are set to $172.5\gev$ and $125\gev$ respectively.
The samples are normalised using the {\sc Top++} cross section predictions discussed above.
 
\begin{figure*}[tbp]
\centering
\subfloat[]{\includegraphics[width=0.33\textwidth]{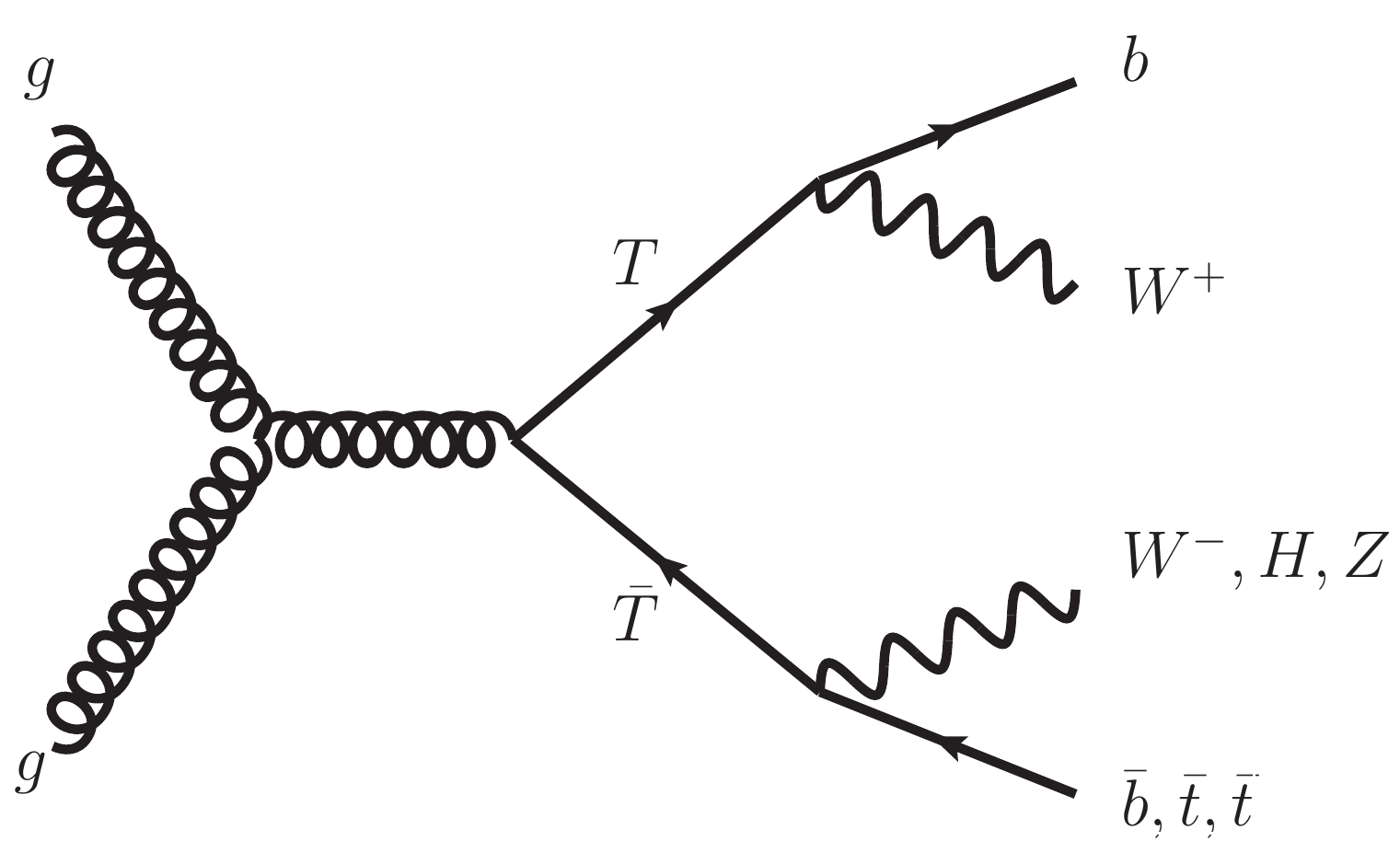}}
\subfloat[]{\includegraphics[width=0.33\textwidth]{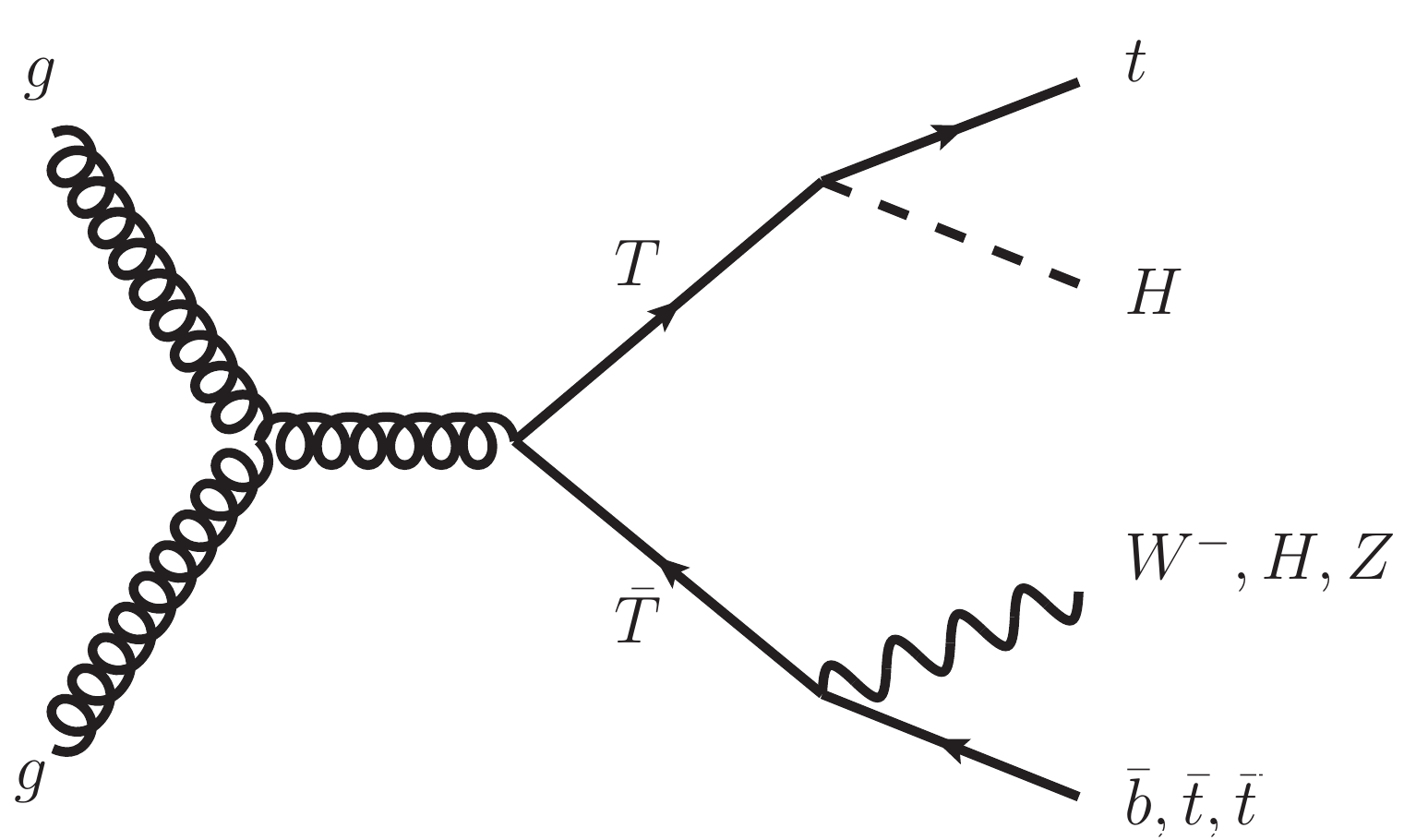}} 
\subfloat[]{\includegraphics[width=0.33\textwidth]{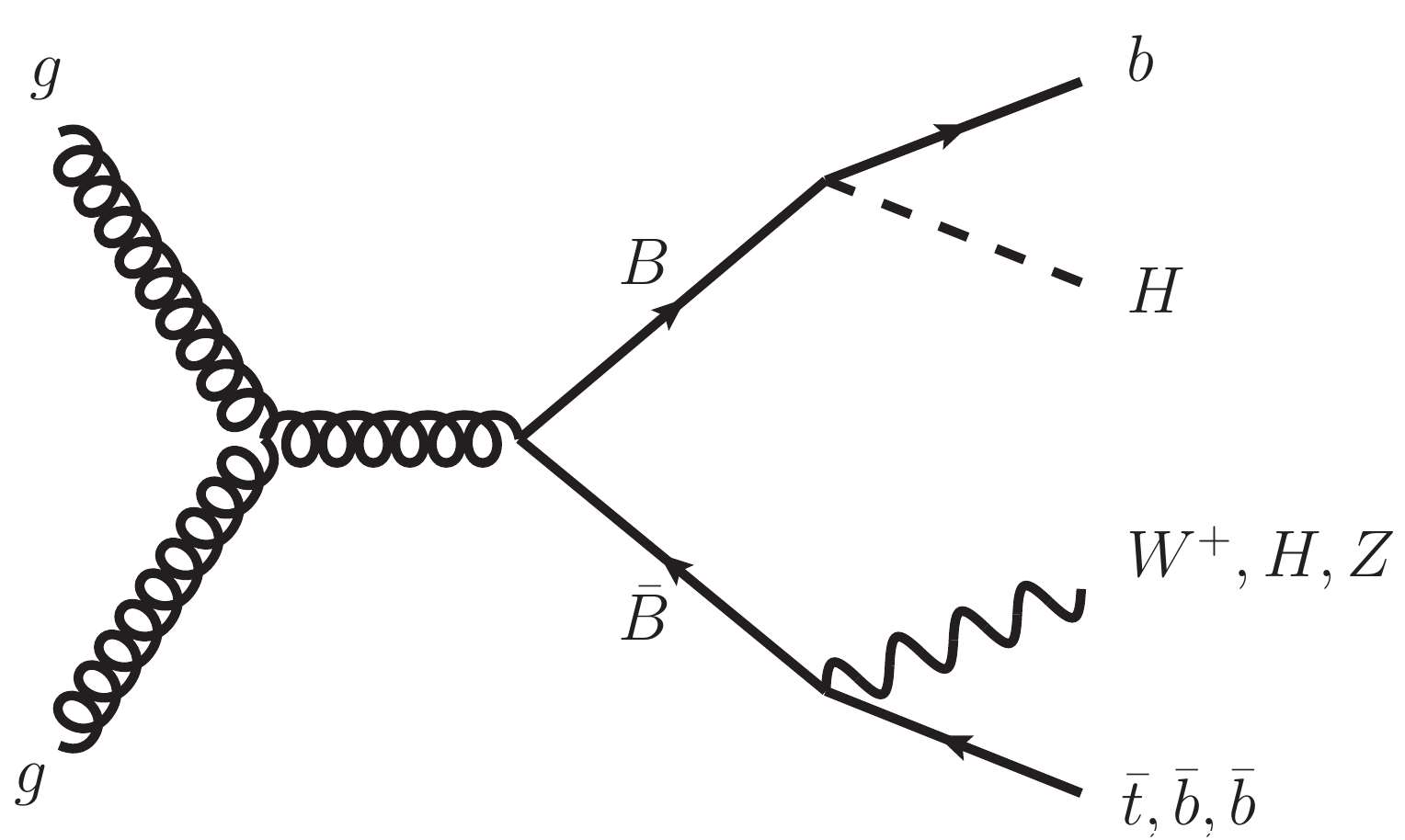}} 
\caption{Representative leading-order Feynman diagrams for $T\bar{T}$ production probed by (a) the $T\bar{T} \to Wb$+X search and
(b) the  $T\bar{T} \to Ht$+X search, and (c) for $B\bar{B}$ production probed by the $B\bar{B} \to Hb$+X search. }
\label{fig:VLQ_FD}
\end{figure*}

\begin{figure*}[tbp]
\centering
\subfloat[]{\includegraphics[width=0.45\textwidth]{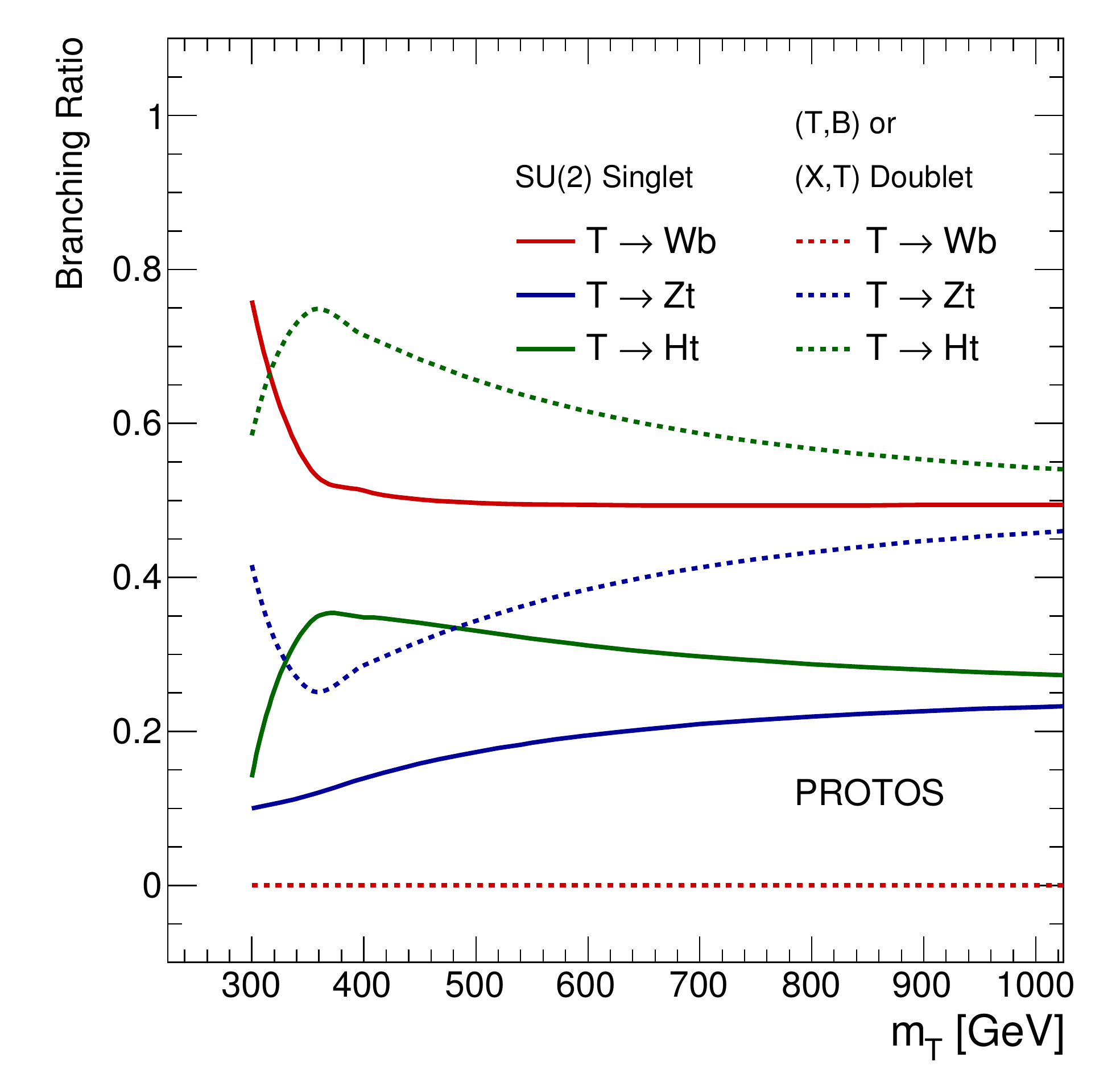}}
\subfloat[]{\includegraphics[width=0.45\textwidth]{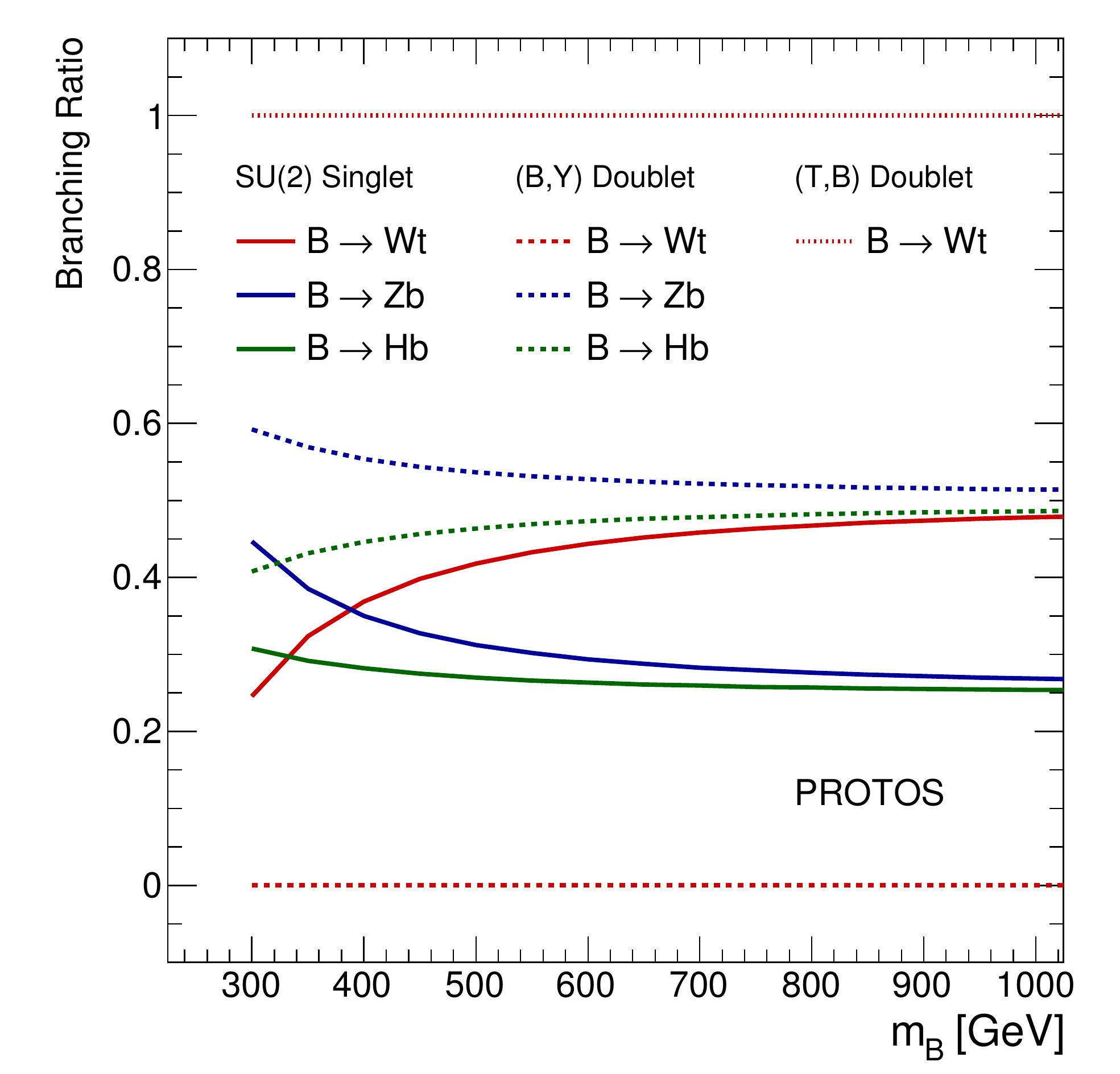}} 
\caption{Branching ratios for the different decay modes as a function of heavy-quark mass in the case of (a) a vector-like $T$ quark and
(b) a vector-like $B$ quark, as computed with {\sc Protos}. In both cases the branching ratios are
provided for an SU(2) singlet and two different SU(2) doublet scenarios. }
\label{fig:VLQ_BR}
\end{figure*}

\subsection{Four-top-quark production}

\begin{figure*}[tbp]
\centering
\begin{tabular}{cc}
\subfloat[]{\includegraphics[width=0.33\textwidth]{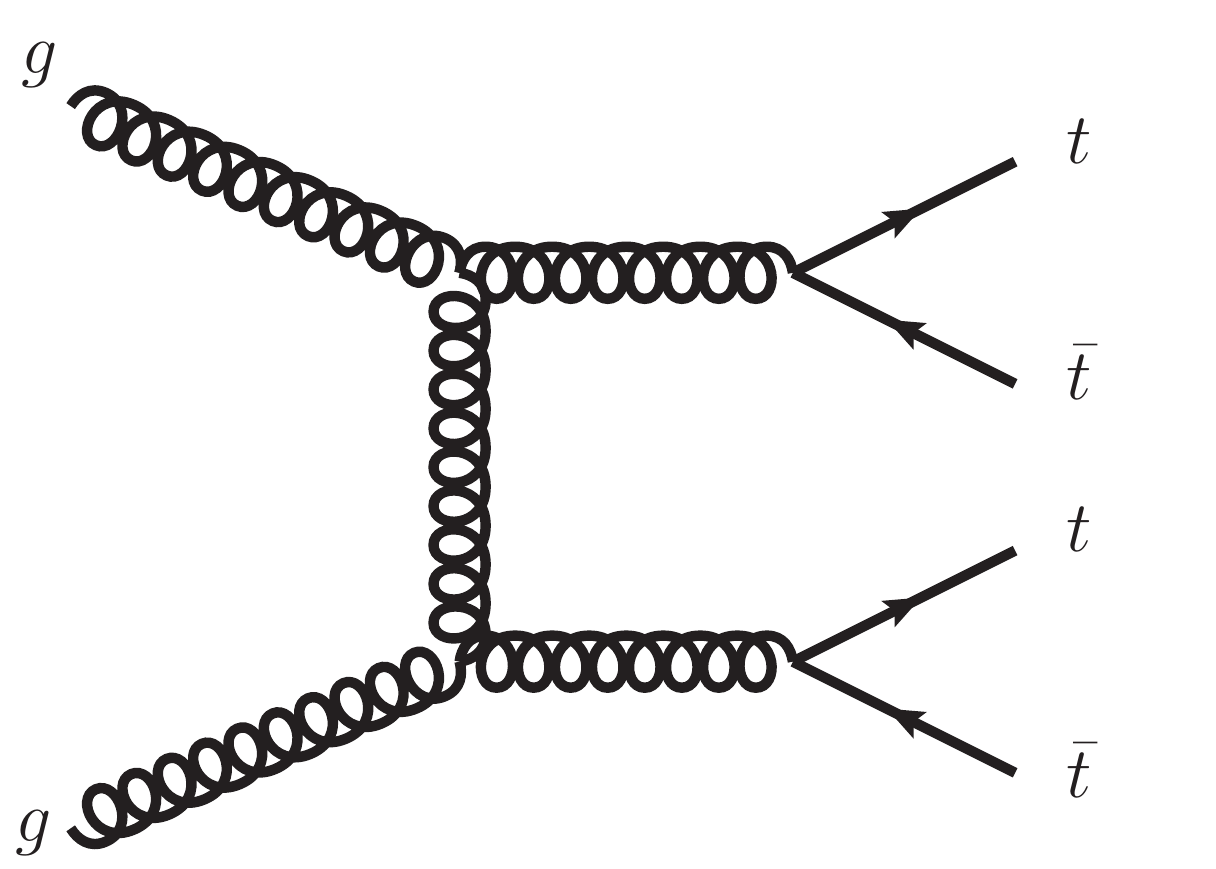}} &
\subfloat[]{\includegraphics[width=0.30\textwidth]{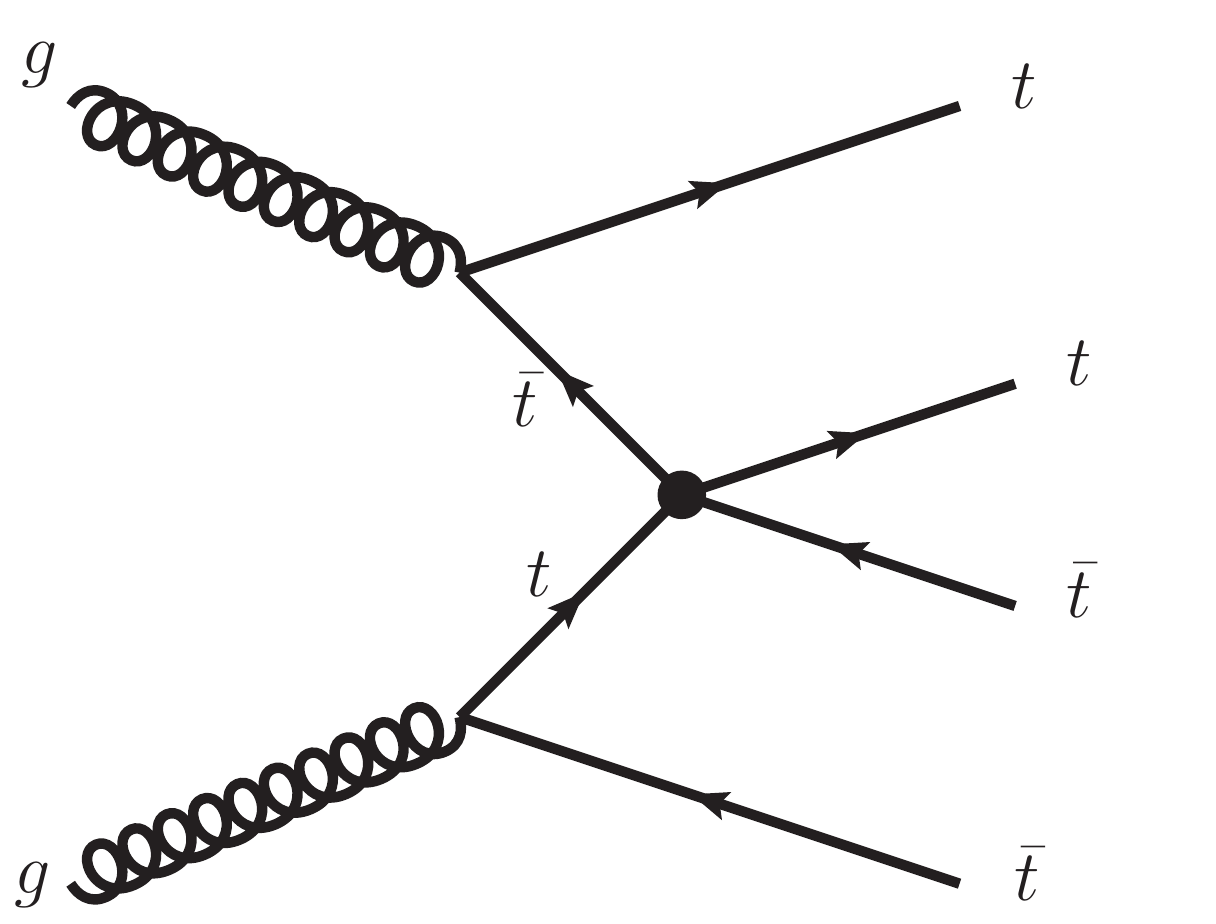}} \\
\subfloat[]{\includegraphics[width=0.33\textwidth]{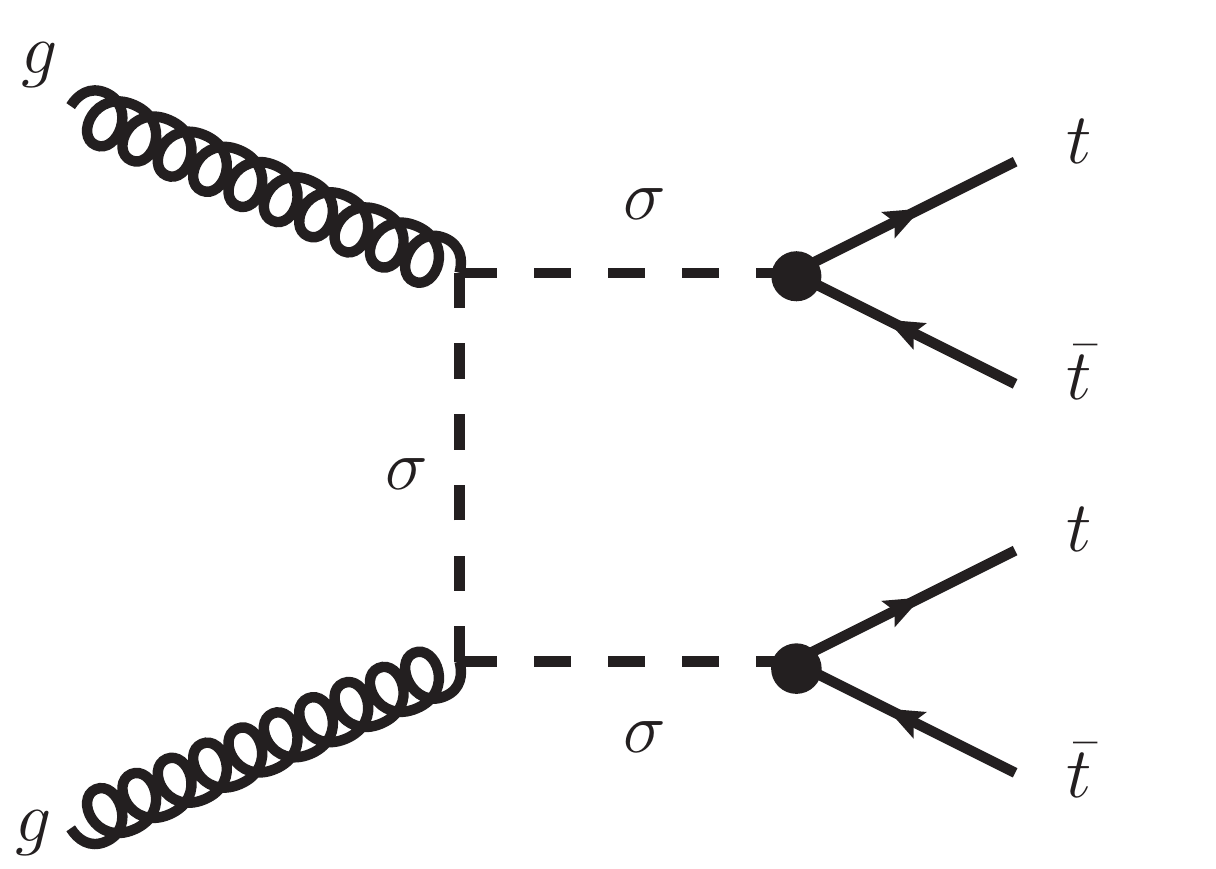}} &
\subfloat[]{\includegraphics[width=0.43\textwidth]{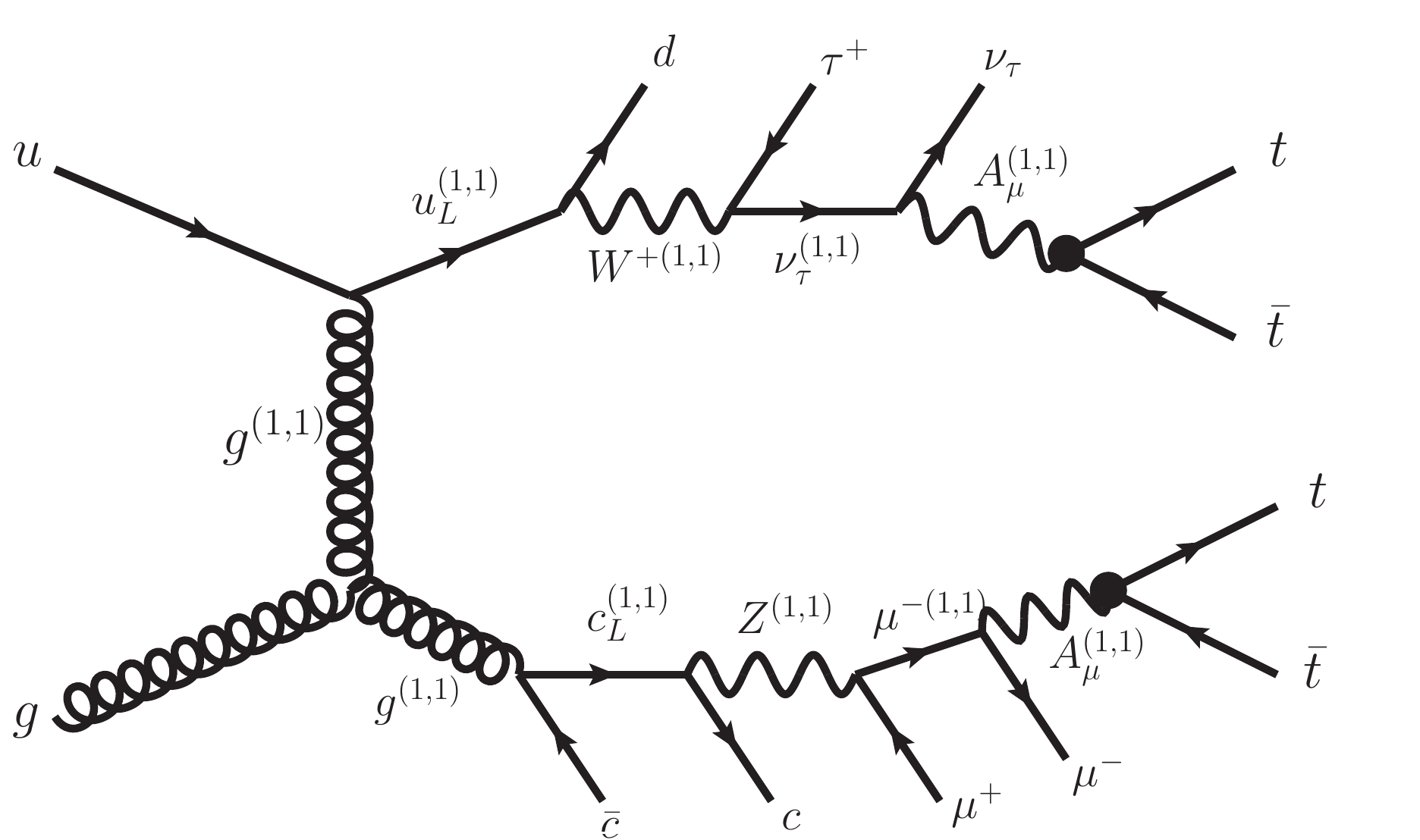}} \\
\end{tabular}
\caption{Representative leading-order Feynman diagrams for four-top-quark production within (a) the SM 
and several beyond-the-SM scenarios (see text for details): (b) via an effective four-top-quark interaction in an effective field theory 
model, (c) via scalar-gluon-pair production, and (d) via cascade decays from Kaluza--Klein excitations in a universal extra dimensions 
model with two extra dimensions compactified using the geometry of the real projective plane. }
\label{fig:fourtop_FD}
\end{figure*}

The production cross section for four-top-quark events in the SM is very small ($\sigma_{\fourtop} \simeq 1$~fb at $\sqrt{s}=8\tev$)~\cite{Barger:1991vn,Barger:2010uw}, 
but it can be significantly enhanced in several BSM scenarios. Figure~\ref{fig:fourtop_FD} depicts representative LO Feynman diagrams
for four-top-quark production within the SM and the different BSM scenarios considered in this paper. A class of models involving new heavy vector 
particles strongly coupled to the right-handed top quark, such as top quark compositeness~\cite{Pomarol:2008bh,Lillie:2007hd,Kumar:2009vs} 
or Randall--Sundrum extra dimensions~\cite{Guchait:2007jd},
can be described via an effective field theory (EFT) involving a four-fermion contact interaction~\cite{Degrande:2010kt} (figure~\ref{fig:fourtop_FD}(b)). 
The Lagrangian assumed is 

\begin{equation}
{\cal L}_{4t} = \frac{|C_{4t}|}{\Lambda^2} (\tbar_{\rm R} \gamma^\mu t_{\rm R}) (\tbar_{\rm R} \gamma_\mu t_{\rm R}),
\end{equation}

\noindent where $t_{\rm R}$ is the right-handed top quark spinor,  $\gamma_\mu$ are the Dirac matrices, $C_{4t}$ is the coupling constant, 
and $\Lambda$ is the energy scale of new physics.
Only the contact interaction operator with right-handed top quarks is considered, since left-handed operators are already strongly
constrained by the precision electroweak measurements~\cite{Georgi:1994ha}. 

In addition, two specific models are considered involving new heavy particles: scalar gluon (sgluon) pair production and 
a Universal Extra Dimensions (UED) model. Sgluons are colour-adjoint scalars, denoted by $\sigma$, that
appear in several extensions of the SM, both supersymmetric~\cite{Plehn:2008ae,Choi:2008ub} and 
non-supersymmetric~\cite{Kilic:2009mi,Kilic:2008pm,Burdman:2006gy,Calvet:2012rk}. The dominant production
mode at the LHC is in pairs via the strong interaction, $gg \to \sigma \sigma$. For sgluon masses above
twice the top quark mass, the dominant decay mode is into $\ttbar$, giving rise to a four-top-quark final
state (figure~\ref{fig:fourtop_FD}(c)). The UED model considered has two extra dimensions that are
compactified using the geometry of the real projective plane (2UED/RPP)~\cite{Cacciapaglia:2009pa}, leading to a
discretisation of the momenta along their directions. A tier of Kaluza--Klein towers is labelled by two 
integers, $k$ and $\ell$, referred to as ``tier $(k,\ell)$''. Within a given tier, the squared masses of the particles 
are given at leading order by  $m^2 = k^2/R_4^2+\ell^2/R_5^2$, where $\pi R_4$ and $\pi R_5$ 
are the size of the two  extra dimensions. The model is parameterised by $R_4$ and $R_5$ or, alternatively, by 
$m_{\KK}=1/R_4$ and $\xi=R_4/R_5$. Four-top-quark production can arise from tier (1,1), where particles
from this tier have to be pair produced because of symmetries of the model. Then they chain-decay to the 
lightest particle of this tier, the heavy photon $A^{(1,1)}$, by emitting SM particles (figure~\ref{fig:fourtop_FD}(d)). 
The branching ratios of $A^{(1,1)}$ into SM particles are not predicted by the model, although the decay 
into $\ttbar$ is expected to be dominant~\cite{Cacciapaglia:2011kz}. Four-top-quark events can also arise from 
tiers (2,0) and (0,2) via a similar mechanism. In this case the expected cross section for four-top-quark production 
is reduced compared to that from tier (1,1) since each state in tiers (2,0) and (0,2) can decay 
directly into a pair of SM particles or into a pair of states in tiers (1,0) or (0,1) via bulk interactions, 
resulting in smaller branching ratios for decay into $t\bar{t}$~\cite{Cacciapaglia:2011kz}.
In the following, when considering four-top-quark production from a given tier, it is 
assumed that the $A$ photon in that tier decays with 100\% branching ratio into $\ttbar$ 
while $A$ photons from other tiers cannot decay into $\ttbar$.
Within this model, observations of dark-matter relic abundance prefer values of $m_{\KK}$ between $600\gev$ and $1200\gev$~\cite{Arbey:2012ke}.

Simulated samples of four-top-quark production within the SM, within an EFT model, and within the 2UED/RPP model, 
are generated with the {\sc Madgraph5} 1.3.33~\cite{Alwall:2011uj}  LO
generator and the MSTW 2008 PDF set, interfaced to {\sc Pythia} 8.1~\cite{Sjostrand:2007gs} and the 
AU2 UE tune~\cite{ATLASUETune3}. In the case of the 2UED/RPP model, samples are
generated for four different values of $m_{\KK}$ (600, 800, 1000 and $1200\gev$) and the {\sc Bridge}~\cite{Meade:2007js} 
generator is used to decay the pair-produced excitations from tier (1,1) generated by {\sc Madgraph5}.
Constraints for tiers (2,0) and (0,2) can be derived from those for tier (1,1) together with the theoretical cross sections.
Samples of four-top-quark production via sgluon pairs are generated with {\sc Pythia} 6.426 with the 
CTEQ6L1~\cite{Nadolsky:2008zw} PDF set and the AUET2B UE tune, for seven different values of the sgluon mass
between $350\gev$ and $1250\gev$, and normalised to the NLO theoretical cross section~\cite{GoncalvesNetto:2012nt}. 

Events from minimum-bias interactions are simulated with the {\sc Pythia} 8.1 generator with the 
MSTW 2008 LO PDF set and the A2 tune~\cite{ATLASUETune3}. They are overlaid on the simulated signal 
events according to the luminosity profile of the recorded data. 
The contributions from these pileup interactions are modelled both within the same bunch crossing as the 
hard-scattering process and in neighbouring bunch crossings.  Finally, the generated samples are processed 
through a simulation~\cite{Aad:2010ah} of the detector geometry and response using {\sc Geant4}~\cite{Agostinelli:2002hh} 
with a fast simulation of the calorimeter response~\cite{Aad:2010ah}. 
All samples are processed through the same reconstruction software as the data. 
Simulated events are corrected so that the object identification efficiencies, energy
scales and energy resolutions match those determined from data control samples.

\section{Background modelling}
\label{sec:bkg_model}

After event preselection, the main background is $\ttbar$+jets
production, with the production of a $W$ boson in association with
jets ($W$+jets) and multijet events contributing to a lesser extent.
Small contributions arise from single top quark, $Z$+jets and diboson
($WW,WZ,ZZ$) production, as well as from the associated production of a vector boson $V$ ($V=W,Z$)
or a Higgs boson and a $\ttbar$ pair ($\ttbar V$ and $\ttbar H$). Multijet events contribute 
to the selected sample via the misidentification of a jet or a photon as an
electron or via the presence of a non-prompt lepton, e.g.~from a
semileptonic $b$- or $c$-hadron decay; the corresponding yield is estimated via data-driven methods.  
The rest of the background contributions are estimated from 
simulation and normalised to their theoretical cross sections. 
In the case of the $\ttbar$+jets and $W/Z$+jets background predictions, 
further corrections are applied to improve agreement between the data and simulation,
as discussed in sections~\ref{sec:bkgmodel_ttbar} and~\ref{sec:bkgmodel_vjets} respectively.

All simulated background samples utilise {\sc Photos 2.15}~\cite{Golonka:2005pn} to simulate 
photon radiation and {\sc Tauola 1.20}~\cite{Jadach:1990mz} to simulate $\tau$ decays.
Similarly to the signal samples, they also include a simulation of pileup interactions, and are
processed through a full {\sc Geant4} detector simulation and the same reconstruction software as the data.
Further details about the modelling of each of the backgrounds are provided below.

\subsection{$\ttbar$+jets background}
\label{sec:bkgmodel_ttbar}
Simulated samples of $\ttbar$+jets events are generated with the next-to-leading-order (NLO) generator 
{\sc Powheg-Box} 2.0~\cite{Frixione:2007nw,Nason:2004rx,Frixione:2007vw,Alioli:2010xd}  using the CT10 PDF set~\cite{Lai:2010vv}. 
The nominal sample is interfaced to 
{\sc Pythia} 6.425~\cite{Sjostrand:2006za} with the CTEQ6L1 PDF set and the 
Perugia2011C UE tune~\cite{Skands:2010ak}.  An alternative sample, used to study 
the uncertainty related to the fragmentation model, is interfaced to {\sc Herwig} v6.520~\cite{Corcella:2000bw} 
with the CTEQ6L1 PDF set and {\sc Jimmy} v4.31~\cite{Butterworth:1996zw} to simulate the UE.
The $\ttbar$+jets samples are normalised 
to the theoretical cross section obtained with {\sc Top++}, performed at NNLO in QCD and including resummation 
of NNLL soft gluon terms.

The $\ttbar$+jets samples are generated inclusively, but events are categorised depending
on the flavour content of additional particle jets in the event (i.e. jets not originating from
the decay of the $\ttbar$ system). Particle jets are reconstructed with the anti-$k_t$
algorithm with a radius parameter $R=0.4$ and are required to have $\pt>15\gev$ and
$|\eta|<2.5$. Events where at least one such particle jet is matched within $\Delta R<0.4$ to a $b$-hadron
with $\pt>5\gev$ not originating from a top quark decay are labelled as $\ttbb$ events.
Similarly, events where at least one such particle jet is matched within $\Delta R<0.4$ to a $c$-hadron
with $\pt>5\gev$ not originating from a $W$ boson decay, that are not labelled already as $\ttbb$, 
are labelled as $\ttcc$ events. Events labelled as either $\ttbb$  or
$\ttcc$ are generically referred to below as $\ttbar$+HF events, where HF stands for ``heavy flavour''.
The remaining events are labelled as $\ttbar$+light-jet events, including those with no additional jets. 
In {\sc Powheg}+{\sc Pythia} the modelling of $\ttbar$+HF is via the parton-shower evolution. To study 
uncertainties related to this simplified description, an alternative $\ttbar$+jets sample is generated
with {\sc Madgraph5} 1.5.11 using the CT10 PDF set. It includes tree-level diagrams with up to three 
additional partons (including $b$- and $c$-quarks) and is interfaced to {\sc Pythia} 6.425.

Since the best possible modelling of the $\ttbar$+jets background is a key aspect of these searches, 
a correction is applied to simulated $\ttbar$ events in {\sc Powheg}+{\sc Pythia}
based on the ratio of the differential cross sections measured in data and simulation at $\sqrt{s}=7\tev$ 
as a function of top quark $\pt$ and $\ttbar$ system $\pt$~\cite{Aad:2014zka}.
The stability of the ratio between $\sqrt{s}=7\tev$ and $\sqrt{s}=8\tev$ was studied to support
the usage of $\sqrt{s}=7\tev$ data to correct the simulation at $\sqrt{s}=8\tev$.
This correction significantly improves agreement between simulation and data in 
distributions such as the jet multiplicity and the \pt of decay products of the $\ttbar$ system.
This correction is applied only to $\ttbar$+light-jets and $\ttcc$ events.
The modelling of the $\ttbb$ background, particularly important for the $Ht/Hb$+X searches,
is improved by reweighting the {\sc Powheg}+{\sc Pythia} prediction to an NLO prediction 
of $\ttbb$ including parton showering~\cite{Cascioli:2013era}, based on 
{\sc Sherpa+OpenLoops}~\cite{Gleisberg:2008ta, Cascioli:2011va} using the CT10 PDF set.  This reweighting is performed
for different topologies of $\ttbb$ in such a way that the inter-normalisation of each of the 
categories and the relevant kinematic distributions are at NLO accuracy. 
More details about the modelling of the $\ttbar$+jets background can be found in ref.~\cite{Aad:2015gra}.

\subsection{$W/Z$+jets background}
\label{sec:bkgmodel_vjets}
Samples of $W/Z$+jets events are generated with up to five additional partons using the {\sc Alpgen} v2.14~\cite{Mangano:2002ea} 
LO generator and the CTEQ6L1 PDF set, interfaced to {\sc Pythia} v6.426 for parton showering and fragmentation.
To avoid double-counting of partonic configurations generated by both the matrix-element  calculation and the parton shower, 
a parton--jet matching scheme (``MLM matching'')~\cite{Mangano:2001xp} is employed. 
The $W$+jets samples are generated separately for $W$+light-jets, $Wb\bar{b}$+jets, $Wc\bar{c}$+jets, and $Wc$+jets. 
The $Z$+jets samples are generated separately for $Z$+light-jets, $Zb\bar{b}$+jets, and $Zc\bar{c}$+jets. Overlap between 
$VQ\bar{Q}$+jets ($V=W,Z$ and $Q=b,c$)  events generated from the matrix-element calculation and those generated from parton-shower 
evolution in the $W/Z$+light-jets samples is avoided via an algorithm based on the angular separation between the extra heavy quarks:
if $\Delta R(Q,\bar{Q})>0.4$, the matrix-element prediction is used, otherwise the parton-shower prediction is used. 
Both the $W$+jets and $Z$+jets background contributions are normalised to their inclusive NNLO theoretical cross sections~\cite{Melnikov:2006kv}.
Further corrections are applied to $W/Z$+jets events in order to better describe data in the preselected sample.
Scale factors for each of the $W$+jets categories ($Wb\bar{b}$+jets, $Wc\bar{c}$+jets, $Wc$+jets and $W$+light-jets) are derived
for events with one lepton and at least four jets by simultaneously analysing six different event categories, defined by the
$b$-tag multiplicity (0, 1 and $\geq$2) and the sign of the lepton charge. The $b$-tag multiplicity provides information 
about the heavy-flavour composition of the $W$+jets background, while the lepton charge is used to determine the normalisation 
of each component, exploiting the expected charge asymmetry for $W$+jets production in $pp$ collisions as predicted by {\sc Alpgen}.
In the case of $Z$+jets events, a correction to the heavy-flavour fraction was derived to reproduce the relative rates of $Z$+2-jets
events with zero and one $b$-tagged jets observed in data. In addition, the $Z$ boson $\pt$ spectrum was compared between
data and the simulation in $Z$+2-jets events, and a reweighting function was derived in order to improve the modelling.

\subsection{Other simulated background}

Samples of single-top-quark backgrounds corresponding to the $t$-channel, $s$-channel and $Wt$ production mechanisms 
are generated with {\sc Powheg-Box} 2.0~\cite{Alioli:2009je,Re:2010bp} 
using the CT10 PDF set and  interfaced  to {\sc Pythia} 6.425 with the CTEQ6L1 PDF set and the Perugia2011C UE tune.  
Overlaps between the \ttbar\ and $Wt$ final states are removed using the ``diagram removal'' scheme~\cite{Frixione:2005vw}.
The single-top-quark samples are normalised to the approximate NNLO theoretical cross sections~\cite{Kidonakis:2011wy,Kidonakis:2010ux,Kidonakis:2010tc}
calculated using the MSTW 2008 NNLO PDF set. 

The $WW/WZ/ZZ$+jets samples are generated with up to three additional partons using {\sc Alpgen} v2.13 and the 
CTEQ6L1 PDF set, interfaced to {\sc Herwig} v6.520 and {\sc Jimmy} v4.31 for parton showering, fragmentation and UE modelling. 
The MLM parton--jet matching scheme is used. The $WW$+jets samples require at least
one of the $W$ bosons to decay leptonically, while the $WZ/ZZ$+jets samples require one $Z$ boson to decay leptonically,
with the other boson decaying inclusively. Additionally, $WZ$+jets samples requiring the $W$ and $Z$ bosons to decay
leptonically and hadronically respectively, are generated with up to three additional partons 
(including massive $b$- and $c$-quarks) using {\sc Sherpa} v1.4.1 and the CT10 PDF set.
All diboson samples are normalised to their NLO theoretical cross sections~\cite{Campbell:1999ah}

Samples of $\ttbar V$ events, including $\ttbar WW$, are generated with up to two additional partons using {\sc Madgraph5} 1.3.28 
with the CTEQ6L1 PDF set, and interfaced to {\sc Pythia} 6.425 with the AUET2B UE tune.
A sample of $\ttbar H$ events is generated with the {\sc PowHel} framework~\cite{Garzelli:2011vp}, which combines the {\sc Powheg-Box} generator and
NLO matrix elements obtained from the HELAC-Oneloop package~\cite{Bevilacqua:2011xh}. The sample is generated
using the CT10nlo PDF set~\cite{Lai:2010vv}. Showering is performed with
{\sc Pythia} 8.1 using the CTEQ6L1 PDF set and the AU2 UE tune~\cite{ATLASUETune1,ATLASUETune2}. 
Inclusive decays of the Higgs boson are assumed in the generation of the $\ttbar H$ sample.
The $\ttbar V$ samples are normalised to the NLO cross section predictions~\cite{Garzelli:2012bn}.
The $\ttbar H$ sample is normalised using the NLO cross section~\cite{Dawson:2003zu,Beenakker:2002nc,Beenakker:2001rj}  
and the Higgs decay branching ratios~\cite{Djouadi:1997yw,Bredenstein:2006rh,Actis:2008ts,Denner:2011mq} collected in ref.~\cite{Dittmaier:2011ti}.

\subsection{Multijet background}
Multijet events can enter the selected data sample through several
production and misreconstruction mechanisms.  In the electron
channel, the multijet background consists of non-prompt electrons
as well as misidentified photons (e.g.~with a conversion into an $e^+e^-$ pair)
or jets with a high fraction of their energy deposited in the EM calorimeter.  
In the muon channel, the background contributed by multijet events is predominantly 
due to final states with non-prompt muons, such as those from semileptonic $b$-
or $c$-hadron decays.  

The multijet background normalisation and shape are estimated directly
from data by using the ``matrix method''
technique~\cite{Aad:2010ey}.  The matrix method exploits differences in lepton-identification-related 
properties between prompt, isolated leptons from
$W$ and $Z$ boson decays (referred to as ``real leptons'' below) and those
where the leptons are either non-isolated or result from the
misidentification of photons or jets (referred to as ``fake leptons'' below). 
For this purpose, two samples are defined after imposing the final kinematic 
selection criteria, differing only in the lepton identification criteria: a ``tight''
sample and a ``loose'' sample, the former being a subset of the
latter.  The tight selection employs the complete set of lepton identification criteria 
used in the analysis. For the loose selection the lepton isolation requirements are omitted.
The method assumes that the number of
selected events in each sample ($\nl$ and $\nt$) can be expressed as a
linear combination of the numbers of events with real and fake
leptons, so that the number of multijet events in the tight sample 
is given by 

\begin{equation}
N^{\mathrm{tight}}_{\mathrm{MJ}} = \frac{\epsf}{\epsr-\epsf}(\epsr \nl - \nt),
\end{equation}

\noindent where $\epsr$ ($\epsf$) represents the probability for a real (fake) lepton that satisfies
the loose criteria to also satisfy the tight ones. Both probabilities are measured
in data control samples. 
To measure $\epsr$, samples enriched in real leptons from $W$ bosons decays
are selected by requiring high $\met$ or $\mtw$. The average $\epsr$ is $\sim$0.75 ($\sim$0.98) in the
electron (muon) channel. To measure $\epsf$, samples enriched in multijet background are selected
by requiring either low $\met$ (electron channel) or high impact parameter significance for the lepton track (muon channel).
The average $\epsf$ value is $\sim$0.35 ($\sim$0.20) in the electron (muon) channel. Dependencies 
of $\epsr$ and $\epsf$ on quantities such as lepton $\pt$ and $\eta$, $\Delta R$ between the
lepton and the closest jet, or number of $b$-tagged jets, are parameterised in order to obtain
a more accurate estimate.

\section{Search for $T\bar{T} \to Wb$+X production}
\label{sec:search_wbx}

This search is sensitive to $T\bar{T}$ production where at least one of the $T$ quarks decays into a $W$ boson and a $b$-quark, although
it is particularly optimised for $T\bar{T} \to W^+ b W^- \bar{b}$ events. One of the $W$ bosons present in the final state is then required to decay leptonically.
After the preselection described in section~\ref{sec:data_presel}, further background suppression is achieved by applying requirements aimed at exploiting the distinct kinematic 
features of the signal. The large $T$ quark mass results in energetic $W$ bosons and $b$-quarks in the final state with large 
angular separation between them, while the decay products from the boosted $W$ bosons have small angular separation. 
The combination of these properties is very effective in distinguishing the dominant
$t\bar{t}$ background since $t\bar{t}$ events with boosted $W$ boson configurations are rare and are typically
characterised by a small angular separation between the $W$ boson and the $b$-quark from the top quark decay.

To take advantage of these properties, it is necessary to identify the 
hadronically decaying $W$ boson ($W_{\rm had}$) as well as the $b$-jets in the event.
The candidate $b$-jets are defined as the two jets with the highest $b$-tag discriminant value,
although only one of them is explicitly required to be $b$-tagged in the event selection.
Two types of $W_{\rm had}$ candidates are defined, $W_{\rm had}^{\rm type\;I}$ and $W_{\rm had}^{\rm type\;II}$,
depending on the angular separation between their decay products. $W_{\rm had}^{\rm type\;I}$ candidates correspond to 
boosted $W$ bosons, where the quarks from the $W$-boson decay emerge with small angular separation and are reconstructed as a single jet.
Alternatively, $W_{\rm had}^{\rm type\;II}$ candidates are characterised by two reconstructed jets.
In the construction of both types of $W_{\rm had}$ candidates, the two candidate $b$-jets are not considered.

A $W_{\rm had}^{\rm type\;I}$ candidate is defined as a single jet with $\pt>400\gev$, which is the typical $\pt$ above which the decay
products from a $W$ boson would have an angular separation $\Delta R \leq R_\mathrm{cone}=0.4$.
A $W_{\rm had}^{\rm type\;II}$ candidate is defined as a dijet system with $\pt>250\gev$, angular separation 
$\Delta R(j,j)<0.8$ and mass within the range of $60$--$120\gev$. 
The asymmetric window about the $W$-boson mass value is chosen in order to increase the acceptance for
hadronically decaying $Z$ bosons from $T\bar{T} \to Wb Zt$ events.  
Any jets satisfying the $W_{\rm had}^{\rm type\;I}$ requirements are excluded from consideration when forming $W_{\rm had}^{\rm type\;II}$ candidates.
The leptonically decaying $W$ boson ($W_{\rm lep}$) is reconstructed using the lepton and $\met$, which is taken as a measurement of the neutrino $\pt$. 
Requiring that the invariant mass of the lepton--neutrino system equals the nominal $W$ boson mass allows reconstruction 
of the longitudinal momentum of the neutrino up to a two-fold ambiguity. If two solutions exist, they are both considered.
If no real solution exists, the pseudorapidity of the neutrino is set equal to that of the lepton, since in the kinematic regime of interest  
the decay products of the $W$ boson tend to be collinear.

\begin{table}[h!]
\begin{center}
\begin{tabular}{ll}
\toprule\toprule
Selection & Requirements \\
\midrule
Preselection & Exactly one electron or muon  \\
                       & $\met >20\gev$, $\met +\mtw>60\gev$ \\
                 	     & $\geq$4 jets, $\geq$1 $b$-tagged jets \\
\midrule
Loose selection & Preselection  \\
                  & $\geq$1~$W_{\rm had}$ candidate (type I or type II) \\
                  & $\HT>800\gev$ \\
                  & $\pt(b_1) > 160\gev$, $\pt(b_2) >110\gev$ (type I) or $\pt(b_2) >80\gev$ (type II)\\
                  & $\Delta R(\ell,\nu)<0.8$ (type I) or $\Delta R(\ell,\nu)<1.2$ (type II)\\
\midrule
Tight  selection & Loose selection \\
		      	      &  $\min(\Delta R(\ell, b_{1,2}))>1.4$,  $\min(\Delta R(W_{\rm had}, b_{1,2}))>1.4$ \\    
			      &  $\Delta R(b_1,b_2)>1.0$ (type I) or $\Delta R(b_1,b_2)>0.8$ (type II)\\   
			      & $\Delta m <250\gev$ (type I) [see text for definition]\\                
\bottomrule\bottomrule
\end{tabular}
\caption{\small{Summary of event selection requirements for the $T\bar{T} \to Wb$+X analysis (see text for details).}}
\label{tab:selection}
\end{center}
\end{table}

\begin{figure*}[tbp]
\centering
\subfloat[]{\includegraphics[width=0.45\textwidth]{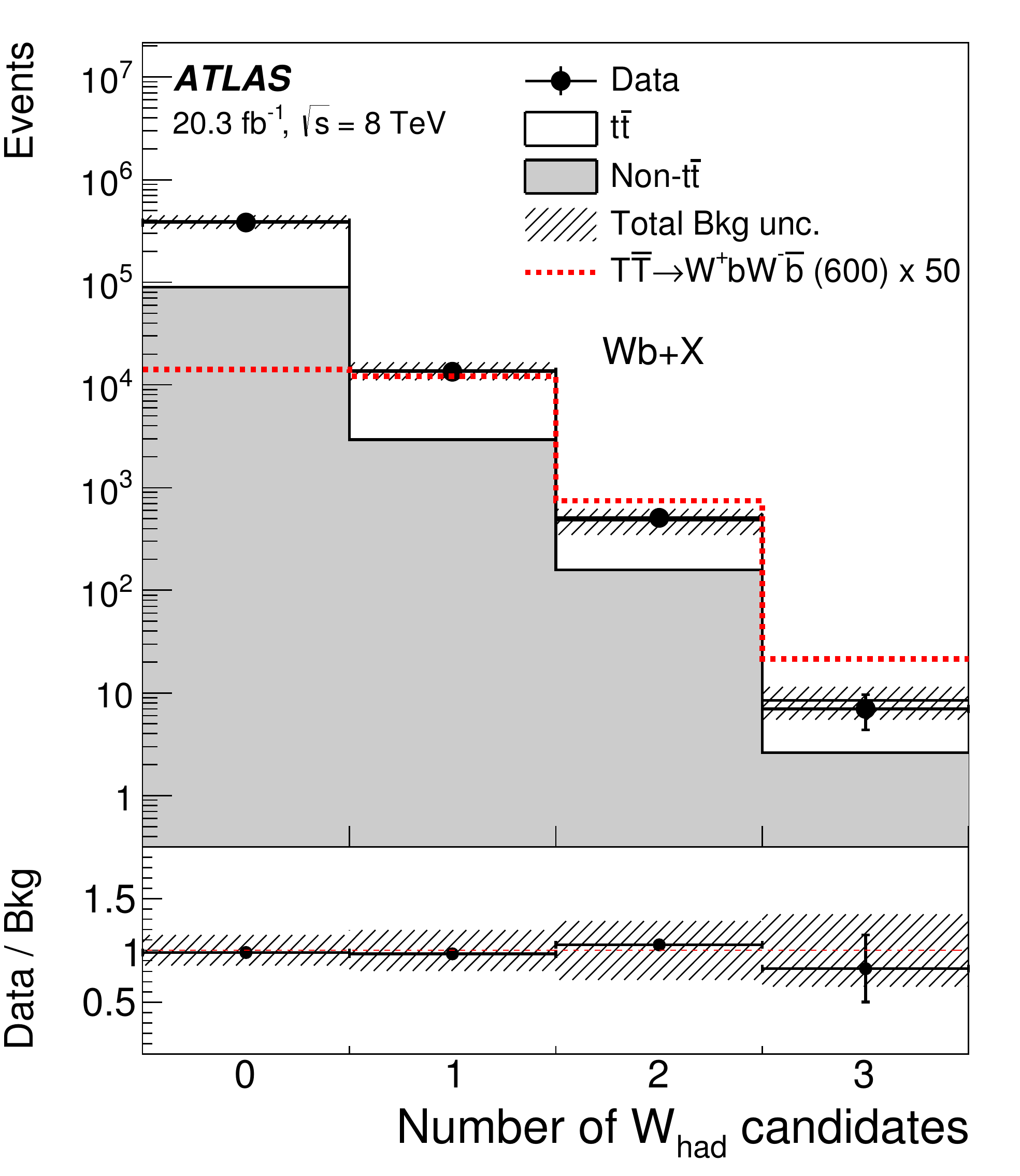}}
\subfloat[]{\includegraphics[width=0.45\textwidth]{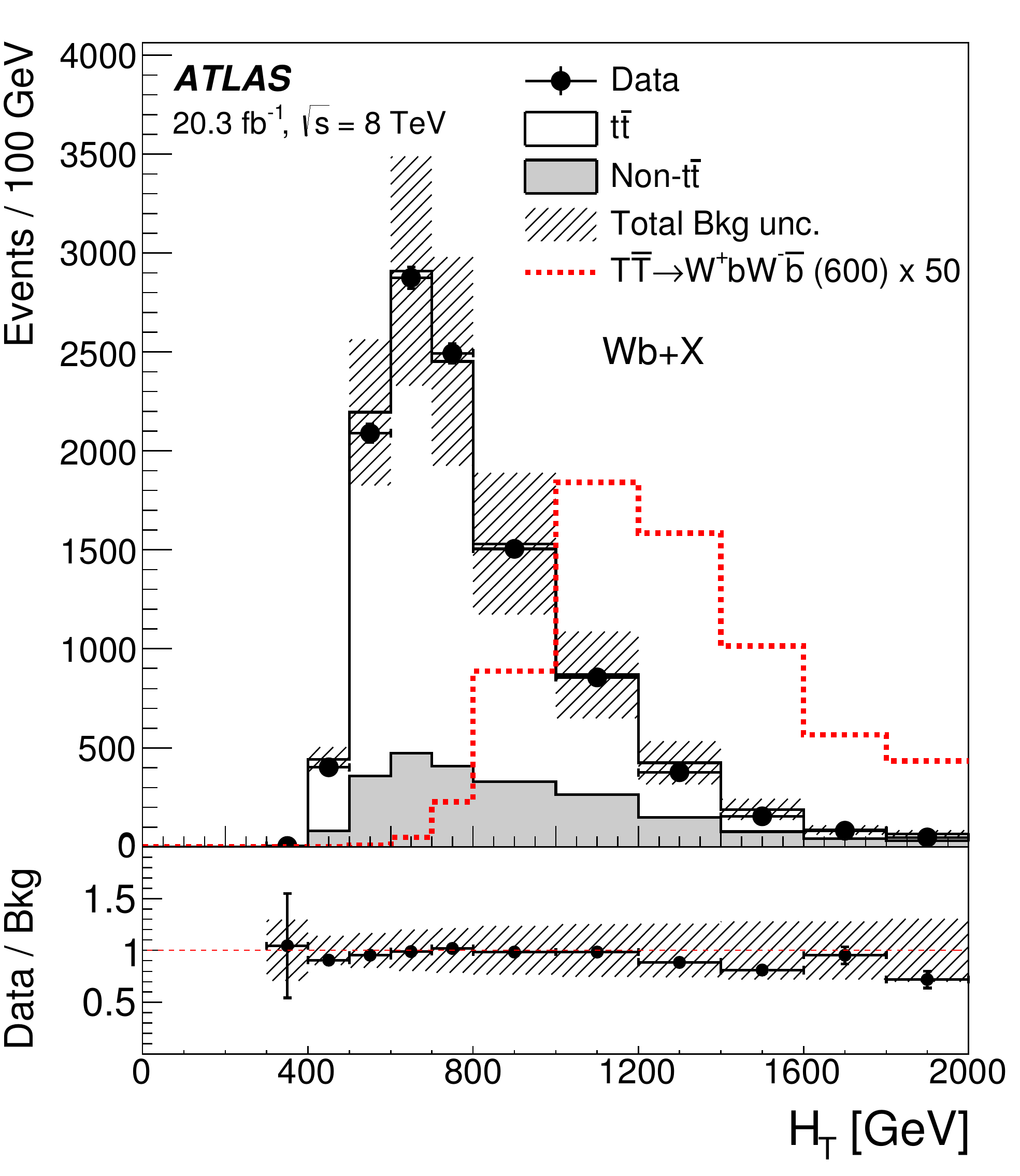}}
\caption{$T\bar{T} \to Wb$+X search: distribution of (a) the number of hadronically decaying $W$ boson ($W_{\rm had}$) candidates after preselection requirements, 
and (b) the scalar sum ($\HT$) of the transverse momenta of the lepton, the selected jets and the 
missing transverse momentum after preselection and $\geq$1~$W_{\rm had}$ candidate requirements.  
The data (solid black points) are compared to the SM prediction (stacked histograms). 
The contributions from backgrounds other than $\ttbar$ are combined into a
single background source referred to as ``Non-$\ttbar$''.
The total uncertainty on the background estimation is shown as a black hashed band.
The expected contribution from a vector-like $T$ quark with mass $m_{T}=600\gev$ under the assumption $\BR(T \to Wb)=1$, multiplied by a factor of 50, 
is also shown (red dashed histogram).
The lower panel shows the ratio of data to the SM prediction. The last bin contains the overflow.}
\label{fig:cutflow_nWhad_HTAll}
\end{figure*}

Table~\ref{tab:selection} summarises the event selection requirements.
Two selections, ``loose'' and ``tight'', are defined, with the latter being more restrictive than the former and representing the final selection. 
As discussed below, the loose selection is used to validate the background modelling in a kinematic regime close to the final selection.
The loose selection considers preselected events with at least one $W_{\rm had}^{\rm type\;I}$ or $W_{\rm had}^{\rm type\;II}$ candidate.
If multiple $W_{\rm had}$ candidates are found in a given event, the one with the highest $\pt$ is chosen.
Figure~\ref{fig:cutflow_nWhad_HTAll}(a) shows the distribution of the number of $W_{\rm had}$ candidates after preselection.
The events must satisfy $\HT>800\gev$, where $\HT$ is the scalar sum of the lepton $\pt$, $\met$ and the $\pt$
of the selected jets. The $\HT$ distribution peaks at 
$\sim$$2 m_{T}$ for signal events, which makes the $\HT>800\gev$ requirement particularly efficient
for signal with $m_{T}\gtrsim 400\gev$, while rejecting a large fraction of the background. 
Figure~\ref{fig:cutflow_nWhad_HTAll}(b) shows the distribution of $\HT$ after the requirement of $\geq$1~$W_{\rm had}$ candidate and prior to the $\HT>800\gev$  requirement.
In addition, the highest-$\pt$ $b$-jet candidate ($b_1$) and the next-to-highest-$\pt$ $b$-jet candidate ($b_2$) are
required to have $\pt(b_1) > 160\gev$ and $\pt(b_2) >110\,(80)\gev$ respectively, in the case of a  $W_{\rm had}^{\rm type\;I}$ ($W_{\rm had}^{\rm type\;II}$) candidate.
Finally, the angular separation between the lepton and the reconstructed neutrino is required to satisfy $\Delta R(\ell,\nu)<0.8\,(1.2)$ in 
case of a  $W_{\rm had}^{\rm type\;I}$ ($W_{\rm had}^{\rm type\;II}$) candidate.
Figure~\ref{fig:cutflow_DRLepMet_MinDRlb}(a) shows the distributions of $\Delta R(\ell,\nu)$ after all previous requirements and prior to the $\Delta R(\ell,\nu)$ requirement.

The tight selection adds further requirements that are particularly effective at suppressing $t\bar{t}$ background. 
First, a large angular separation between the $W$ bosons and the $b$-jets from the top quark decay is required:
$\min(\Delta R(\ell, b_{1,2}))>1.4$ and $\min(\Delta R(W_{\rm had}, b_{1,2}))>1.4$. 
Figure~\ref{fig:cutflow_DRLepMet_MinDRlb}(b) shows the distributions of $\min(\Delta R(\ell, b_{1,2}))$ after loose selection 
and prior to the $\min(\Delta R(\ell, b_{1,2}))>1.4$ requirement. 
Finally, additional requirements are made on $\Delta R(b_1,b_2)>1.0\,(0.8)$ in the case of a $W_{\rm had}^{\rm type\;I}$ ($W_{\rm had}^{\rm type\;II}$) candidate and
$\Delta m<250\gev$ only in the case of a $W_{\rm had}^{\rm type\;I}$ candidate, where
$\Delta m = \min (|m_{\rm reco}^{\rm lep} - m_{\rm reco}^{\rm had}|)$ is the smallest absolute difference between the reconstructed heavy-quark masses
obtained by pairing the $W_{\rm lep}$ and $W_{\rm had}$ candidates with the two $b$-jet candidates as described in the following.
The reconstruction of the $W_{\rm lep}$ candidate usually yields two solutions, and there
are two possible ways to pair the $b$-jet candidates with the $W$ boson candidates to form the heavy quarks. 
Among all possible combinations, the one yielding the smallest $\Delta m$ is chosen.
The main discriminating variable used in this search is the reconstructed heavy-quark mass ($m_{\rm reco}$), built 
from the $W_{\rm had}$ candidate and one of the two $b$-jet candidates.
The resulting $m_{\rm reco}$ distributions for the loose and tight selections are shown in figure~\ref{fig:mreco}
for the sum of $W_{\rm had}^{\rm type\;I}$ and $W_{\rm had}^{\rm type\;II}$ events.
The tight selection has the better expected sensitivity, and only this selection is 
chosen to derive the final result of the search. The loose selection, displaying a significant $t\bar{t}$ background
at low $m_{\rm reco}$ which is in good agreement with the expectation, provides further confidence in the 
background modelling prior to the application of $b$-jet isolation requirements in the tight selection.

\begin{figure*}[tbp]
\centering
\subfloat[]{\includegraphics[width=0.45\textwidth]{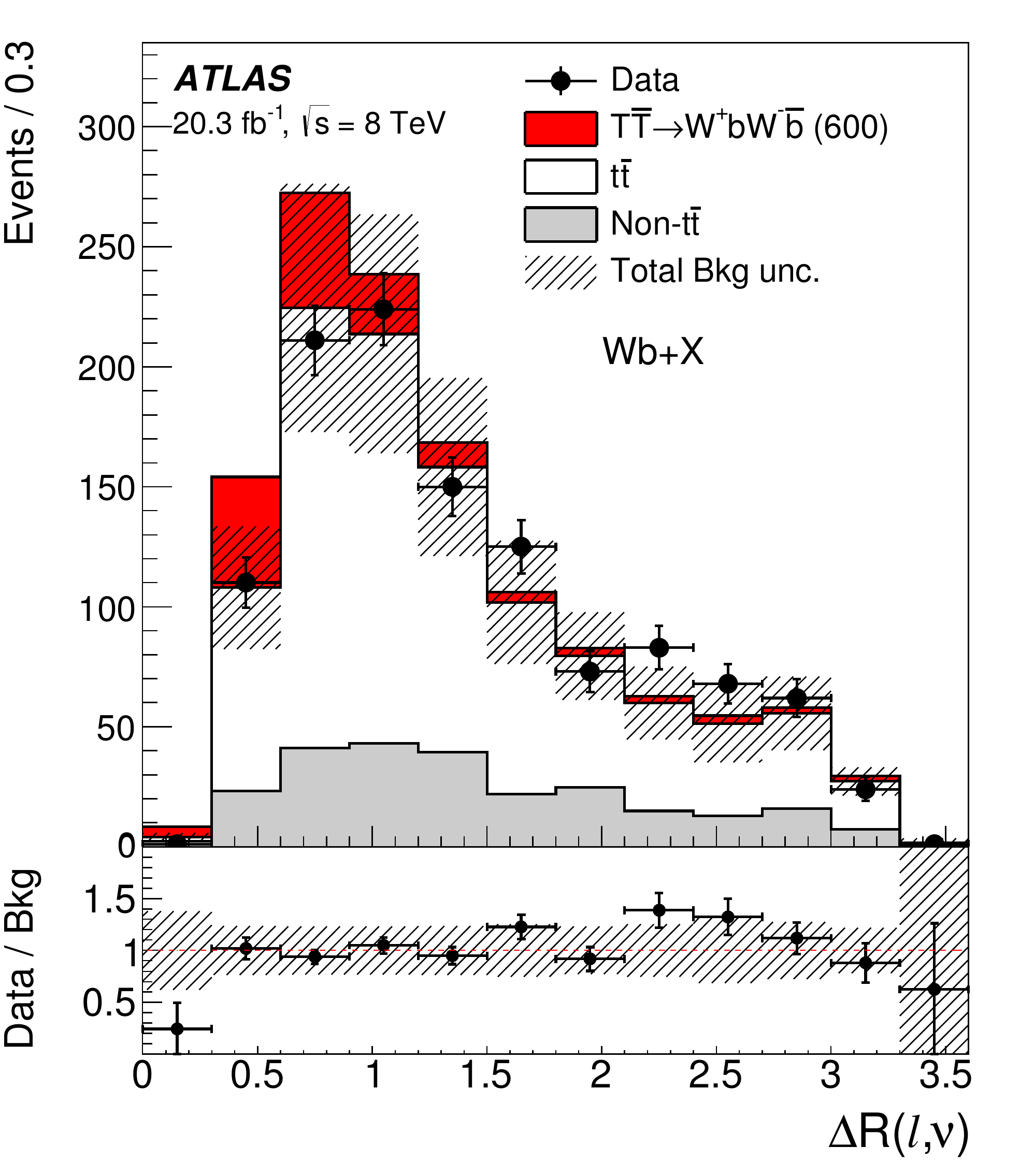}}
\subfloat[]{\includegraphics[width=0.45\textwidth]{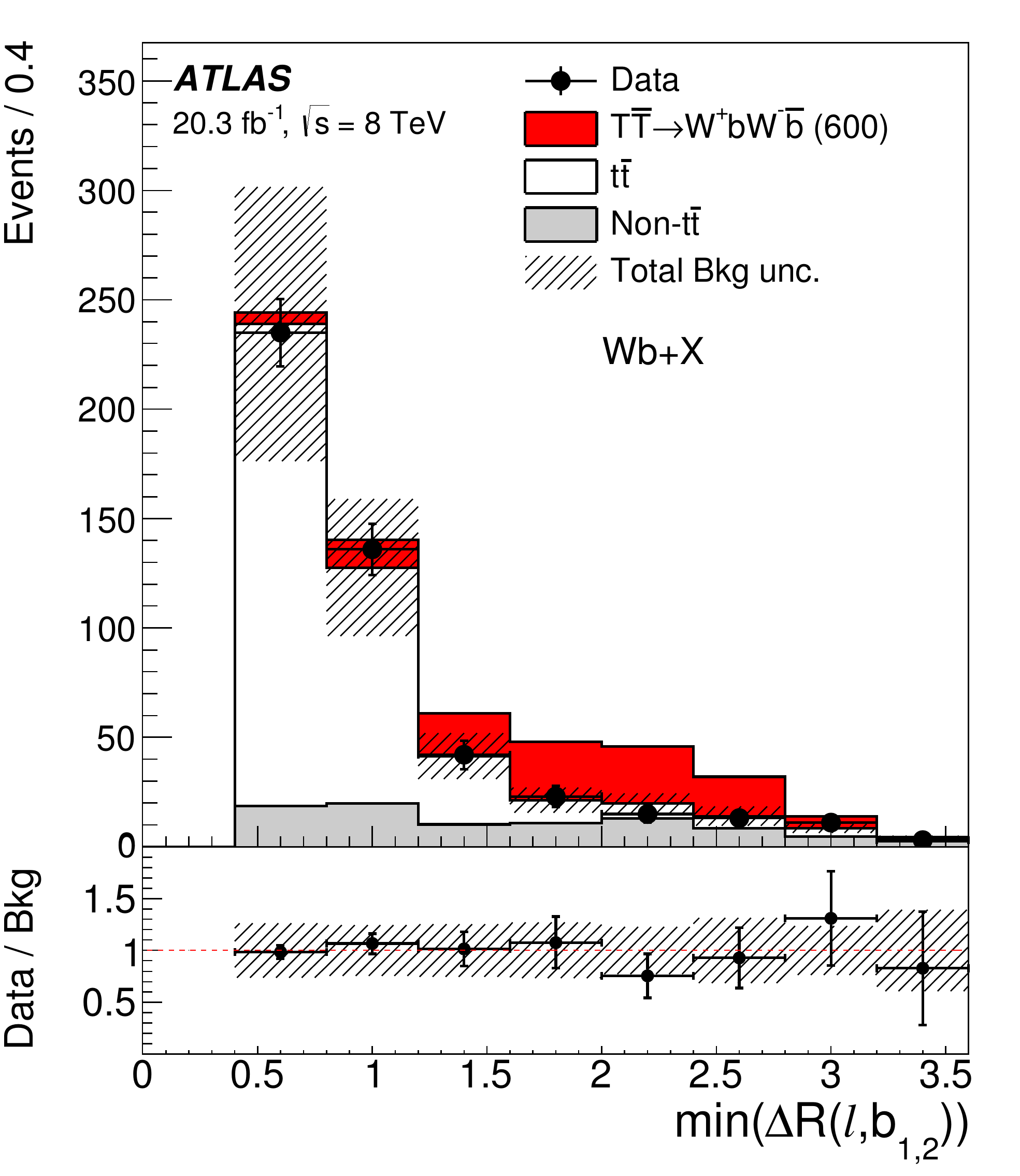}}
\caption{$T\bar{T} \to Wb$+X search: distribution of (a) the angular separation between the lepton and the reconstructed neutrino ($\Delta R(\ell,\nu)$), 
and (b) the minimum angular separation between the lepton and the two candidate $b$-jets ($\min(\Delta R(\ell, b_{1,2}))$).
The selections made include all previous requirements except for the requirement on each of these variables (see text for details).  
The data (solid black points) are compared to the SM prediction (stacked histograms). 
The contributions from backgrounds other than $\ttbar$ are combined into a
single background source referred to as ``Non-$\ttbar$''.
The total uncertainty on the background estimation is shown as a black hashed band.
The expected contribution from a vector-like $T$ quark with mass $m_{T}=600\gev$ under the assumption $\BR(T \to Wb)=1$ is 
also shown (red histogram), stacked on top of the SM background.
The lower panel shows the ratio of data to the SM prediction. The last bin contains the overflow.}
\label{fig:cutflow_DRLepMet_MinDRlb}
\end{figure*}

\begin{figure*}[tbp]
\centering
\subfloat[]{\includegraphics[width=0.45\textwidth]{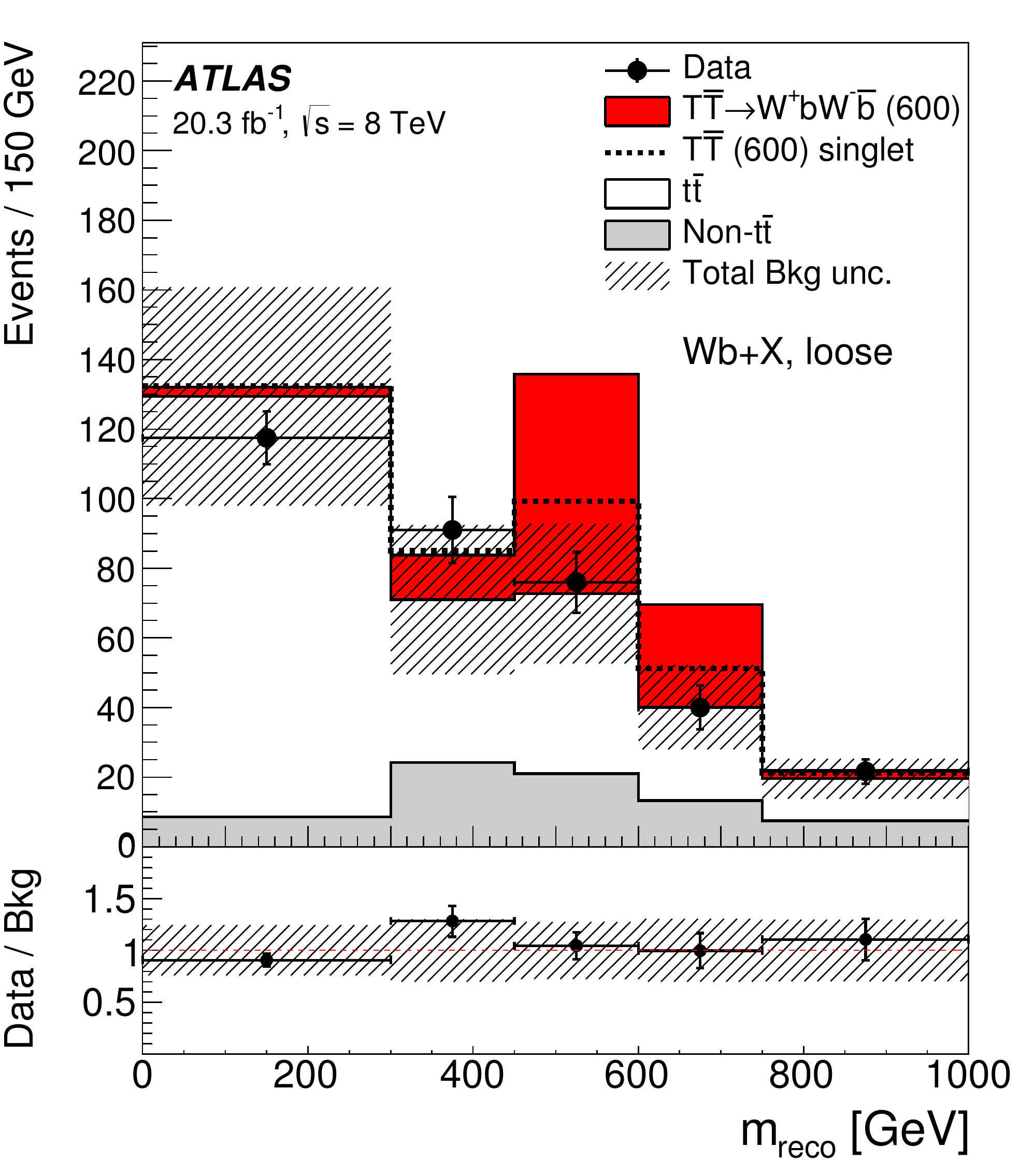}}
\subfloat[]{\includegraphics[width=0.45\textwidth]{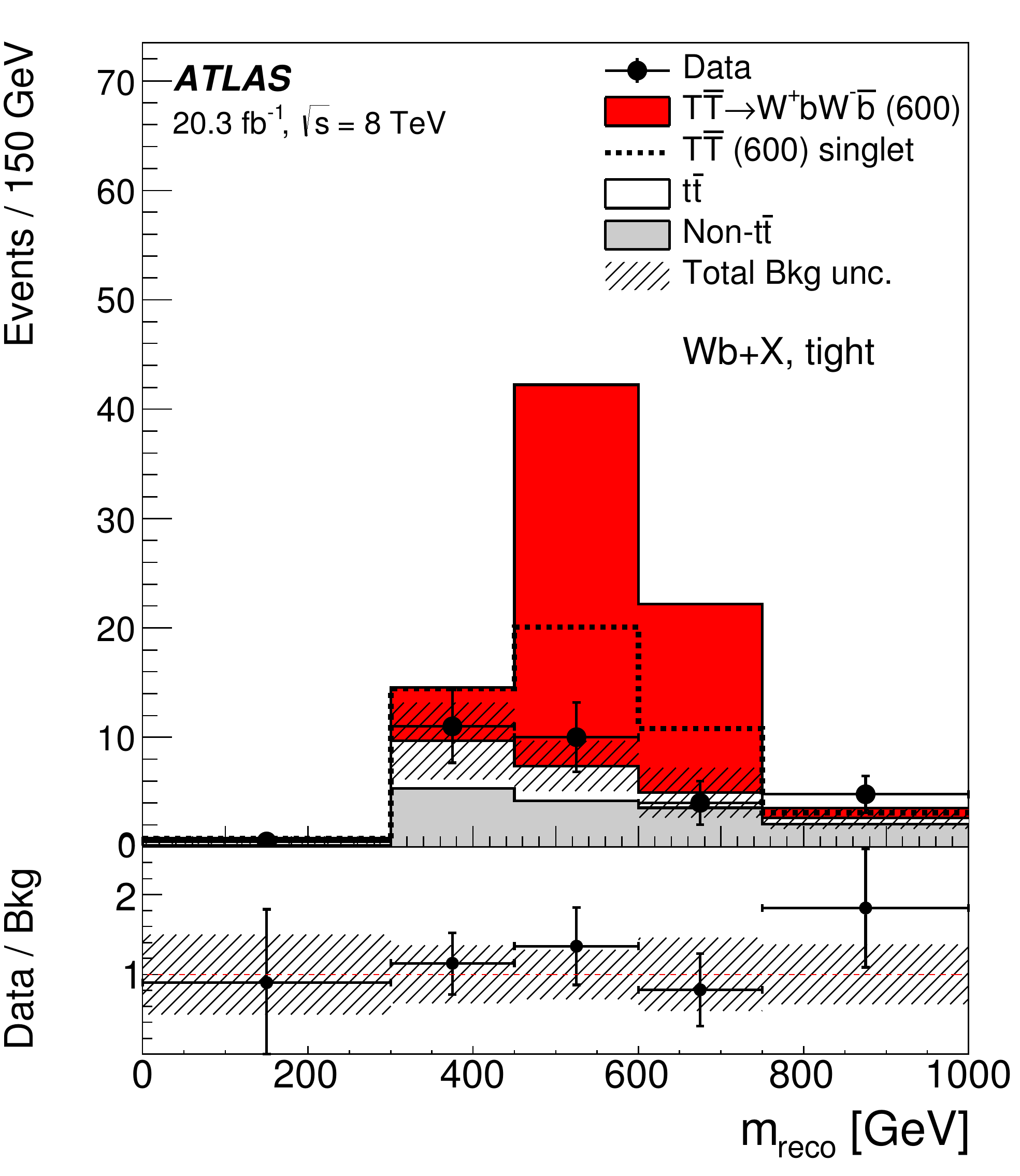}}
\caption{$T\bar{T} \to Wb$+X search: distribution of the reconstructed heavy-quark mass ($m_{\rm reco}$) after (a) the loose selection
and (b) the tight selection, for the sum of $W_{\rm had}^{\rm type\;I}$ and $W_{\rm had}^{\rm type\;II}$ events. 
The data (solid black points) are compared to the SM prediction (stacked histograms). 
The contributions from backgrounds other than $\ttbar$ are combined into a
single background source referred to as ``Non-$\ttbar$''.
The total uncertainty on the background estimation is shown as a black hashed band.
The expected contributions from a vector-like $T$ quark with mass $m_{T}=600\gev$ in two scenarios, 
$\BR(T \to Wb)=1$ (red histogram) and singlet (dashed black histogram), are also shown stacked on top of the SM background.
The lower panel shows the ratio of data to the SM prediction. The last bin contains the overflow.}
\label{fig:mreco}
\end{figure*}

Table~\ref{tab:yields} presents a summary of the background estimates 
for the loose and tight selections, as well as a
comparison of the total predicted and observed yields. 
The quoted uncertainties include both the statistical and systematic contributions.
The latter are discussed in section~\ref{sec:systematics}.
The predicted and observed yields are in agreement within these uncertainties.

\begin{table*}
\begin{center}
\begin{tabular}{lD{,}{\,\pm\,}{-1}D{,}{\,\pm\,}{-1}}
\toprule\toprule
 &  \multicolumn{1}{c}{Loose selection} & \multicolumn{1}{c}{Tight selection}  \\
\midrule
$T\bar{T}$ ($m_{T}=600\gev$)\\
$\BR(T \to Wb)=1$  &  115 , 10 & 58.9 , 5.9 \\
Singlet & 60.3 , 5.1 & 24.5 , 2.3 \\
\midrule
$t\bar{t}$ & 390 , 110 & 10.7 , 4.3 \\
$t\bar{t}V$   &  6.5 , 2.5 & 0.4 , 0.2 \\
$t\bar{t}H$   &  1.6 , 0.4 & 0.10 , 0.03 \\
$W$+jets   &  38 , 19 & 11.4 , 6.2\\
$Z$+jets   &  1.5 , 1.2 & 0.4 , 0.4 \\
Single top   &  36 , 17 & 2.2 , 1.5  \\
Diboson &  5.6 , 1.4 & 1.5 , 0.6 \\
Multijet &  0.3 , 1.6 & 0.8 , 0.7 \\
\midrule
Total background &  480 , 120 & 27.6 , 8.6 \\
\midrule
Data& \multicolumn{1}{c}{$\;\;\;\,478$} & \multicolumn{1}{c}{$\;\;\;\,34$}  \\
\bottomrule\bottomrule
\end{tabular}
\caption{\small{$T\bar{T} \to Wb$+X search: number of observed events, integrated over the whole mass spectrum, compared to the SM expectation 
after the loose and tight selections.
The expected signal yields in two different scenarios for a vector-like $T$ quark with $m_{T}=600\gev$, $\BR(T \to Wb)=1$ and singlet, are also shown.
The quoted uncertainties include both the statistical and systematic contributions.} }
\label{tab:yields}
\end{center}
\end{table*}

\section{Search for $T\bar{T} \to Ht$+X and $\fourtop$ production}
\label{sec:search_htx}

This search is focused on $T\bar{T}$ production where at least one of the $T$ quarks decays into a Higgs boson and 
a top quark resulting from the following processes:  $T\bar{T} \to HtH\bar{t}$, $ZtHt$ and 
$WbHt$.\footnote{In the following $ZtHt$ is used to denote both $ZtH\bar{t}$ and its charge conjugate, $HtZ\bar{t}$.
Similar notation is used for other processes, as appropriate.}
For the dominant $H\to b\bar{b}$ decay mode, the final-state signature is characterised by high jet and $b$-tag
multiplicities, which provide a powerful experimental handle to suppress the background. Similarly, this search
is also sensitive to $T\bar{T} \to ZtZ\bar{t}$ and $WbZt$, with $Z\to b\bar{b}$.
High jet and $b$-tag multiplicities are also characteristic of $\fourtop$ events (both within the SM and in BSM
extensions), which makes this search also sensitive to this process. 
Figure~\ref{fig:shape_njet_nbtag}(a) compares the jet multiplicity distribution
after preselection (described in section~\ref{sec:data_presel}) between the total background and several signal scenarios. 
Signal events have, on average, higher jet multiplicity than the background.   
The higher $b$-quark content of signal events results in a higher $b$-tag multiplicity than
for the background, as illustrated in figure~\ref{fig:shape_njet_nbtag}(b) for events with $\geq$6 jets.
Therefore, after preselection, the final selection requirements are $\geq$5 jets of which $\geq$2 jets
are $b$-tagged, leaving a sample completely dominated by $t\bar{t}$+jets background. 
In order to ensure a non-overlapping analysis sample and to facilitate the combination of results, 
events accepted by the $Wb$+X search are rejected. 
This veto only removes about 2\% of the events with $\geq$6 jets and $\geq$4 $b$-tagged jets in data.

\begin{figure*}[tbp]
\centering
\subfloat[]{\includegraphics[width=0.48\textwidth]{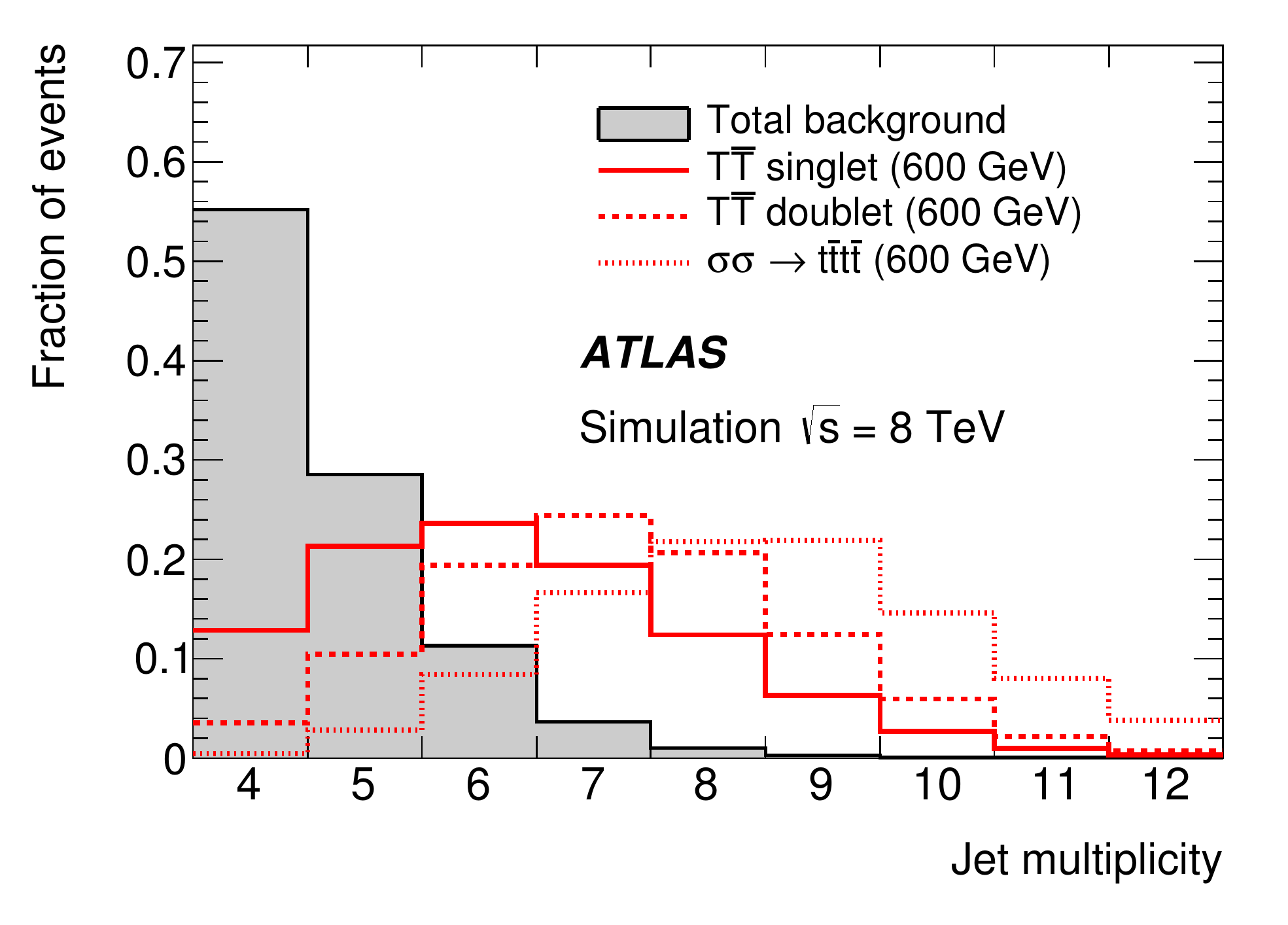}}
\subfloat[]{\includegraphics[width=0.48\textwidth]{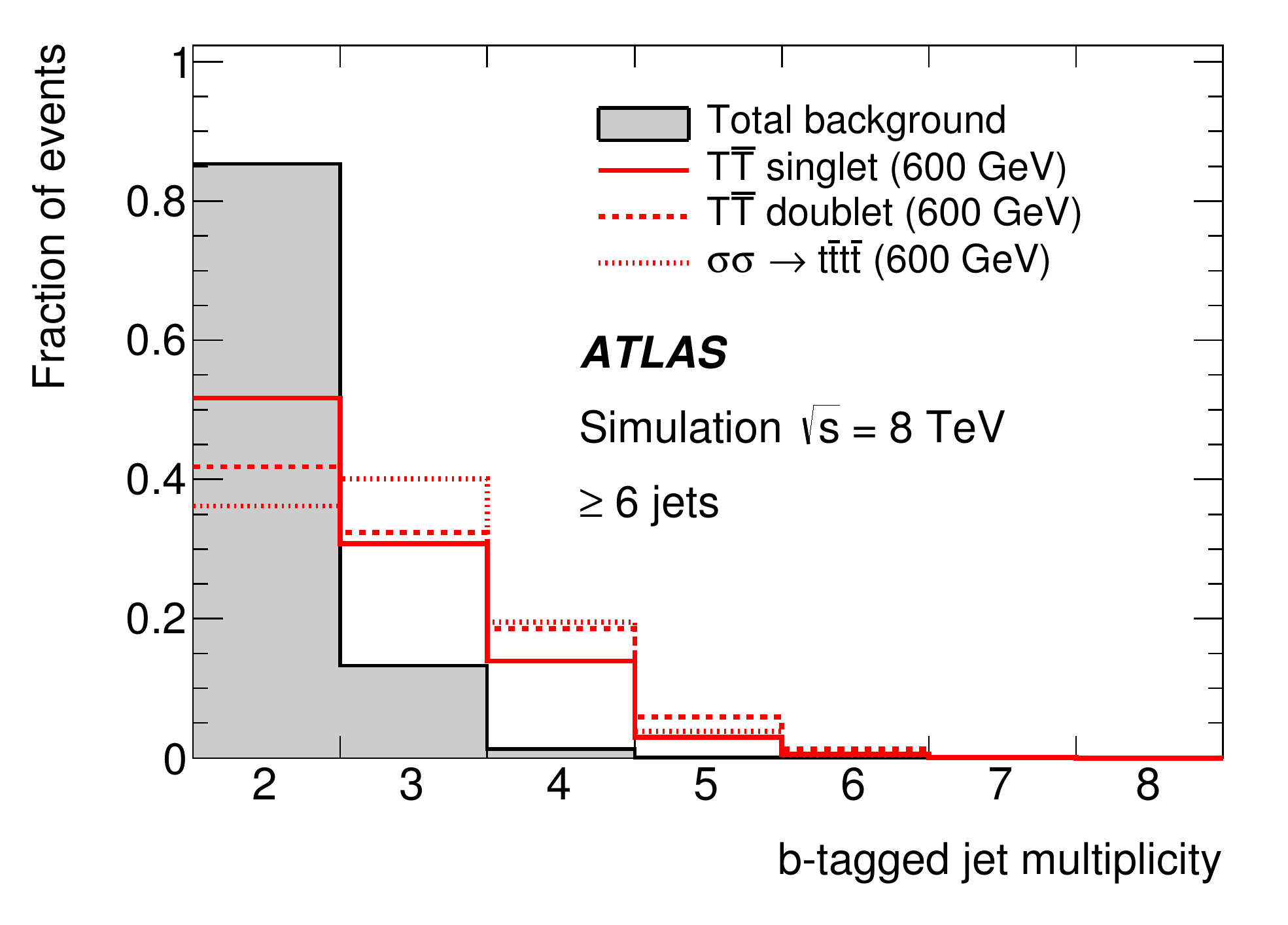}}
\caption{$T\bar{T} \to Ht$+X search (simulated events): comparison of (a) the jet multiplicity distribution after preselection, and (b) the $b$-tag multiplicity distribution 
after the requirement of $\geq$6 jets, between the total background (shaded histogram) and several signal scenarios considered in this search:
$T\bar{T}$ production in the $T$ quark singlet (red solid histogram) and doublet (red dashed histogram) cases, and sgluon pair production giving a four-top-quark 
final state (red dotted histogram). A mass of $600\gev$ is assumed for the $T$ quark and the sgluon.}
\label{fig:shape_njet_nbtag}
\end{figure*}

In order to optimise the sensitivity of the search, the selected events are categorised into different channels
depending on the number of jets (5 and $\geq$6) and on the number of $b$-tagged jets (2, 3 and $\geq$4).
The channel with $\geq$6 jets and $\geq$4 $b$-tagged jets has the largest signal-to-background ratio and 
therefore drives the sensitivity of the search. 
The channels with two and three $b$-tagged jets have significantly lower  signal-to-background ratio. 
These are particularly useful to calibrate the $t\bar{t}$+jets background prediction and constrain the  
related systematic uncertainties. In the case of the channel with $\geq$6 jets and $\geq$4 $b$-tagged jets the background
uncertainty is dominated by uncertainties on the $b$-tagging, jet energy calibration and physics modelling,  
including the $t\bar{t}$+HF content.
A detailed discussion of the systematic uncertainties considered is given in section~\ref{sec:systematics}.
In addition, events with $\geq$6 jets and 3 or $\geq$4 $b$-tagged jets are split into two channels each depending on the
value of the invariant mass of the two $b$-tagged jets with lowest $\Delta R$ separation: $M_{bb}^{{\rm min}\Delta R}<100\gev$ (``low $M_{bb}^{{\rm min}\Delta R}$'')
and $M_{bb}^{{\rm min}\Delta R}>100\gev$ (``high $M_{bb}^{{\rm min}\Delta R}$''). For high values of $m_{T}$, the Higgs boson from the $T \to Ht$ decay 
has high $\pt$, and the $b\bar{b}$ pair from the Higgs boson decay has smaller angular separation than other 
pairs resulting from combinatorial background. As shown in figure~\ref{fig:shape_Mbb_HTAll}(a), the $M_{bb}^{{\rm min}\Delta R}$ variable 
provides a good approximation to the reconstructed $H\to b\bar{b}$ invariant mass and allows the separation of these channels into channels depleted or enriched in 
$T\to Ht$, $H\to b\bar{b}$ decays, the latter having a higher signal-to-background ratio.
Therefore, the total number of analysis channels considered in this search is eight:
(5 j, 2 b), (5 j, 3 b), (5 j, $\geq$4 b), ($\geq$6 j, 2 b), ($\geq$6 j, 3 b, low $M_{bb}^{{\rm min}\Delta R}$), 
($\geq$6 j, 3 b, high $M_{bb}^{{\rm min}\Delta R}$), ($\geq$6 j, $\geq$4 b, low $M_{bb}^{{\rm min}\Delta R}$), 
and ($\geq$6 j, $\geq$4 b, high $M_{bb}^{{\rm min}\Delta R}$),
where ($n$ j, $m$ b) indicates $n$ selected jets and $m$ $b$-tagged jets.

\begin{figure*}[tbp]
\centering
\subfloat[]{\includegraphics[width=0.48\textwidth]{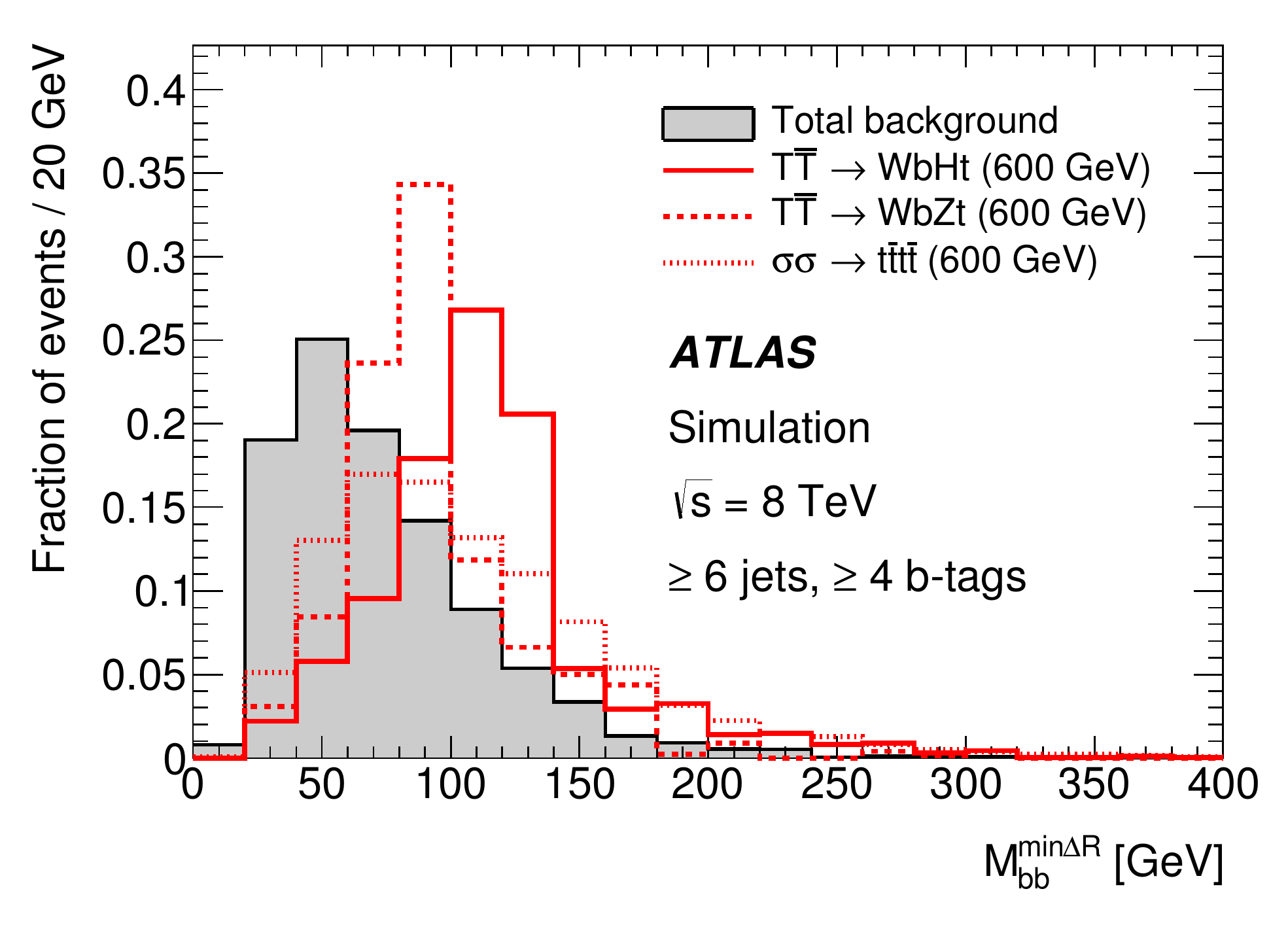}}
\subfloat[]{\includegraphics[width=0.48\textwidth]{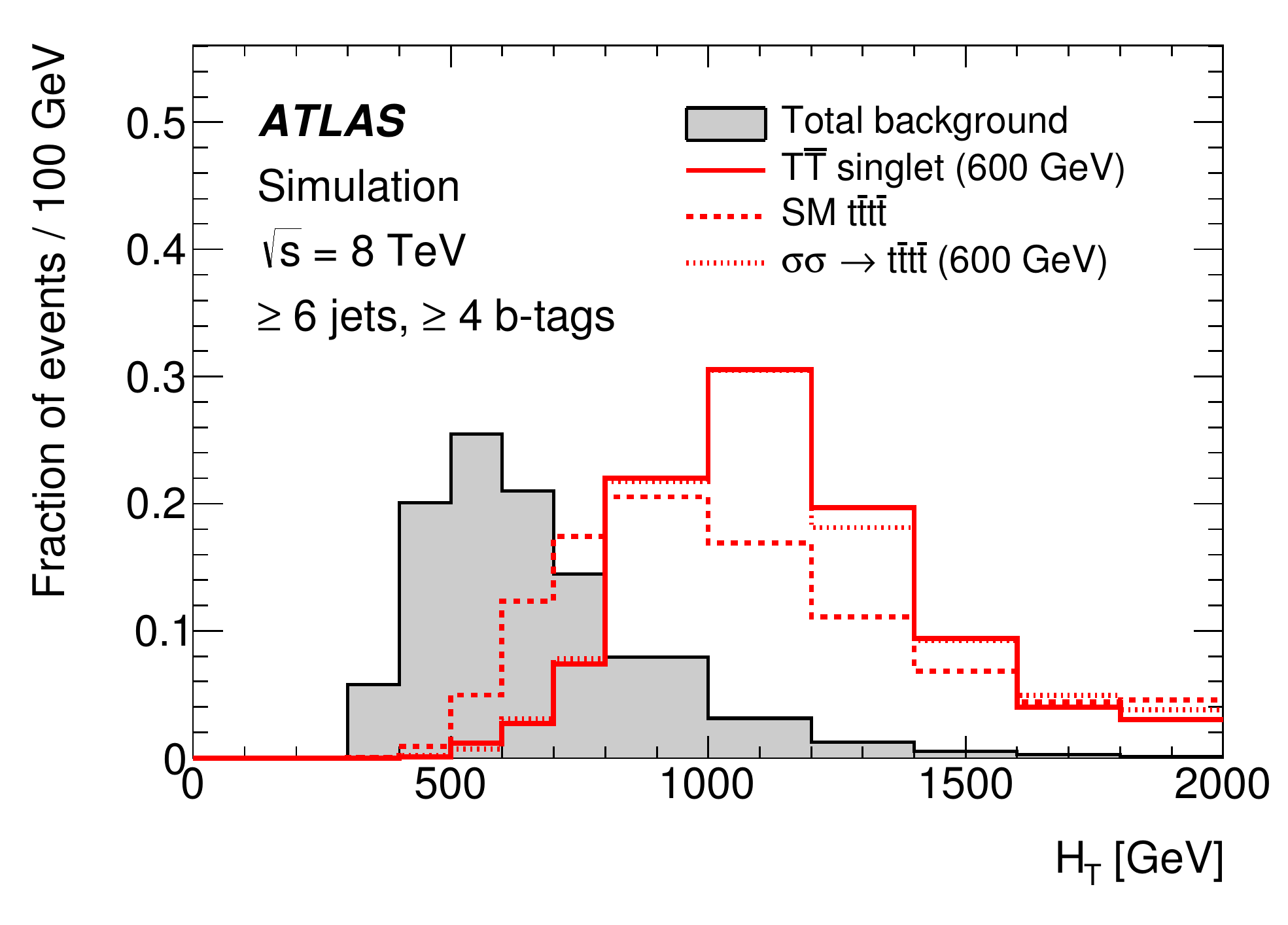}}
\caption{$T\bar{T} \to Ht$+X search (simulated events): comparison of the distributions of (a) the invariant mass of the two $b$-tagged jets with lowest $\Delta R$ separation
($M_{bb}^{{\rm min}\Delta R}$), and (b) the scalar sum of the transverse momenta of the lepton, the selected jets and the missing transverse 
momentum ($\HT$), between the total background (shaded histogram) and several signal scenarios considered in this search:
$T\bar{T} \to WbHt$ (red solid histogram),  $T\bar{T} \to WbZt$ or SM $\fourtop$ production (red dashed histograms), and sgluon pair production 
giving a $\fourtop$ final state (red dotted histogram). A mass of $600\gev$ is assumed for the $T$ quark and the sgluon.
The selection used in both (a) and (b) corresponds to events satisfying the preselection requirements and with $\geq$6 jets and $\geq$4 $b$-tagged jets.}
\label{fig:shape_Mbb_HTAll}
\end{figure*}

To further improve the separation between signal and background,
the distinct kinematic features of the signal are exploited. In particular, the large $T$ quark mass results
in energetic leptons and jets in the final state, and $\HT$ provides a suitable discriminating variable between
signal and background. Figure~\ref{fig:shape_Mbb_HTAll}(b) compares the $\HT$ distribution between signal and 
background for events with $\geq$6 jets and $\geq$4 $b$-tagged jets. The $\HT$ distribution is quite similar for 
different signal scenarios corresponding to pair production of exotic particles with
the same mass ($600\gev$ in this case), and significantly different from that of the background.
The discrimination between signal and background increases with mass.

Figures~\ref{fig:prefit_HtX_unblinded_1} and~\ref{fig:prefit_HtX_unblinded_2} show the comparison of data and prediction for the $\HT$ distributions in each of the analysis
channels considered. The corresponding predicted and observed yields per channel can be found in table~\ref{tab:Prefit_Yields_HtX_unblind}.
Following the statistical procedure outlined in section~\ref{sec:stat_analysis}, a fit to the observed $\HT$ distributions in data in the eight analysis channels is performed. 
This provides an improved background prediction with smaller uncertainties, and hence improved sensitivity to a signal. 
The results are presented in section~\ref{sec:result}.

\begin{figure*}[htbp]
\begin{center}
\subfloat[]{\includegraphics[width=0.45\textwidth]{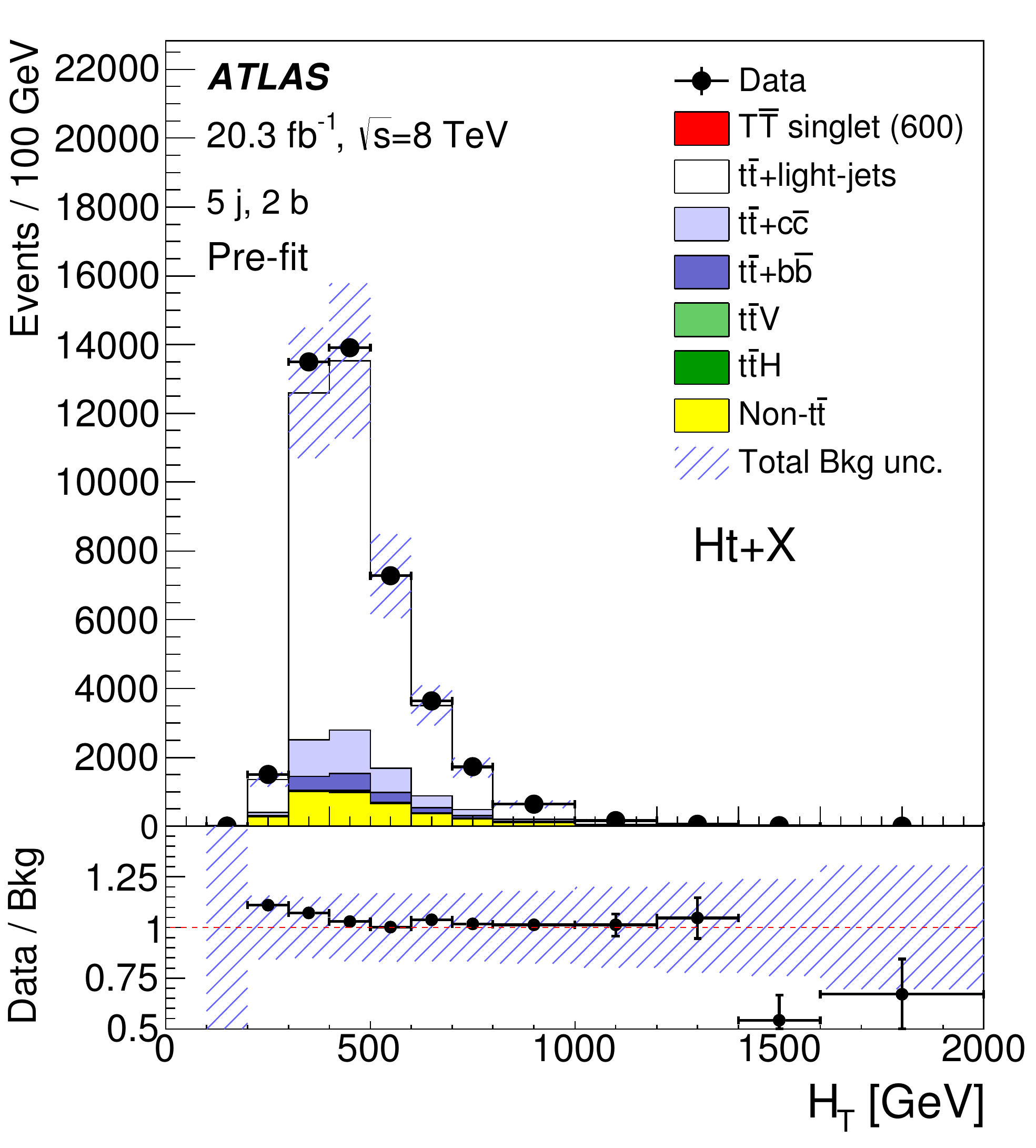}}
\subfloat[]{\includegraphics[width=0.45\textwidth]{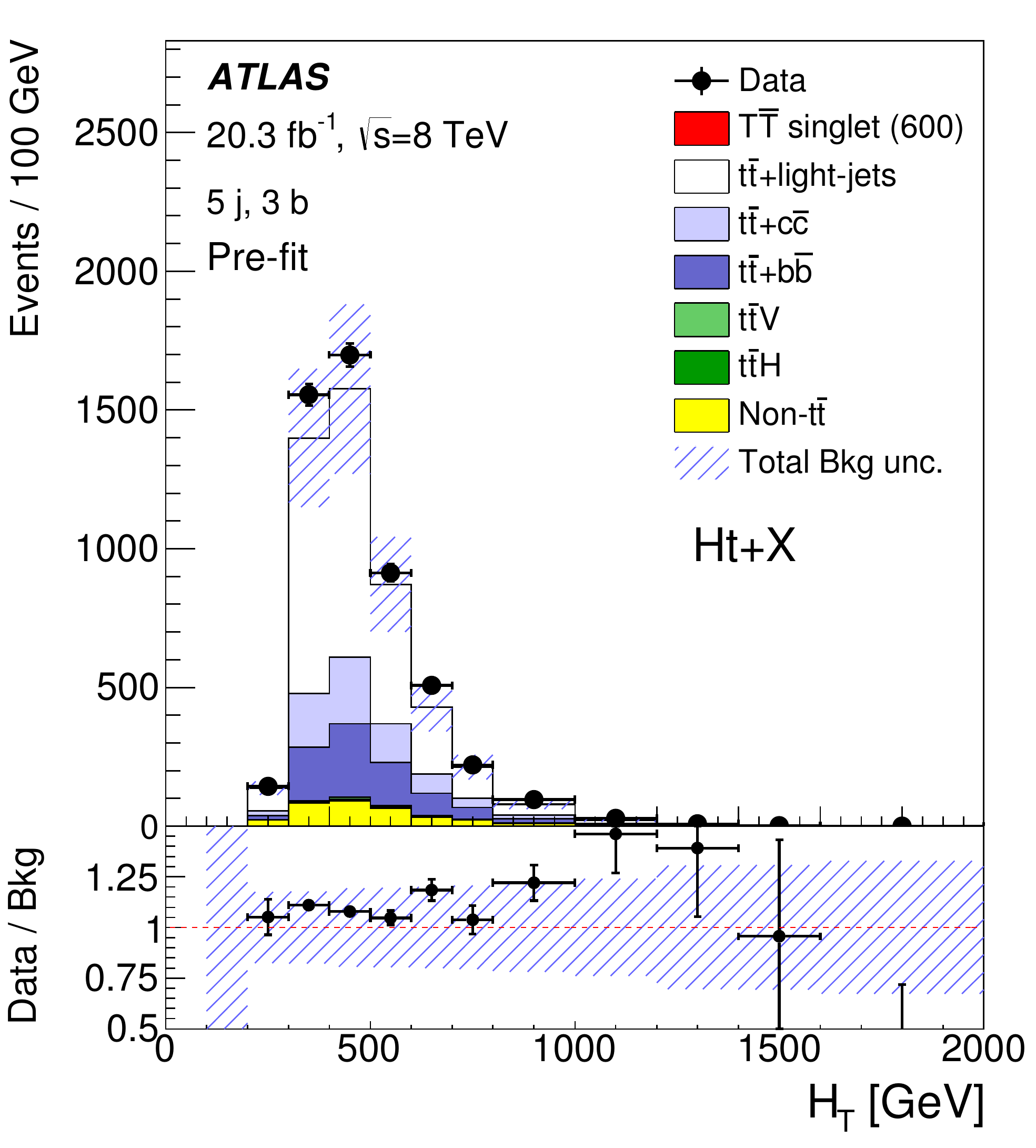}} \\
\subfloat[]{\includegraphics[width=0.45\textwidth]{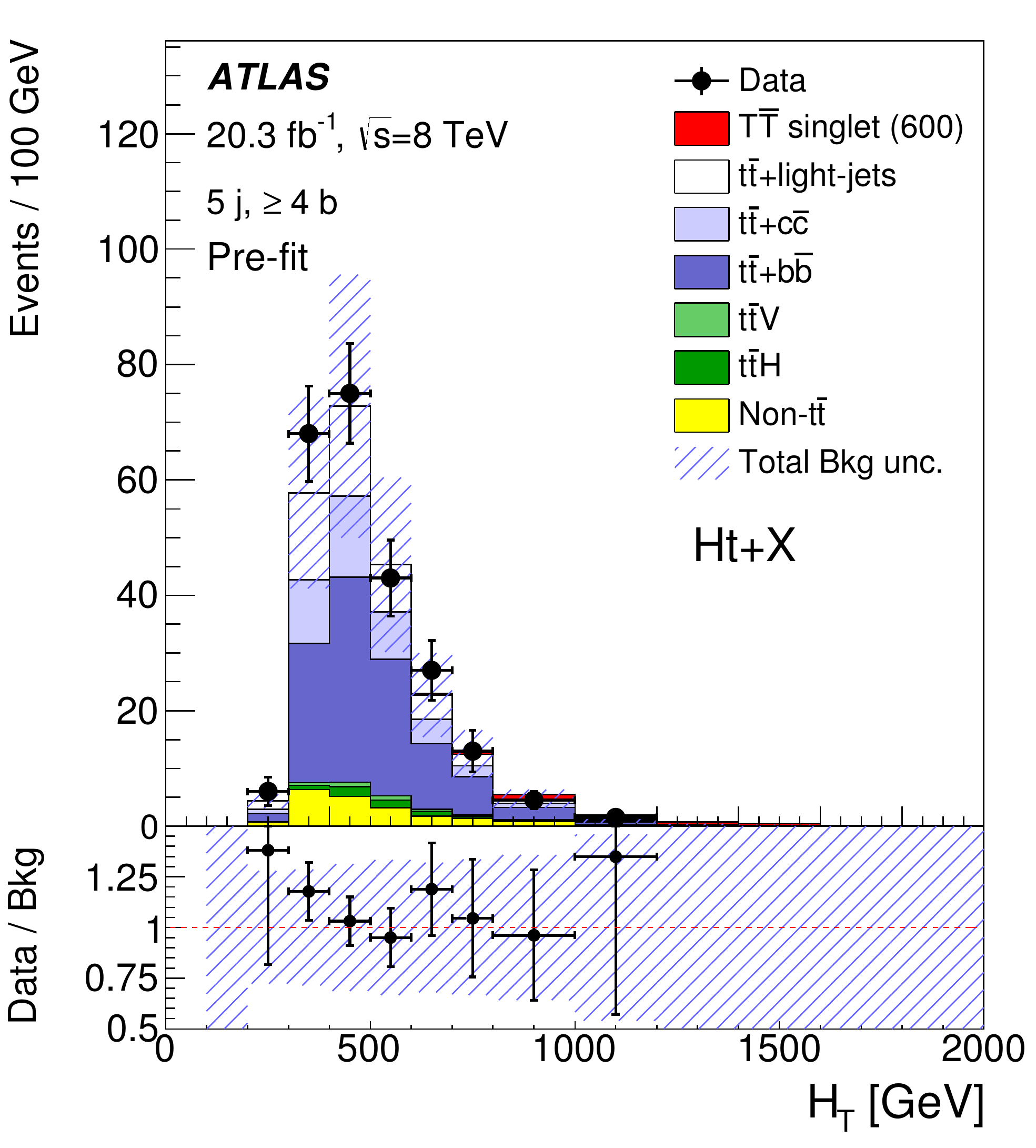}} 
\subfloat[]{\includegraphics[width=0.45\textwidth]{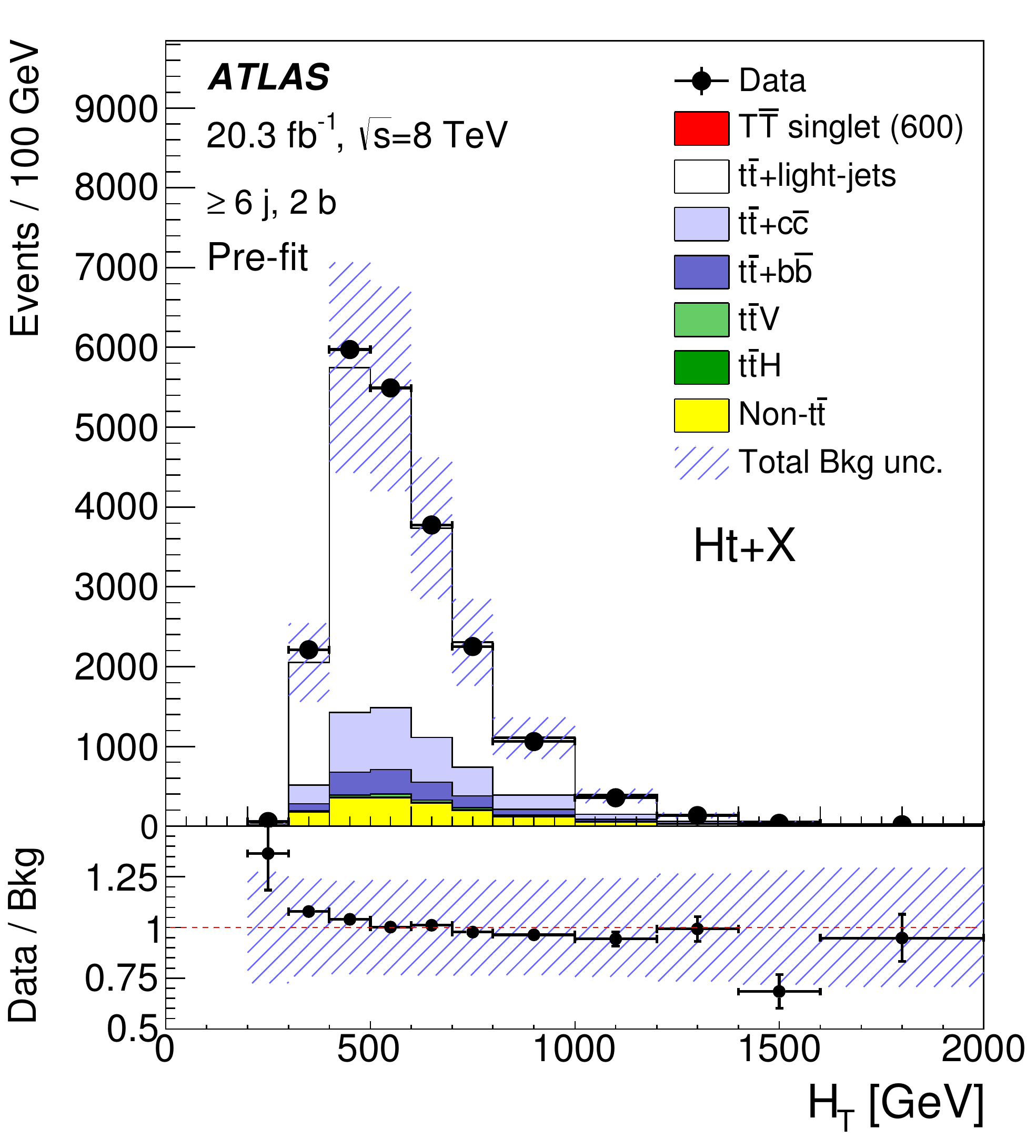}} \\
\caption{$T\bar{T} \to Ht$+X search: comparison between data and prediction for the distribution of
the scalar sum ($\HT$) of the transverse momenta of the lepton, the selected jets and the missing transverse 
momentum in each of the analysed channels after final selection:
(a) (5 j, 2 b), (b) (5 j, 3 b), (c) (5 j, $\geq$4 b), and (d) ($\geq$6 j, 2 b). 
The background prediction is shown before the fit to data. 
The contributions from $W/Z$+jets,  single top, diboson and multijet backgrounds are 
combined into a single background source referred to as ``Non-$\ttbar$''.
Also shown is the expected signal contribution from a singlet vector-like $T$ quark with mass $m_{T}=600\gev$.
The last bin in all figures contains the overflow. The bottom panel displays the ratio of
data to the total background prediction. The hashed area represents the total uncertainty on the background.}
\label{fig:prefit_HtX_unblinded_1} 
\end{center}
\end{figure*}

\begin{figure*}[htbp]
\begin{center}
\subfloat[]{\includegraphics[width=0.45\textwidth]{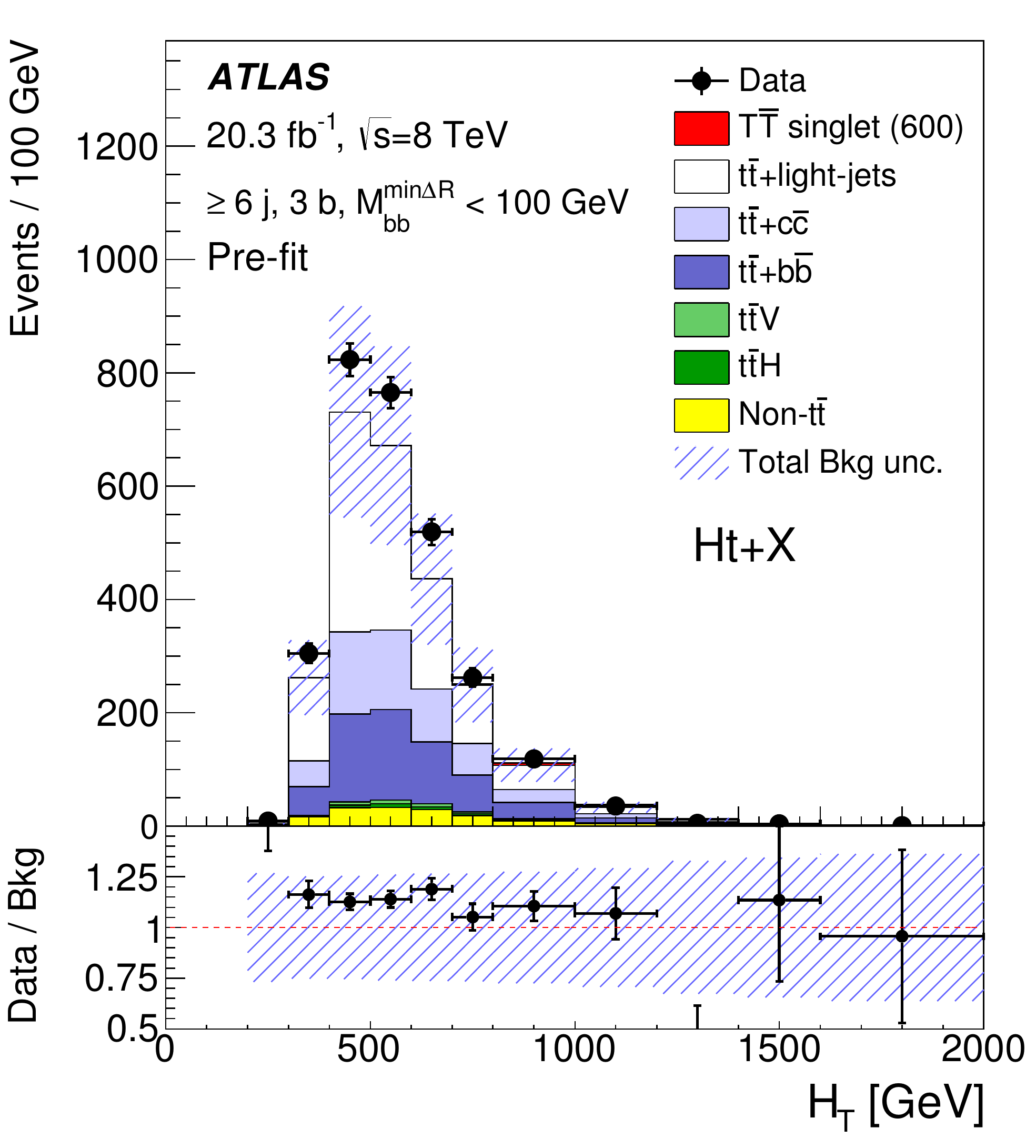}} 
\subfloat[]{\includegraphics[width=0.45\textwidth]{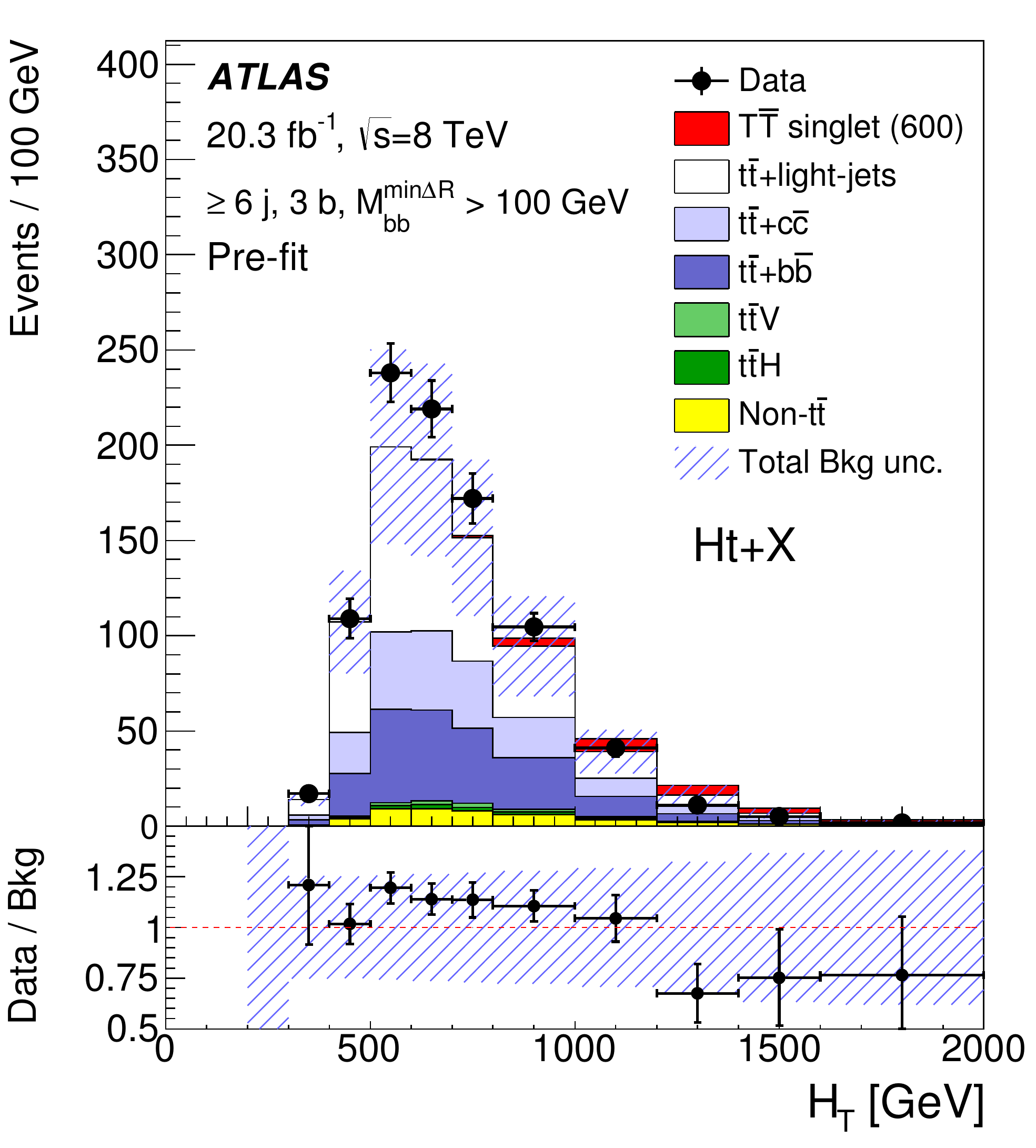}} \\
\subfloat[]{\includegraphics[width=0.45\textwidth]{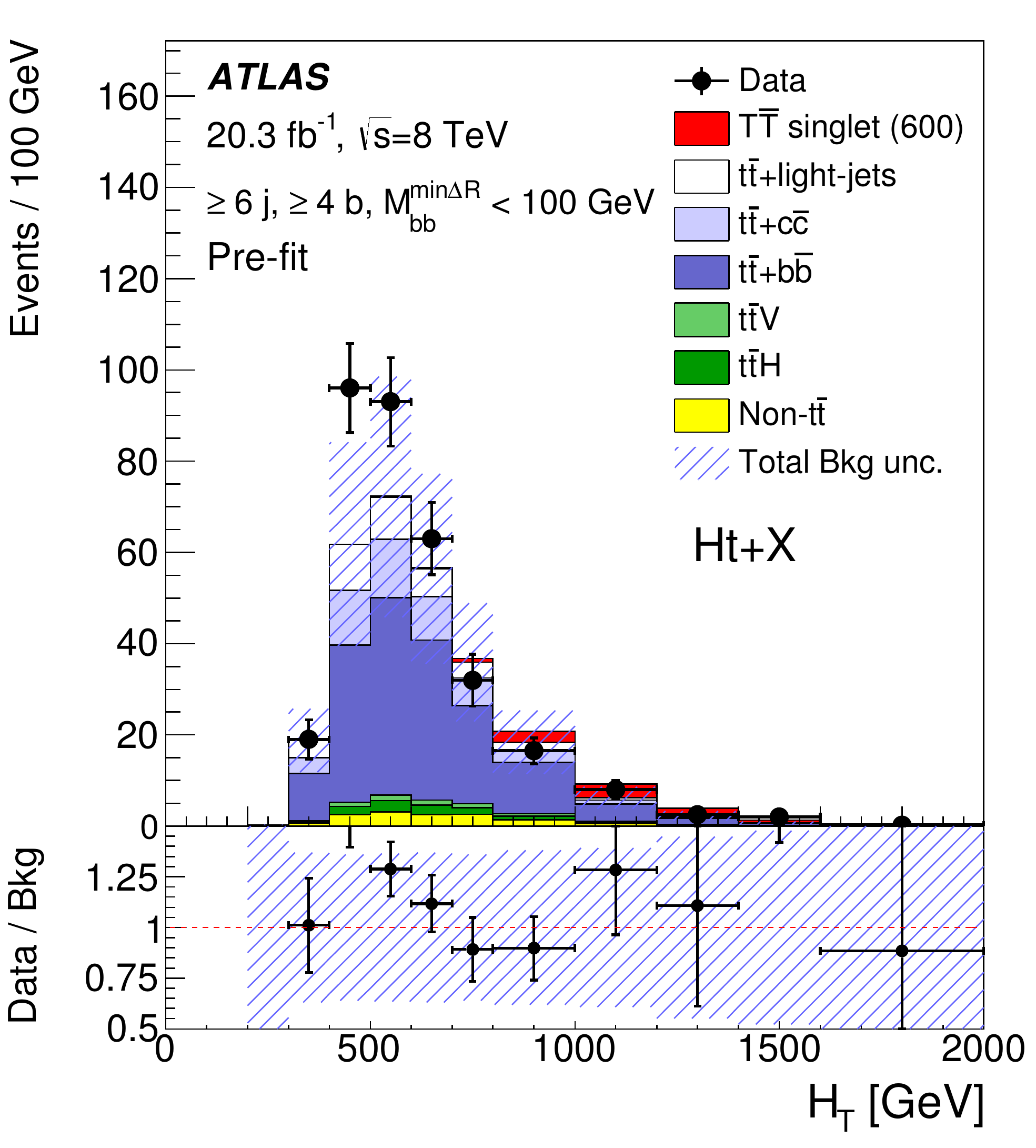}} 
\subfloat[]{\includegraphics[width=0.45\textwidth]{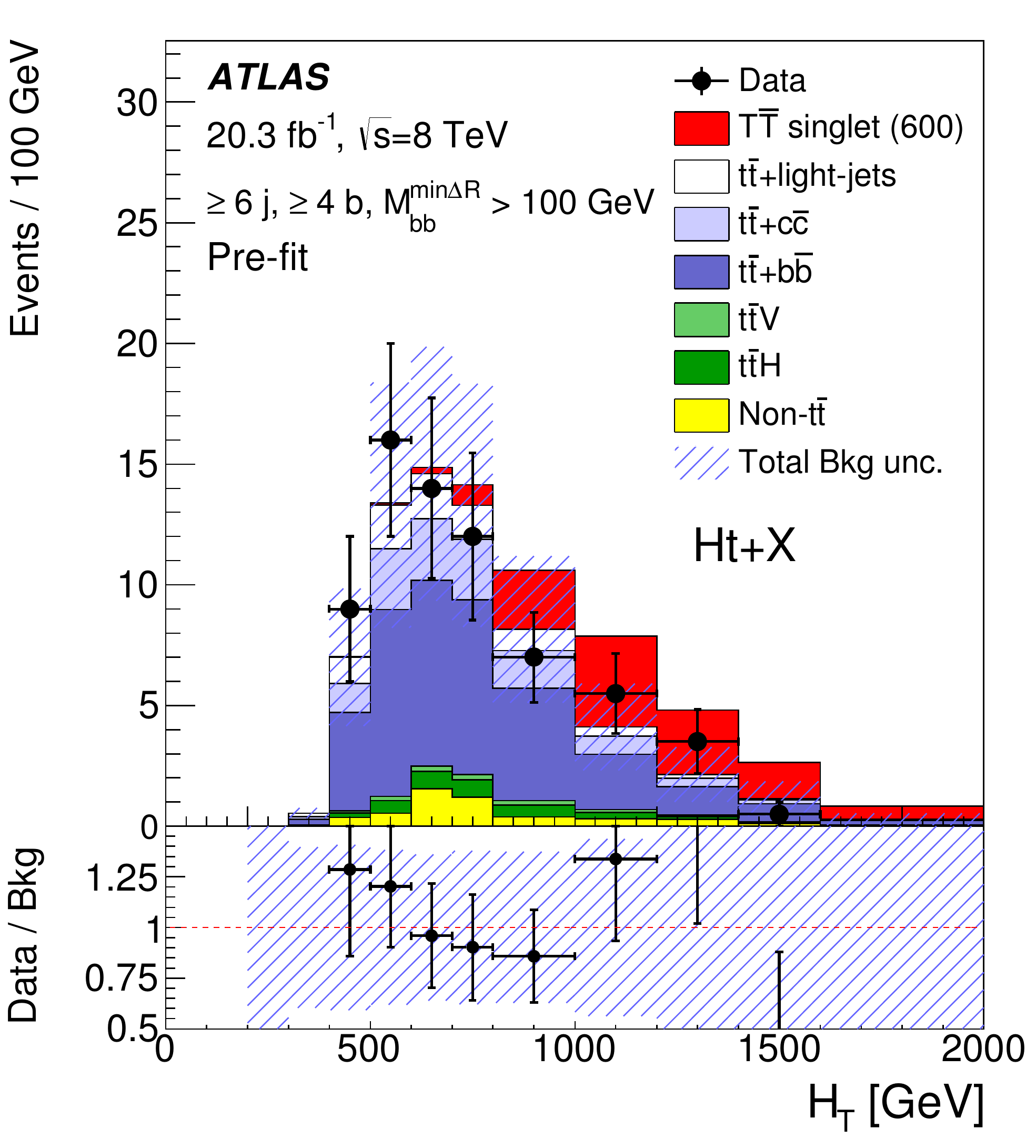}} 
\caption{$T\bar{T} \to Ht$+X search: comparison between data and prediction for the distribution of
the scalar sum ($\HT$) of the transverse momenta of the lepton, the selected jets and the missing transverse 
momentum in each of the analysed channels after final selection:
(a) ($\geq$6 j, 3 b, low $M_{bb}^{{\rm min}\Delta R}$), (b) ($\geq$6 j, 3 b, high $M_{bb}^{{\rm min}\Delta R}$), 
(c) ($\geq$6 j, $\geq$4 b, low $M_{bb}^{{\rm min}\Delta R}$), and (d) ($\geq$6 j, $\geq$4 b, high $M_{bb}^{{\rm min}\Delta R}$). 
The background prediction is shown before the fit to data. 
The contributions from $W/Z$+jets,  single top, diboson and multijet backgrounds are 
combined into a single background source referred to as ``Non-$\ttbar$''.
Also shown is the expected signal contribution from a singlet vector-like $T$ quark with mass $m_{T}=600\gev$.
The last bin in all figures contains the overflow. The bottom panel displays the ratio of
data to the total background prediction. The hashed area represents the total uncertainty on the background.}
\label{fig:prefit_HtX_unblinded_2} 
\end{center}
\end{figure*}

\begin{table}
\begin{center}
\begin{tabular}{l*{4}{c}}
\toprule\toprule
 & 5 j, 2 b & 5 j, 3 b & 5 j, $\geq$4 b & $\geq$6 j, 2 b\\
\midrule
$T\bar{T}$ ($m_{T}=600\gev$) \\
{\,} Singlet  & $52.5 \pm 4.2$ & $19.0 \pm 2.3$ & $5.8 \pm 1.2$ & $123.3 \pm 6.2$\\
{\,} $(T,B)$ or $(X,T)$ doublet& $25.8 \pm 2.0$ & $14.0 \pm 1.4$ & $5.0 \pm 1.0$ & $154.1 \pm 6.4$\\
$\sigma\sigma \to \fourtop$ ($m_{\sigma}=800\gev$) & $2.0 \pm 0.3$ & $1.4 \pm 0.3$ & $0.3 \pm 0.1$ & $64.8 \pm 4.6$\\
$\fourtop$+X (Tier (1,1), $m_{\KK}=800\gev$) & $1.0 \pm 0.4$ & $0.6 \pm 0.3$ & $0.06 \pm 0.05$ & $180 \pm 29$\\
\midrule
$t\bar{t}$+light-jets & $32400 \pm 5300$ & $2930 \pm 520$ & $48 \pm 12$ & $16200 \pm 4000$\\
$t\bar{t}+c\bar{c}$ & $3800 \pm 2100$ & $730 \pm 410$ & $42 \pm 24$ & $3300 \pm 1800$\\
$t\bar{t}+b\bar{b}$ & $1530 \pm 800$ & $800 \pm 420$ & $108 \pm 58$ & $1300 \pm 700$\\
$t\bar{t}V$ & $140 \pm 46$ & $24.9 \pm 8.1$ & $2.9 \pm 1.0$ & $172 \pm 56$\\
$t\bar{t}H$ & $39.2 \pm 1.7$ & $20.8 \pm 1.6$ & $5.6 \pm 0.7$ & $60.2 \pm 4.5$\\
$W$+jets & $1600 \pm 1000$ & $111 \pm 71$ & $5.0 \pm 3.4$ & $770 \pm 530$\\
$Z$+jets & $360 \pm 120$ & $24.8 \pm 8.4$ & $1.2 \pm 0.5$ & $185 \pm 67$\\
Single top & $1630 \pm 320$ & $169 \pm 36$ & $7.0 \pm 1.0$ & $730 \pm 200$\\
Diboson & $85 \pm 27$ & $7.3 \pm 2.5$ & $0.4 \pm 0.2$ & $45 \pm 15$\\
Multijet & $133 \pm 48$ & $33 \pm 12$ & $6.9 \pm 2.6$ & $56 \pm 20$\\
\midrule
Total background & $41700 \pm 6400$          & $4840 \pm 900$          & $228 \pm 69$          & $22800 \pm 5200$         \\
\midrule
Data & $43319$ & $5309$ & $244$ & $23001$\\
\bottomrule\bottomrule     \\
\end{tabular}
\vspace{0.1cm}

\begin{tabular}{l*{4}{c}}
\toprule\toprule
 & \begin{tabular}{@{}c@{}}$\geq$6 j, 3 b\\ low $M_{bb}^{{\rm min}\Delta R}$\end{tabular} & \begin{tabular}{@{}c@{}}$\geq$6 j, 3 b\\ high $M_{bb}^{{\rm min}\Delta R}$\end{tabular} & \begin{tabular}{@{}c@{}}$\geq$6 j, $\geq$4 b\\ low $M_{bb}^{{\rm min}\Delta R}$\end{tabular} & \begin{tabular}{@{}c@{}}$\geq$6 j, $\geq$4  b\\ high $M_{bb}^{{\rm min}\Delta R}$\end{tabular}\\
\midrule
$T\bar{T}$ ($m_{T}=600\gev$) \\
{\,} Singlet & $29.5 \pm 2.0$ & $44.0 \pm 3.6$ & $17.7 \pm 1.9$ & $24.1 \pm 3.7$\\
{\,} $(T,B)$ or $(X,T)$ doublet & $50.2 \pm 2.5$ & $68.9 \pm 4.1$ & $41.0 \pm 3.9$ & $53.8 \pm 7.3$\\
$\sigma\sigma \to \fourtop$ ($m_{\sigma}=800\gev$) & $22.5 \pm 1.6$ & $50.7 \pm 3.5$ & $9.3 \pm 1.0$ & $16.2 \pm 2.6$\\
$\fourtop$+X (Tier (1,1), $m_{\KK}=800\gev$) & $33.6 \pm 2.8$ & $132.5 \pm 5.9$ & $27.7 \pm 2.3$ & $75 \pm 13$\\
\midrule
$t\bar{t}$+light-jets & $1280 \pm 350$ & $440 \pm 110$ & $38 \pm 14$ & $9.3 \pm 3.9$\\
$t\bar{t}+c\bar{c}$ & $550 \pm 320$ & $220 \pm 120$ & $53 \pm 31$ & $14.7 \pm 9.0$\\
$t\bar{t}+b\bar{b}$ & $620 \pm 330$ & $250 \pm 140$ & $178 \pm 95$ & $46 \pm 25$\\
$t\bar{t}V$ & $28.7 \pm 9.2$ & $12.5 \pm 4.2$ & $6.2 \pm 2.0$ & $1.5 \pm 0.5$\\
$t\bar{t}H$ & $24.9 \pm 1.9$ & $11.6 \pm 1.3$ & $10.6 \pm 1.2$ & $4.1 \pm 0.6$\\
$W$+jets & $68 \pm 46$ & $16 \pm 10$ & $6.6 \pm 4.8$ & $0.6 \pm 0.4$\\
$Z$+jets & $15.7 \pm 6.3$ & $3.3 \pm 1.3$ & $1.6 \pm 0.6$ & $0.3 \pm 0.1$\\
Single top & $74 \pm 22$ & $32 \pm 12$ & $7.8 \pm 2.2$ & $2.1 \pm 1.3$\\
Diboson & $4.2 \pm 1.6$ & $1.2 \pm 0.5$ & $0.4 \pm 0.1$ & $0.2 \pm 0.1$\\
Multijet & $1.9 \pm 0.8$ & $4.8 \pm 2.1$ & $<0.01$ & $2.8 \pm 1.0$\\
\midrule
Total background & $2670 \pm 680$          & $990 \pm 260$          & $300 \pm 110$          & $81 \pm 30$         \\
\midrule
Data & $3015$ & $1085$ & $362$ & $84$\\
\bottomrule\bottomrule     \\
\end{tabular}
\vspace{0.1cm}

\end{center}
\vspace{-0.5cm}
\caption{$T\bar{T} \to Ht$+X search: predicted and observed yields in each of the analysis channels considered.
The background prediction is shown before the fit to data. Also shown are the signal predictions for different benchmark scenarios considered.
The quoted uncertainties are the sum in quadrature of statistical and systematic uncertainties on the yields.}
\label{tab:Prefit_Yields_HtX_unblind}
\end{table}

\section{Search for $B\bar{B} \to Hb$+X production}
\label{sec:search_hbx}

\begin{figure*}[tbp!]
\centering
\subfloat[]{\includegraphics[width=0.48\textwidth]{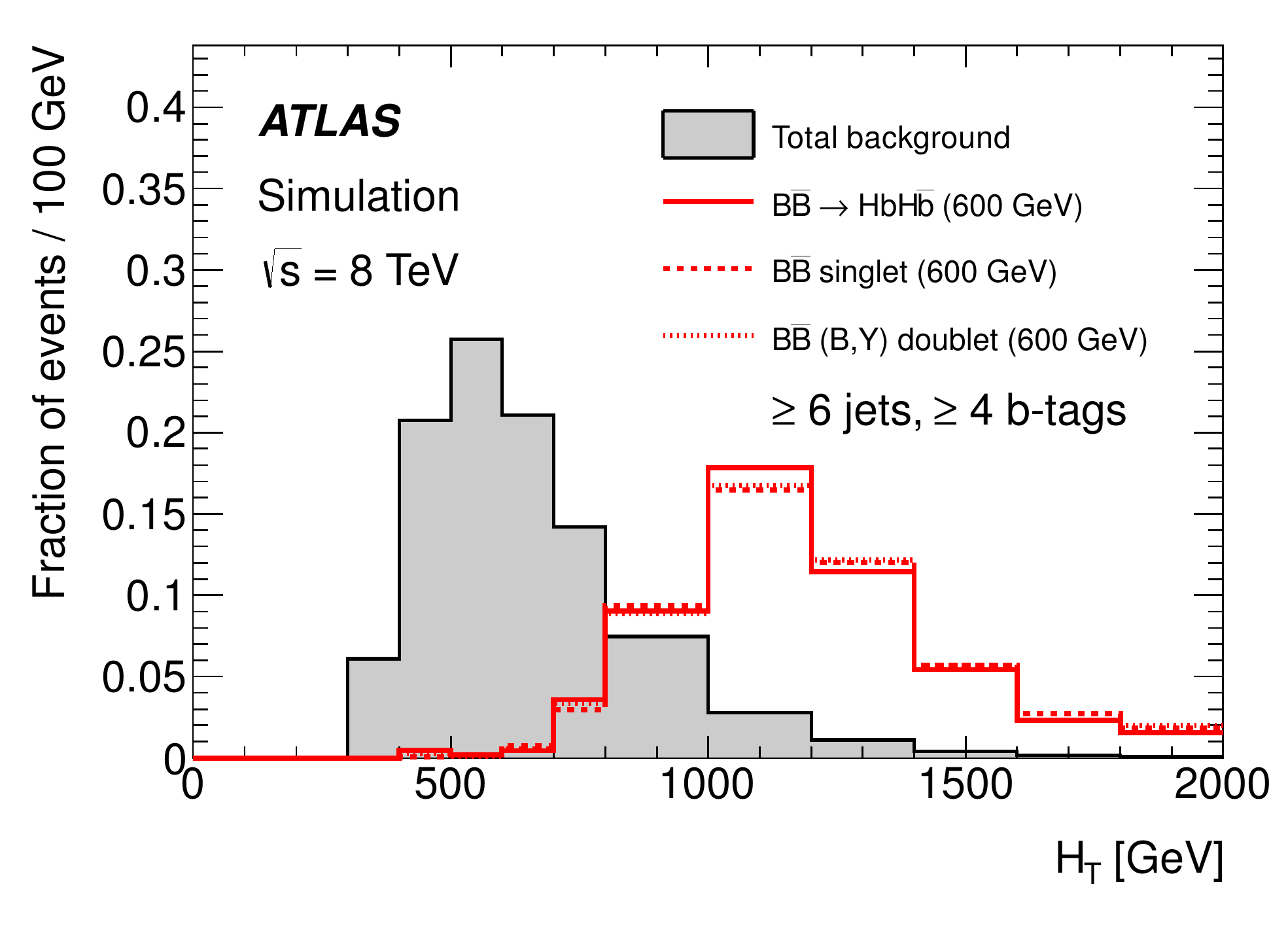}}
\subfloat[]{\includegraphics[width=0.48\textwidth]{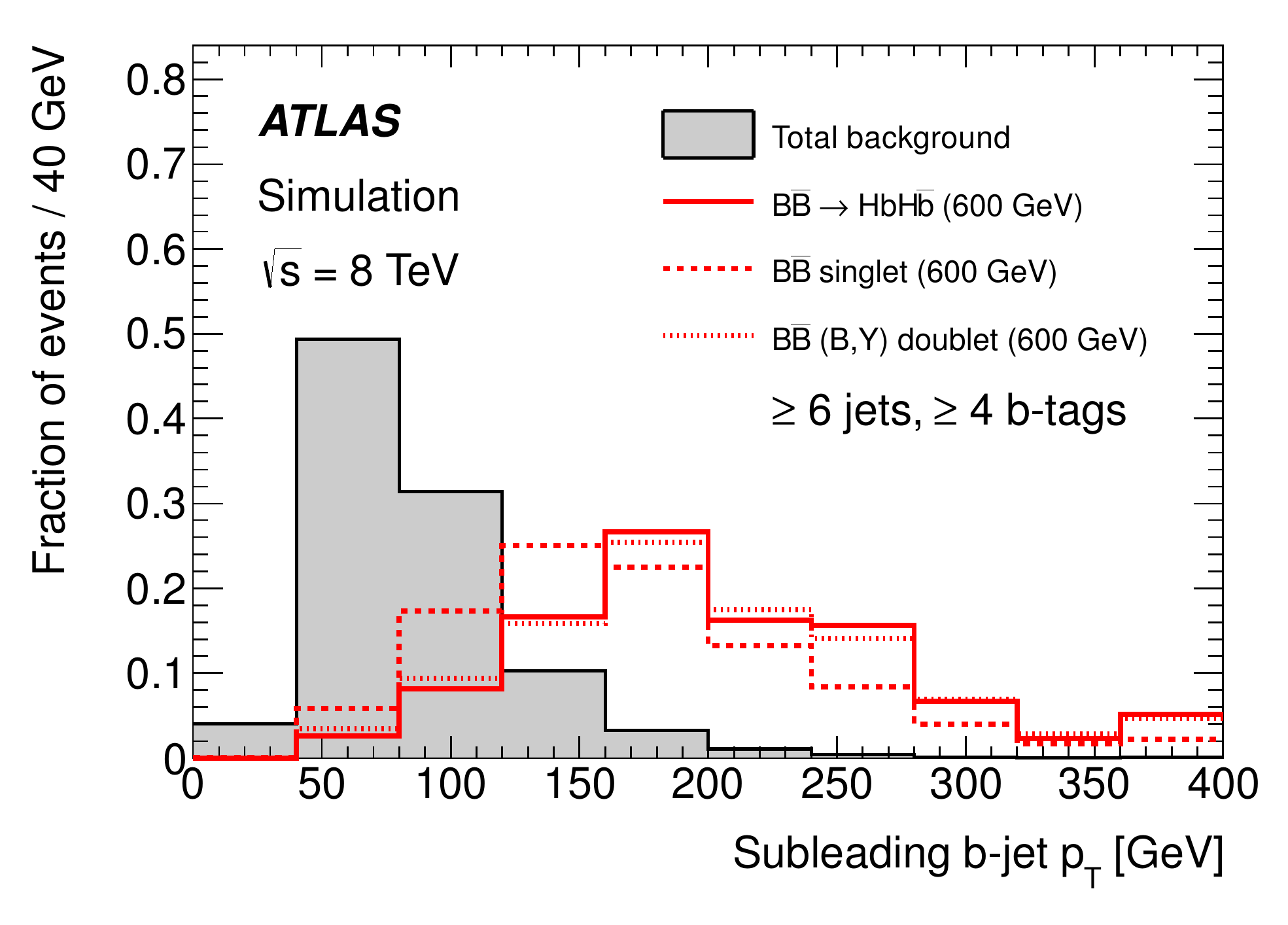}}
\caption{$B\bar{B} \to Hb$+X search (simulated events): comparison of the distributions of (a) the scalar sum of the transverse 
momenta of the lepton, the selected jets and the missing transverse momentum ($\HT$), and 
(b) the transverse momentum of the next-to-highest-transverse-momentum $b$-jet,
between the total background (shaded histogram) and several $B\bar{B}$ signal scenarios considered in this search:
$\BR(B \to Hb)=1$ (red solid histogram), $B$ quark singlet (red dashed histogram), and $B$ quark from a $(B,Y)$ doublet 
(red dotted histogram). In all cases a mass of $600\gev$ is assumed for the $B$ quark.
The selection used in both (a) and (b) corresponds to events satisfying the preselection requirements and with $\geq$6 jets and $\geq$4 $b$-tagged jets.}
\label{fig:shape_ht_ptb2}
\end{figure*}

This search is focused on $B\bar{B}$ production where at least one of the $B$ quarks decays into a Higgs boson and a $b$ quark, 
a decay mode that was omitted from previous searches~\cite{Aad:2014efa,Aad:2015gdg,Aad:2015mba}.
In particular, the $B\bar{B} \to HbH\bar{b}$ final state is the least covered one because the most-common Higgs boson decay mode, $H \to \bbbar$, leads 
to a challenging final state with six $b$-jets and no leptons. In contrast, cleaner experimental signatures involving leptons 
tend to be suppressed by the small decay branching ratios. However, a sizeable signal rate results from 
the mixed decay mode where one of the Higgs bosons decays into $W^+W^-$, while the other Higgs boson decays into $\bbbar$:
$B\bar{B} \to HbH\bar{b} \to (W^+W^-)b(\bbbar)\bbar$. When one of the $W$ bosons decays leptonically, this leads to the final-state signature 
considered in this search, involving one lepton and high jet and $b$-tag multiplicities, analogous to the signature exploited by the $T\bar{T} \to Ht$+X search.

Consequently, this search considers the same discriminating variable, $\HT$, and the same eight analysis channels as the $T\bar{T} \to Ht$+X search.
Figure~\ref{fig:shape_ht_ptb2}(a) illustrates the good separation between signal and background in the $\HT$ distribution for 
events passing the preselection requirements and with $\geq 6$ jets and $\geq 4$ $b$-tagged jets.
A peculiarity of the $B \to Hb$ decay mode is that the $b$-jet originating (directly) from the $B$-quark decay
can have very high transverse momentum in the case of a heavy $B$ quark. To exploit this feature, the event selection is tightened relative to that used
in the $T\bar{T} \to Ht$+X search by raising the minimum $\pt$ requirement on the two highest-$\pt$ (leading) $b$-tagged jets to $\pt >150\gev$.
Figure~\ref{fig:shape_ht_ptb2}(b) shows the distribution of the subleading $b$-jet $\pt$ for 
events passing the preselection requirements and with $\geq 6$ jets and $\geq 4$ $b$-tagged jets.
The tighter requirement on the subleading $b$-jet $\pt$ rejects about 90\% of the $\ttbar$ background while retaining a large acceptance for the $B\bar{B} \to Hb$+X signal.
This search is also sensitive to other $B\bar{B}$ final states, such as  $B\bar{B} \to HbWt$, that typically do not
involve multilepton final states in the topologies usually searched for (opposite-sign dileptons with a $Z\to \ell^+\ell^-$ candidate, 
same-sign dileptons, and trileptons), and is thus complementary to previous searches~\cite{Aad:2014efa,Aad:2015gdg,Aad:2015mba}.

Figures~\ref{fig:prefit_HbX_unblinded_1} and~\ref{fig:prefit_HbX_unblinded_2} show the comparison of data and prediction for the $\HT$ distributions in each of the analysis
channels considered. The corresponding predicted and observed yields per channel can be found in table~\ref{tab:Prefit_Yields_HbX_unblind}.
The results of the fit to the data to improve the background prediction, as in the $T\bar{T} \to Ht$+X search, are presented in section~\ref{sec:result}.

\begin{figure*}[htbp]
\begin{center}
\subfloat[]{\includegraphics[width=0.45\textwidth]{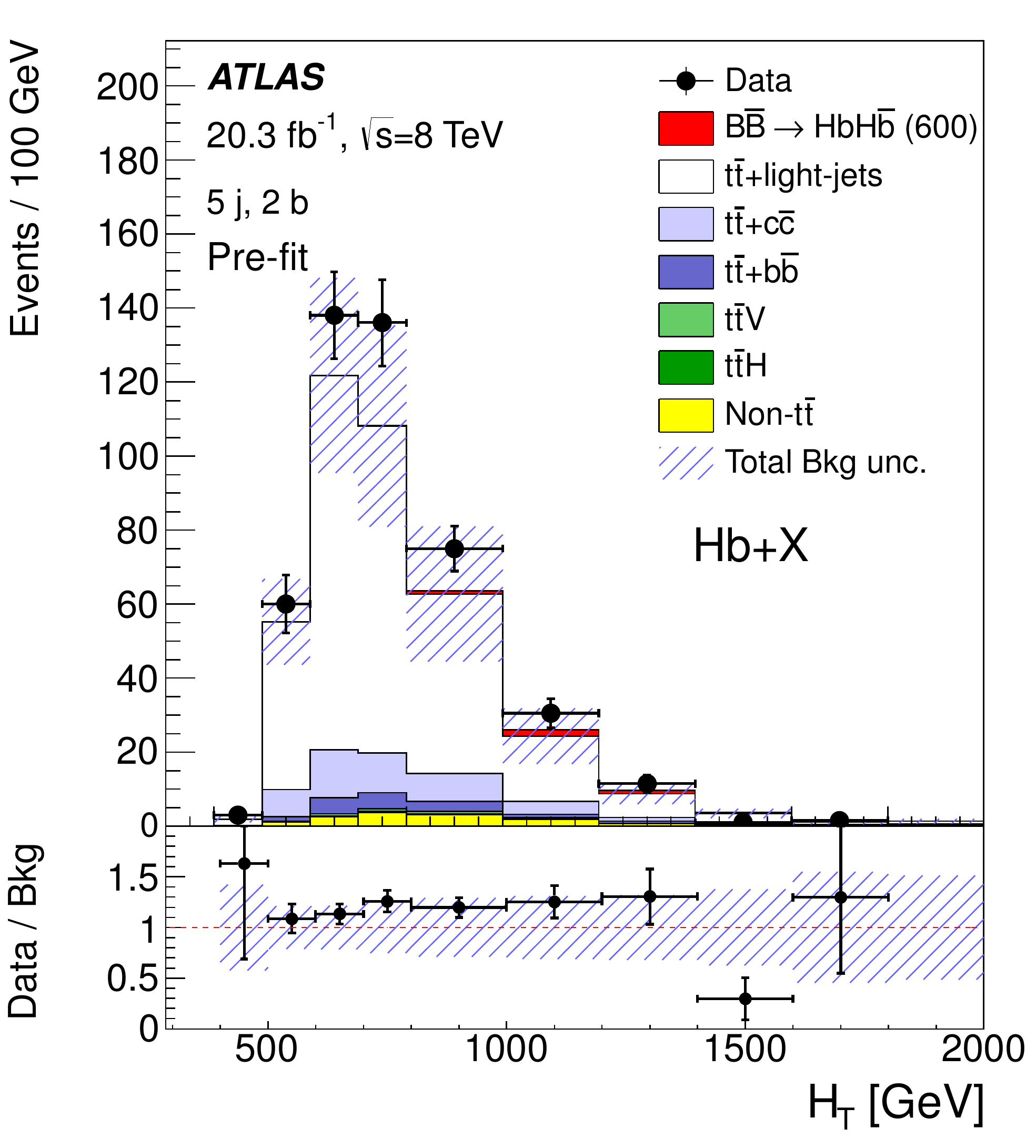}}
\subfloat[]{\includegraphics[width=0.45\textwidth]{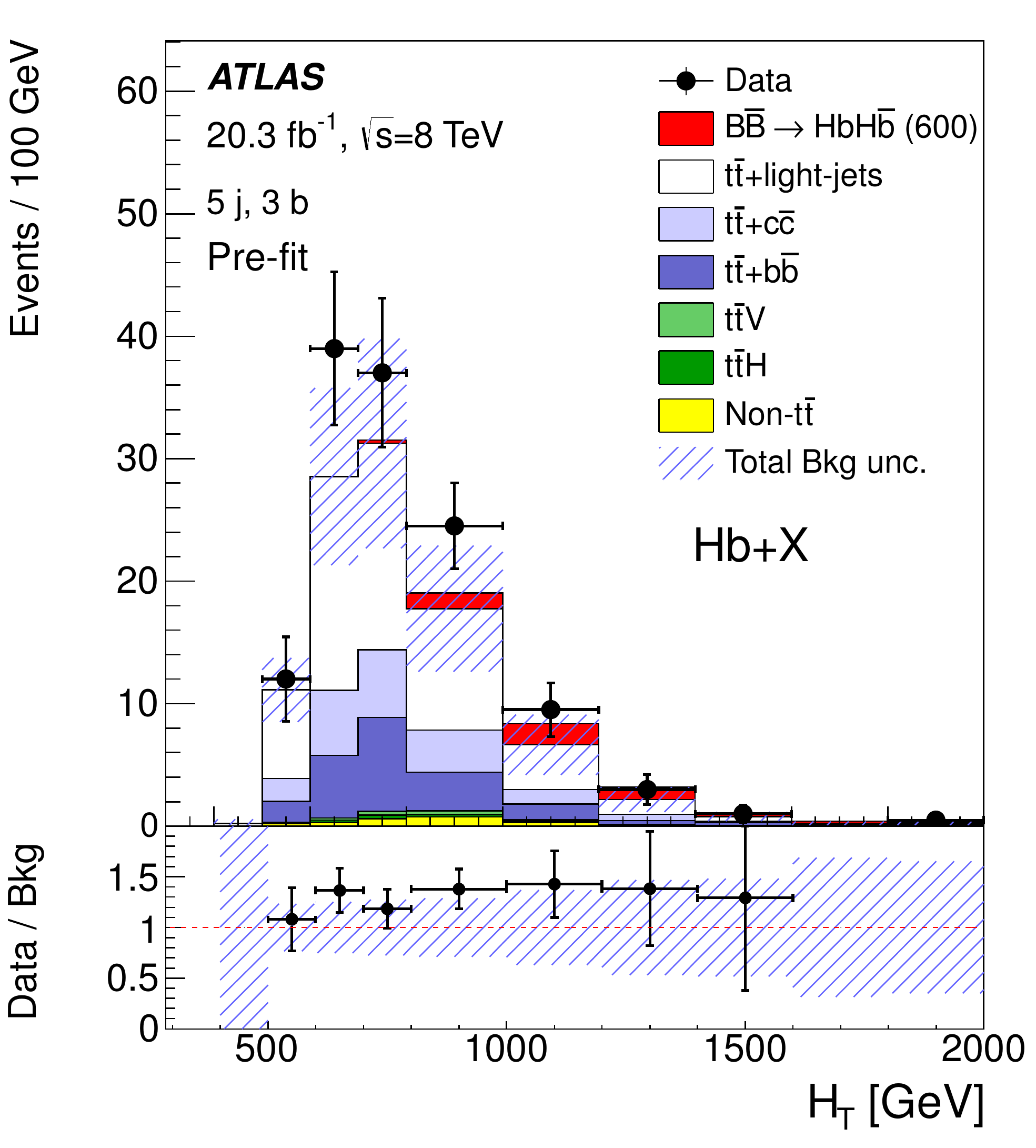}} \\
\subfloat[]{\includegraphics[width=0.45\textwidth]{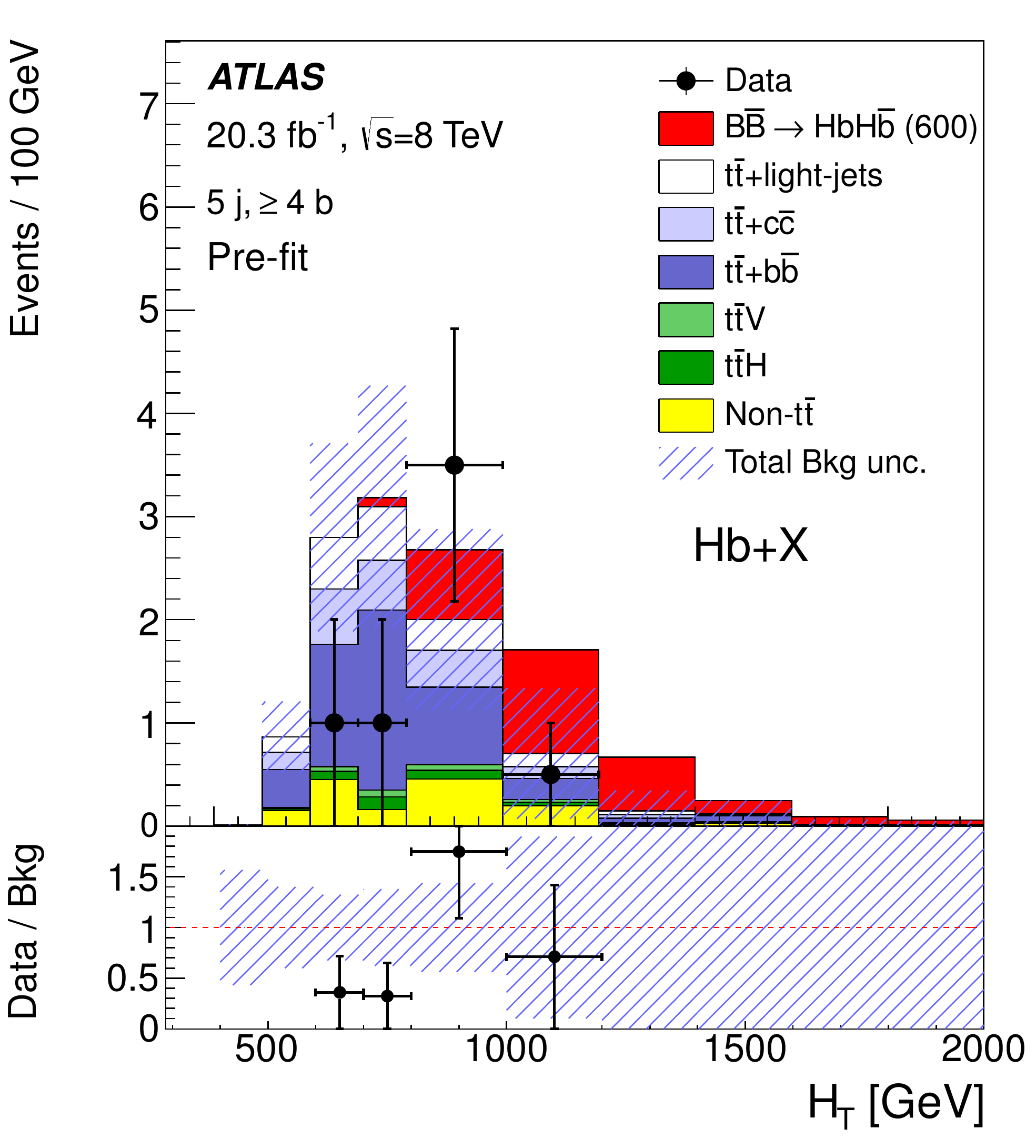}} 
\subfloat[]{\includegraphics[width=0.45\textwidth]{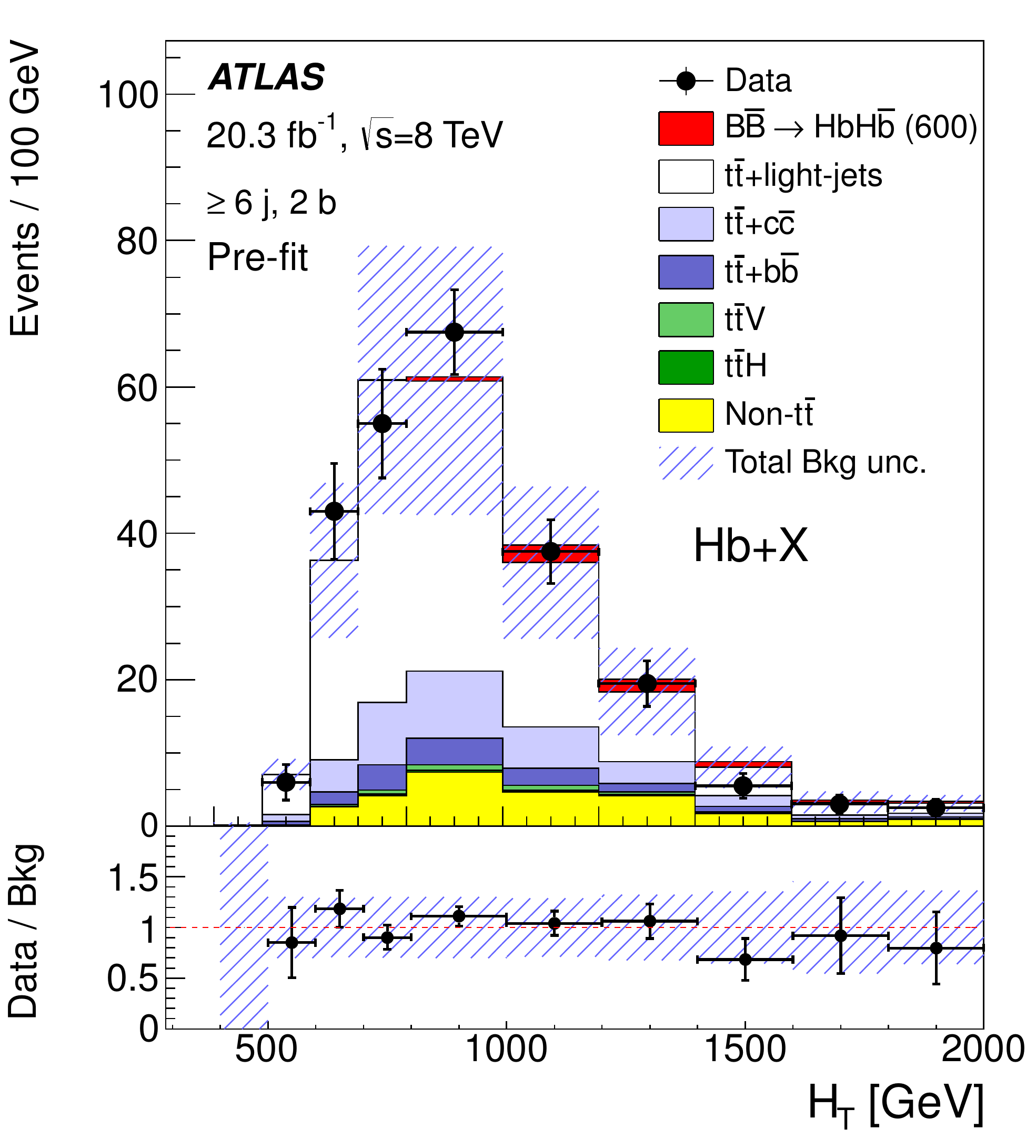}} 
\caption{$B\bar{B} \to Hb$+X search: comparison between data and prediction for the distribution of
the scalar sum ($\HT$) of the transverse momenta of the lepton, the selected jets and the missing transverse 
momentum in each of the analysed channels after final selection:
a) (5 j, 2 b), (b) (5 j, 3 b), (c) (5 j, $\geq$4 b), and (d) ($\geq$6 j, 2 b). 
The background prediction is shown before the fit to data. 
The contributions from $W/Z$+jets,  single top, diboson and multijet backgrounds are 
combined into a single background source referred to as ``Non-$\ttbar$''.
Also shown is the expected signal contribution from a vector-like $B$ quark with mass $m_{B}=600\gev$ under the assumption $\BR(B \to Hb)=1$.
The last bin in all figures contains the overflow. The bottom panel displays the ratio of
data to the total background prediction. The hashed area represents the total uncertainty on the background.}
\label{fig:prefit_HbX_unblinded_1} 
\end{center}
\end{figure*}

\begin{figure*}[htbp]
\begin{center}
\subfloat[]{\includegraphics[width=0.45\textwidth]{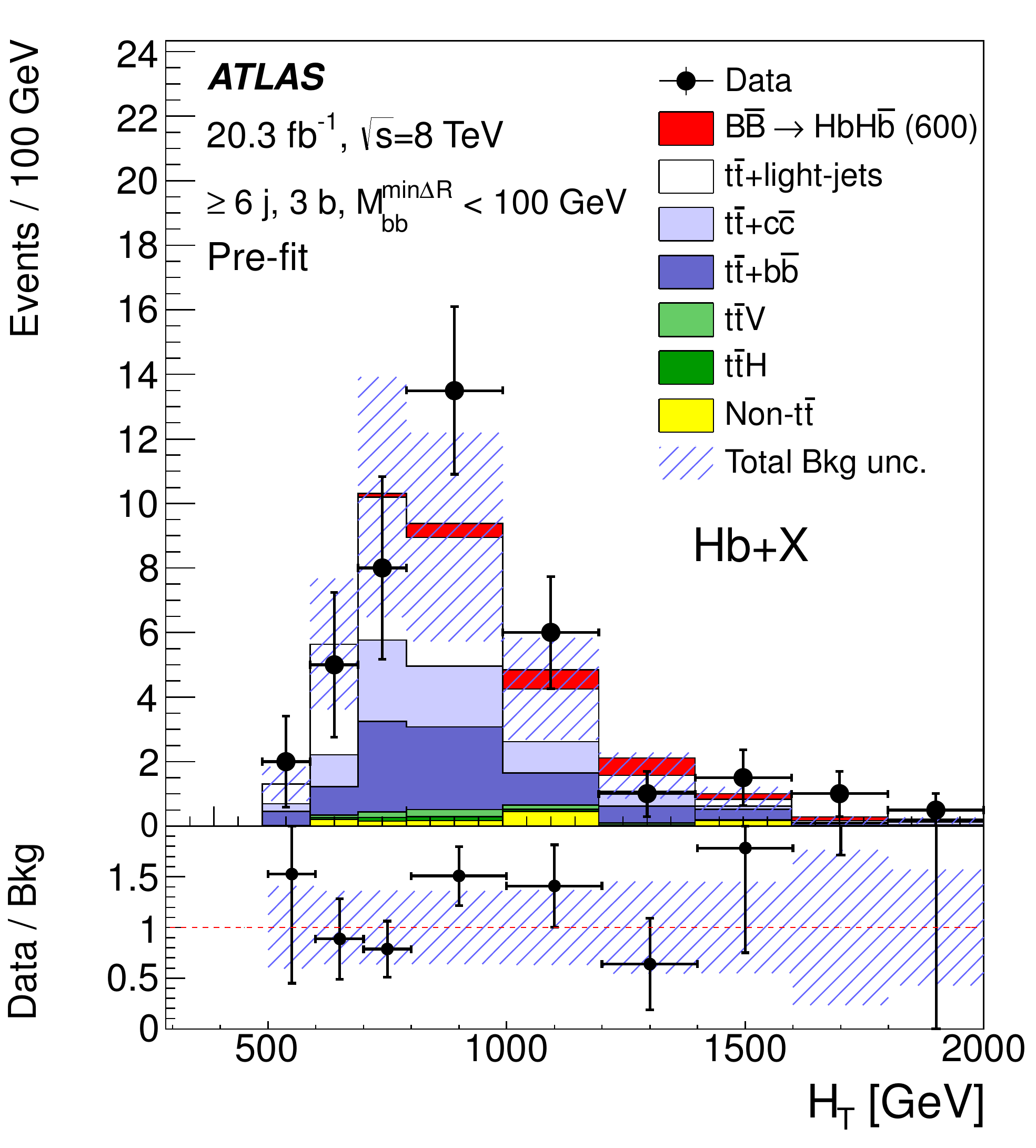}} 
\subfloat[]{\includegraphics[width=0.45\textwidth]{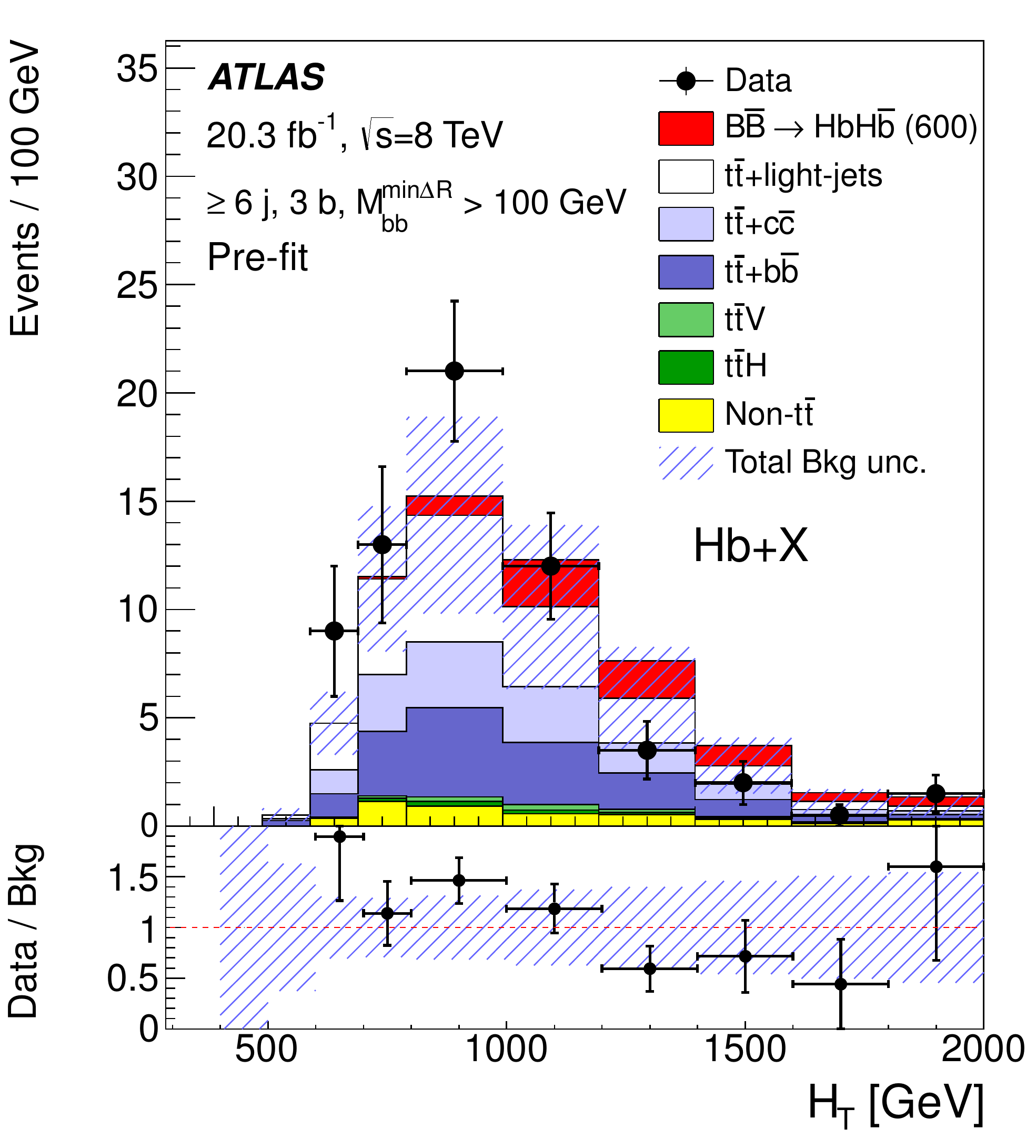}} \\
\subfloat[]{\includegraphics[width=0.45\textwidth]{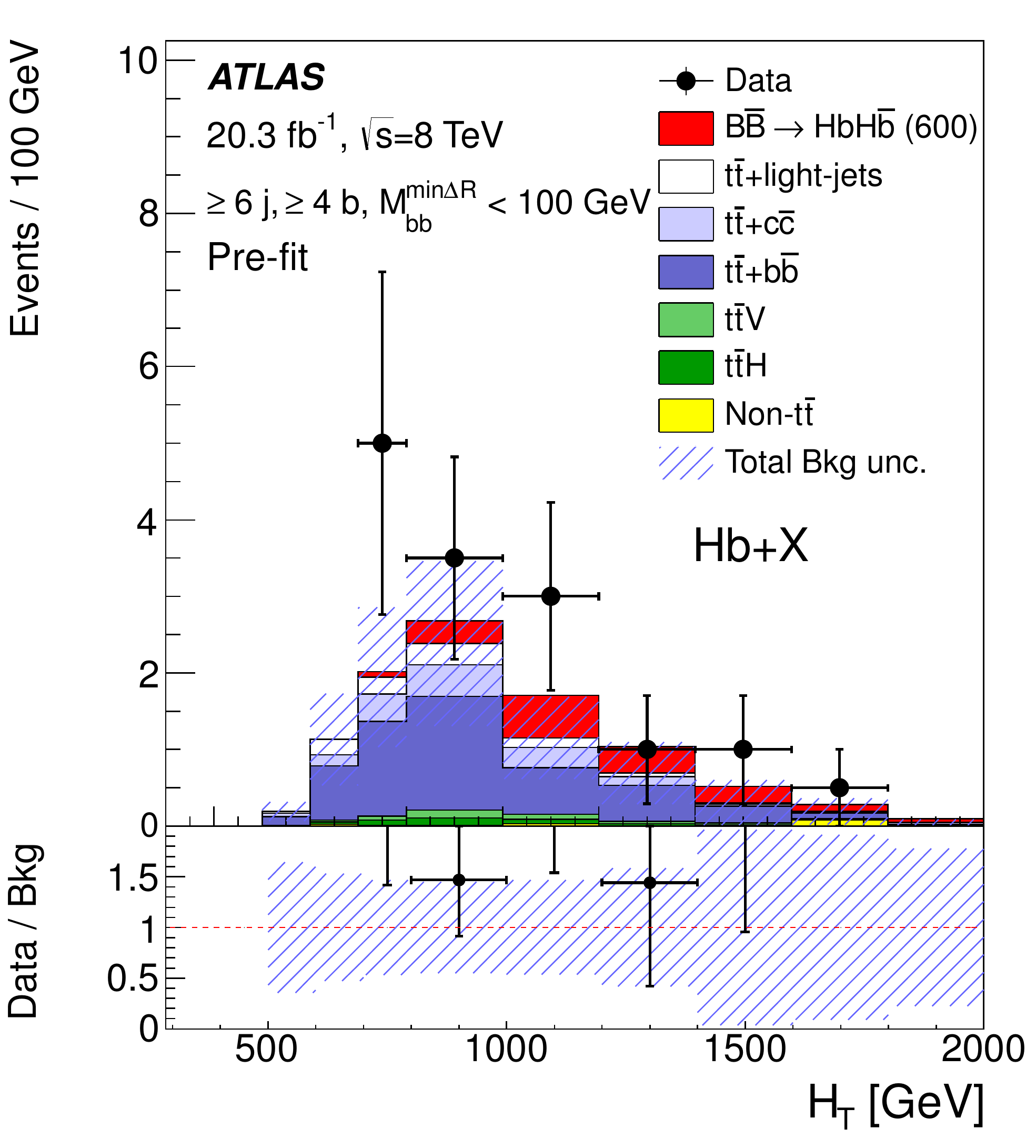}} 
\subfloat[]{\includegraphics[width=0.45\textwidth]{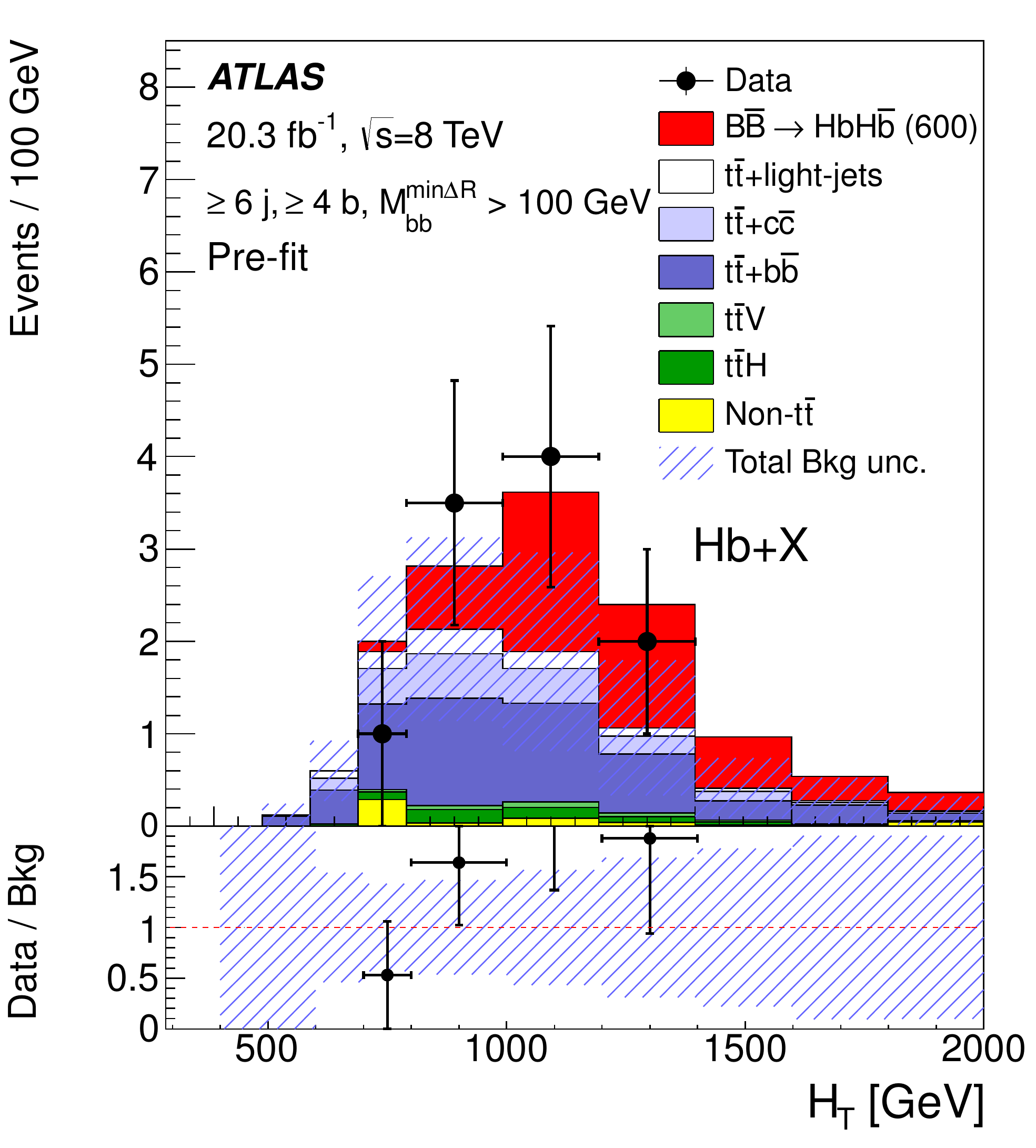}} 
\caption{$B\bar{B} \to Hb$+X search: comparison between data and prediction for the distribution of
the scalar sum ($\HT$) of the transverse momenta of the lepton, the selected jets and the missing transverse 
momentum in each of the analysed channels after final selection:
(a) ($\geq$6 j, 3 b, low $M_{bb}^{{\rm min}\Delta R}$), (b) ($\geq$6 j, 3 b, high $M_{bb}^{{\rm min}\Delta R}$), 
(c) ($\geq$6 j, $\geq$4 b, low $M_{bb}^{{\rm min}\Delta R}$), and (d) ($\geq$6 j, $\geq$4 b, high $M_{bb}^{{\rm min}\Delta R}$). 
The background prediction is shown before the fit to data. 
The contributions from $W/Z$+jets,  single top, diboson and multijet backgrounds are 
combined into a single background source referred to as ``Non-$\ttbar$''.
Also shown is the expected signal contribution from a vector-like $B$ quark with mass $m_{B}=600\gev$ under the assumption $\BR(B \to Hb)=1$.
The last bin in all figures contains the overflow. The bottom panel displays the ratio of
data to the total background prediction. The hashed area represents the total uncertainty on the background.}
\label{fig:prefit_HbX_unblinded_2} 
\end{center}
\end{figure*}

\begin{table}
\begin{center}
\begin{tabular}{l*{4}{c}}
\toprule\toprule
 & 5 j, 2 b & 5 j, 3 b & 5 j, $\geq$4 b & $\geq$6 j, 2 b\\
\midrule
$B\bar{B}$ ($m_{B}=600\gev$) \\
$\BR(B \to Hb)=1$ & $8.6 \pm 1.1$ & $9.3 \pm 2.2$ & $5.0 \pm 1.4$ & $11.9 \pm 3.0$\\
Singlet  & $12.2 \pm 1.9$ & $8.8 \pm 1.7$ & $3.4 \pm 0.8$ & $27.4 \pm 4.3$\\
$(B,Y)$ doublet & $8.5 \pm 1.1$ & $5.8 \pm 1.4$ & $2.8 \pm 0.8$ & $10.9 \pm 2.1$\\
\midrule
$t\bar{t}$+light-jets & $389 \pm 93$ & $72 \pm 18$ & $2.1 \pm 0.7$ & $234 \pm 74$\\
$t\bar{t}+c\bar{c}$ & $56 \pm 42$ & $23 \pm 15$ & $2.2 \pm 1.5$ & $55 \pm 40$\\
$t\bar{t}+b\bar{b}$ & $19 \pm 14$ & $25 \pm 14$ & $5.5 \pm 3.2$ & $22 \pm 15$\\
$t\bar{t}V$ & $4.2 \pm 1.4$ & $1.6 \pm 0.5$ & $0.3 \pm 0.1$ & $5.1 \pm 1.7$\\
$t\bar{t}H$ & $1.0 \pm 0.1$ & $1.1 \pm 0.2$ & $0.5 \pm 0.1$ & $1.5 \pm 0.2$\\
$W$+jets & $21 \pm 12$ & $3.5 \pm 2.1$ & $0.6 \pm 0.5$ & $12.5 \pm 7.9$\\
$Z$+jets & $8.2 \pm 3.3$ & $2.8 \pm 2.8$ & $0.5 \pm 0.5$  & $4.3 \pm 4.1$\\
Single top & $41.3 \pm 7.2$ & $8.8 \pm 1.9$ & $0.6 \pm 0.1$ & $28.0 \pm 6.8$\\
Diboson & $1.9 \pm 0.9$ & $0.5 \pm 0.3$ & $0.07 \pm 0.07$  & $1.2 \pm 0.7$\\
Multijet & $<0.01$ & $<0.01$ & $0.4 \pm 0.2$ & $0.2 \pm 0.1$\\
\hline
Total background& $540 \pm 120$          & $139 \pm 35$          & $12.8 \pm 4.9$          & $360 \pm 100$         \\
\hline
Data & $576$ & $165$ & $10$ & $375$\\
\bottomrule\bottomrule     \\
\end{tabular}
\vspace{0.1cm}

\begin{tabular}{l*{4}{c}}
\toprule\toprule
 & \begin{tabular}{@{}c@{}}$\geq$6 j, 3 b\\ low $M_{bb}^{{\rm min}\Delta R}$\end{tabular} & \begin{tabular}{@{}c@{}}$\geq$6 j, 3 b\\ high $M_{bb}^{{\rm min}\Delta R}$\end{tabular} & \begin{tabular}{@{}c@{}}$\geq$6 j, $\geq$4 b\\ low $M_{bb}^{{\rm min}\Delta R}$\end{tabular} & \begin{tabular}{@{}c@{}}$\geq$6 j, $\geq$4 b\\ high $M_{bb}^{{\rm min}\Delta R}$\end{tabular}\\
\midrule
$B\bar{B}$ ($m_{B}=600\gev$) \\
$\BR(B \to Hb)=1$ & $3.8 \pm 0.6$ & $13.1 \pm 1.8$ & $3.2 \pm 0.7$ & $9.6 \pm 2.0$\\
Singlet & $7.1 \pm 0.9$ & $15.8 \pm 2.5$ & $4.6 \pm 0.9$ & $7.5 \pm 1.5$\\
$(B,Y)$ doublet & $2.7 \pm 0.3$ & $7.0 \pm 1.3$ & $2.3 \pm 0.6$ & $3.9 \pm 0.9$\\
\midrule
$t\bar{t}$+light-jets & $21.3 \pm 9.0$ & $32.8 \pm 9.5$ & $1.4 \pm 0.5$ & $1.5 \pm 0.6$\\
$t\bar{t}+c\bar{c}$ & $10.8 \pm 7.5$ & $20 \pm 15$ & $2.2 \pm 1.6$ & $2.9 \pm 2.2$\\
$t\bar{t}+b\bar{b}$ & $13.1 \pm 8.5$ & $24 \pm 16$ & $7.8 \pm 4.8$ & $8.1 \pm 5.3$\\
$t\bar{t}V$ & $1.1 \pm 0.4$ & $1.6 \pm 0.6$ & $0.6 \pm 0.2$ & $0.4 \pm 0.2$\\
$t\bar{t}H$ & $0.7 \pm 0.1$ & $1.4 \pm 0.2$ & $0.5 \pm 0.1$ & $0.9 \pm 0.2$\\
$W$+jets & $2.0 \pm 1.3$ & $1.1 \pm 0.8$ & $0.3 \pm 0.3$ & $0.05 \pm 0.05$         \\
$Z$+jets & $0.11 \pm 0.07$ & $0.2 \pm 0.1$ & $<0.01$ & $<0.01$\\
Single top & $3.2 \pm 0.6$ & $5.1 \pm 2.2$ & $0.8 \pm 0.2$ & $0.3 \pm 0.2$\\
Diboson & $0.2 \pm 0.1$ & $0.09 \pm 0.03$ & $0.02 \pm 0.01$ & $<0.01$\\
Multijet & $<0.01$ & $0.6 \pm 0.2$ & $<0.01$ & $0.4 \pm 0.1$\\
\hline
Total background & $53 \pm 18$          & $87 \pm 30$          & $13.7 \pm 5.9$          & $14.5 \pm 7.3$         \\
\hline
Data & $62$ & $103$ & $23$ & $20$\\
\bottomrule\bottomrule    \\
\end{tabular}
\vspace{0.1cm}

\end{center}
\vspace{-0.5cm}
\caption{$B\bar{B} \to Hb$+X search: predicted and observed yields in each of the analysis channels considered.
The background prediction is shown before the fit to data. Also shown are the signal predictions for different benchmark scenarios considered.
The quoted uncertainties are the sum in quadrature of statistical and systematic uncertainties on the yields.}
\label{tab:Prefit_Yields_HbX_unblind}
\end{table}

\FloatBarrier

\section{Systematic uncertainties}
\label{sec:systematics}
				   
Several sources of systematic uncertainty are considered that can affect the normalisation of signal 
and background and/or the shape of their corresponding final discriminant distributions.  
Individual sources of systematic uncertainty are considered uncorrelated.  Correlations of a given 
systematic uncertainty are maintained across processes and channels.  
Table~\ref{tab:SystSummary} presents a list of all systematic uncertainties considered in the analyses 
and indicates whether they are taken to be normalisation-only, or to affect both shape and normalisation.
\begin{table}[ht!]
\centering
\begin{tabular}{lcc}
\toprule\toprule
Systematic uncertainty & Type  & Components \\
\midrule
Luminosity                  &  N & 1\\\midrule\midrule
{\bf Reconstructed Objects}                 &   & \\
Electron                  & SN & 5 \\
Muon                      &  SN & 6 \\\midrule
Jet reconstruction      & SN & 1\\ 
Jet vertex fraction         & SN    & 1\\
Jet energy scale            & SN & 22\\
Jet energy resolution       & SN & 1\\
Missing transverse momentum  & SN & 2\\ \midrule
$b$-tagging efficiency      & SN & 6\\
$c$-tagging efficiency      & SN & 4\\
Light-jet tagging efficiency    & SN & 12\\ 
High-\pt\ tagging   & SN & 1 \\ \midrule\midrule
{\bf Background Model}                 &   & \\
$t\bar{t}$ cross section    &  N & 1\\
$t\bar{t}$ modelling: $\pt$ reweighting   & SN & 9\\
$t\bar{t}$ modelling: parton shower & SN & 3\\
$t\bar{t}$+HF: normalisation & N & 2 \\
$t\bar{t}$+$c\bar{c}$: HF reweighting  & SN & 2 \\
$t\bar{t}$+$c\bar{c}$: generator & SN & 4 \\
$t\bar{t}$+$b\bar{b}$: NLO Shape & SN & 8 \\\midrule
$W$+jets normalisation      &  N & 3\\
$Z$+jets normalisation      &  N & 3\\
Single top cross section    &  N & 1\\
Single top model            &  SN & 1\\
Diboson normalisation  &  N & 1\\
$t\bar{t}V$ cross section   &  N & 1\\
$t\bar{t}V$ model           &  SN & 1\\ 
$t\bar{t}H$ cross section & N & 1 \\
$t\bar{t}H$ model       & SN & 2 \\ 
Multijet normalisation  &  N & 2\\
\bottomrule\bottomrule
\end{tabular}
\caption{\label{tab:SystSummary} List of systematic uncertainties considered. 
An ``N" means that the uncertainty is taken as normalisation-only for all 
processes and channels affected, whereas ``SN" means that the uncertainty is 
taken on both shape and normalisation.
Some of the systematic uncertainties are split into several components for a more
accurate treatment.}
\end{table}

%
Table~\ref{tab:SystSummary_WbX} presents a summary of the systematic uncertainties for the $T\bar{T} \to Wb$+X search and
their impact on the normalisation of signal and backgrounds. A similar summary is presented for the $T\bar{T} \to Ht$+X and $B\bar{B} \to Hb$+X
searches in tables~\ref{tab:SystSummary_HtX} and~\ref{tab:SystSummary_HbX} respectively, restricted to the highest-sensitivity channel
and displaying only the signal and the $\ttbar$+jets background categories.
Tables~\ref{tab:SystSummary_HtX} and~\ref{tab:SystSummary_HbX} also show the impact of the systematic uncertainties before and after the fit to data.
\begin{table*}[h!]
\centering
\begin{tabular}{l*{4}{c}}
\toprule\toprule
 & Signal & $t\bar{t}$ & Non-$t\bar{t}$ & Total background \\
\midrule
Luminosity & $\pm 2.8 $  & $\pm 2.8 $  & $\pm 2.8 $  & $\pm 2.8 $ \\ 
Lepton efficiencies & $\pm 1.6 $  & $\pm 1.6 $  & $\pm 1.5 $  & $\pm 1.6 $ \\ 
Jet energy scale & $+3.4$/$-7.2$  & $\pm 16$  & $+19$/$-9$  & $+17$/$-12$ \\ 
Jet efficiencies & $\pm 1.5 $  & $\pm 1.6 $  & $\pm 1.6 $  & $\pm 1.6 $ \\ 
Jet energy resolution & $\pm 1.1 $  & $\pm 0.6 $  & $\pm 2.6 $  & $\pm 1.8 $ \\ 
$b$-tagging efficiency & $\pm 5.0$  & $\pm 0.7$  & $\pm 2.9 $  & $\pm 2.0 $ \\ 
$c$-tagging efficiency & $\pm 0.4 $  & $\pm 1.2 $  & $\pm 2.3 $  & $\pm 1.9 $ \\ 
Light-jet tagging efficiency & $\pm 0.2 $  & $\pm 1.3 $  & $\pm 1.6 $  & $\pm 1.4 $ \\ 
High-\pt tagging efficiency & $\pm 3.2 $  & $\pm 1.3 $  & $\pm 0.8 $  & $\pm 1.1 $ \\ 
Missing transverse momentum & --  & $\pm 2.6 $  & --  & $\pm 1.0 $\\ 
$t\bar{t}$: reweighting & --  & $\pm 15 $  & --  & $\pm 5.9 $ \\ 
$t\bar{t}$: parton shower & --  & $\pm 9.3 $  & --  & $\pm 3.6 $ \\ 
$t\bar{t}$+HF: normalisation & --  & $+12.0$/$-5.5$  & --  & $+4.5$/$-2.1$ \\ 
$t\bar{t}$+HF: modelling & --  & $\pm 30 $  & --  & $\pm 11 $ \\ 
Theoretical cross sections & --  & $\pm 6.0$ & $\pm 33$  & $\pm 20$\\
Multijet normalisation & --  & --  & $\pm 2.9 $  & $\pm 1.8 $ \\  
Non-$t\bar{t}$ modelling & --  & --  & $\pm 2.3 $  & $\pm 1.4 $ \\ 
\midrule
Total  & $+7.7$/$-10.0$  & $\pm 40$  & $\pm 35 $  & $\pm 29$ \\ 
\bottomrule\bottomrule
\end{tabular}
\caption{$T\bar{T} \to Wb$+X search: summary of the systematic uncertainties considered and their impact (in \%) on the normalisation of signal 
and backgrounds. Only sources of systematic uncertainty resulting in a normalisation change of at least 0.5\% are displayed.
The signal shown corresponds to a vector-like $T$ quark with mass $m_{T}=600\gev$ and $\BR(T \to Wb)=1$.}
\label{tab:SystSummary_WbX}
\end{table*}

\begin{table*}[h!]
\centering
\begin{tabular}{l | c c c c | c c c}
\multicolumn{8}{c}{$\geq$6 j, $\geq$4 b, high $M_{bb}^{{\rm min}\Delta R}$}\\
\toprule\toprule
 & \multicolumn{4}{c|}{Pre-fit} & \multicolumn{3}{c}{Post-fit} \\ 
 &  Signal & $t\bar{t}$+light-jets & $t\bar{t}+c\bar{c}$ & $t\bar{t}+b\bar{b}$ &  $t\bar{t}$+light-jets & $t\bar{t}+c\bar{c}$ & $t\bar{t}+b\bar{b}$ \\
\midrule
Luminosity  & $\pm 2.8 $  & $\pm 2.8 $  & $\pm 2.8 $  & $\pm 2.8 $  & $\pm 2.6 $  & $\pm 2.6 $  & $\pm 2.6 $ \\ 
Lepton efficiencies  & $\pm 1.5 $  & $\pm 1.5 $  & $\pm 1.5 $  & $\pm 1.5 $  & $\pm 1.5 $  & $\pm 1.4 $  & $\pm 1.5 $ \\ 
Jet energy scale  & $\pm 4.4 $  & $\pm 15 $  & $\pm 11 $  & $\pm 12 $  & $\pm 8.7 $  & $\pm 6.4 $  & $\pm 6.7 $ \\ 
Jet efficiencies  & --  & $\pm 4.0 $  & $\pm 2.2 $  & $\pm 1.9 $  & $\pm 2.7 $  & $\pm 1.5 $  & $\pm 1.3 $ \\ 
Jet energy resolution  & $\pm 0.1 $  & $\pm 4.4 $  & $\pm 3.8 $  & $\pm 0.5 $  & $\pm 3.1 $  & $\pm 2.6 $  & $\pm 0.4 $ \\ 
$b$-tagging efficiency  & $\pm 13 $  & $\pm 5.6 $  & $\pm 5.4 $  & $\pm 9.3 $  & $\pm 4.6 $  & $\pm 4.6 $  & $\pm 6.6 $ \\ 
$c$-tagging efficiency  & $\pm 1.6 $  & $\pm 5.8 $  & $\pm 12 $  & $\pm 3.1 $  & $\pm 5.6 $  & $\pm 11 $  & $\pm 2.9 $ \\ 
Light-jet tagging efficiency  & $\pm 0.6 $  & $\pm 20 $  & $\pm 5.7 $  & $\pm 2.0 $  & $\pm 17 $  & $\pm 5.1 $  & $\pm 1.8 $ \\ 
High-\pt tagging efficiency  & $\pm 4.8 $  & $\pm 0.7 $  & $\pm 1.7 $  & $\pm 1.6 $  & $\pm 0.6 $  & $\pm 1.3 $  & $\pm 1.2 $ \\ 
$t\bar{t}$: reweighting  & --  & $\pm 13 $  & $\pm 15 $  & --  & $\pm 10 $  & $\pm 10 $  & -- \\ 
$t\bar{t}$: parton shower  & --  & $\pm 28 $  & $\pm 17 $  & $\pm 6.2 $  & $\pm 13 $  & $\pm 11 $  & $\pm 4.0 $ \\ 
$t\bar{t}$+HF: normalisation  & --  & --  & $\pm 50 $  & $\pm 50 $  & --  & $\pm 32 $  & $\pm 18 $ \\ 
$t\bar{t}$+HF: modelling  & --  & --  & $\pm 17 $  & $\pm 12 $  & --  & $\pm 16 $  & $\pm 10 $ \\ 
Theoretical cross sections  & --  & $\pm 6.3 $  & $\pm 6.3 $  & $\pm 6.3 $  & $\pm 4.6 $  & $\pm 4.6 $  & $\pm 4.6 $ \\ 
\midrule
Total   & $\pm 15 $  & $\pm 42 $  & $\pm 61 $  & $\pm 55 $  & $\pm 22 $  & $\pm 30 $  & $\pm 15 $ \\ 
\bottomrule\bottomrule
\end{tabular}
\caption{$T\bar{T} \to Ht$+X search: summary of the systematic uncertainties considered in the ($\geq$6 j, $\geq$4 b, high $M_{bb}^{{\rm min}\Delta R}$)  
channel and their impact (in \%) on the normalisation of signal and backgrounds, before and after the fit to data.
Only sources of systematic uncertainty resulting in a normalisation change of at least 0.5\% are displayed.
The signal shown corresponds to a singlet vector-like $T$ quark with mass $m_{T}=600\gev$. 
The total post-fit uncertainty can be different from the sum in quadrature of individual sources due to the anti-correlations between them resulting from the fit to the data.}
\label{tab:SystSummary_HtX}
\end{table*}

\begin{table*}[h!]
\centering
\begin{tabular}{l | c c c c | c c c}
\multicolumn{8}{c}{$\geq$6 j, $\geq$4 b, high $M_{bb}^{{\rm min}\Delta R}$}\\
\toprule\toprule
 & \multicolumn{4}{c|}{Pre-fit} & \multicolumn{3}{c}{Post-fit} \\ 
 &  Signal & $t\bar{t}$+light-jets & $t\bar{t}+c\bar{c}$ & $t\bar{t}+b\bar{b}$ &  $t\bar{t}$+light-jets & $t\bar{t}+c\bar{c}$ & $t\bar{t}+b\bar{b}$ \\
\midrule
Luminosity  & $\pm 2.8 $  & $\pm 2.8 $  & $\pm 2.8 $  & $\pm 2.8 $  & $\pm 2.7 $  & $\pm 2.7 $  & $\pm 2.7 $ \\ 
Lepton efficiencies  & $\pm 1.6 $  & $\pm 1.4 $  & $\pm 1.5 $  & $\pm 1.7 $  & $\pm 1.4 $  & $\pm 1.5 $  & $\pm 1.6 $ \\ 
Jet energy scale  & $\pm 5.6 $  & $\pm 14 $  & $\pm 14 $  & $\pm 11 $  & $\pm 13 $  & $\pm 14 $  & $\pm 11 $ \\ 
Jet efficiencies  & $\pm 3.1 $  & $\pm 3.3 $  & $\pm 1.0 $  & $\pm 0.9 $  & $\pm 3.2 $  & $\pm 0.9 $  & $\pm 0.8 $ \\ 
Jet energy resolution  & $\pm 0.1 $  & $\pm 6.0 $  & $\pm 1.1 $  & $\pm 1.9 $  & $\pm 4.5 $  & $\pm 0.9 $  & $\pm 1.5 $ \\ 
$b$-tagging efficiency  & $\pm 16 $  & $\pm 7.6 $  & $\pm 9.2 $  & $\pm 16 $  & $\pm 3.9 $  & $\pm 5.2 $  & $\pm 7.5 $ \\ 
$c$-tagging efficiency  & $\pm 1.0 $  & $\pm 6.1 $  & $\pm 15 $  & $\pm 3.0 $  & $\pm 5.8 $  & $\pm 14 $  & $\pm 2.8 $ \\ 
Light-jet tagging efficiency  & --  & $\pm 19 $  & $\pm 6.3 $  & $\pm 2.4 $  & $\pm 18 $  & $\pm 5.8 $  & $\pm 2.3 $ \\ 
High-\pt tagging efficiency  & $\pm 11 $  & $\pm 2.7 $  & $\pm 5.3 $  & $\pm 5.0 $  & $\pm 1.9 $  & $\pm 3.8 $  & $\pm 3.6 $ \\ 
$t\bar{t}$: reweighting  & --  & $\pm 15 $  & $\pm 16 $  & --  & $\pm 14 $  & $\pm 15 $  & -- \\ 
$t\bar{t}$: parton shower  & --  & $\pm 22 $  & $\pm 35 $  & $\pm 26 $  & $\pm 14 $  & $\pm 33 $  & $\pm 24 $ \\ 
$t\bar{t}$+HF: normalisation  & --  & --  & $\pm 50 $  & $\pm 50 $  & --  & $\pm 44 $  & $\pm 30 $ \\ 
$t\bar{t}$+HF: modelling  & --  & --  & $\pm 27 $  & $\pm 24 $  & --  & $\pm 28 $  & $\pm 21 $ \\ 
Theoretical cross sections  & --  & $\pm 6.3 $  & $\pm 6.2 $  & $\pm 6.3 $  & $\pm 5.9 $  & $\pm 5.9 $  & $\pm 5.9 $ \\ 
\midrule
Total   & $\pm 21 $  & $\pm 38 $  & $\pm 73 $  & $\pm 65 $  & $\pm 24 $  & $\pm 46 $  & $\pm 27 $ \\
\bottomrule\bottomrule
\end{tabular}
\caption{$B\bar{B} \to Hb$+X search: summary of the systematic uncertainties considered in the ($\geq$6 j, $\geq$4 b, high $M_{bb}^{{\rm min}\Delta R}$)  
channel and their impact (in \%) on the normalisation of signal and backgrounds, before and after the fit to data.
Only sources of systematic uncertainty resulting in a normalisation change of at least 0.5\% are displayed.
The signal shown corresponds to a vector-like $B$ quark with mass $m_{B}=600\gev$ and $\BR(B \to Hb)=1$.
The total post-fit uncertainty can be different from the sum in quadrature of individual sources due to the anti-correlations between them resulting from the fit to the data.}
\label{tab:SystSummary_HbX}
\end{table*}

In the case of the $T\bar{T} \to Wb$+X search, the total systematic uncertainty in the background normalisation is
approximately 29\%, with the dominant contributions originating from the normalisation of the $W$+jets
background (20\%), jet energy scale ($+17\%$/$-12\%$) and the $\ttbar$+HF normalisation (11\%).
The total systematic uncertainty in the signal normalisation is $+8\%$/$-10\%$, with comparable contributions
from jet energy scale and $b$-tagging uncertainties.

The leading sources of systematic uncertainty in the  $T\bar{T} \to Ht$+X and $B\bar{B} \to Hb$+X searches
vary depending on the analysis channel considered, but they typically originate from $\ttbar$+jets modelling
(including $\ttbar$+HF), jet energy scale and $b$-tagging. For example, the total systematic uncertainty 
in the background normalisation in the highest-sensitivity channel  ($\geq$6 j, $\geq$4 b, high $M_{bb}^{{\rm min}\Delta R}$) 
of the $T\bar{T} \to Ht$+X search is approximately 37\%, 
with the largest contributions originating from $\ttbar$+HF normalisation (23\%), jet energy scale (10\%) and
$b$-tagging (9\%). However, as discussed previously, the  fit to data in the eight analysis channels in these
searches allows the overall background uncertainty to be reduced significantly, to approximately 5\% in the case
of the $T\bar{T} \to Ht$+X search. More details about the fit to data can be found in section~\ref{sec:result_fits}.
The total systematic uncertainty on the signal normalisation is approximately 15\%, almost all due to $b$-tagging uncertainties.

The following sections describe each of the systematic uncertainties considered in the analyses. 

\subsection{Luminosity}
\label{sec:syst_lumi}
The uncertainty on the integrated luminosity is 2.8\%, affecting the overall normalisation of
all processes estimated from the simulation. It is derived following the same methodology as that 
detailed in ref.~\cite{Aad:2013ucp}.

\subsection{Reconstructed objects}
\label{sec:syst_objects}

\subsubsection{Leptons}
Uncertainties associated with leptons arise from the reconstruction,
identification and trigger, as well as the lepton momentum scale and
resolution.
The reconstruction and identification efficiency of electrons and
muons, as well as the efficiency of the trigger used to record the
events, differ slightly between data and simulation.  Scale factors are derived using
tag-and-probe techniques on $Z\to \ell^+\ell^-$ ($\ell=e,\mu$) data
and simulated samples, and are applied to the simulation to correct for differences.
Additional sources of uncertainty originate from the corrections
applied to adjust the lepton momentum scale and resolution in
the simulation to match those in data, measured using reconstructed 
distributions of the $Z\to \ell^+\ell^-$ and $J/\psi \to \ell^+\ell^-$ masses,
as well as the measured $E/p$ in $W\to e\nu$ events, where $E$ and $p$ are
the electron energy and momentum, as measured by the calorimeter and the tracker respectively.
The combined effect of all these uncertainties results in an overall normalisation 
uncertainty on the signal and background of approximately 1.5\%.

\subsubsection{Jets and missing transverse momentum}
Uncertainties associated with jets arise from the efficiency of jet reconstruction 
and identification based on the JVF variable, as well as the jet energy scale
and resolution. The uncertainty associated with the jet reconstruction efficiency
is assessed by randomly removing 0.2\% of the jets with $\pt$ below $30\gev$,
which is the level of disagreement between data and the simulation, and has a 
negligible impact in the analysis. The per-jet efficiency to satisfy the JVF requirement
is measured in $Z(\to \ell^+\ell^-)$+1-jet events in data and simulation,
selecting separately events enriched in hard-scatter jets and events
enriched in jets from pileup, and good agreement is found. The associated uncertainty
is estimated by changing the nominal JVF cut value by $\pm 0.1$ 
and repeating the analysis using the modified cut value, resulting in normalisation
uncertainties in the range of 1--5\%, depending on the jet multiplicity under consideration
and the $\pt$ spectra of the jets.
The jet energy scale and its uncertainty were derived by combining
information from test-beam data, LHC collision data and
simulation~\cite{Aad:2014bia}.  The jet energy scale uncertainty is split into 22
uncorrelated sources with their respective jet $\pt$ and $\eta$
dependences and are treated independently in this analysis. 
It represents one of the leading sources of uncertainty associated with reconstructed 
objects, affecting the normalisations of signal and backgrounds by approximately 
5\% and 15\% respectively, in the most signal-rich channels considered. 
The jet energy resolution was measured in data and simulation
as a function of jet $\pt$ and rapidity using dijet events. They are found to agree
within 10\%, and the corresponding uncertainty is assessed by smearing the jet
$\pt$ in the simulation.

The $\met$ reconstruction is affected by uncertainties associated with leptons and jet 
energy scales and resolutions, which are propagated to $\met$ and thus are included 
under the corresponding uncertainty categories in tables~\ref{tab:SystSummary_WbX}--\ref{tab:SystSummary_HbX}. 
Additional small uncertainties associated with the modelling of the underlying event, in particular its impact on
the $\pt$ scale and resolution of unclustered energy, are also taken into account
and are displayed in tables~\ref{tab:SystSummary_WbX}--\ref{tab:SystSummary_HbX}
under the category of ``Missing transverse momentum".

\subsubsection{Heavy- and light-flavour tagging}
Efficiencies to tag jets from $b$- and $c$-quarks in the simulation are 
corrected to match the efficiencies in data by $\pt$-dependent factors 
in the approximate ranges 0.9--1.0 and 0.9--1.1 respectively,  
whereas the light-jet efficiency is scaled by $\pt$- and $\eta$-dependent
scale factors in the range 1.2--1.5~\cite{BTaggingEfficiency,CLTaggingEfficiency}.
Uncertainties on these scale factors include a total of six independent sources
affecting  $b$-jets and four independent sources affecting $c$-jets. 
Each of these uncertainties has different jet $\pt$ dependence.
Twelve uncertainties are considered for the light-jets tagging, which depend on the jet \pt\ and \eta\ regions. 
These systematic uncertainties are taken as uncorrelated between $b$-jets, 
$c$-jets, and light-jets. An additional uncertainty is included due to the extrapolation of the 
$b$-, $c$-, and light-jet-tagging scale factors for jets with $\pt$ beyond the kinematic reach 
of the data calibration samples used:  $\pt>300\gev$ for $b$- and $c$-jets, and
$\pt>750\gev$ for light-jets. This uncertainty is evaluated in the simulation by comparing 
the tagging efficiencies while varying e.g. the fraction of tracks with shared hits in the silicon detectors 
or the fraction of fake tracks resulting from random combinations of hits, both of which typically increase at high $\pt$ due to
growing track multiplicity and density of hits within the jet. 
These uncertainties are taken to be correlated among the three jet flavours.
As an example, the uncertainties on the tagging efficiencies for $b$-jets and $c$-jets with $300\gev \leq \pt < 500\gev$ are 14\% and 23\%
respectively.

\subsection{Background modelling}
\label{sec:syst_bkgmodeling}

\subsubsection{$\ttbar$+jets}

A number of systematic uncertainties affecting the modelling of $t\bar{t}$+jets 
are considered. These include the uncertainty on the theoretical prediction for the inclusive cross section, 
uncertainties associated with the reweighting procedure applied to $\ttbar$+light-jets 
and $\ttcc$ processes,  uncertainties affecting the modelling of $t\bar{t}$+HF-jets 
production, and uncertainties associated with the choice of parton shower and 
hadronisation model. A summary of these uncertainties can be found below. Additional
details can be found in ref.~\cite{Aad:2015gra}.

An uncertainty of $+5\%$/$-6\%$ is assumed for the inclusive $t\bar{t}$ production
cross section~\cite{Czakon:2011xx}, including contributions from varying the factorisation and renormalisation 
scales and uncertainties arising from the PDF, $\alpha_{\rm S}$ and the top quark mass.
The PDF and $\alpha_{\rm S}$ uncertainties were calculated using the PDF4LHC 
prescription.

Uncertainties associated with the reweighting procedure applied to $\ttbar$+light-jets 
and $\ttcc$ processes include the nine leading sources of uncertainty in the differential cross
section measurement at $\sqrt{s}=7\tev$~\cite{Aad:2014zka}, dominated by the modelling of 
initial- and final-state radiation and the choice of event generator for $\ttbar$ production. 

Uncertainties affecting the modelling of $\ttbb$ production include those associated with
the NLO prediction from {\sc Sherpa}+{\sc OpenLoops}, which is used for reweighting of the default {\sc Powheg} $\ttbb$ prediction.
These include three different scale variations, including changing the functional form of the renormalisation scale, 
changing the functional form of the factorisation and resummation scales, and varying the renormalisation scale by a
factor of two up and down. In addition,  a different shower recoil model scheme and 
two alternative PDF sets (MSTW and NNPDF) are considered. A fraction of the $\ttbb$ background
predicted by {\sc Powheg+Pythia} originates from multiple parton interactions or final-state radiation from top decay products. 
Such backgrounds are not part of the NLO prediction, and these two categories are kept separate and
subject to additional normalisation uncertainties.
The NLO corrections and associated systematic uncertainties are adjusted so that the overall normalisation of the
$\ttbb$ background at the particle level is fixed, i.e. effectively only migrations across categories and distortions
to the shape of the kinematic distributions are considered.
Detailed comparisons of $\ttbb$ between {\sc Powheg+Pythia} and {\sc Sherpa+OpenLoops} 
show that the cross sections agree to better than 50\%, which is taken as a normalisation uncertainty
for $\ttbb$. 

Beyond the uncertainties associated with the reweighting procedure, additional uncertainties are assigned
to the modelling of the $\ttcc$ component of the background, which again is not part of the NLO prediction 
used for $\ttbb$. These include two uncertainties taken as the full difference between applying and not applying the
reweightings of the top quark and $\ttbar$ $\pt$ spectra. In addition, four uncertainties are considered associated 
with the choice of LO generator: the full difference between {\sc Powheg+Pythia} and {\sc Madgraph5}+{\sc Pythia} simulations,
as well as variations in generator parameters (factorisation and renormalisation scales, matching threshold and $c$-quark mass),
which are derived using {\sc Madgraph5}+{\sc Pythia} simulations and applied to the {\sc Powheg+Pythia} simulation.
Analogously to the procedure used in the $\ttbb$ background estimate, these uncertainties are adjusted so that the overall 
normalisation of the $\ttcc$ background at the particle level is fixed.
Finally, an overall normalisation uncertainty of 50\% is also assigned to the $\ttcc$ component, taken
as uncorrelated with the same normalisation uncertainty applied to $\ttbb$, since only the $\ttbb$ process
is normalised to a NLO prediction.

An uncertainty due to the choice of parton shower and hadronisation model 
is derived by comparing events produced by {\sc Powheg} interfaced to {\sc Pythia} 
or {\sc Herwig}. In the case of $\ttbar$+light-jets and $\ttcc$, a reweighting of the top quark and $\ttbar$ $\pt$ spectra
is also applied to the {\sc Powheg}+{\sc Herwig} samples to ensure reliable modelling
of the top quark kinematics. The corresponding correction factors were recalculated for {\sc Powheg}+{\sc Herwig} in order
to match the differential cross section measurements at $\sqrt{s}=7\tev$. 
In the case of $\ttbb$, the various HF categories and the corresponding
partonic kinematics in {\sc Powheg}+{\sc Herwig} are reweighted to match the NLO prediction
of  {\sc Sherpa}+{\sc OpenLoops}, so that only the effect of changing the 
hadronisation model is propagated. 
Given the different effect of this uncertainty on the $\ttbar$+light-jets, $\ttcc$ and $\ttbb$, it is treated as
uncorrelated between the three processes. 
This treatment prevents an undue reduction of this systematic uncertainty on $\ttcc$ and $\ttbb$ 
by constraining it for $\ttbar$+light-jets via the fit to data in the highly populated channels with two $b$-tagged jets.

\subsubsection{$W/Z$+jets}
\label{sec:vjets_systs}
Uncertainties affecting the modelling of the $W/Z$+jets background include 5\%  
from their respective normalisations to the theoretical NNLO cross sections~\cite{Melnikov:2006kv}, 
as well as an additional 24\% normalisation uncertainty added in quadrature for each additional 
inclusive parton multiplicity bin, based on a comparison among different algorithms 
for merging LO matrix elements and parton showers~\cite{Alwall:2007fs}.
The above uncertainties are taken as uncorrelated between $W$+jets and $Z$+jets. 

\subsubsection{Other simulated background}
Uncertainties affecting the modelling of the single-top-quark background include a 
$+5\%$/$-4\%$ uncertainty on the total cross section estimated as a weighted average 
of the theoretical uncertainties on $t$-, $Wt$- and $s$-channel production~\cite{Kidonakis:2011wy,Kidonakis:2010ux,Kidonakis:2010tc},
as well as a systematic uncertainty on $Wt$-channel production concerning the separation 
between $t\bar{t}$ and $Wt$ at NLO~\cite{Frixione:2008yi}.  The latter is estimated by comparing
the nominal sample, which uses the so-called ``diagram subtraction'' scheme, with an alternative sample
using the ``diagram removal'' scheme.

Uncertainties on the diboson background normalisation include 5\% from the
NLO theoretical cross sections~\cite{Campbell:1999ah} added in quadrature 
to an uncertainty of 24\% due to the extrapolation to the high jet multiplicity channels,
following the procedure discussed in section~\ref{sec:vjets_systs}.

Uncertainties on the $\ttbar V$ and $\ttbar H$ normalisations are 30\% and $+9\%$/$-12\%$ respectively,
from the uncertainties on their respective NLO theoretical cross sections~\cite{Campbell:2012dh,Garzelli:2012bn,Dittmaier:2011ti}. 
Additional small uncertainties arising from scale variations, which change the amount of initial-state radiation and thus the event
kinematics, are also included.

\subsubsection{Multijet}
Uncertainties on the data-driven multijet background estimate receive
contributions from the limited sample size in data, particularly at high jet and $b$-tag multiplicities, as 
well as from the uncertainty on the rate of fake leptons, estimated in 
different control regions (e.g. selected with either an upper $\met$ or $\mtw$ requirement). 
A combined normalisation uncertainty of 50\% due 
to all these effects is assigned, which is taken as correlated across jet
and $b$-tag multiplicity bins, but uncorrelated between electron and muon channels. 
No explicit shape uncertainty is assigned since the large statistical uncertainties associated with
the multijet background prediction, which are uncorrelated 
bin-to-bin in the final discriminating variable, effectively cover all possible shape uncertainties.

\section{Statistical analysis}
\label{sec:stat_analysis}

For a given search, the distributions of the final discriminating variables in each of the analysis channels considered 
are combined to test for the presence of a signal. 
The statistical analysis is based on a binned likelihood function ${\cal L}(\mu,\theta)$ constructed as
a product of Poisson probability terms over all bins considered in the analysis. This function depends
on the signal-strength parameter $\mu$,  a multiplicative factor to the theoretical signal production cross section,
and $\theta$, a set of nuisance parameters that encode the effect of systematic uncertainties
on the signal and background expectations and are implemented in the likelihood function as Gaussian or
log-normal priors. Therefore, the total number of expected events in a given bin depends on $\mu$ and
$\theta$. The nuisance parameters $\theta$ allow variations of the expectations for signal and background
according to the corresponding systematic uncertainties, and their fitted values correspond to the deviations from
the nominal expectations that globally provide the best fit to the data.
This procedure allows a reduction of the impact of systematic uncertainties on 
the search sensitivity by taking advantage of the highly populated background-dominated channels included in the likelihood fit.
It requires a good understanding of the systematic effects affecting the shapes of the discriminant distributions.
Detailed validation studies of the fitting procedure have been performed using the simulation.
To verify the improved background prediction, fits are performed under the background-only hypothesis.
Differences between the data and the background prediction are checked relative to the smaller post-fit uncertainties
in kinematic variables other than the ones used in the fit.

The test statistic $q_\mu$ is defined as the profile likelihood ratio: 
$q_\mu = -2\ln({\cal L}(\mu,\hat{\hat{\theta}}_\mu)/{\cal L}(\hat{\mu},\hat{\theta}))$,
where $\hat{\mu}$ and $\hat{\theta}$ are the values of the parameters that
maximise the likelihood function (with the constraint $0\leq \hat{\mu} \leq \mu$), and $\hat{\hat{\theta}}_\mu$ are the values of the
nuisance parameters that maximise the
likelihood function for a given value of $\mu$. 
Statistical uncertainties in each bin of the discriminant distributions are also taken into account via dedicated parameters in the fit.     
The test statistic $q_\mu$ is implemented in the {\sc RooFit} package~\cite{Verkerke:2003ir,RooFitManual} and
is used to measure the compatibility of the observed data with the background-only hypothesis (i.e. the discovery test)  
setting $\mu=0$ in the profile likelihood ratio: $q_0 = -2\ln({\cal L}(0,\hat{\hat{\theta}}_0)/{\cal L}(\hat{\mu},\hat{\theta}))$.
The $p$-value (referred to as $p_0$) representing the compatibility of the data with the background-only hypothesis is estimated by integrating
the distribution of $q_0$ from background-only pseudo-experiments, approximated using the asymptotic formulae given in ref.~\cite{Cowan:2010js,ErratumCowan:2010js}, 
above the observed value of $q_0$. Some model dependence exists
in the estimation of the $p_0$-value, as a given signal scenario needs to be assumed in the calculation of the denominator of $q_\mu$, even 
if the overall signal normalisation is left floating and fitted to data. The observed $p_0$-value is checked for each explored signal scenario.
In the absence of any significant excess above the background expectation, upper limits on the signal production cross section for each of the 
signal scenarios considered are derived by using $q_\mu$ in the CL$_{\rm{s}}$ method~\cite{Junk:1999kv,Read:2002hq}.
For a given signal scenario, values of the production cross section (parameterised by $\mu$) yielding CL$_{\rm{s}}$$<$0.05, 
where CL$_{\rm{s}}$ is computed using the asymptotic approximation~\cite{Cowan:2010js,ErratumCowan:2010js}, are excluded at $\geq$95\% CL.

\section{Results}
\label{sec:result}

This section presents the results obtained from the searches discussed in sections~\ref{sec:search_wbx}--\ref{sec:search_hbx},
following the statistical analysis discussed in section~\ref{sec:stat_analysis}.

\subsection{Likelihood fits to data}
\label{sec:result_fits}

The consideration of high-statistics background-dominated channels in the analysis allows an improved background prediction 
with significantly reduced systematic uncertainties to be obtained during the statistical analysis, as discussed in section~\ref{sec:stat_analysis}.
This is the strategy adopted in the $T\bar{T} \to Ht$+X and $B\bar{B} \to Hb$+X searches. In contrast, the small number of data events in the $T\bar{T} \to Wb$+X 
search results in virtually the same background prediction and uncertainties both pre-fit and post-fit.
Figures~\ref{fig:postfit_HtX_unblinded_1} and~\ref{fig:postfit_HtX_unblinded_2} show the comparison of data and the post-fit background 
prediction for the $\HT$ distributions in each of the analysis channels considered in the $T\bar{T} \to Ht$+X search. The corresponding
comparisons for the $B\bar{B} \to Hb$+X search can be found in figures~\ref{fig:postfit_HbX_unblinded_1} and~\ref{fig:postfit_HbX_unblinded_2}. 
The fit to the data is performed under the background-only hypothesis. 
Tables with the corresponding predicted and observed yields per channel can be found in appendix~\ref{sec:postfit_yields_appendix}.

Compared to the pre-fit distributions shown in sections~\ref{sec:search_htx} and~\ref{sec:search_hbx}, the total background 
uncertainty is significantly reduced after the fit, not only in the background-dominated channels, but also in the signal-rich channels. The reduced 
uncertainty results from the significant constraints provided by the data on some systematic uncertainties, as well as the anti-correlations 
among sources of systematic uncertainty resulting from the fit to the data. For example, the uncertainty in the $\ttbb$ background in the
highest-sensitivity channel ($\geq$6 j, $\geq$4 b, high $M_{bb}^{{\rm min}\Delta R}$) is reduced from about 60\% prior to the fit 
to about 15\% and 30\% in the $T\bar{T} \to Ht$+X and the $B\bar{B} \to Hb$+X searches, respectively. The larger post-fit uncertainty in the case of 
the $B\bar{B} \to Hb$+X search is partly caused by the smaller number of data events due to the selection requirements being tighter than in the $T\bar{T} \to Ht$+X search.

\begin{figure*}[htbp]
\begin{center}
\subfloat[]{\includegraphics[width=0.45\textwidth]{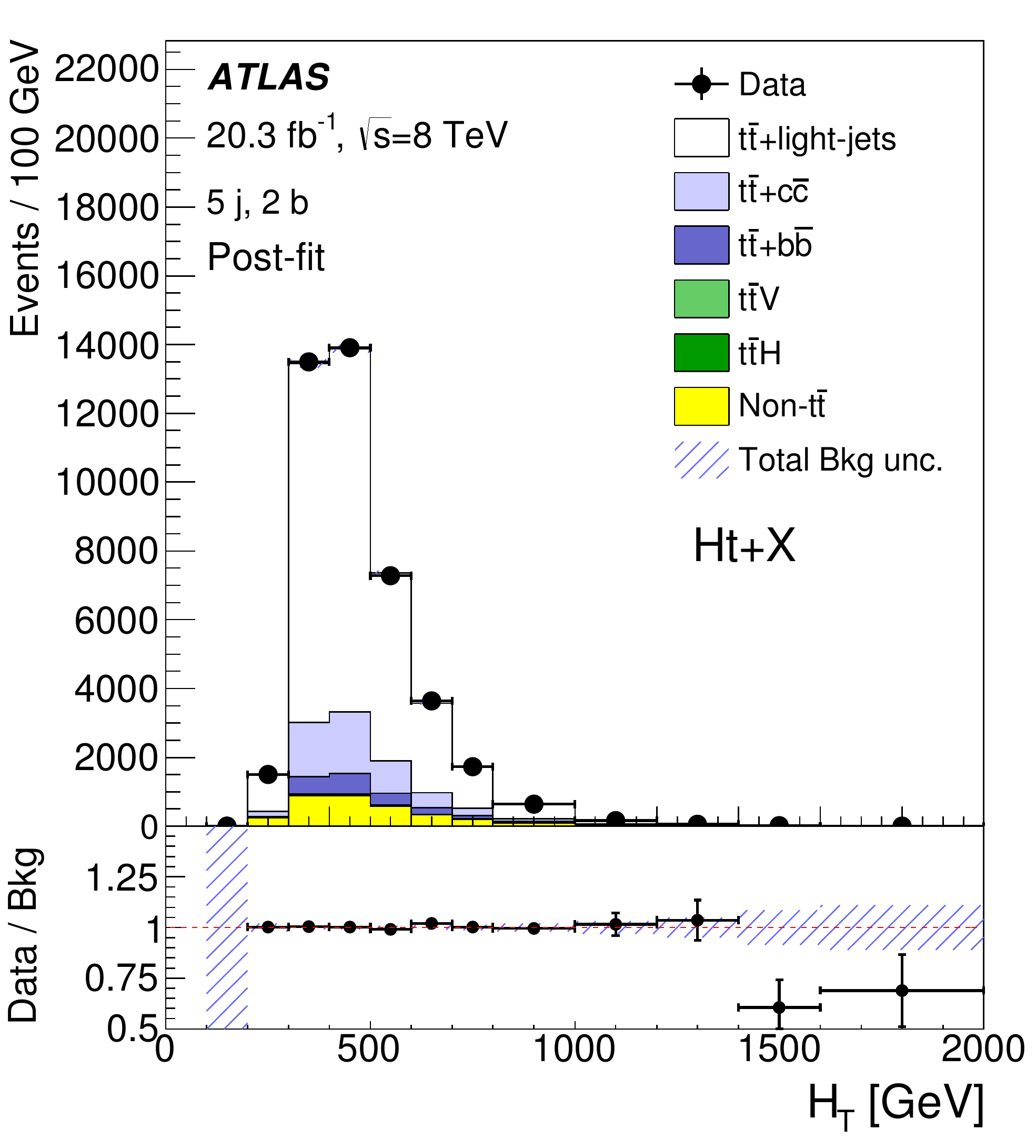}}
\subfloat[]{\includegraphics[width=0.45\textwidth]{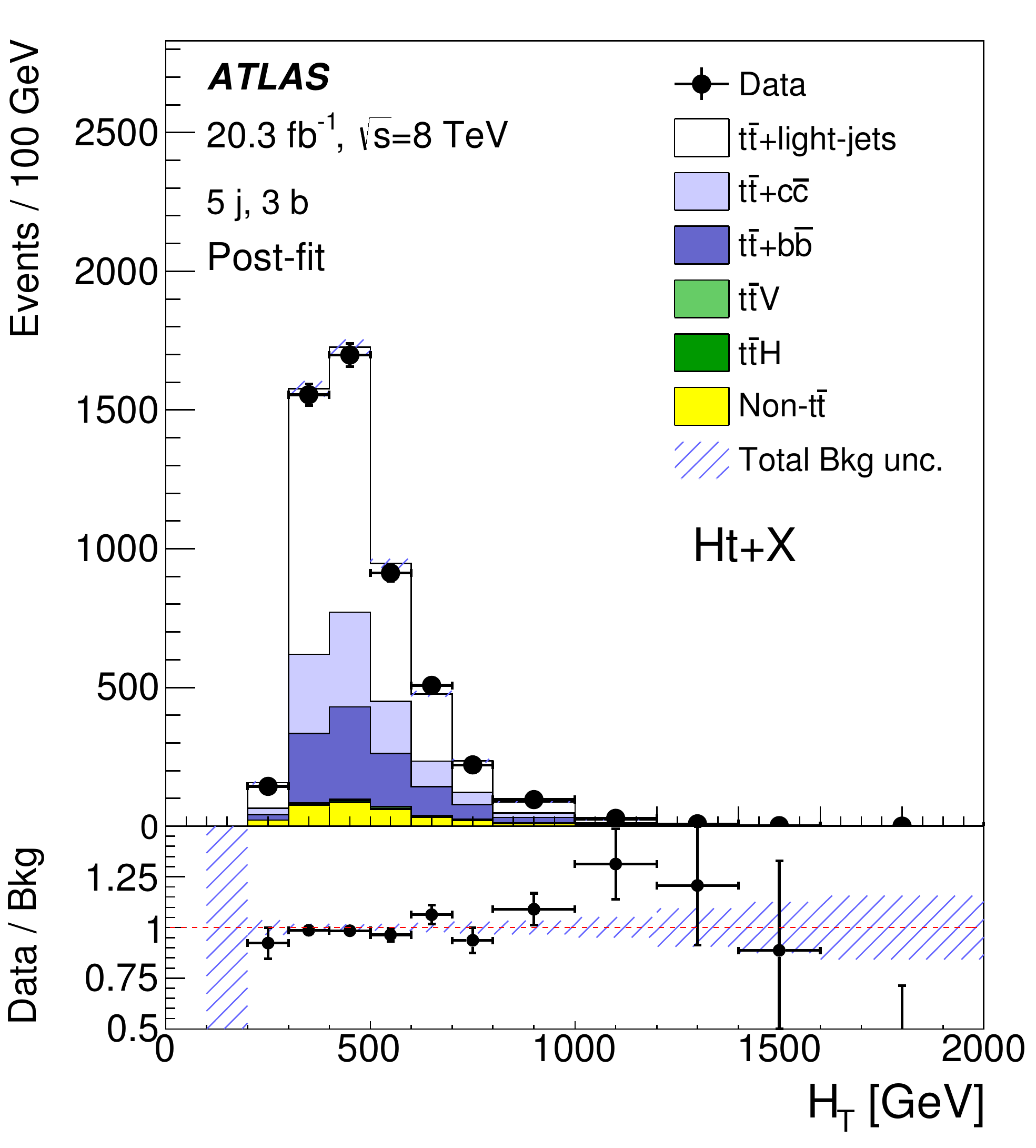}} \\
\subfloat[]{\includegraphics[width=0.45\textwidth]{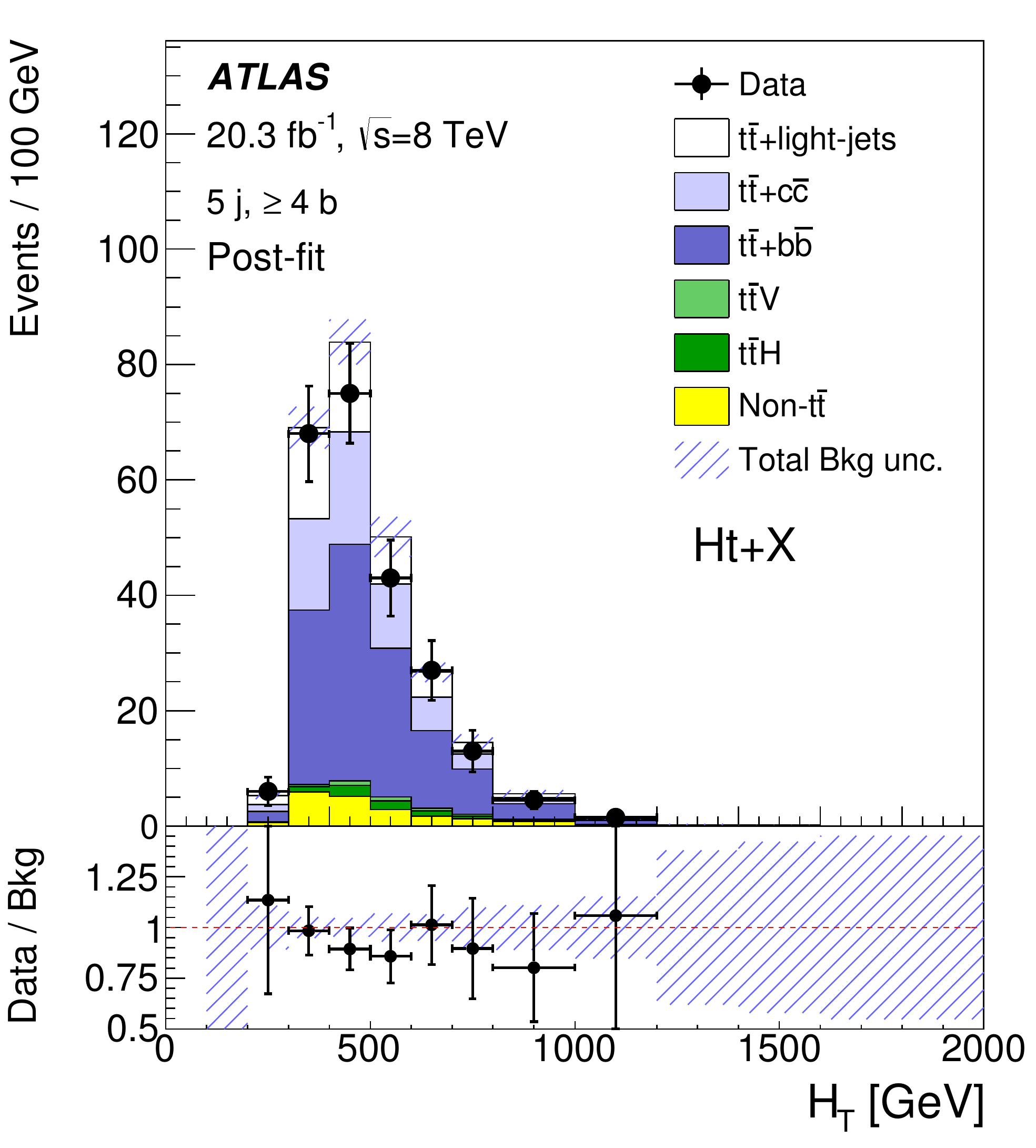}} 
\subfloat[]{\includegraphics[width=0.45\textwidth]{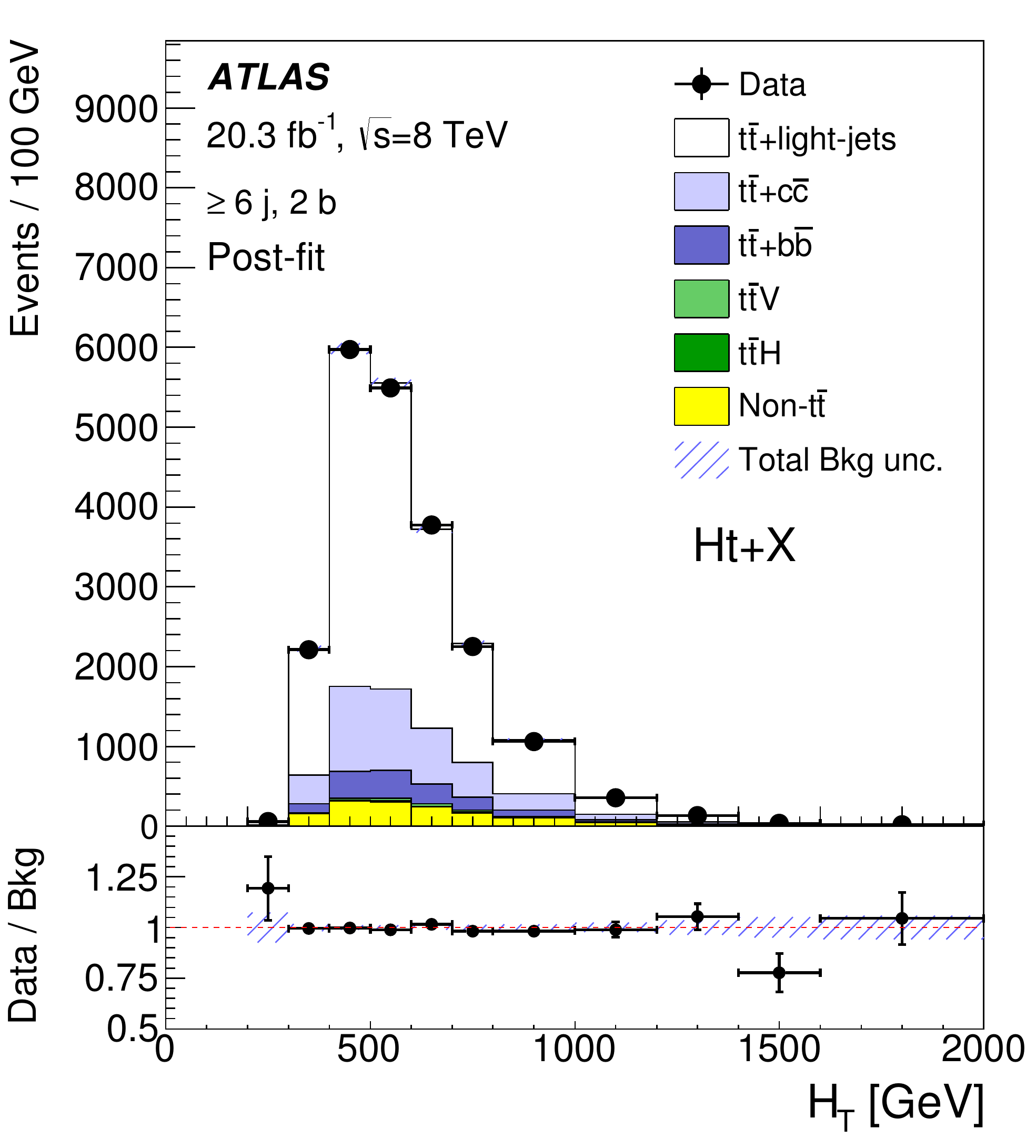}} 
\caption{$T\bar{T} \to Ht$+X search: comparison between data and prediction for the distribution of
the scalar sum ($\HT$) of the transverse momenta of the lepton, the selected jets and the missing transverse 
momentum in each of the analysed channels after final selection:
(a) (5 j, 2 b), (b) (5 j, 3 b), (c) (5 j, $\geq$4 b), and (d) ($\geq$6 j, 2 b). 
The background prediction is shown after the fit to data under the background-only hypothesis.
The small contributions from $W/Z$+jets,  single top, diboson and multijet backgrounds are 
combined into a single background source referred to as ``Non-$\ttbar$''.
The last bin in all figures contains the overflow. The bottom panel displays the ratio of
data to the total background prediction. The hashed area represents the total uncertainty on the background.}
\label{fig:postfit_HtX_unblinded_1} 
\end{center}
\end{figure*}

\begin{figure*}[htbp]
\begin{center}
\subfloat[]{\includegraphics[width=0.45\textwidth]{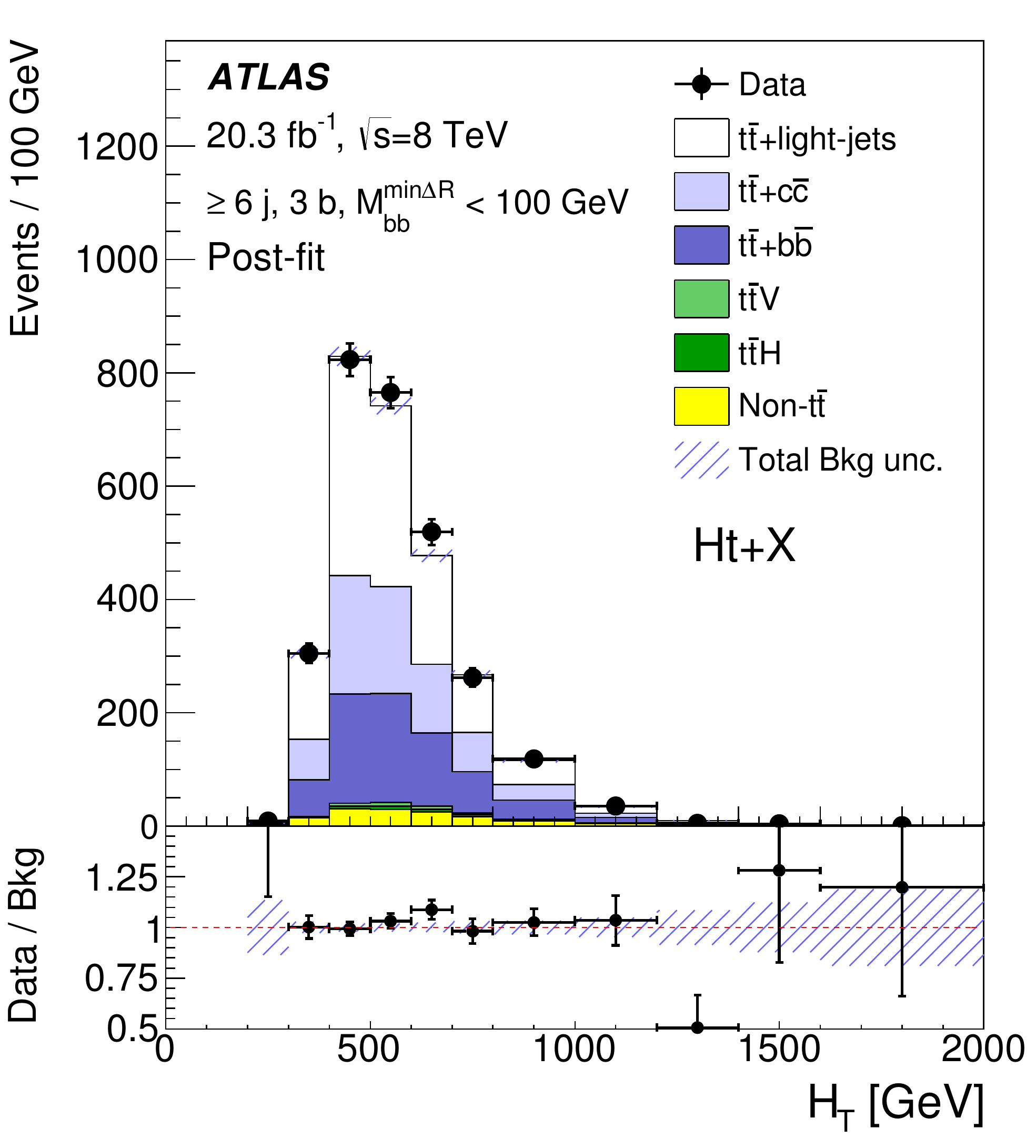}} 
\subfloat[]{\includegraphics[width=0.45\textwidth]{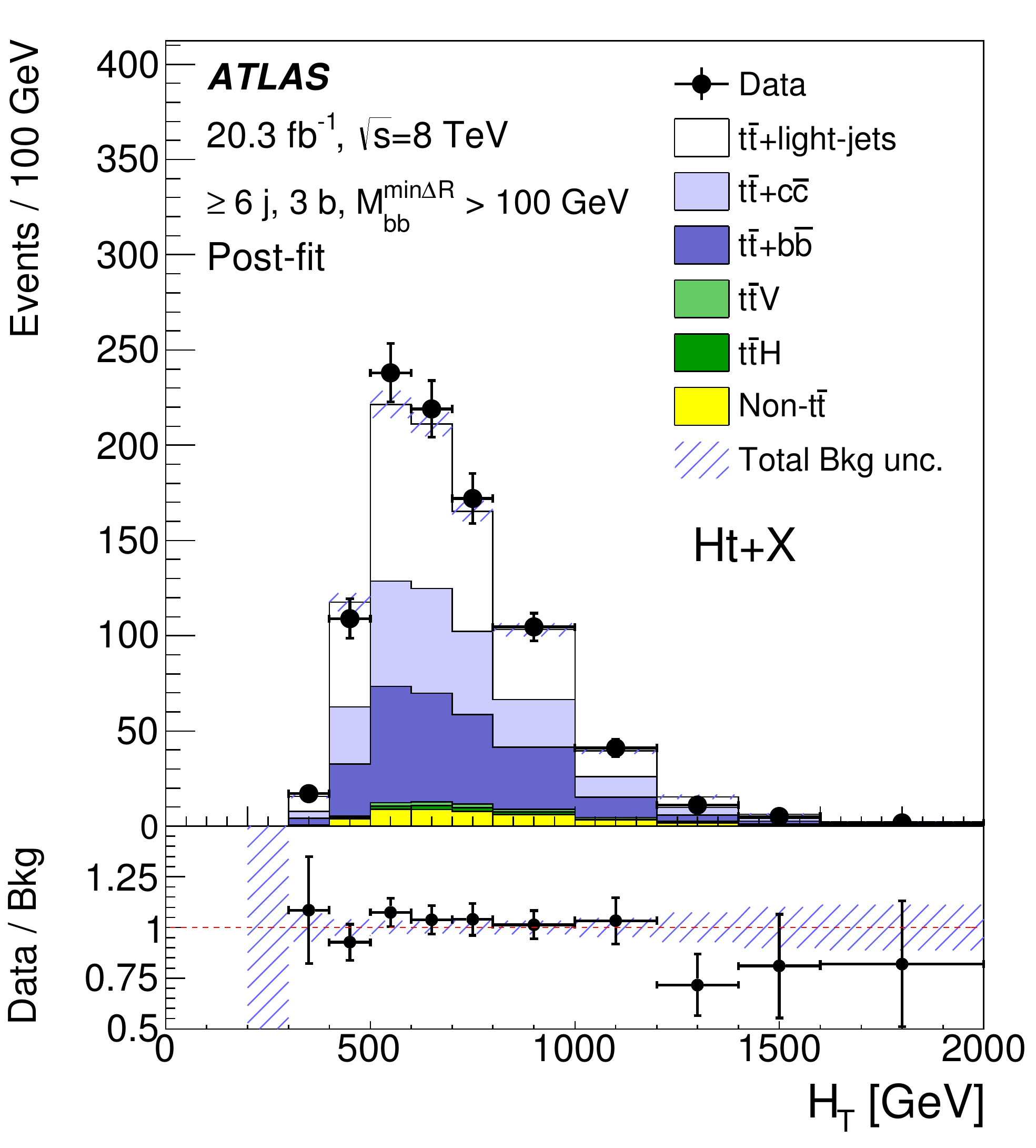}} \\
\subfloat[]{\includegraphics[width=0.45\textwidth]{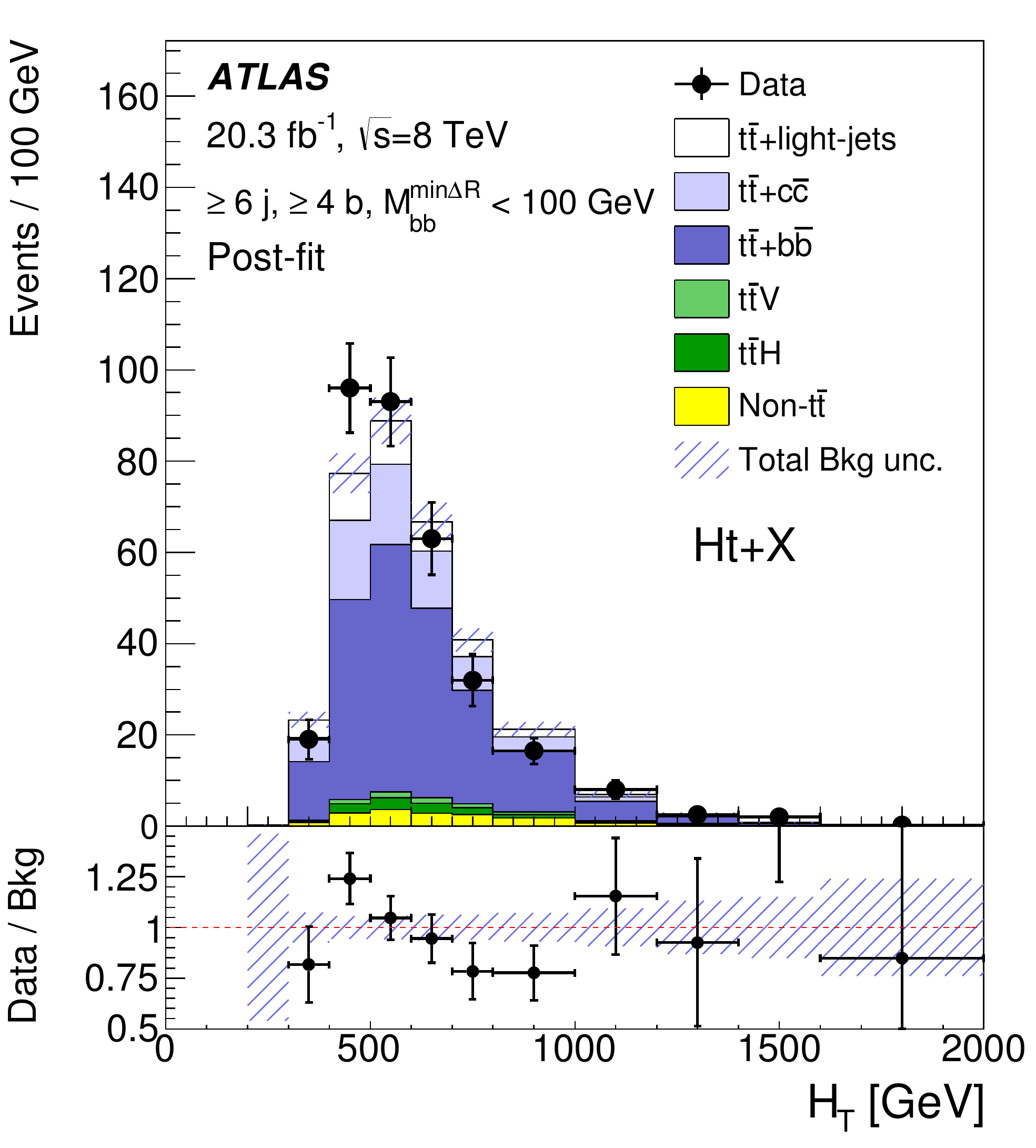}} 
\subfloat[]{\includegraphics[width=0.45\textwidth]{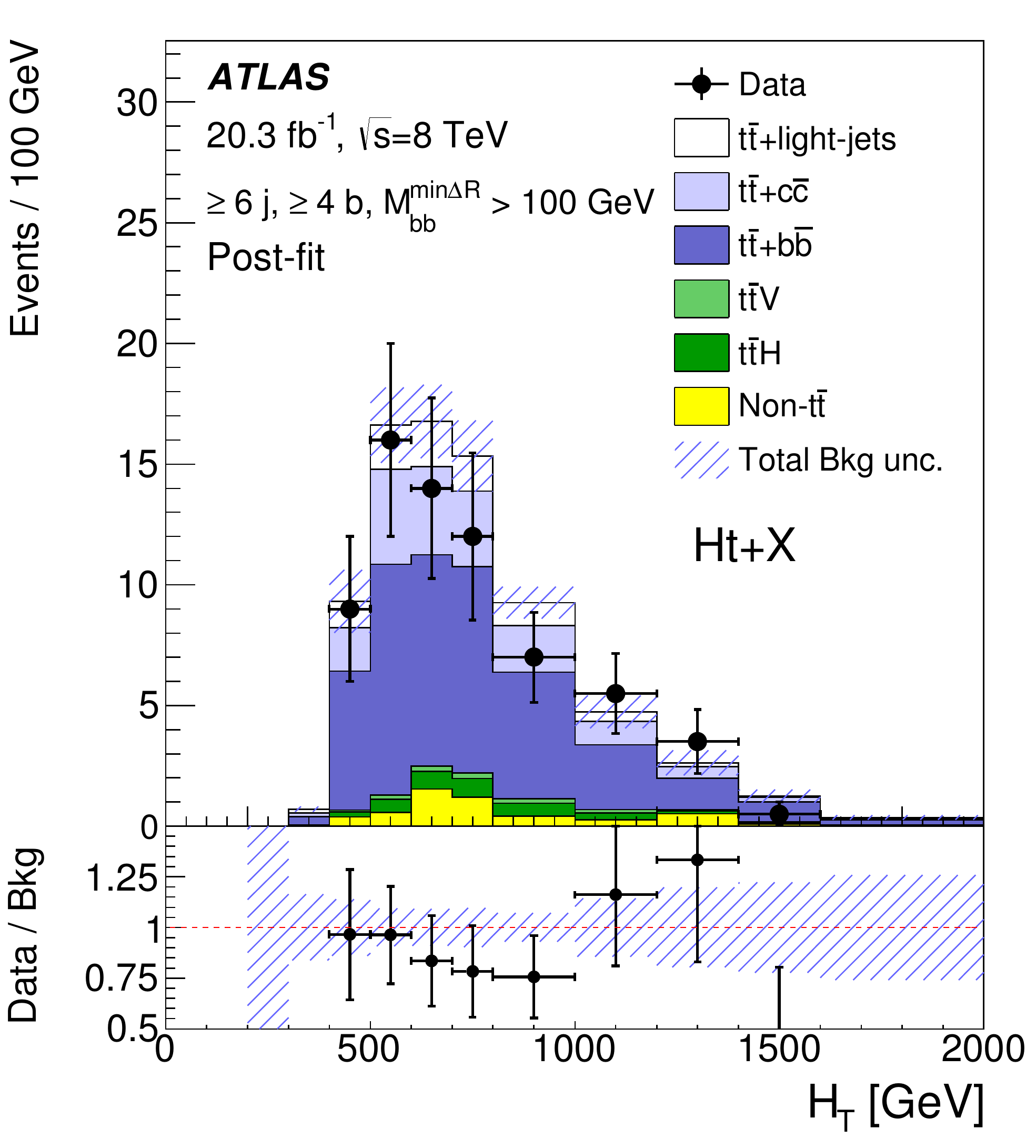}} 
\caption{$T\bar{T} \to Ht$+X search: comparison between data and prediction for the distribution of
the scalar sum ($\HT$) of the transverse momenta of the lepton, the selected jets and the missing transverse 
momentum in each of the analysed channels after final selection:
(a) ($\geq$6 j, 3 b, low $M_{bb}^{{\rm min}\Delta R}$), (b) ($\geq$6 j, 3 b, high $M_{bb}^{{\rm min}\Delta R}$), 
(c) ($\geq$6 j, $\geq$4 b, low $M_{bb}^{{\rm min}\Delta R}$), and (d) ($\geq$6 j, $\geq$4 b, high $M_{bb}^{{\rm min}\Delta R}$). 
The background prediction is shown after the fit to data under the background-only hypothesis.
The small contributions from $W/Z$+jets,  single top, diboson and multijet backgrounds are 
combined into a single background source referred to as ``Non-$\ttbar$''.
The last bin in all figures contains the overflow. The bottom panel displays the ratio of
data to the total background prediction. The hashed area represents the total uncertainty on the background.}
\label{fig:postfit_HtX_unblinded_2} 
\end{center}
\end{figure*}

\begin{figure*}[htbp]
\begin{center}
\subfloat[]{\includegraphics[width=0.45\textwidth]{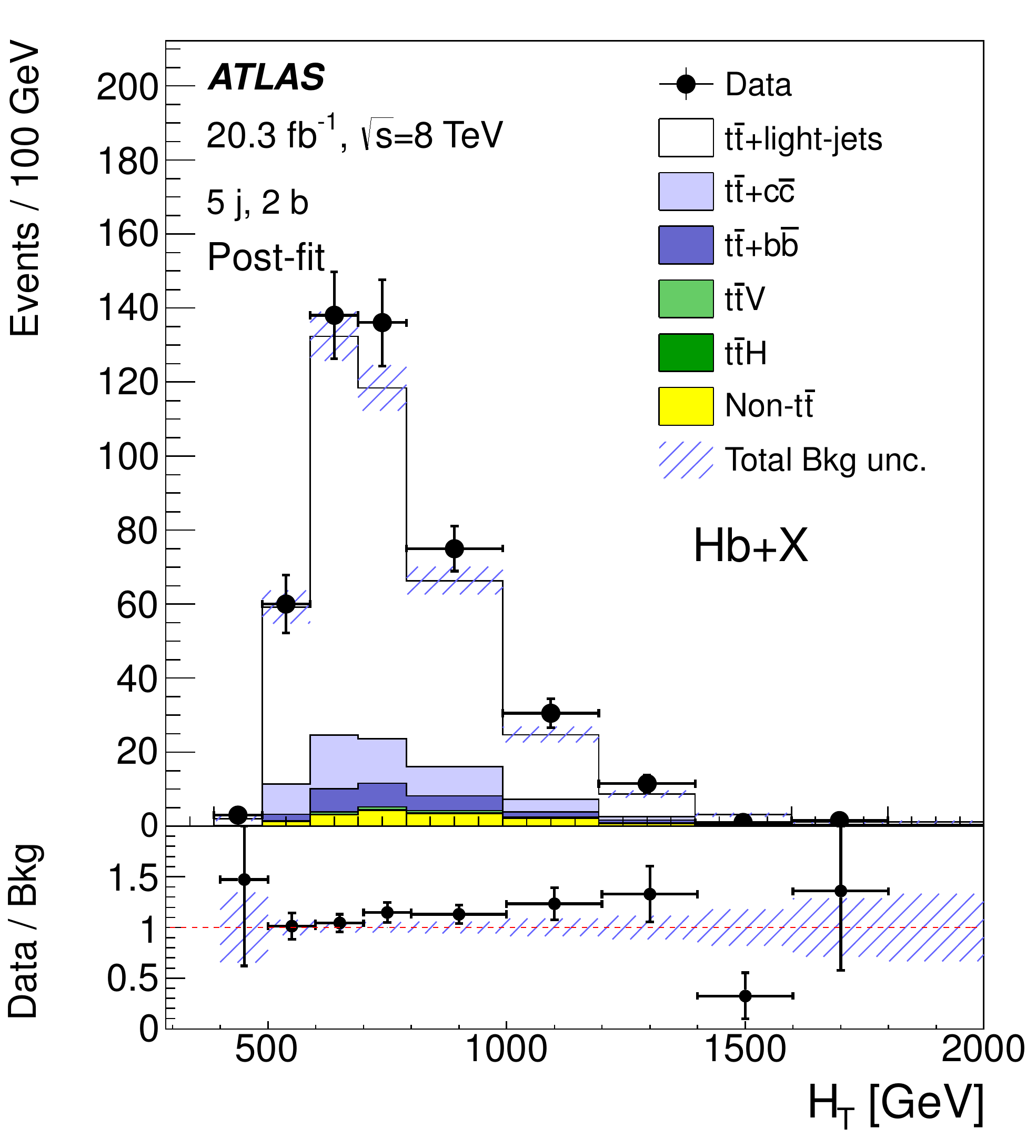}}
\subfloat[]{\includegraphics[width=0.45\textwidth]{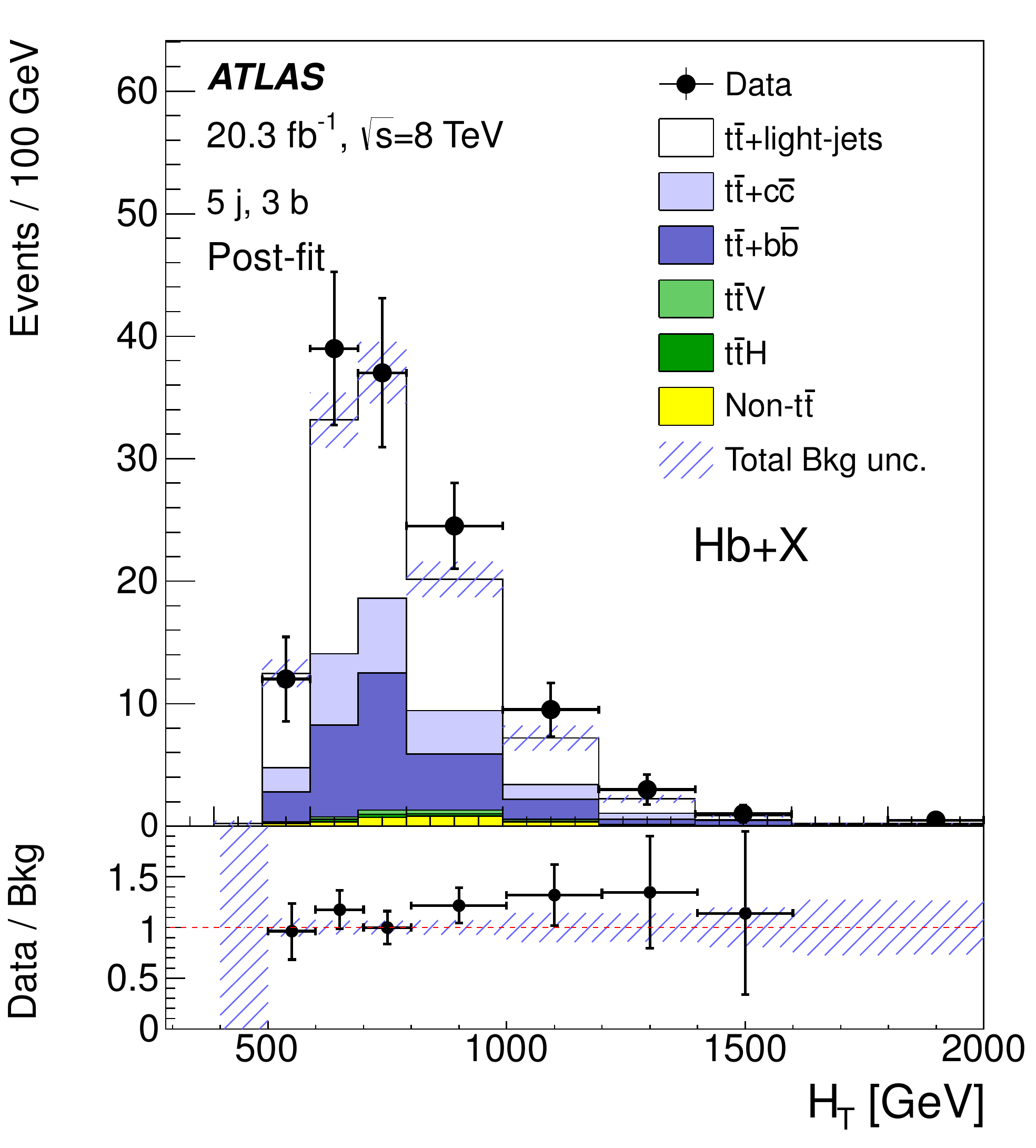}} \\
\subfloat[]{\includegraphics[width=0.45\textwidth]{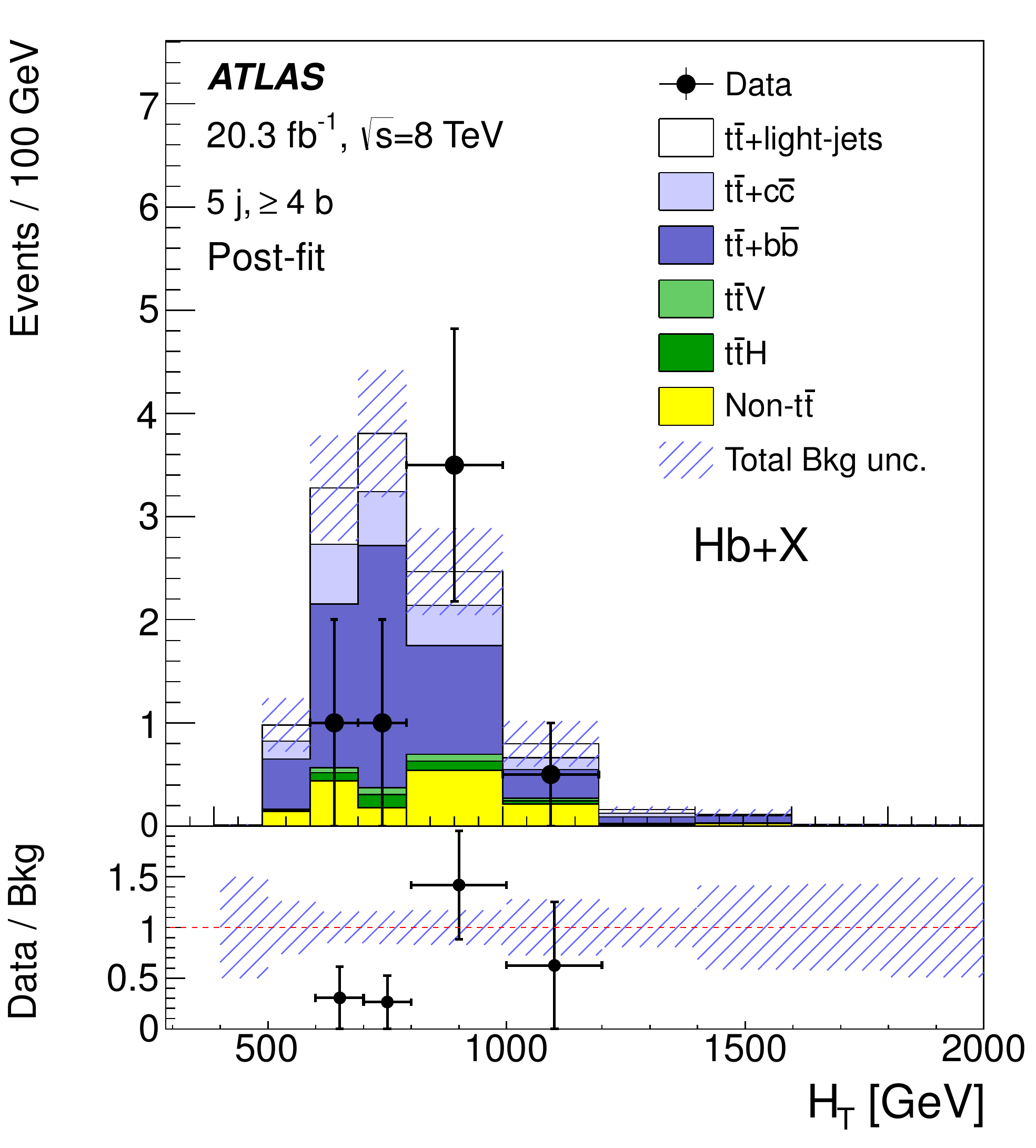}} 
\subfloat[]{\includegraphics[width=0.45\textwidth]{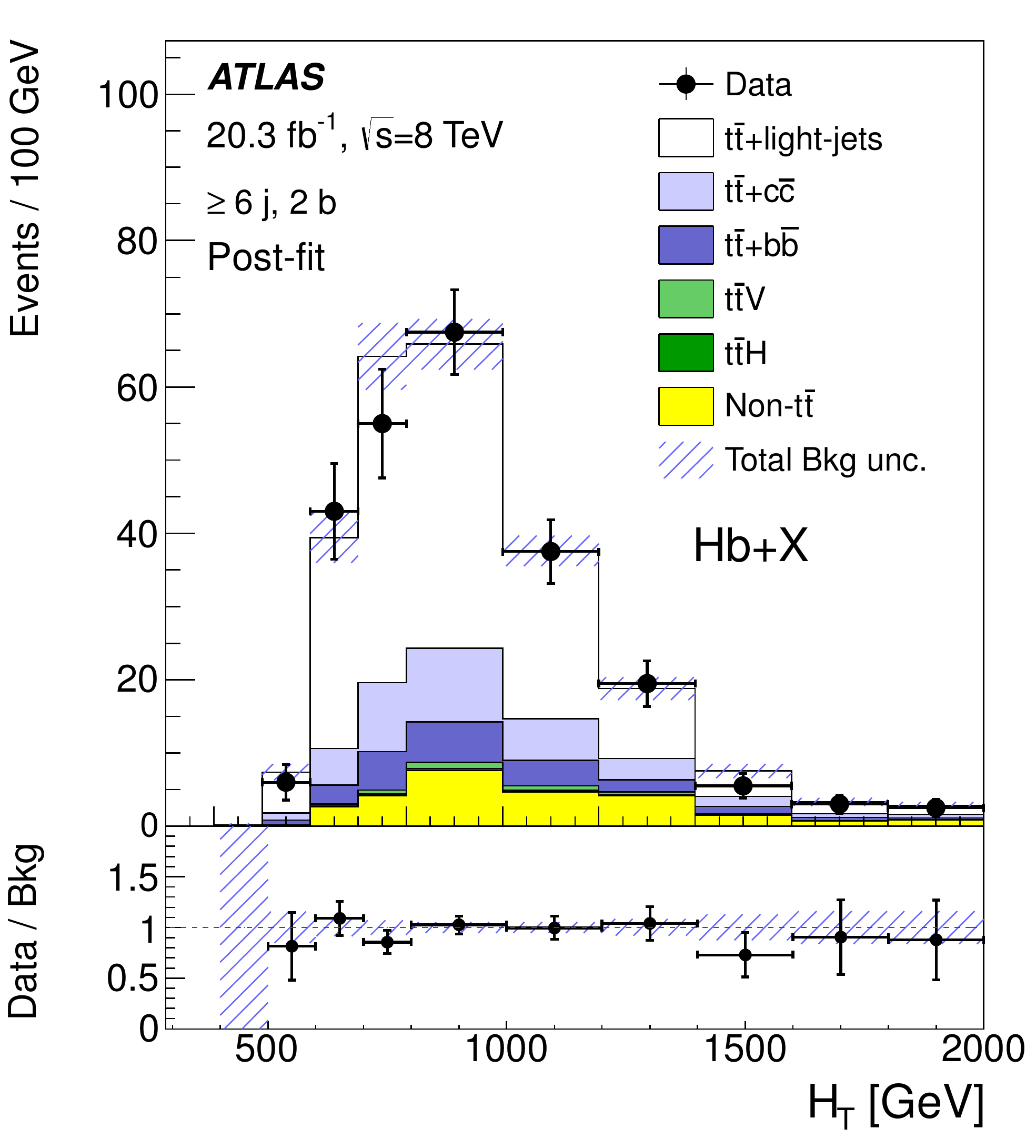}} 
\caption{$B\bar{B} \to Hb$+X search: comparison between data and prediction for the distribution of
the scalar sum ($\HT$) of the transverse momenta of the lepton, the selected jets and the missing transverse 
momentum in each of the analysed channels after final selection:
(a) (5 j, 2 b), (b) (5 j, 3 b), (c) (5 j, $\geq$4 b), and (d) ($\geq$6 j, 2 b). 
The background prediction is shown after the fit to data under the background-only hypothesis.
The small contributions from $W/Z$+jets,  single top, diboson and multijet backgrounds are 
combined into a single background source referred to as ``Non-$\ttbar$''.
The last bin in all figures contains the overflow. The bottom panel displays the ratio of
data to the total background prediction. The hashed area represents the total uncertainty on the background.}
\label{fig:postfit_HbX_unblinded_1} 
\end{center}
\end{figure*}

\begin{figure*}[htbp]
\begin{center}
\subfloat[]{\includegraphics[width=0.45\textwidth]{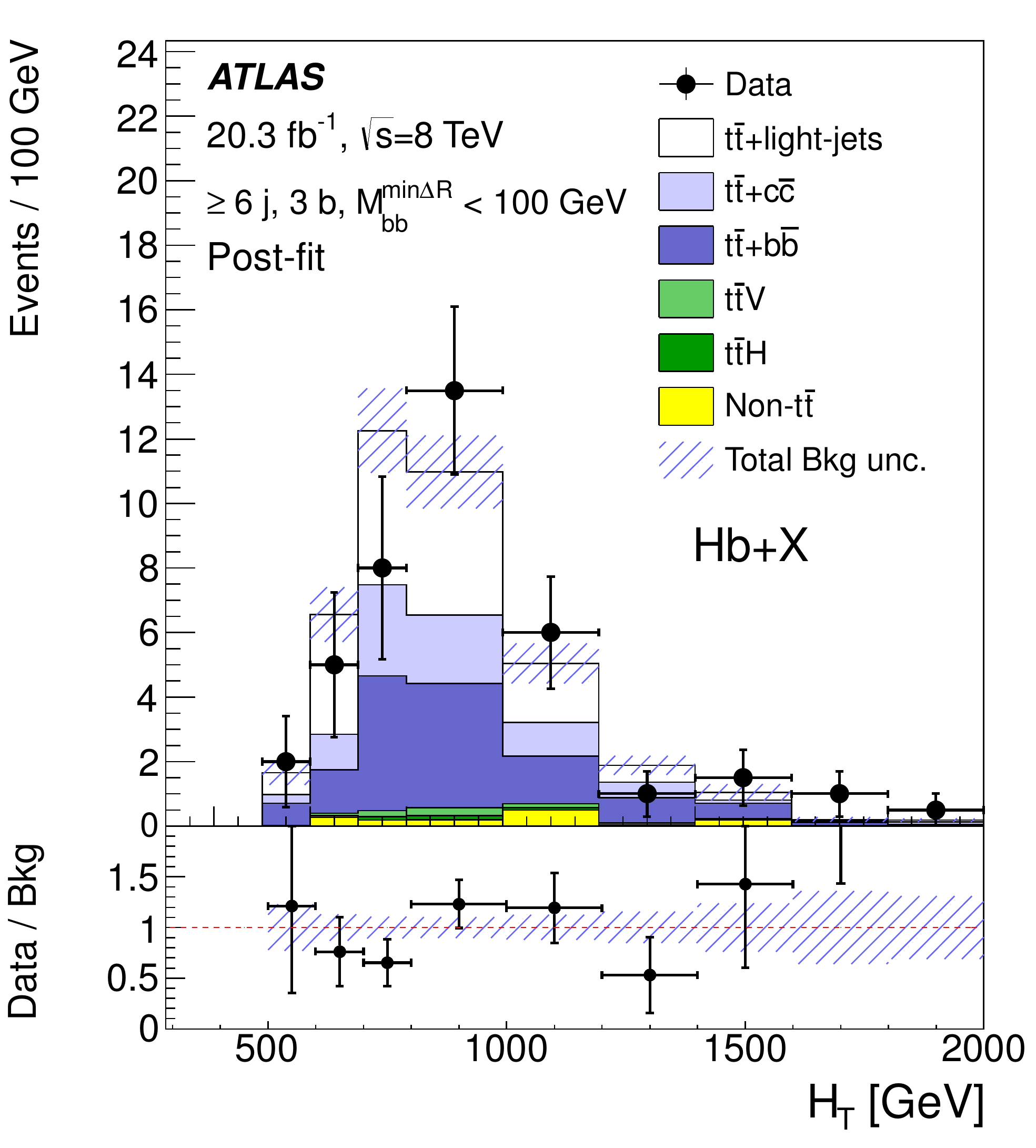}} 
\subfloat[]{\includegraphics[width=0.45\textwidth]{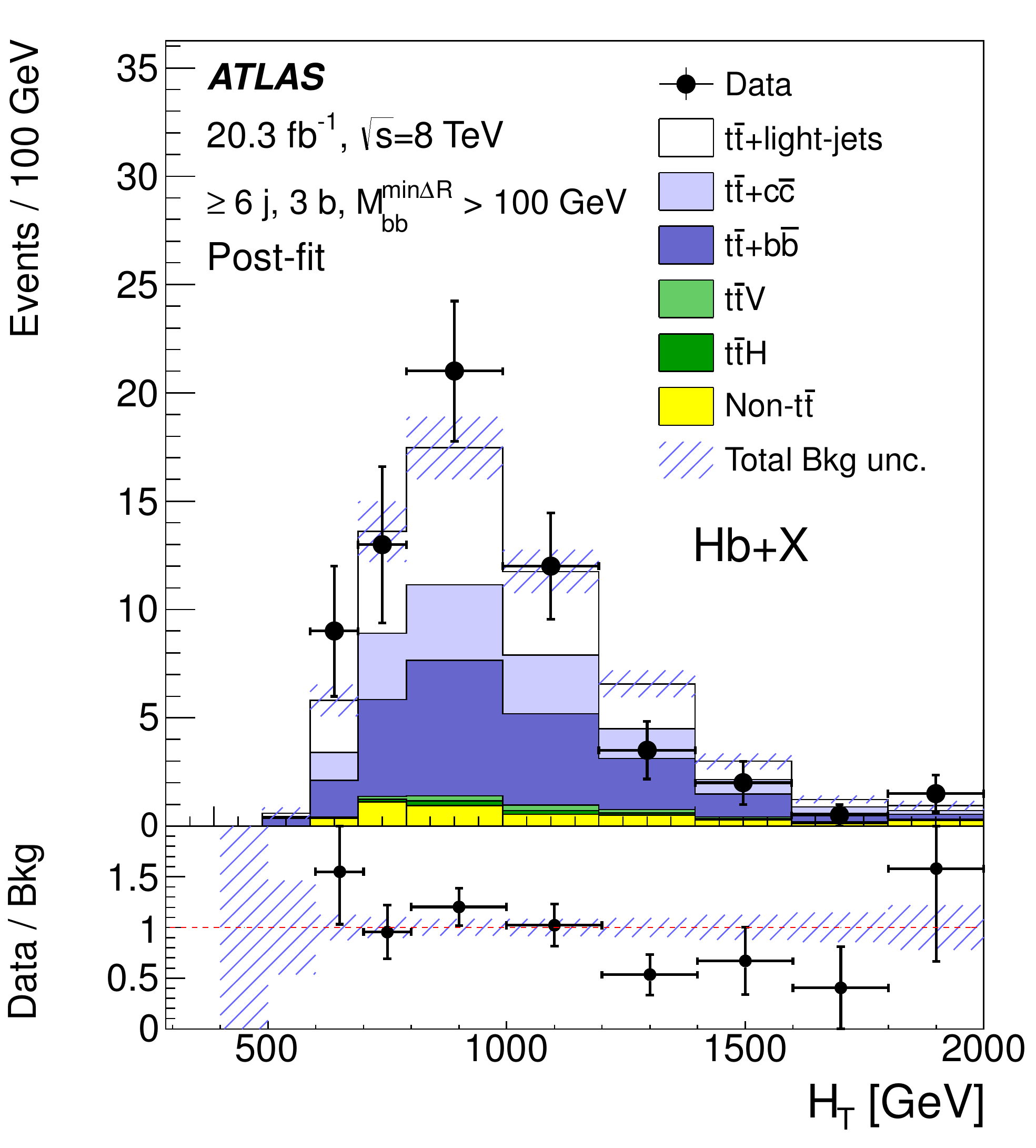}} \\
\subfloat[]{\includegraphics[width=0.45\textwidth]{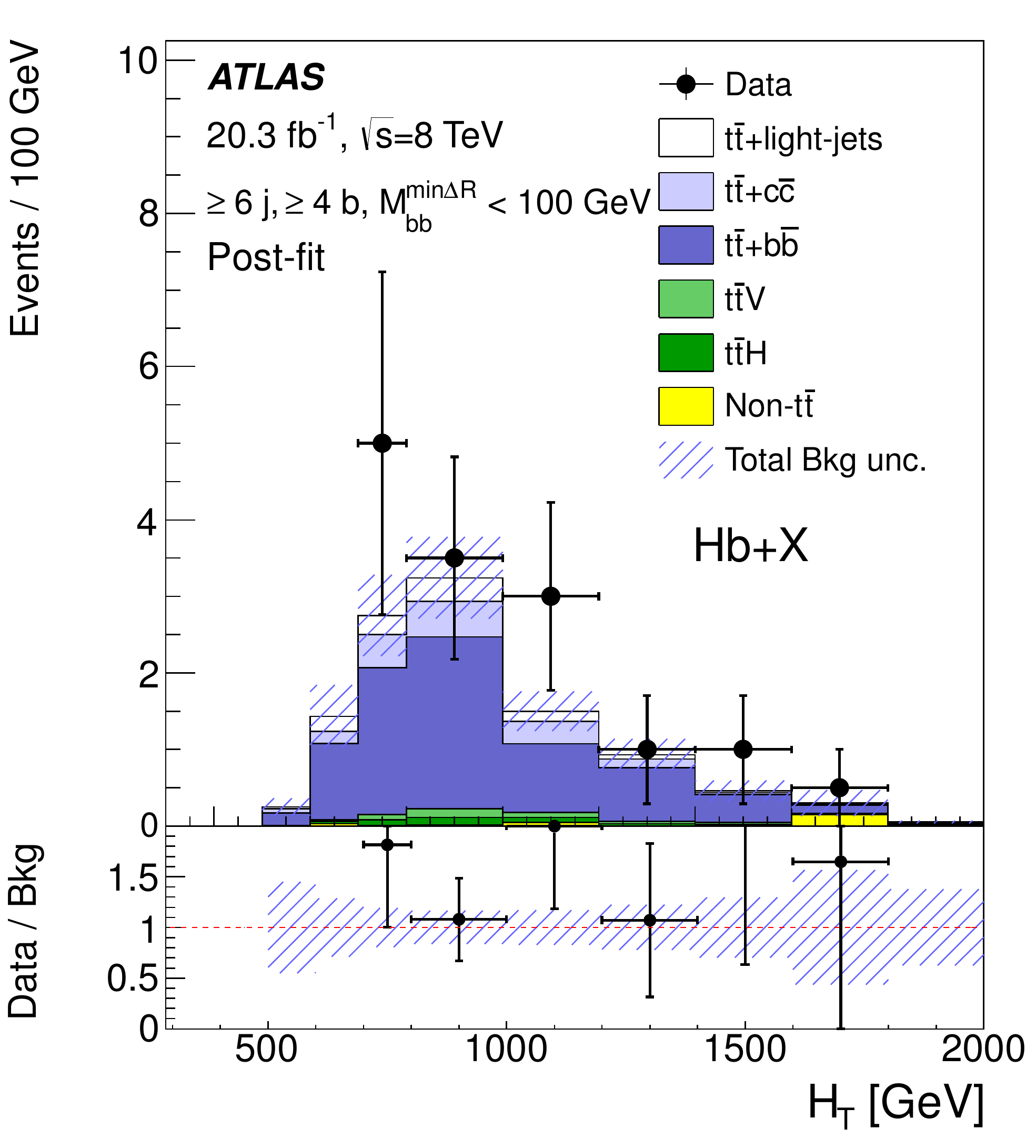}} 
\subfloat[]{\includegraphics[width=0.45\textwidth]{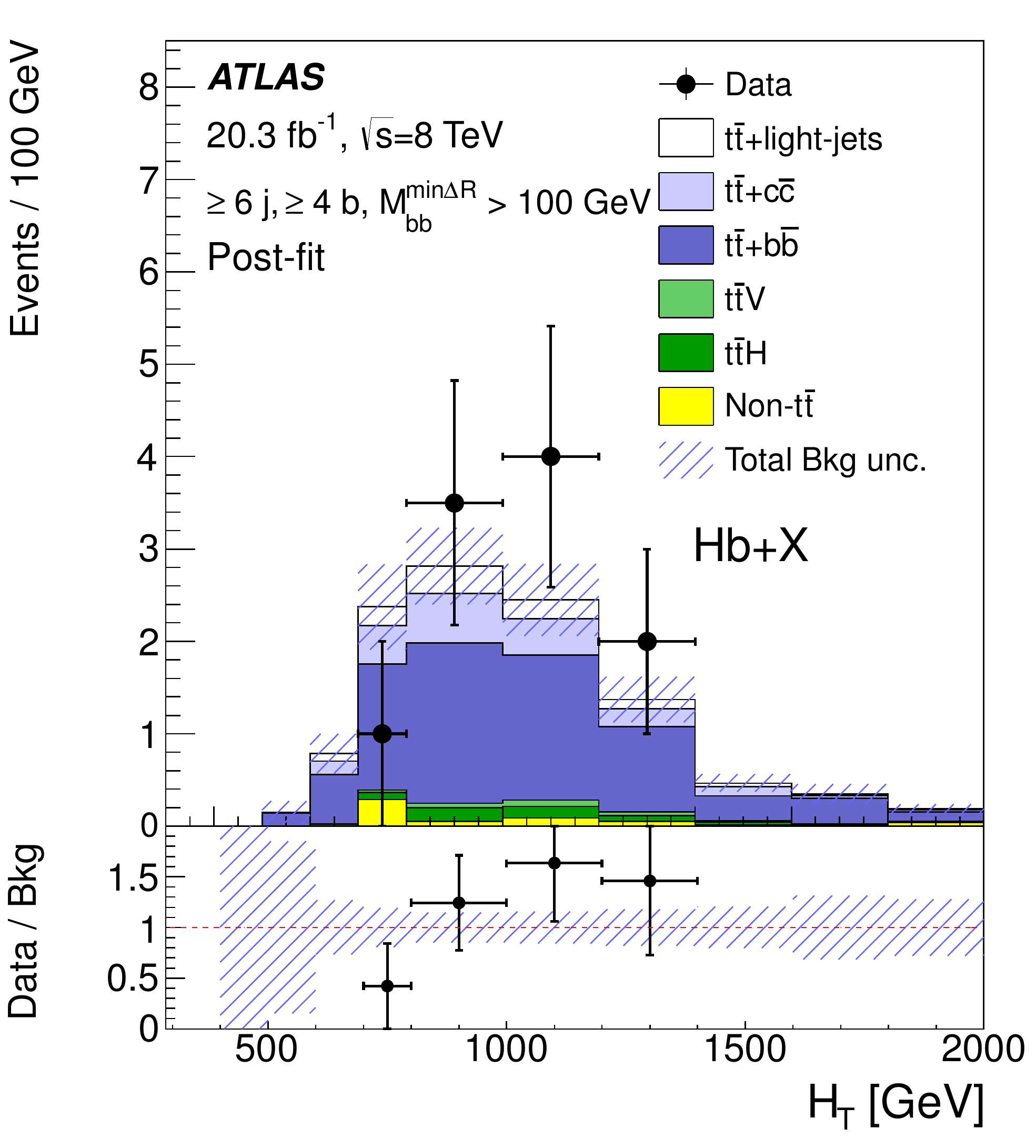}} 
\caption{$B\bar{B} \to Hb$+X search: comparison between data and prediction for the distribution of
the scalar sum ($\HT$) of the transverse momenta of the lepton, the selected jets and the missing transverse 
momentum in each of the analysed channels after final selection:
(a) ($\geq$6 j, 3 b, low $M_{bb}^{{\rm min}\Delta R}$), (b) ($\geq$6 j, 3 b, high $M_{bb}^{{\rm min}\Delta R}$), 
(c) ($\geq$6 j, $\geq$4 b, low $M_{bb}^{{\rm min}\Delta R}$), and (d) ($\geq$6 j, $\geq$4 b, high $M_{bb}^{{\rm min}\Delta R}$). 
The background prediction is shown after the fit to data under the background-only hypothesis.
The small contributions from $W/Z$+jets,  single top, diboson and multijet backgrounds are 
combined into a single background source referred to as ``Non-$\ttbar$''.
The last bin in all figures contains the overflow. The bottom panel displays the ratio of
data to the total background prediction. The hashed area represents the total uncertainty on the background.}
\label{fig:postfit_HbX_unblinded_2} 
\end{center}
\end{figure*}

\subsection{Limits on $T\bar{T}$ production}

The compatibility of the data with the background prediction is assessed by computing the $p_0$-value for each 
signal scenario considered, defined by the assumed values for the heavy quark mass (see Sect.~\ref{sec:vlq_model}) 
and the three decay branching ratios, which are varied in steps of 0.05 requiring that  they add up to unity.
In the case of the $T\bar{T} \to Wb$+X search alone,  the smallest $p_0$-value found, 0.023, is obtained for $m_{T}=600\gev$, 
$\BR(T \to Wb)=0.30$ and $\BR(T \to Ht)=0.65$ [$\BR(T \to Zt)=1-\BR(T \to Wb)-\BR(T \to Ht)=0.05$],
and corresponds to a local significance of 2.0 standard deviations above the background-only prediction.
In the case of the $T\bar{T} \to Ht$+X search, 
the smallest $p_0$-value found, $0.44$, is obtained for $m_{T}=600\gev$, 
$\BR(T \to Wb)=0.0$, $\BR(T \to Ht)=0.0$, and $\BR(T \to Zt)=1.0$,
and corresponds to a local significance of $0.2$ standard deviations above the background-only prediction.
Thus, no significant excess above the background expectation is found in either of the two searches.

Since the two searches have complementary sensitivity to different decay modes of a vector-like $T$ quark, they are combined 
in a single likelihood function taking into account the correlation of systematic uncertainties.
Upper limits at 95\% CL on the $T\bar{T}$ production cross section are set in several benchmark scenarios as a function of the
$T$ quark mass $m_{T}$ and are compared to the theoretical prediction from {\sc Top++}, as shown in figure~\ref{fig:limits1D_TT}. 
The resulting lower limits on $m_{T}$ correspond to the central value of the theoretical cross section.
The scenarios considered involve different assumptions on the decay branching ratios: $\BR(T \to Wb)=1$, singlet and doublet.
Only the $T\bar{T} \to Wb$+X search is sensitive to a $T$ quark with $\BR(T \to Wb)=1$, 
yielding an observed (expected) 95\% CL lower limit of $m_{T}>770\,(795)\gev$. 
This represents the most stringent limit to date, and is also applicable to a $Y$ vector-like quark with electric charge of
$-4/3$ and decaying into a $W^-$ boson and a $b$ quark.
Both searches are sensitive to a vector-like singlet $T$ quark. The $T\bar{T} \to Wb$+X and $T\bar{T} \to Ht$+X searches yield 
observed (expected) 95\% CL  limits of $m_{T}>660\,(670)\gev$ and $m_{T}>765\,(720)\gev$ respectively.
The combination of both analyses results in a slight improvement over the $T\bar{T} \to Ht$+X search alone, yielding
$m_{T}>800\,(755)\gev$. 
Finally, only the $T\bar{T} \to Ht$+X search is sensitive to a vector-like doublet $T$ quark, 
yielding an observed (expected) 95\% CL lower limit of $m_{T}>855\,(820)\gev$.

\begin{figure*}[tbp]
\centering
\subfloat[]{\includegraphics[width=0.45\textwidth]{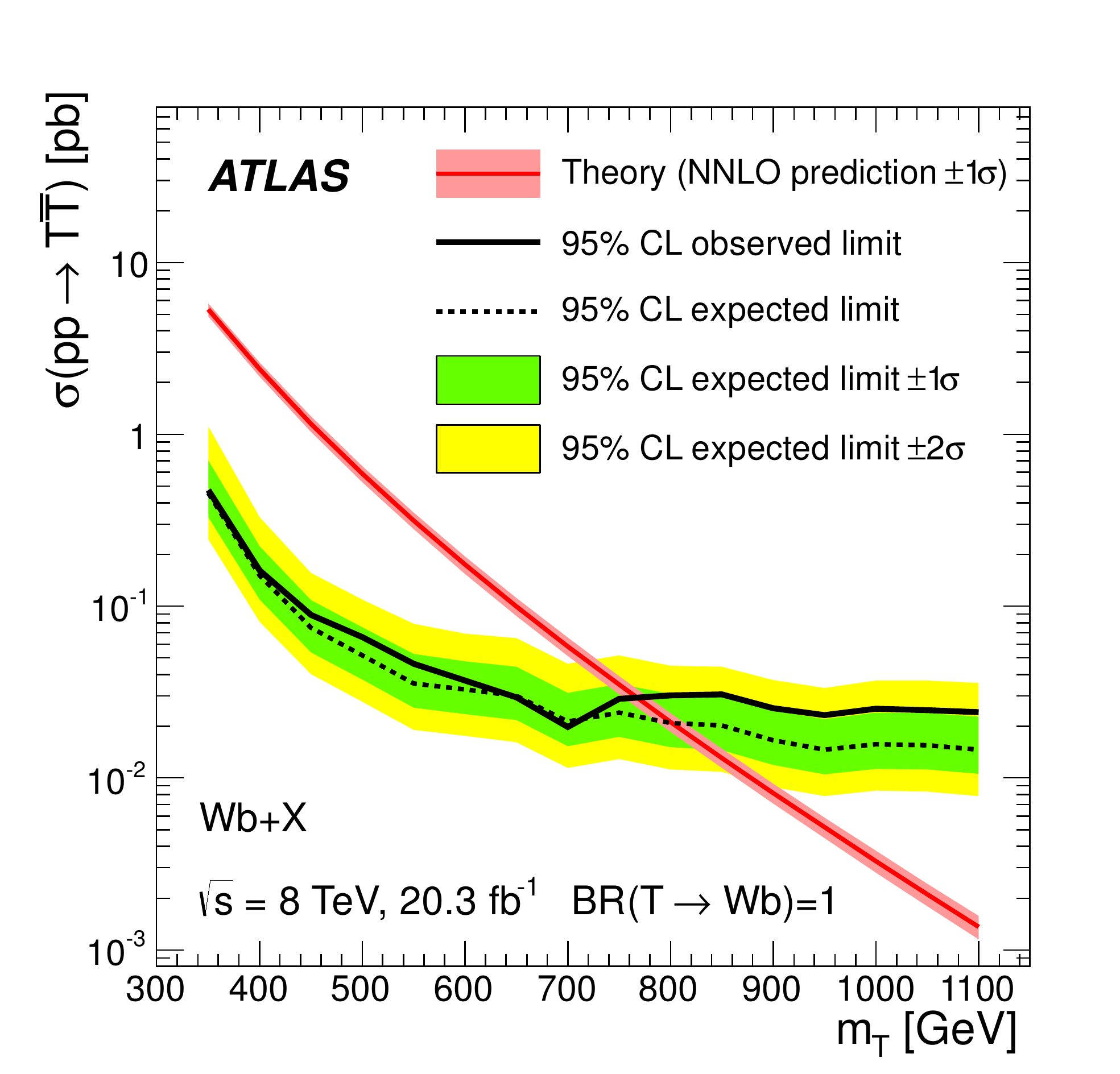}}
\subfloat[]{\includegraphics[width=0.45\textwidth]{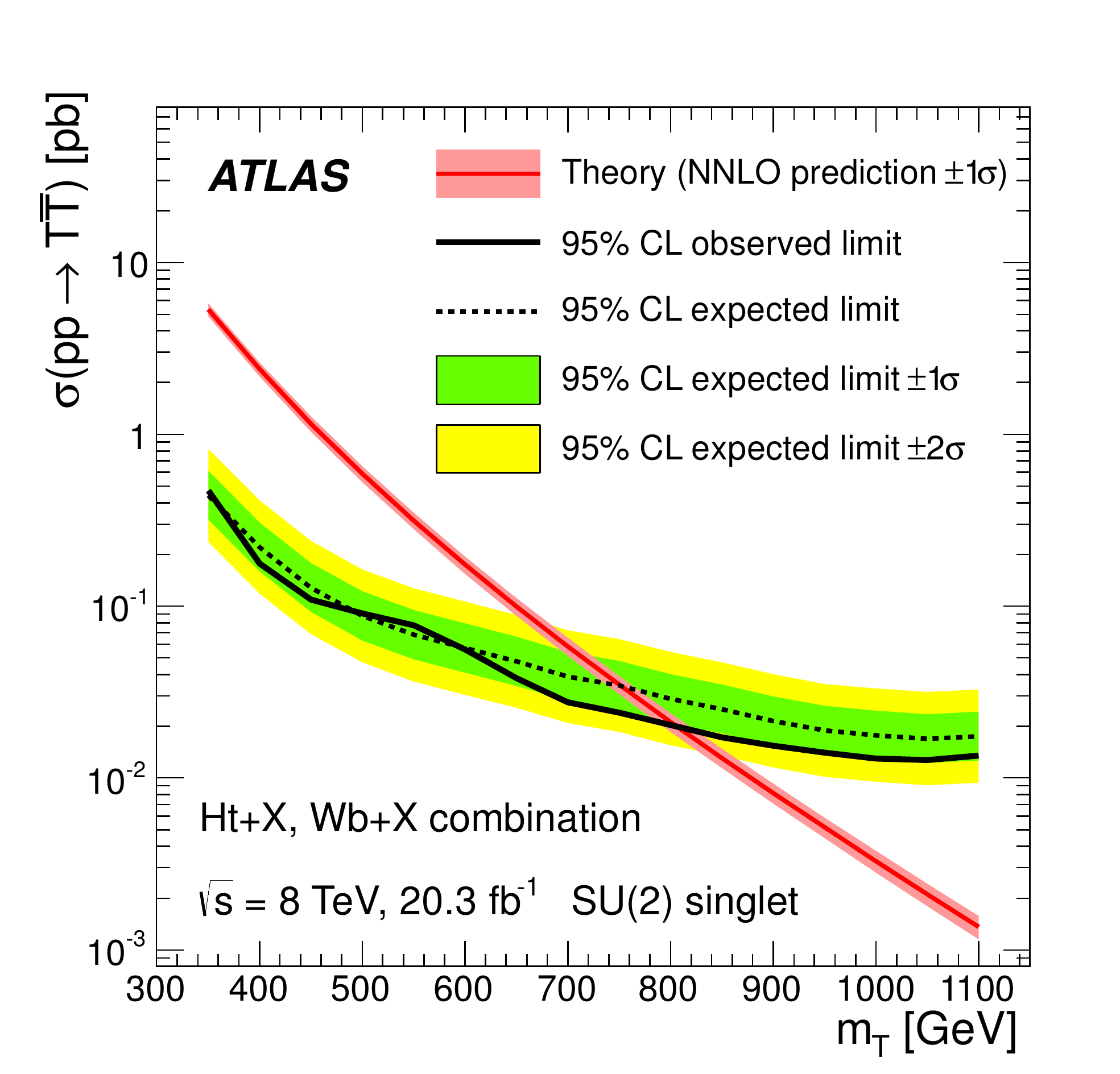}} \\
\subfloat[]{\includegraphics[width=0.45\textwidth]{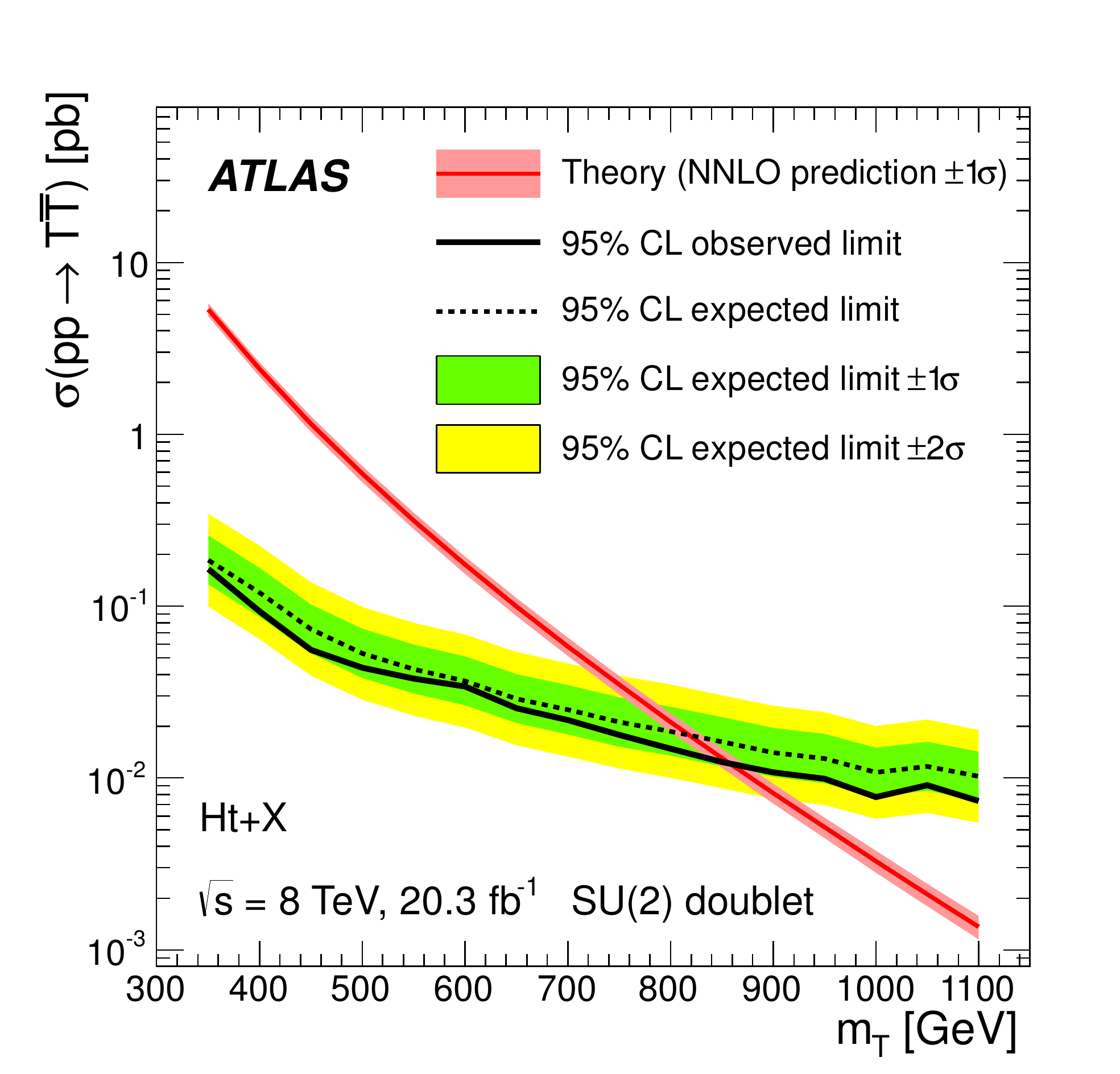}}
\caption{Observed (solid line) and expected (dashed line) 95\% CL upper limits on the $T\bar{T}$ cross section as a function of the $T$ quark mass 
(a) under the assumption $\BR(T \to Wb)=1$, (b) for a $T$ quark singlet, and (c) for a $T$ quark doublet.
The surrounding shaded bands correspond to $\pm1$ and $\pm2$ standard deviations around the expected limit. 
The thin red line and band show the theoretical prediction and its $\pm1$ standard deviation uncertainty.}
\label{fig:limits1D_TT}
\end{figure*}

The same searches are used to derive exclusion limits on vector-like $T$ quark production for different 
values of $m_{T}$ and as a function of $\BR(T\to W b)$ and $\BR(T\to Ht)$.
To probe this branching ratio plane, the signal samples are reweighted by the ratio
of the desired branching ratio to the original branching ratio in {\sc Protos}, and the complete analysis is repeated.
The resulting 95\% CL exclusion limits are shown in figure~\ref{fig:limits2D_TT} for the combination of the $T\bar{T} \to Wb$+X and $T\bar{T} \to Ht$+X searches,
for different values of $m_{T}$.  Figure~\ref{fig:limits2D_TT_temp} presents the corresponding 
observed and expected $T$ quark mass limits in the plane of $\BR(T \to Ht)$ versus $\BR(T \to Wb)$,
obtained by linear interpolation of the estimated CL$_{\rm{s}}$ versus $m_{T}$.

The combined results set observed lower limits on the $T$ quark mass ranging between $715\gev$ and $950\gev$  
for all possible values of the branching ratios into the three decay modes. This implies that any branching ratio scenario is
excluded at 95\% CL for a $T$ quark with mass below $715\gev$. The corresponding range
of expected lower limits is between $675\gev$ and $885\gev$.
The exclusion limits for the individual searches can be
found in appendix~\ref{sec:limits_appendix}. These figures illustrate the complementarity of these searches
and how their combination improves over simply taking the most sensitive search for each assumed branching ratio scenario, 
leading to large regions in the branching ratio plane being excluded.

In addition to the combined $T\bar{T} \to Wb$+X and $T\bar{T} \to Ht$+X result discussed in this paper, the ATLAS Collaboration has performed
searches for $T\bar{T}$ production in several multilepton final states: same-sign dileptons and trileptons~\cite{Aad:2015gdg} 
and opposite-sign dileptons and trileptons with a $Z$ boson candidate~\cite{Aad:2014efa} (referred to as the $Zb/t$+X search).
These searches have overlapping selections and have not been combined.  
Figure~\ref{fig:limits2D_T_Summary_temp} summarises the most restrictive observed and expected $T$ quark mass limits 
in the plane of $\BR(T \to Ht)$ versus $\BR(T \to Wb)$, set by any of these searches.
The observed lower limits on the $T$ quark mass range between $730\gev$ and $950\gev$  
for all possible values of the branching ratios into the three decay modes, representing an improvement
over previous results~\cite{Chatrchyan:2013uxa}.  The corresponding range of expected lower limits is between 
$715\gev$ and $885\gev$.

\begin{figure*}[tbp]
\centering
\includegraphics[width=0.9\textwidth]{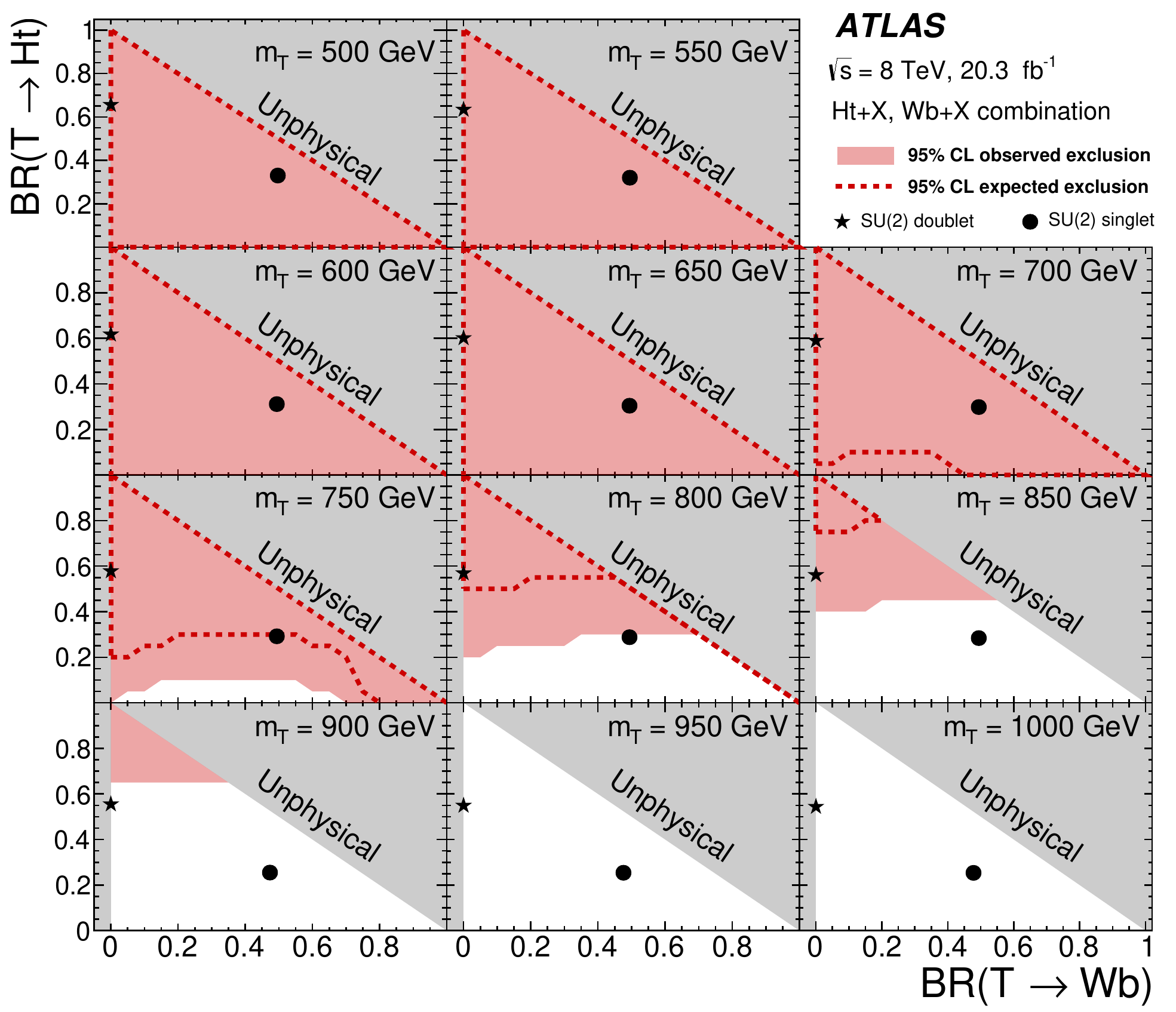}
\caption{
Observed (red filled area) and expected (red dashed line) 95\% CL exclusion in the plane of
$\BR(T \to Wb)$ versus $\BR(T \to Ht)$ from the combination of the $T\bar{T} \to Wb$+X and $T\bar{T} \to Ht$+X searches, 
for different values of the vector-like $T$ quark mass.
The grey (dark shaded) area corresponds to the unphysical region where the sum of branching ratios exceeds unity. 
The default branching ratio values from the {\sc Protos} event generator for the weak-isospin singlet and doublet cases 
are shown as plain circle and star symbols respectively. 
\label{fig:limits2D_TT}}
\end{figure*}

\begin{figure*}[tbp]
\centering
\subfloat[]{\includegraphics[width=0.48\textwidth]{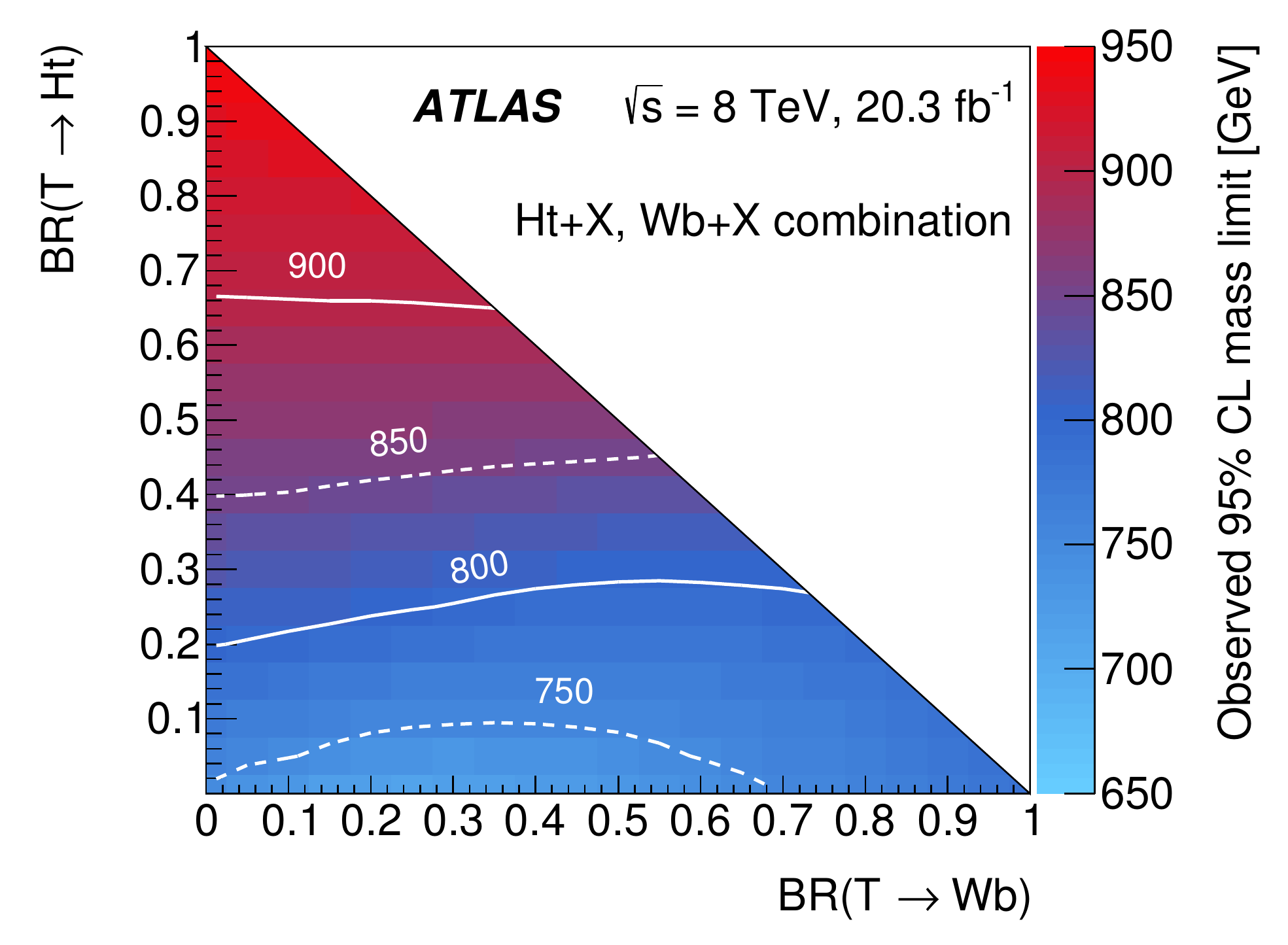}}
\subfloat[]{\includegraphics[width=0.48\textwidth]{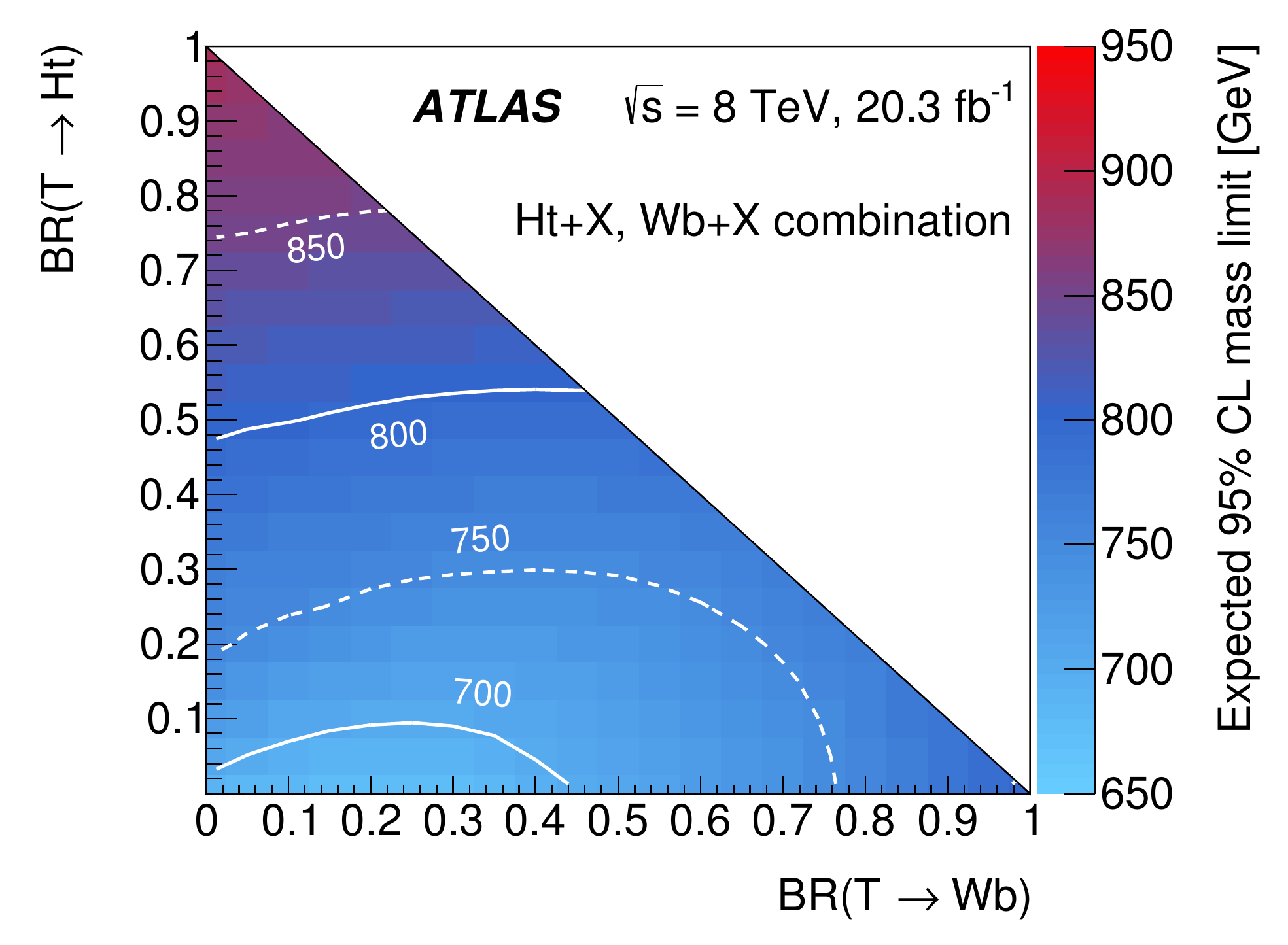}}
\caption{(a) Observed and (b) expected limit (95\% CL) on the mass of the $T$ quark in the plane 
of $\BR(T \to Ht)$ versus $\BR(T \to Wb)$ for the combination of the $T\bar{T} \to Wb$+X and $T\bar{T} \to Ht$+X searches.
Contour lines are provided to guide the eye.}
\label{fig:limits2D_TT_temp}
\end{figure*}

\begin{figure*}[tbp]
\centering
\subfloat[]{\includegraphics[width=0.48\textwidth]{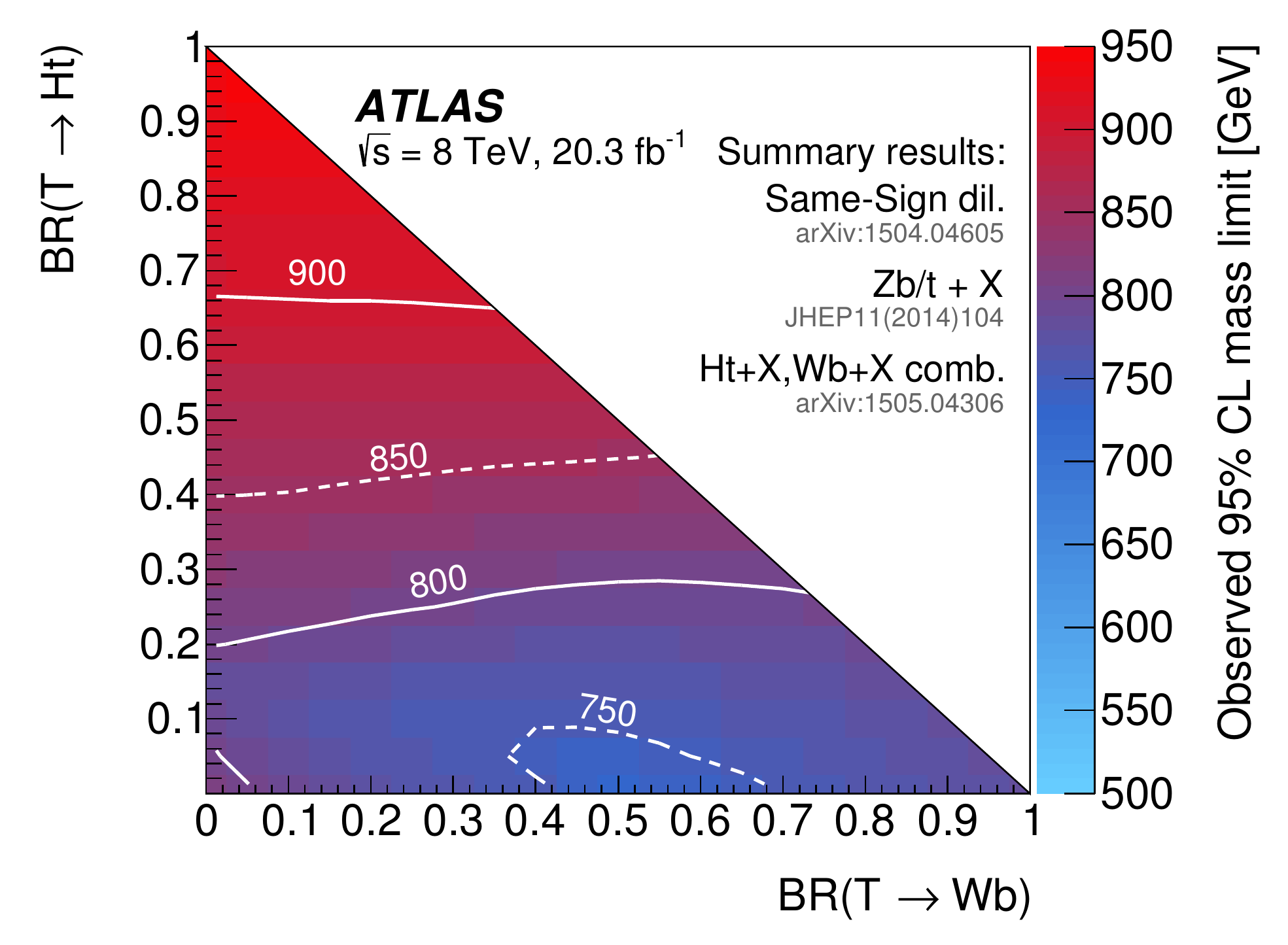}}
\subfloat[]{\includegraphics[width=0.48\textwidth]{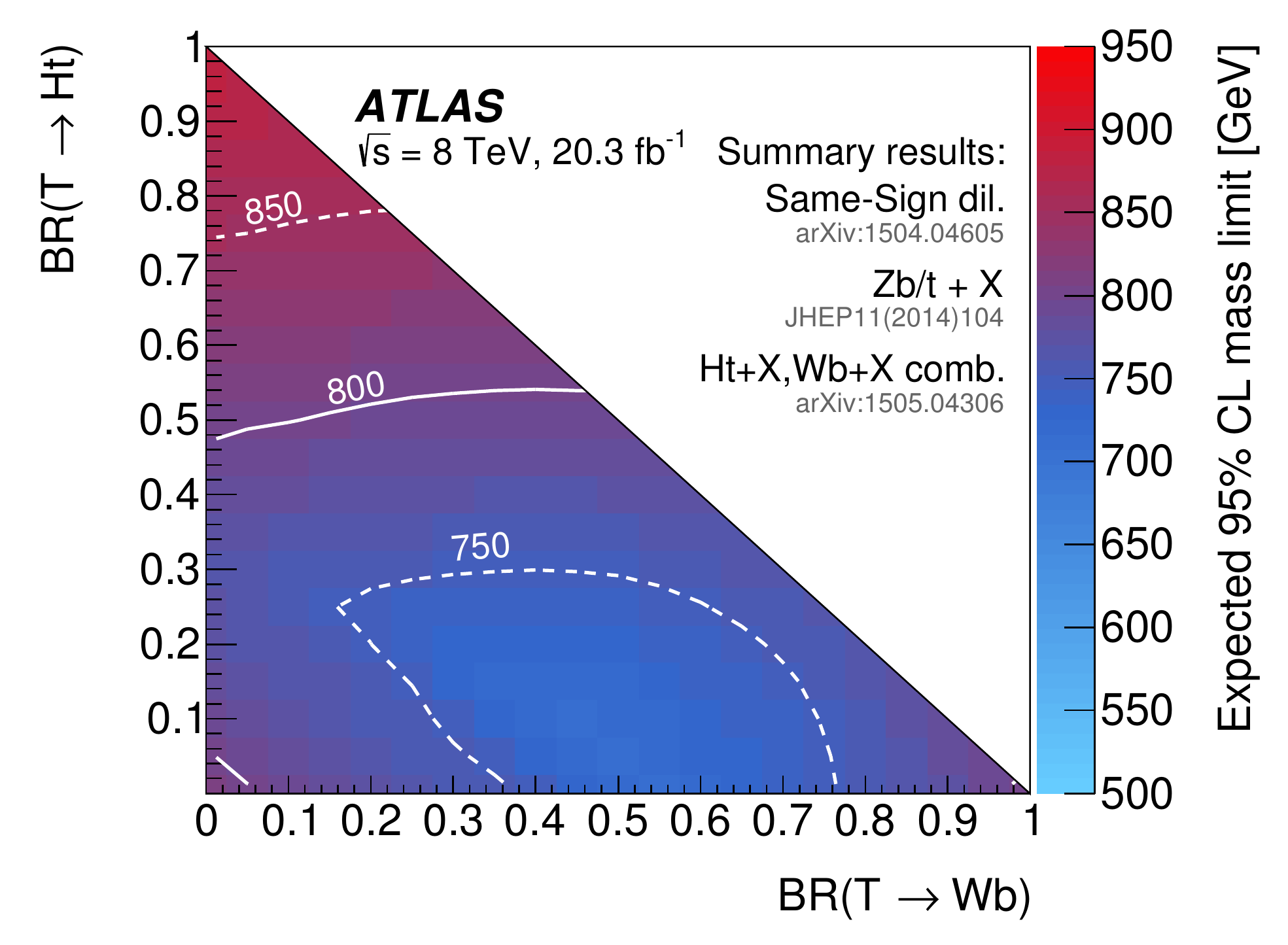}}
\caption{Summary of the most restrictive (a) observed and (b) expected limit (95\% CL) on the mass of the $T$ quark 
in the plane of $\BR(T \to Ht)$ versus $\BR(T \to Wb)$  from all ATLAS searches for $T\bar{T}$ production (see text for details).
Contour lines are provided to guide the eye.}
\label{fig:limits2D_T_Summary_temp}
\end{figure*}

\subsection{Limits on $B\bar{B}$ production}

In the case of the $B\bar{B} \to Hb$+X search, the smallest $p_0$-value found, 0.023, is obtained for $m_{B}=450\gev$, 
$\BR(B \to Wt)=0.0$ and $\BR(B \to Hb)=0.3$ [$\BR(B \to Zb)=1-\BR(B \to Wt)-\BR(B \to Hb)=0.7$),  
and corresponds to a local significance of 2.0 standard deviations above the background-only prediction.

Upper limits at 95\% CL on the $B\bar{B}$ production cross section are set for two benchmark scenarios
as a function of the $B$ quark mass, as shown in figure~\ref{fig:limits1D_BB}. 
Assuming $\BR(B \to Hb) = 1$, the intervals $350 < m_{B} < 580\gev$ and $635 < m_{B} < 700\gev$ are excluded at 95\% CL. The expected
exclusion is $m_{B}>625\gev$ at 95\% CL. For branching ratios corresponding to a $B$ singlet, the observed (expected) 95\% CL
limit is $m_{B}>735 \,(635) \gev$.
Exclusion limits are set for values of $m_{B}$ and as a function of $\BR(B \to Wt)$ and $\BR(B \to Hb)$, shown
in figure~\ref{fig:limits2D_BB}. The search is particularly sensitive at large $\BR(B \to Hb)$, and also at large $\BR(B \to Wt)$.
Figure~\ref{fig:limits2D_BB_temp} presents the corresponding 
observed and expected $B$ quark mass limits in the plane of $\BR(B \to Hb)$ versus $\BR(B \to Wt)$.

Beyond the $B\bar{B} \to Hb$+X search presented in this paper, which focuses on the $B \to Hb$ decay, the
ATLAS Collaboration has performed several other searches for $B\bar{B}$ production that are complementary to each other.
A search in the lepton-plus-jets final state~\cite{Aad:2015mba}, referred to as $B\bar{B} \to Wt$+X,
and the search in same-sign dilepton and multilepton events~\cite{Aad:2015gdg},
probe primarily the $B \to Wt$ decay mode.
The $Zb/t$+X search~\cite{Aad:2014efa} is most sensitive to $B \to Zb$ production.
Figure~\ref{fig:limits2D_B_Summary_temp} summarises the most restrictive observed and expected $B$ quark mass limits 
in the plane of $\BR(B \to Hb)$ versus $\BR(B \to Wt)$, set by any of these searches.
The observed lower limits on the $B$ quark mass range between $575\gev$ and $813\gev$  
for all possible values of the branching ratios into the three decay modes. 
The corresponding range of expected lower limits is between $615\gev$ and $800\gev$.

\begin{figure*}[tbp]
\centering
\subfloat[]{\includegraphics[width=0.45\textwidth]{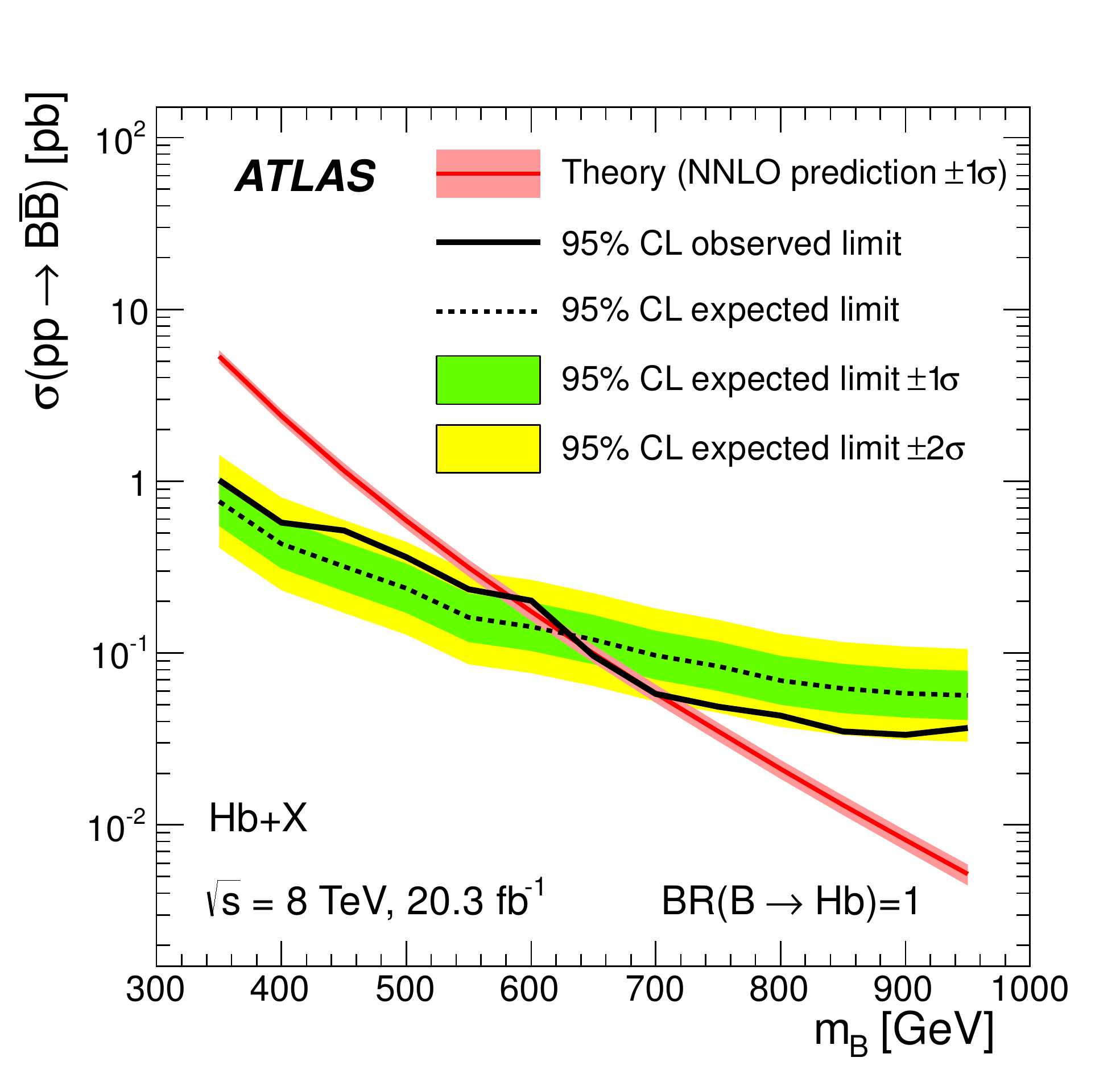}}
\subfloat[]{\includegraphics[width=0.45\textwidth]{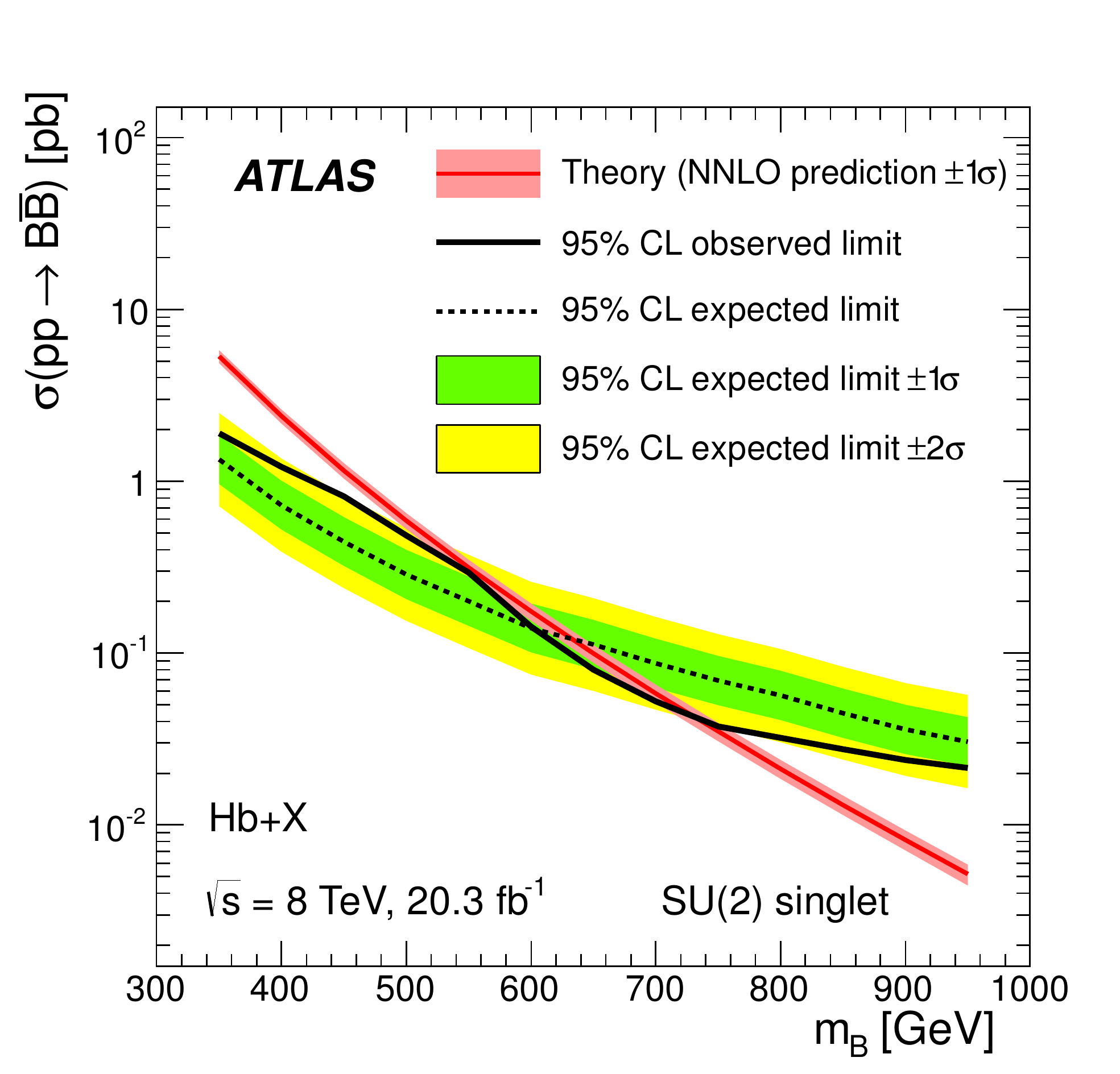}}
\caption{Observed (solid line) and expected (dashed line) 95\% CL upper limits on the $B\bar{B}$ cross section
as a function of the $B$ quark mass (a) under the assumption $\BR(B \to Hb)=1$ and (b) for a $B$ quark singlet. 
The surrounding shaded bands correspond to $\pm1$ and $\pm2$ standard deviations around the expected limit. 
The thin red line and band show the theoretical prediction and its $\pm1$ standard deviation uncertainty.}
\label{fig:limits1D_BB}
\end{figure*}

\begin{figure*}[tbp]
\centering
\includegraphics[width=0.9\textwidth]{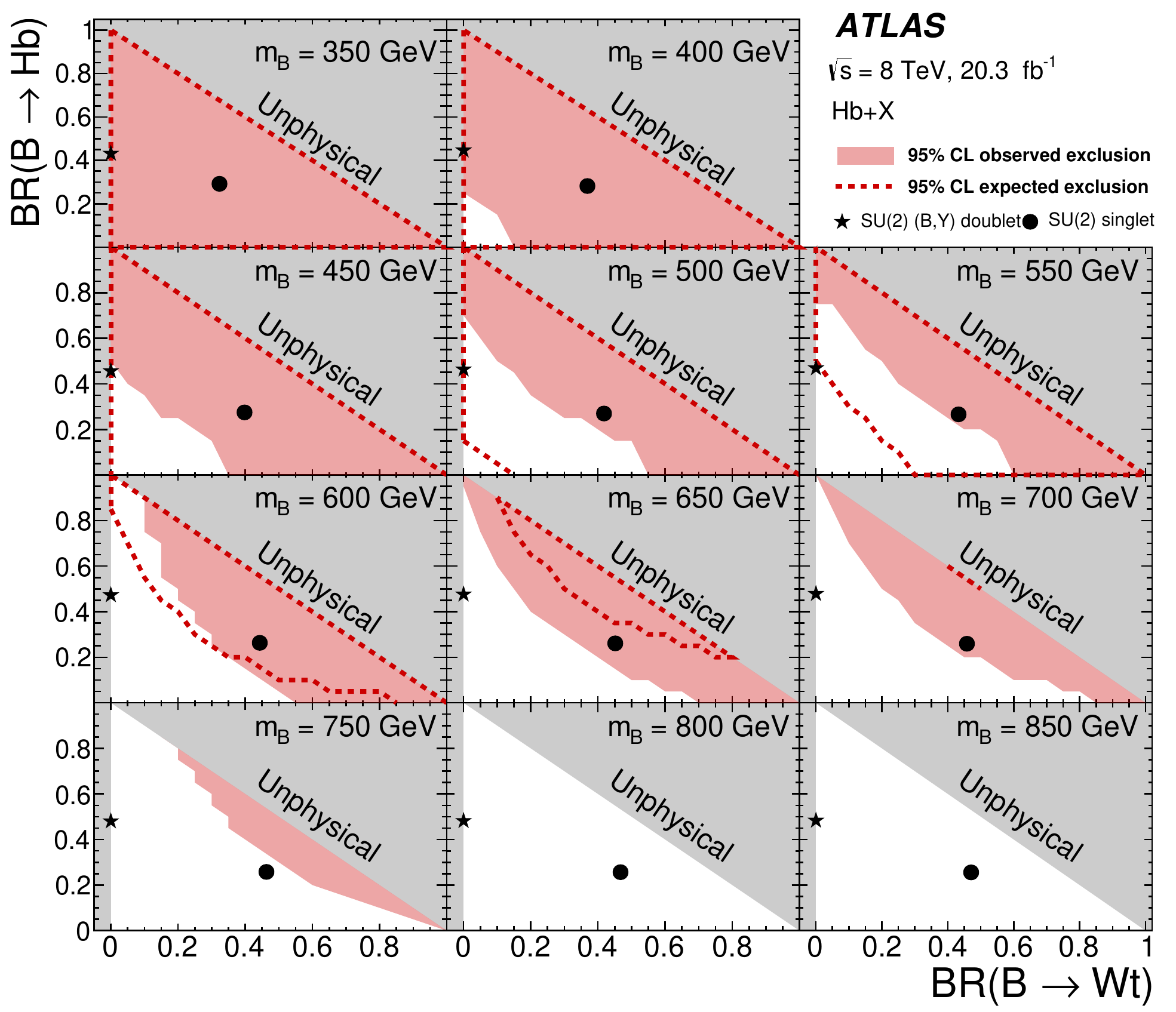}
\caption{
Observed (red filled area) and expected (red dashed line) 95\% CL exclusion in the plane of
$\BR(B \to Wt)$ versus $\BR(B \to Hb)$ from the $B\bar{B} \to Hb$+X search, 
for different values of the vector-like $B$ quark mass.
The grey (dark shaded) area corresponds to the unphysical region where the sum of branching ratios exceeds unity. 
The default branching ratio values from the {\sc Protos} event generator for the weak-isospin singlet and $(B,Y)$ doublet cases 
are shown as plain circle and star symbols respectively.
\label{fig:limits2D_BB}}
\end{figure*}

\begin{figure*}[tbp]
\centering
\subfloat[]{\includegraphics[width=0.48\textwidth]{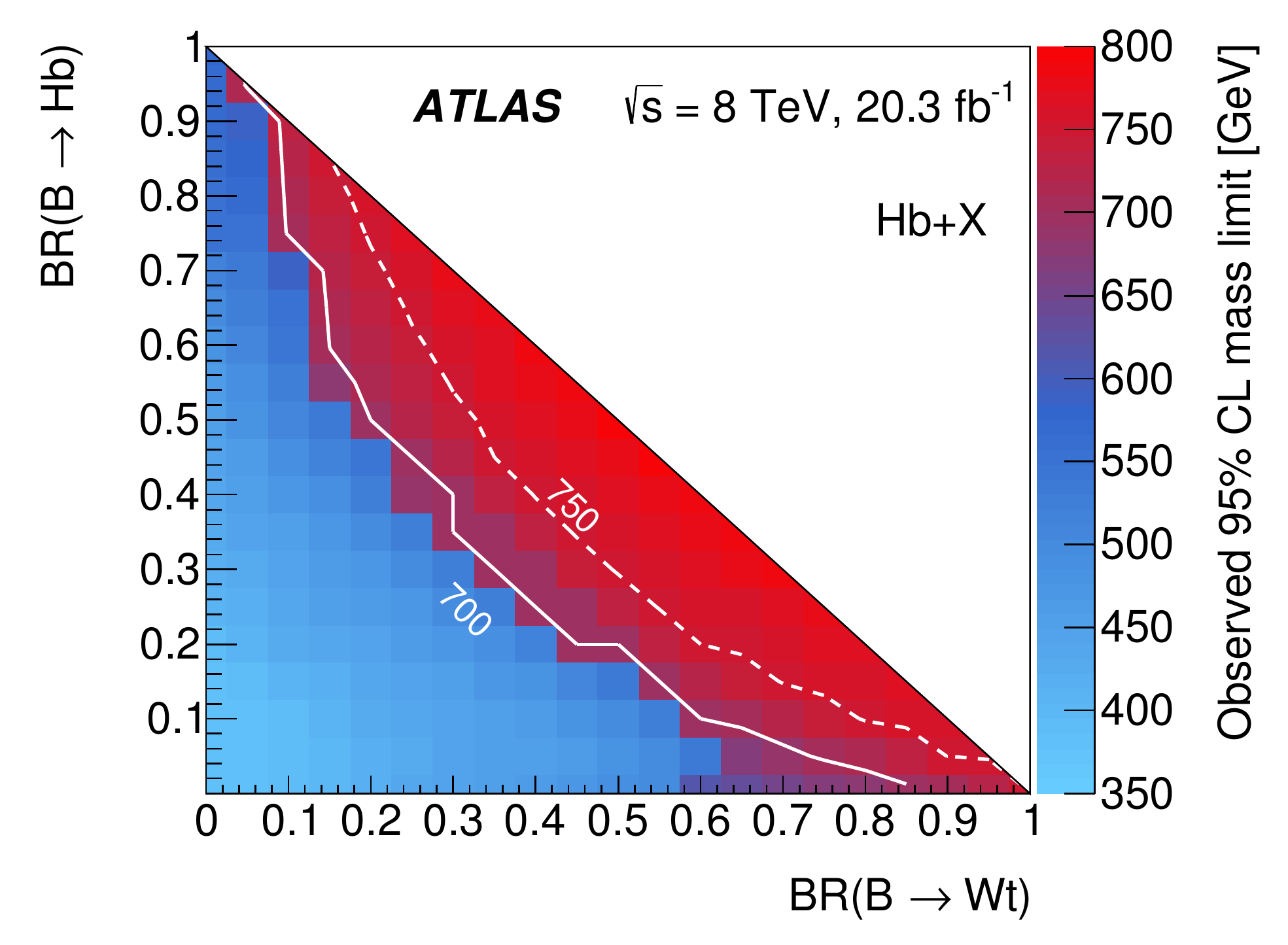}}
\subfloat[]{\includegraphics[width=0.48\textwidth]{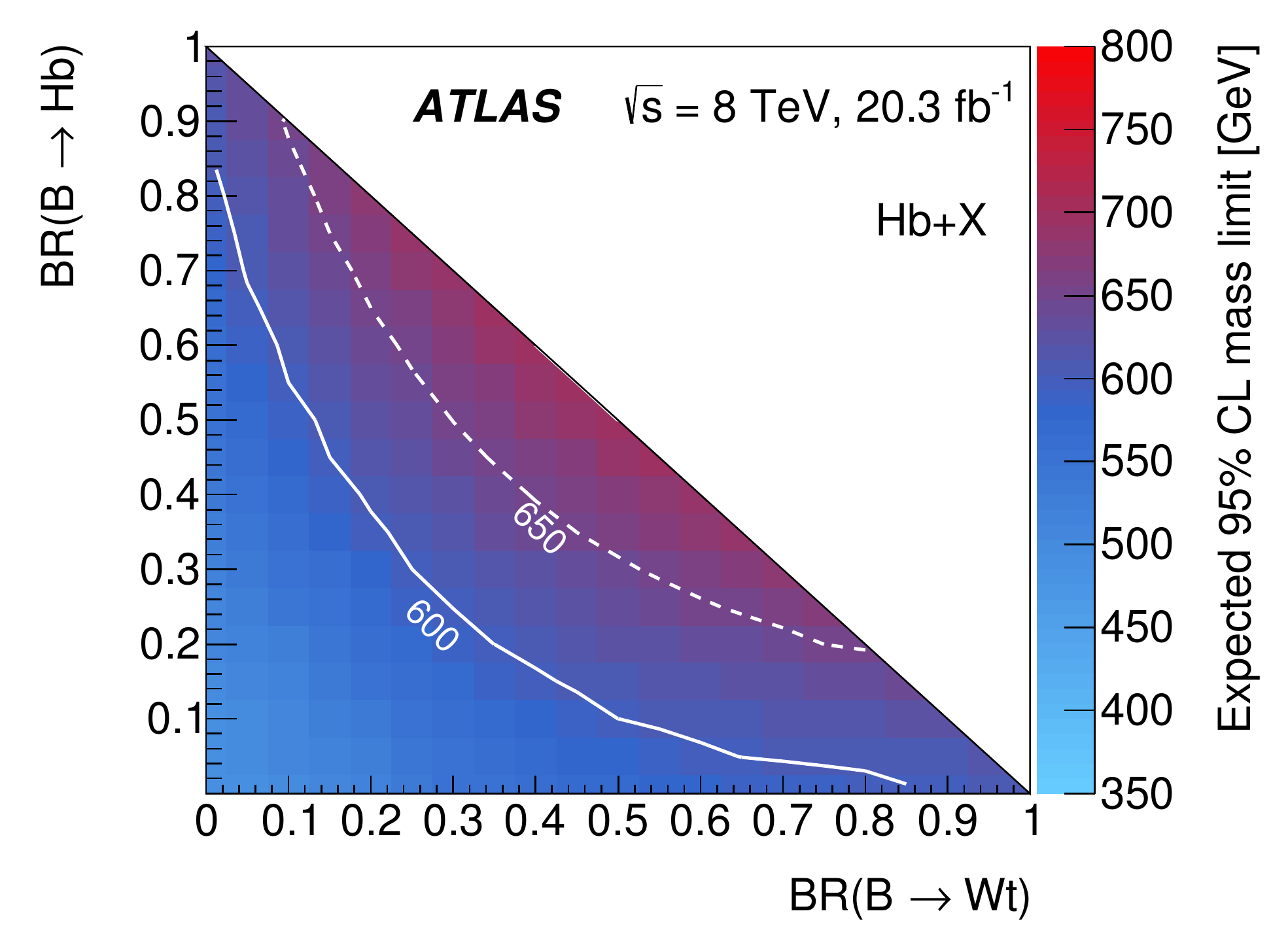}}
\caption{(a) Observed and (b) expected limit (95\% CL) on the mass of the $B$ quark in the plane 
of $\BR(B \to Hb)$ versus $\BR(B \to Wt)$ for the $B\bar{B} \to Hb$+X search.
Contour lines are provided to guide the eye.}
\label{fig:limits2D_BB_temp}
\end{figure*}

\begin{figure*}[tbp]
\centering
\subfloat[]{\includegraphics[width=0.48\textwidth]{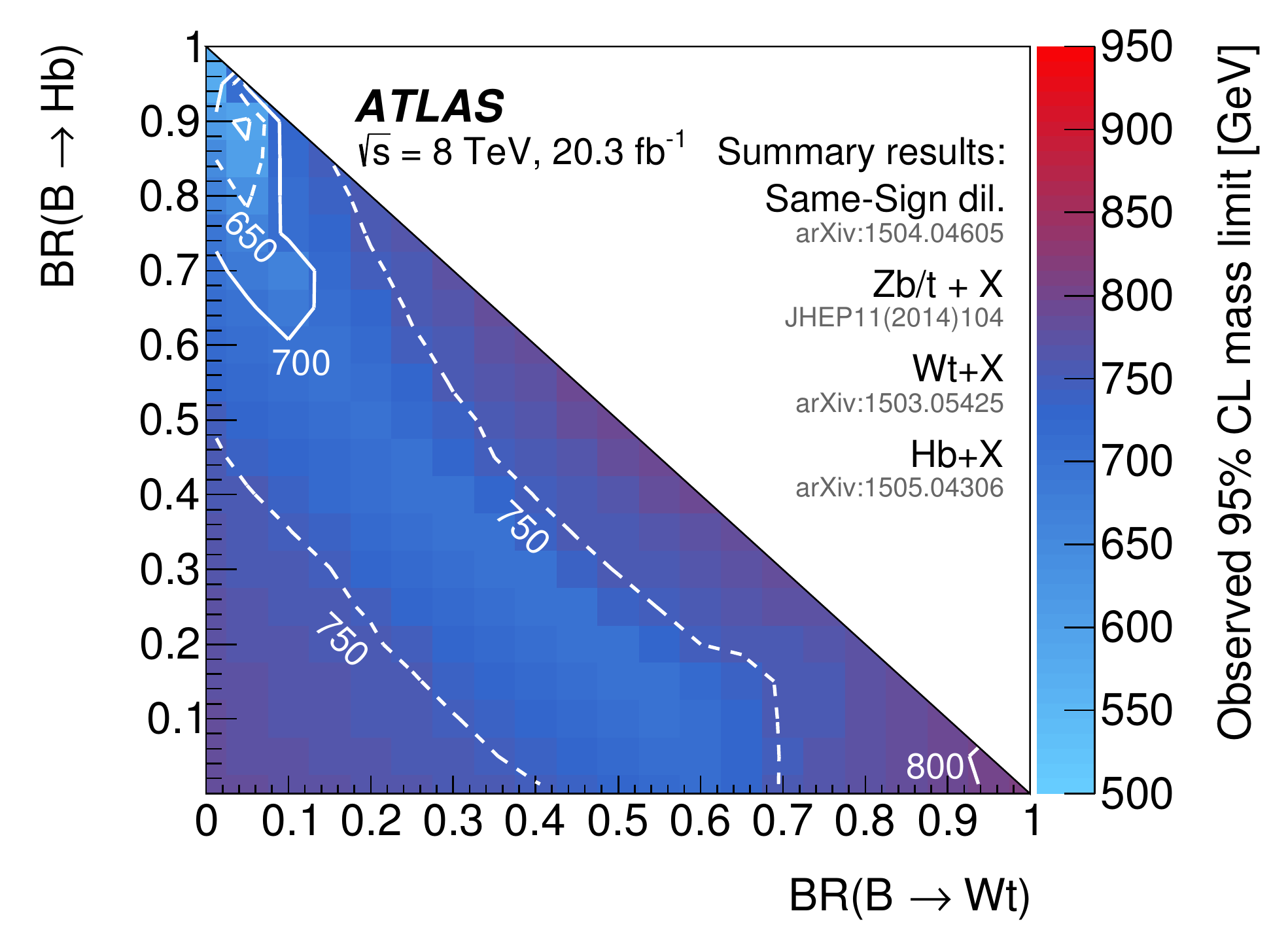}}
\subfloat[]{\includegraphics[width=0.48\textwidth]{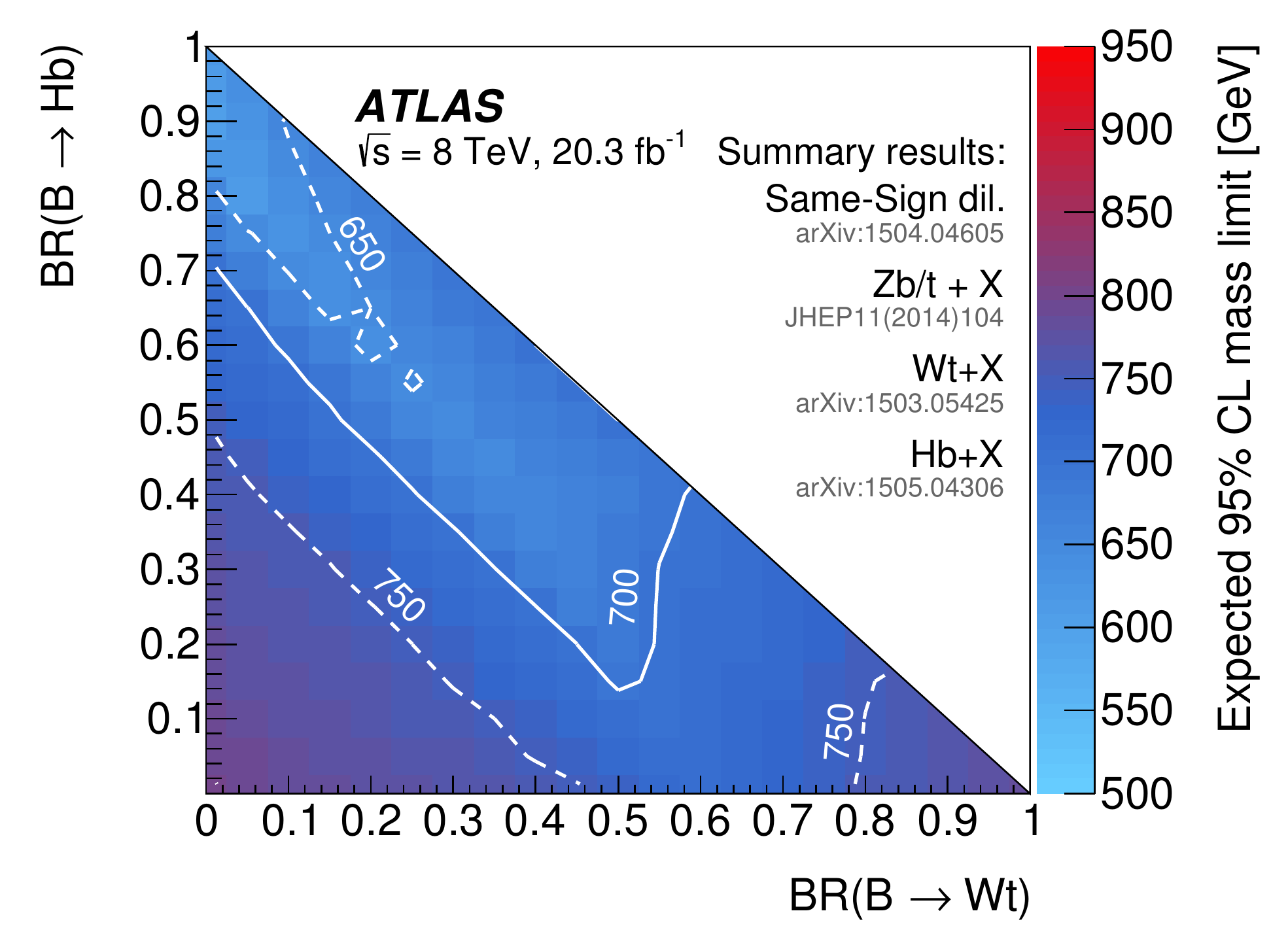}}
\caption{Summary of the most restrictive (a) observed and (b) expected limit (95\% CL) on the mass of the $B$ quark 
in the plane of $\BR(B \to Hb)$ versus $\BR(B \to Wt)$ from all ATLAS searches for $B\bar{B}$ production (see text for details).
Contour lines are provided to guide the eye.}
\label{fig:limits2D_B_Summary_temp}
\end{figure*}

\subsection{Limits on $\fourtop$ production}

The $Ht$+X analysis is also used to set limits on four-top-quark production considering different signal benchmark
scenarios: SM-like $\fourtop$, $\fourtop$ via an EFT model with a four-top-quark contact interaction, sgluon pair production with
decay into $t\bar{t}$, and $\fourtop$+X via the 2UED/RPP model. Except for the case of SM-like $\fourtop$ production,
for which the ATLAS multilepton search~\cite{Aad:2015gdg} achieves the best expected sensitivity, in all other
benchmark scenarios this analysis achieves the most restrictive expected bounds.

In the case of $\fourtop$ production with the SM kinematics, the observed (expected) 95\% CL upper limit on the
production cross section is 23 fb (32 fb), or 34 (47) times the SM prediction. In this scenario the expected
sensitivity of this analysis is comparable to that of previous searches~\cite{Khachatryan:2014sca,Aad:2015gdg}.
In the case of $\fourtop$ production via an EFT model, the observed (expected) 95\% CL upper limit on the
production cross section is 12 fb (16 fb). 
The improved sensitivity in the case of the EFT model results from the harder $\HT$ spectrum compared to that of SM $\fourtop$ production.
The upper limit on the production cross section can be translated into an 
observed (expected) limit on the free parameter of the model, $|C_{4t}|/\Lambda^2<6.6\tev^{-2}\;(7.7\tev^{-2})$. 

The resulting observed and expected upper limits on the sgluon pair production cross section 
times branching ratio are shown in figure~\ref{fig:limits_sgluon} as a function of the sgluon mass 
and are compared to the theoretical prediction. The observed (expected) 95\%  CL  limit  
on the sgluon mass is $1.06\tev$ ($1.02\tev$). 

Finally, in the context of the 2UED/RPP model, the observed and expected upper limits on the production cross section 
times branching ratio are shown in figure~\ref{fig:limits_ued_10} as a function of $m_{\KK}$ 
for the symmetric case ($\xi=R_4/R_5=1$), assuming production by tier (1,1) alone.
The comparison to the LO theoretical cross section translates into an observed (expected) 95\%  CL  limit  
on $m_{\KK}$ of $1.12\tev$ ($1.10\tev$). Four-top-quark events can also arise from 
tiers (2,0) and (0,2). In those tiers the theoretical production cross sections can be 
calculated, leading to more robust results (i.e. there is no need to assume a particular branching ratio). 
The dependence of the tier kinematics on the tier mass also allows the extrapolation of constraints on tier (1,1) to tiers (2,0) and (0,2). 
Excluding a given production cross section for tier (1,1) at a given $m_{\KK}$ is equivalent to
excluding this production cross section for tier (2,0) alone at $m_{\KK}/\sqrt{2}$ and for tier (0,2) at $m_{\KK}/\sqrt{2}\xi$. 
The contribution of tier (0,2) vanishes as $\xi$ increases (highly asymmetric case).
Figure~\ref{fig:limits_ued_20} presents the observed and expected upper limits on the production cross section 
times branching ratio as  function of $m_{\KK}$ for two scenarios: tiers (2,0)+(0,2) alone in the symmetric case,
and tier (2,0) alone in the highly asymmetric case.  In both cases a branching ratio of $A^{(1,1)}\to t\bar{t}$ of 0\% is assumed.
The corresponding observed (expected) 95\%  CL  limits on $m_{\KK}$ are $0.61\tev$ ($0.60\tev$) and $0.57\tev$ ($0.55\tev$) respectively. 

\begin{figure*}[tbp]
\centering
\includegraphics[width=0.45\textwidth]{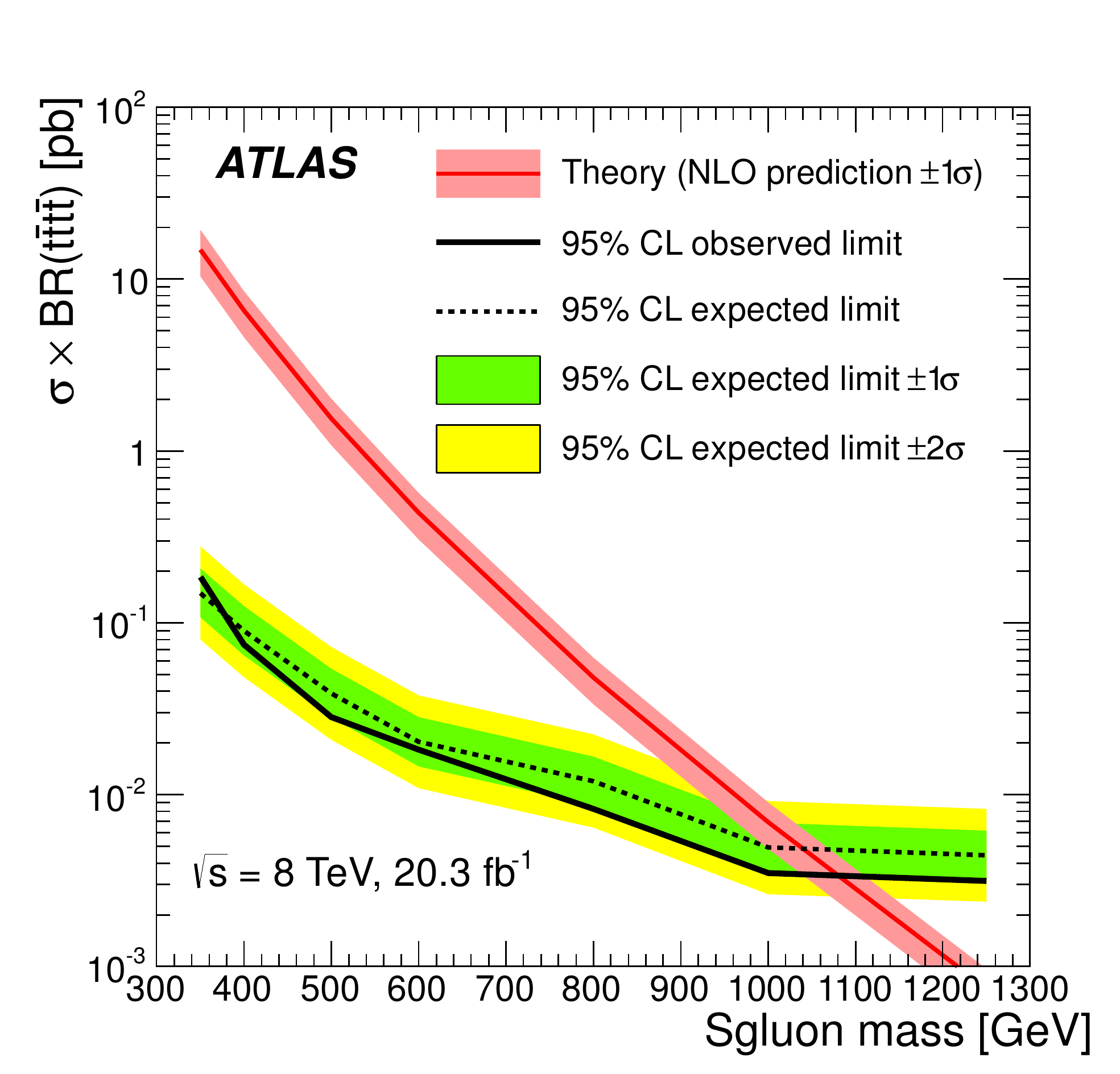}
\caption{
Observed (solid line) and expected (dashed line) 95\% CL upper limits on the sgluon pair production cross section times 
branching ratio as a function of the sgluon mass. 
The surrounding shaded bands correspond to $\pm1$ and $\pm2$ standard deviations around the expected limit. 
The thin red line and band show the theoretical prediction and its $\pm1$ standard deviation uncertainty.
\label{fig:limits_sgluon}}
\end{figure*}

\begin{figure*}[tbp]
\centering
\includegraphics[width=0.45\textwidth]{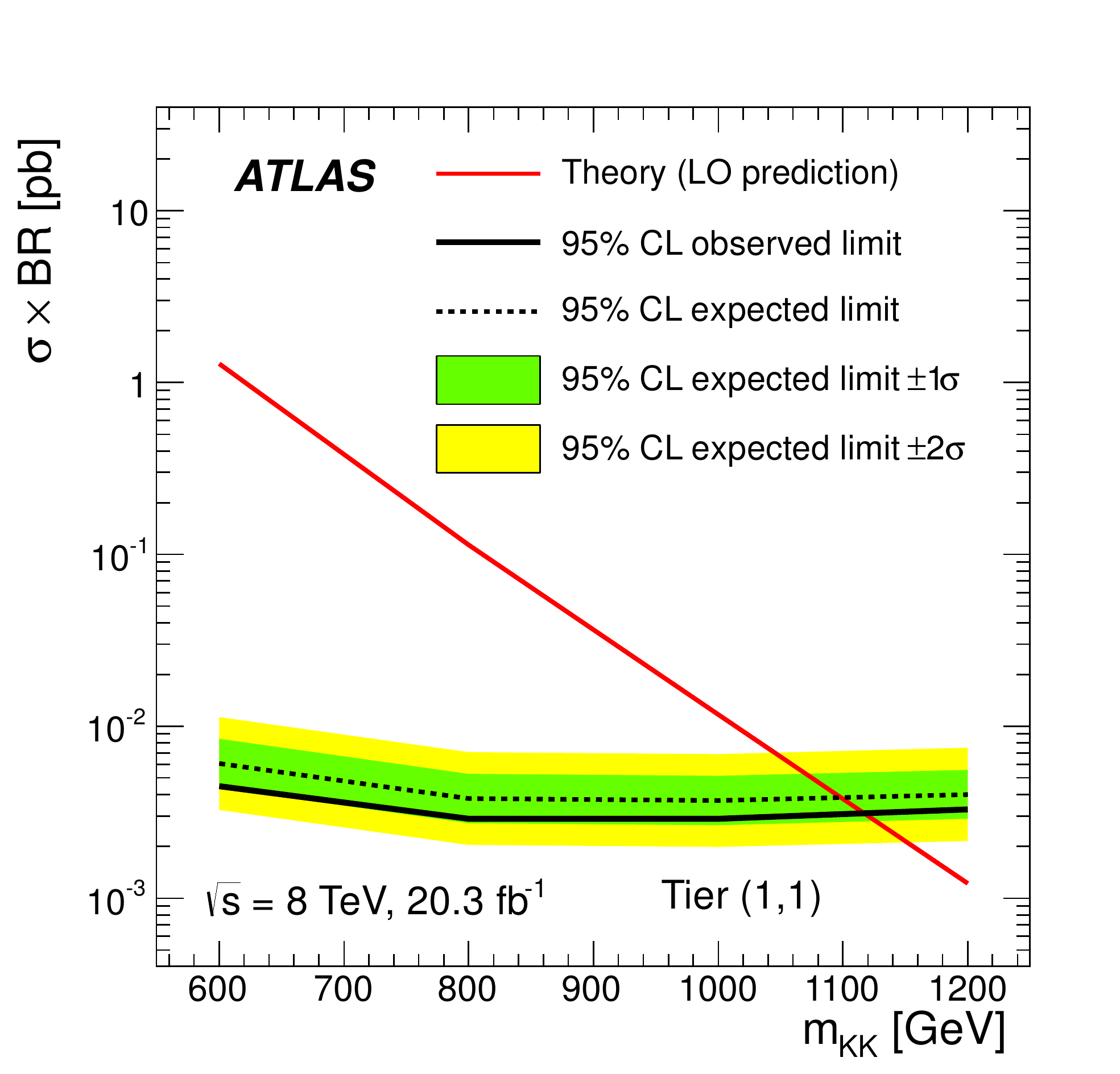}
\caption{
Observed (solid line) and expected (dashed line) 95\% CL upper limits on the production cross section times branching ratio
of four-top-quark events as a function of the Kaluza--Klein mass ($m_{\KK}$) from tier (1,1) in the symmetric case. 
The surrounding shaded bands correspond to $\pm1$ and $\pm2$ standard deviations around the expected limit. 
The thin red line shows the theoretical prediction for the production cross section of four-top-quark events by 
tier (1,1) assuming $\BR(A^{(1,1)}\to t\bar{t})=1$, where $A^{(1,1)}$ is the lightest particle of this tier.
\label{fig:limits_ued_10}}
\end{figure*}

\begin{figure*}[tbp]
\centering
\subfloat[]{\includegraphics[width=0.45\textwidth]{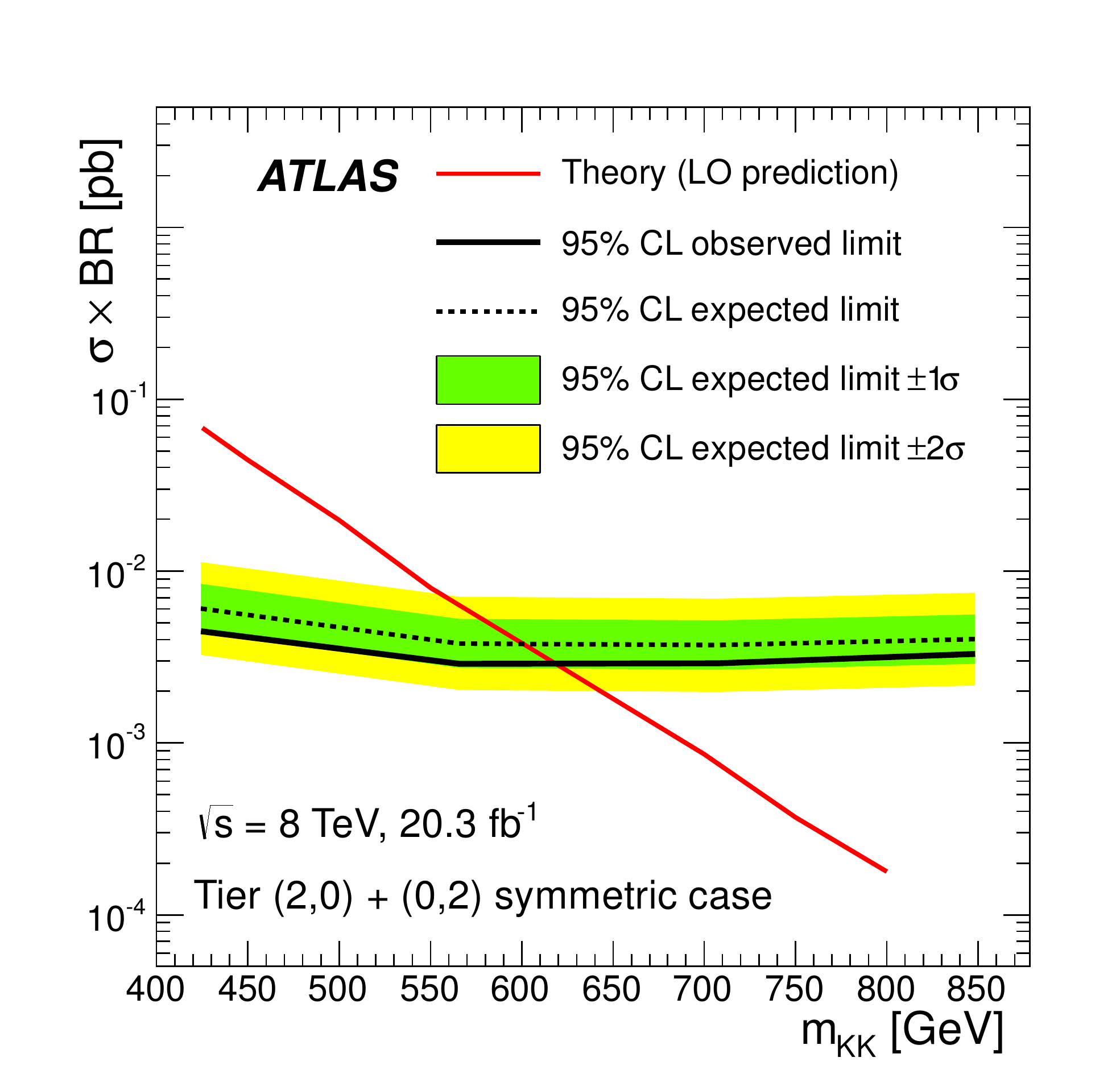}}
\subfloat[]{\includegraphics[width=0.45\textwidth]{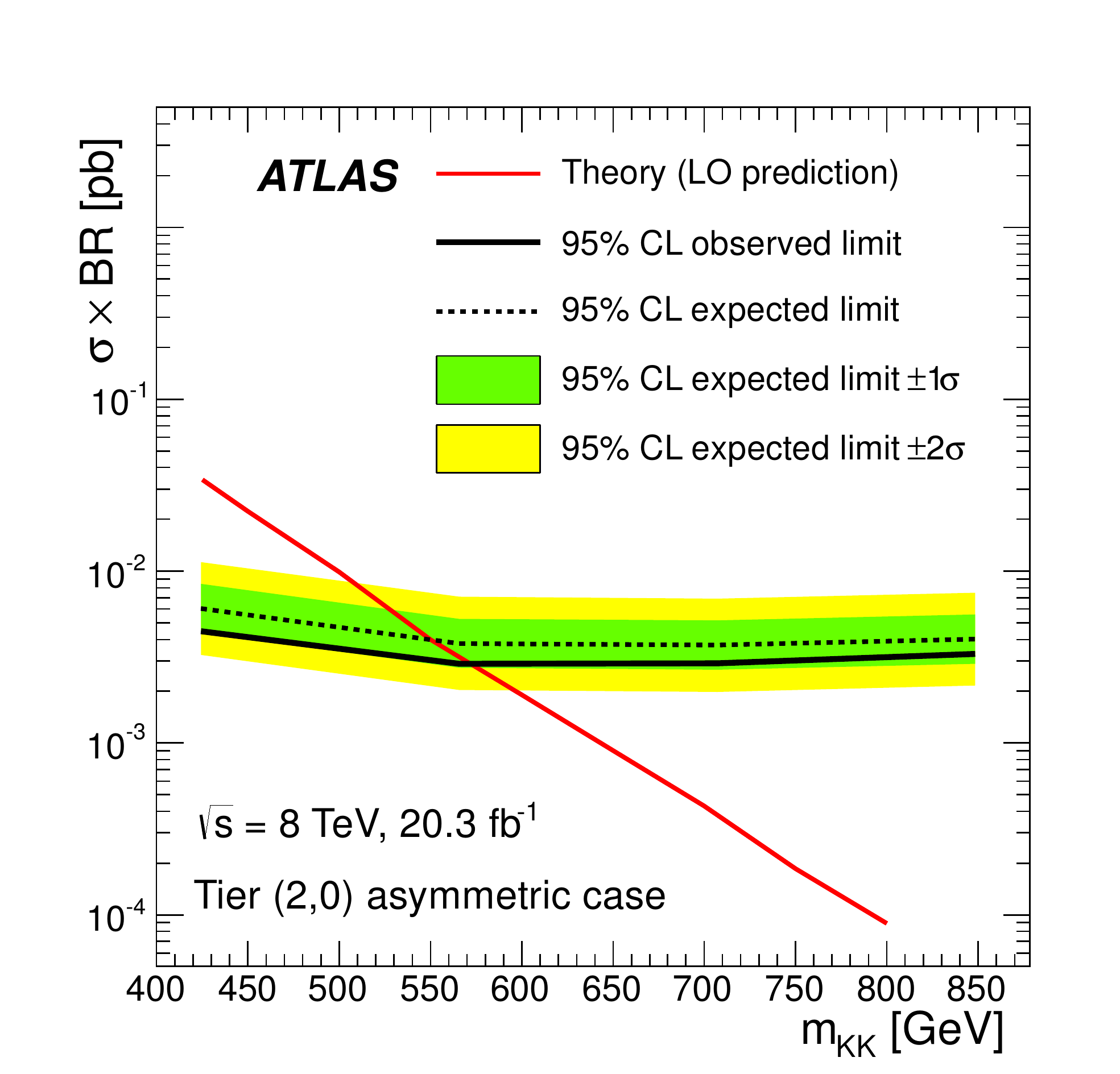}}
\caption{
Observed (solid line) and expected (dashed line) 95\% CL upper limits on the production cross section times branching ratio
of four-top-quark events as a function of the Kaluza--Klein mass ($m_{\KK}$) from (a) tiers (2,0)+(0,2) alone in the symmetric case 
and (b) tier (2,0) alone in the highly asymmetric case.
The surrounding shaded bands correspond to $\pm1$ and $\pm2$ standard deviations around the expected limit. 
The thin red line shows the theoretical prediction for the production cross section of four-top-quark events.
\label{fig:limits_ued_20}}
\end{figure*}

\FloatBarrier

\section{Conclusion}
\label{sec:conclusion}

A search for pair production of vector-like quarks, both up-type ($T$) and down-type ($B$), as well as four-top-quark production has been performed using
$pp$ collision data at $\sqrt{s}=8\tev$ corresponding to an integrated luminosity of 20.3~fb$^{-1}$ recorded with the ATLAS detector at the CERN Large Hadron Collider.
The final states considered have an isolated electron or muon with high transverse momentum, large missing transverse momentum and at least four jets.
Three different analyses are optimised to reach the best sensitivity to the decay channels $T\bar{T} \to Wb$+X, $T\bar{T} \to Ht$+X and $B\bar{B} \to Hb$+X.

No significant deviation from the Standard Model expectation is observed and
lower limits on the masses of the vector-like $T$ ($B$) quark are derived as a function of the 
branching ratios $\BR(T \to Wb)$, $\BR(T \to Zt)$, and $\BR(T \to Ht)$ (respectively $\BR(B \to Wt)$, $\BR(B \to Zb)$, and $\BR(B \to Hb)$). 
The combination of the $T\bar{T} \to Wb$+X, $T\bar{T} \to Ht$+X analyses yields observed lower limits on the $T$ quark mass ranging 
between $715\gev$ and $950\gev$ for all possible values of the branching ratios into three decay modes, 
and are the most stringent constraints to date. 
The $B\bar{B} \to Hb$+X analysis is the first search to target specifically this decay mode and leads to an observed lower limit on the $B$ quark mass 
of 580~GeV for $\BR(B \to Hb) = 1$. Finally, a summary of all ATLAS vector-like quark pair production searches is given.
For $B\bar{B}$ production, the observed lower limits on the $B$ quark mass range between $575\gev$ and $813\gev$  
for all possible values of the branching ratios into the three decay modes.

The $T\bar{T} \to Ht$+X analysis is also used to set limits on four-top-quark production, both in the Standard Model and in several new physics scenarios,
including a four-fermion contact interaction, sgluon pair production and a universal extra dimensions model.
In the case of Standard Model production, a cross section larger than 23~fb is excluded at the 95\% CL. The most restrictive limits to date are obtained
for four-top-quark production in the various new physics scenarios considered.

\section*{Acknowledgements}


We thank CERN for the very successful operation of the LHC, as well as the
support staff from our institutions without whom ATLAS could not be
operated efficiently.

We acknowledge the support of ANPCyT, Argentina; YerPhI, Armenia; ARC,
Australia; BMWFW and FWF, Austria; ANAS, Azerbaijan; SSTC, Belarus; CNPq and FAPESP,
Brazil; NSERC, NRC and CFI, Canada; CERN; CONICYT, Chile; CAS, MOST and NSFC,
China; COLCIENCIAS, Colombia; MSMT CR, MPO CR and VSC CR, Czech Republic;
DNRF, DNSRC and Lundbeck Foundation, Denmark; EPLANET, ERC and NSRF, European Union;
IN2P3-CNRS, CEA-DSM/IRFU, France; GNSF, Georgia; BMBF, DFG, HGF, MPG and AvH
Foundation, Germany; GSRT and NSRF, Greece; RGC, Hong Kong SAR, China; ISF, MINERVA, GIF, I-CORE and Benoziyo Center, Israel; INFN, Italy; MEXT and JSPS, Japan; CNRST, Morocco; FOM and NWO, Netherlands; BRF and RCN, Norway; MNiSW and NCN, Poland; GRICES and FCT, Portugal; MNE/IFA, Romania; MES of Russia and NRC KI, Russian Federation; JINR; MSTD,
Serbia; MSSR, Slovakia; ARRS and MIZ\v{S}, Slovenia; DST/NRF, South Africa;
MINECO, Spain; SRC and Wallenberg Foundation, Sweden; SER, SNSF and Cantons of
Bern and Geneva, Switzerland; NSC, Taiwan; TAEK, Turkey; STFC, the Royal
Society and Leverhulme Trust, United Kingdom; DOE and NSF, United States of
America.

The crucial computing support from all WLCG partners is acknowledged
gratefully, in particular from CERN and the ATLAS Tier-1 facilities at
TRIUMF (Canada), NDGF (Denmark, Norway, Sweden), CC-IN2P3 (France),
KIT/GridKA (Germany), INFN-CNAF (Italy), NL-T1 (Netherlands), PIC (Spain),
ASGC (Taiwan), RAL (UK) and BNL (USA) and in the Tier-2 facilities
worldwide.

%

\clearpage
\appendix
\part*{Appendix}
\addcontentsline{toc}{part}{Appendix}
\section{Post-fit event yields}
\label{sec:postfit_yields_appendix}

Table~\ref{tab:Postfit_Yields_HtX_unblind} presents the observed and predicted background yields in each of the analysis channels 
for the $T\bar{T} \to Ht$+X search, after the fit to the data under the background-only hypothesis. The corresponding observed and predicted
yields for the $B\bar{B} \to Hb$+X search are summarised in table~\ref{tab:Postfit_Yields_HbX_unblind}.

\begin{table}[h!]
\begin{center}
\begin{tabular}{l*{4}{c}}
\toprule\toprule
 & 5 j, 2 b & 5 j, 3 b & 5 j, $\geq$4 b & $\geq$6 j, 2 b\\
\midrule
$t\bar{t}$+light-jets & $32200 \pm 1500$ & $2940 \pm 220$ & $49.1 \pm 8.8$ & $16000 \pm 1000$\\
$t\bar{t}+c\bar{c}$ & $5600 \pm 1700$ & $1000 \pm 310$ & $61 \pm 17$ & $4300 \pm 1300$\\
$t\bar{t}+b\bar{b}$ & $1820 \pm 360$ & $990 \pm 180$ & $124 \pm 19$ & $1440 \pm 280$\\
$t\bar{t}V$ & $139 \pm 44$ & $25.0 \pm 7.9$ & $3.1 \pm 1.0$ & $164 \pm 52$\\
$t\bar{t}H$ & $39.8 \pm 1.4$ & $22.0 \pm 1.2$ & $6.1 \pm 0.5$ & $58.7 \pm 2.9$\\
$W$+jets & $1200 \pm 580$ & $86 \pm 41$ & $4.3 \pm 2.0$ & $560 \pm 280$\\
$Z$+jets & $390 \pm 120$ & $27.6 \pm 8.7$ & $1.6 \pm 0.5$ & $190 \pm 60$\\
Single top & $1600 \pm 260$ & $172 \pm 31$ & $7.1 \pm 0.8$ & $710 \pm 150$\\
Diboson & $88 \pm 27$ & $7.7 \pm 2.6$ & $0.4 \pm 0.2$ & $43 \pm 13$\\
Multijet & $125 \pm 40$ & $31 \pm 10$ & $6.4 \pm 2.2$ & $52 \pm 16$\\
\midrule
Total background & $43240 \pm 320$          & $5360 \pm 79$          & $263 \pm 10$          & $23100 \pm 240$         \\
\midrule
Data & $43319$ & $5309$ & $244$ & $23001$\\
\bottomrule\bottomrule     \\
\end{tabular}
\vspace{0.1cm}

\begin{tabular}{l*{4}{c}}
\toprule\toprule
 & \begin{tabular}{@{}c@{}}$\geq$6 j, 3 b\\ low $M_{bb}^{{\rm min}\Delta R}$\end{tabular} & \begin{tabular}{@{}c@{}}$\geq$6 j, 3 b\\ high $M_{bb}^{{\rm min}\Delta R}$\end{tabular} & \begin{tabular}{@{}c@{}}$\geq$6 j, $\geq$4 b\\ low $M_{bb}^{{\rm min}\Delta R}$\end{tabular} & \begin{tabular}{@{}c@{}}$\geq$6 j, $\geq$4 b\\ high $M_{bb}^{{\rm min}\Delta R}$\end{tabular}\\
\midrule
$t\bar{t}$+light-jets & $1260 \pm 130$ & $421 \pm 43$ &  $38.3 \pm 8.1$ & $9.5 \pm 2.1$ \\
$t\bar{t}+c\bar{c}$ & $760 \pm 210$ & $278 \pm 79$ & $72 \pm 20$ & $20.4 \pm 6.2$ \\
$t\bar{t}+b\bar{b}$ & $730 \pm 120$ & $285 \pm 51$ &  $211 \pm 29$ & $52.0 \pm 7.9$ \\
$t\bar{t}V$ & $28.1 \pm 8.9$ & $12.3 \pm 3.9$ & $6.3 \pm 2.0$ & $1.5 \pm 0.5$ \\
$t\bar{t}H$ & $25.0 \pm 1.3$ & $11.7 \pm 0.9$ & $11.1 \pm 0.9$ & $4.2 \pm 0.4$ \\
$W$+jets & $50 \pm 25$ & $12.0 \pm 6.1$ & $5.4 \pm 2.9$ & $0.4 \pm 0.2$ \\
$Z$+jets & $16.8 \pm 5.5$ & $3.3 \pm 1.2$ & $1.6 \pm 0.5$ & $0.3 \pm 0.1$ \\
Single top & $76 \pm 17$ & $33 \pm 10$ & $11.3 \pm 3.2$ &  $2.8 \pm 1.5$ \\
Diboson & $4.3 \pm 1.5$\ & $1.4 \pm 0.5$ & $0.4 \pm 0.1$ & $0.2 \pm 0.1$ \\
Multijet & $1.7 \pm 0.7$ & $4.3 \pm 1.8$ & $<0.01$ & $2.6 \pm 0.8$ \\
\midrule
Total background & $2948 \pm 54$          & $1062 \pm 25$          & $357 \pm 16$          & $93.9 \pm 5.0$         \\
\midrule
Data & $3015$ & $1085$ & $362$ & $84$\\
\bottomrule\bottomrule     \\
\end{tabular}
\vspace{0.1cm}

\end{center}
\vspace{-0.5cm}
\caption{$T\bar{T} \to Ht$+X search: predicted and observed yields in each of the analysis channels considered.
The background prediction is shown after the fit to data under the background-only hypothesis.
The quoted uncertainties are the sum in quadrature of statistical and systematic uncertainties on the yields, 
computed taking into account correlations among nuisance parameters and among processes.}
\label{tab:Postfit_Yields_HtX_unblind}
\end{table}

\begin{table}
\begin{center}
\begin{tabular}{l*{4}{c}}
\toprule\toprule
 & 5 j, 2 b & 5 j, 3 b & 5 j, $\geq$4 b & $\geq$6 j, 2 b\\
\midrule
$t\bar{t}$+light-jets & $406 \pm 35$ & $77.8 \pm 8.8$ & $2.3 \pm 0.5$ & $239 \pm 26$\\
$t\bar{t}+c\bar{c}$ & $60 \pm 31$ & $25 \pm 11$ & $2.4 \pm 1.1$ & $58 \pm 26$\\
$t\bar{t}+b\bar{b}$ & $28 \pm 10$ & $35.4 \pm 9.3$ & $7.4 \pm 1.9$ & $33 \pm 11$\\
$t\bar{t}V$ & $4.2 \pm 1.3$ & $1.7 \pm 0.5$ & $0.3 \pm 0.1$ & $5.1 \pm 1.6$\\
$t\bar{t}H$ & $1.0 \pm 0.1$ & $1.2 \pm 0.1$ & $0.5 \pm 0.1$ & $1.5 \pm 0.2$\\
$W$+jets & $23 \pm 12$ & $3.9 \pm 2.0$ & $0.8 \pm 0.5$ & $13.9 \pm 7.5$\\
$Z$+jets & $7.2 \pm 2.7$ & $2.0 \pm 2.2$ & $0.6 \pm 0.5$ & $4.0 \pm 3.1$\\
Single top & $41.5 \pm 4.9$ & $9.1 \pm 1.2$ & $0.6 \pm 0.1$ & $26.8 \pm 4.2$\\
Diboson & $1.9 \pm 0.8$ & $0.5 \pm 0.3$ & $0.05 \pm  0.05$ & $1.2 \pm 0.6$\\
Multijet & $<0.01$ & $<0.01$ & $0.3 \pm 0.2$ & $0.18 \pm 0.01$\\
\midrule
Total background & $573 \pm 20$          & $156.3 \pm 8.5$          & $15.2 \pm 1.9$          & $383 \pm 16$         \\
\midrule
Data & $576$ & $165$ & $10$ & $375$\\
\bottomrule\bottomrule     \\
\end{tabular}
\vspace{0.1cm}

\begin{tabular}{l*{4}{c}}
\toprule\toprule
 & \begin{tabular}{@{}c@{}}$\geq$6 j, 3 b\\ low $M_{bb}^{{\rm min}\Delta R}$\end{tabular} & \begin{tabular}{@{}c@{}}$\geq$6 j, 3 b\\ high $M_{bb}^{{\rm min}\Delta R}$\end{tabular} & \begin{tabular}{@{}c@{}}$\geq$6 j, $\geq$4 b\\ low $M_{bb}^{{\rm min}\Delta R}$\end{tabular} & \begin{tabular}{@{}c@{}}$\geq$6 j, $\geq$4 b\\ high $M_{bb}^{{\rm min}\Delta R}$\end{tabular}\\
\midrule
$t\bar{t}$+light-jets & $23.4 \pm 4.5$ & $34.6 \pm 4.9$ & $1.5 \pm 0.3$ & $1.6 \pm 0.4$\\
$t\bar{t}+c\bar{c}$ & $12.0 \pm 5.2$ & $22 \pm 10$ & $2.5 \pm 1.2$ & $3.1 \pm 1.4$\\
$t\bar{t}+b\bar{b}$ & $19.6 \pm 6.2$ & $36 \pm 11$ & $11.8 \pm 3.0$ & $11.8 \pm 3.1$\\
$t\bar{t}V$ & $1.2 \pm 0.4$ & $1.7 \pm 0.5$ & $0.6 \pm 0.2$ & $0.4 \pm 0.1$\\
$t\bar{t}H$ & $0.7 \pm 0.1$ & $1.4 \pm 0.2$ & $0.5 \pm 0.1$ & $0.9 \pm 0.1$\\
$W$+jets & $2.3 \pm 1.3$ & $1.3 \pm 0.8$ & $0.5 \pm 0.5$ & $0.07 \pm 0.06$\\
$Z$+jets & $0.1 \pm 0.1$ & $0.2 \pm 0.1$ & $<0.01$ & $<0.01$\\
Single top & $3.1 \pm 0.4$ & $4.7 \pm 1.4$ & $0.8 \pm 0.1$ & $0.4 \pm 0.2$\\
Diboson & $0.2 \pm 0.1$ & $0.10 \pm 0.03$ & $0.02 \pm 0.01$ & $0.01 \pm 0.01$\\
Multijet & $<0.01$ & $0.6 \pm 0.2$ & $<0.01$ & $0.4 \pm 0.1$\\
\midrule
Total background & $62.6 \pm 5.3$          & $101.9 \pm 7.3$          & $18.3 \pm 2.6$          & $18.6 \pm 2.6$         \\
\midrule
Data & $62$ & $103$ & $23$ & $20$\\
\bottomrule\bottomrule    \\
\end{tabular}
\vspace{0.1cm}

\end{center}
\vspace{-0.5cm}
\caption{$B\bar{B} \to Hb$+X search: predicted and observed yields in each of the analysis channels considered.
The background prediction is shown after the fit to data under the background-only hypothesis.
The quoted uncertainties are the sum in quadrature of statistical and systematic uncertainties on the yields, 
computed taking into account correlations among nuisance parameters and among processes.}
\label{tab:Postfit_Yields_HbX_unblind}
\end{table}

\clearpage
\section{Limits on $T\bar{T}$ production from individual searches}
\label{sec:limits_appendix}

Figure~\ref{fig:limits1D_singlet} shows 95\% CL upper limits on the $T\bar{T}$ production cross section 
as a function of the $T$ quark mass obtained by the individual $T\bar{T} \to Wb$+X and $T\bar{T} \to Ht$+X searches for the singlet scenario.
The $T\bar{T} \to Wb$+X and $T\bar{T} \to Ht$+X searches yield 
observed (expected) 95\% CL  limits of $m_{T}>660\,(665)\gev$ and $m_{T}>765\,(720)\gev$ respectively.
Figure~\ref{fig:limits2D_WbX} shows the 95\% CL exclusion limits on vector-like $T$ quark production, for different 
values of $m_{T}$ and as a function of the two branching ratios $\BR(T\to W b)$ and $\BR(T\to Ht)$, obtained by
the $T\bar{T} \to Wb$+X search.  Figure~\ref{fig:limits2D_WbX_temp}(a,b) present the corresponding 
expected and observed $T$ quark mass limits respectively, in the plane of $\BR(T \to Ht)$ versus $\BR(T \to Wb)$.
The exclusion limits obtained by the $T\bar{T} \to Ht$+X search can be found in figures~\ref{fig:limits2D_HtX} and~\ref{fig:limits2D_HtX_temp}.
The $T\bar{T} \to Wb$+X search sets observed (expected) lower limits on the $T$ quark mass ranging between $350\gev$ and $760\gev$  
($350\gev$ and $800\gev$) for all possible values of the branching ratios into the three  decay modes. 
The $T\bar{T} \to Ht$+X search sets observed (expected) lower limits on the $T$ quark mass ranging between $510\gev$ and $950\gev$  
($505\gev$ and $885\gev$) for all possible values of the branching ratios into the three  decay modes. 

\begin{figure*}[h!]
\centering
\subfloat[]{\includegraphics[width=0.45\textwidth]{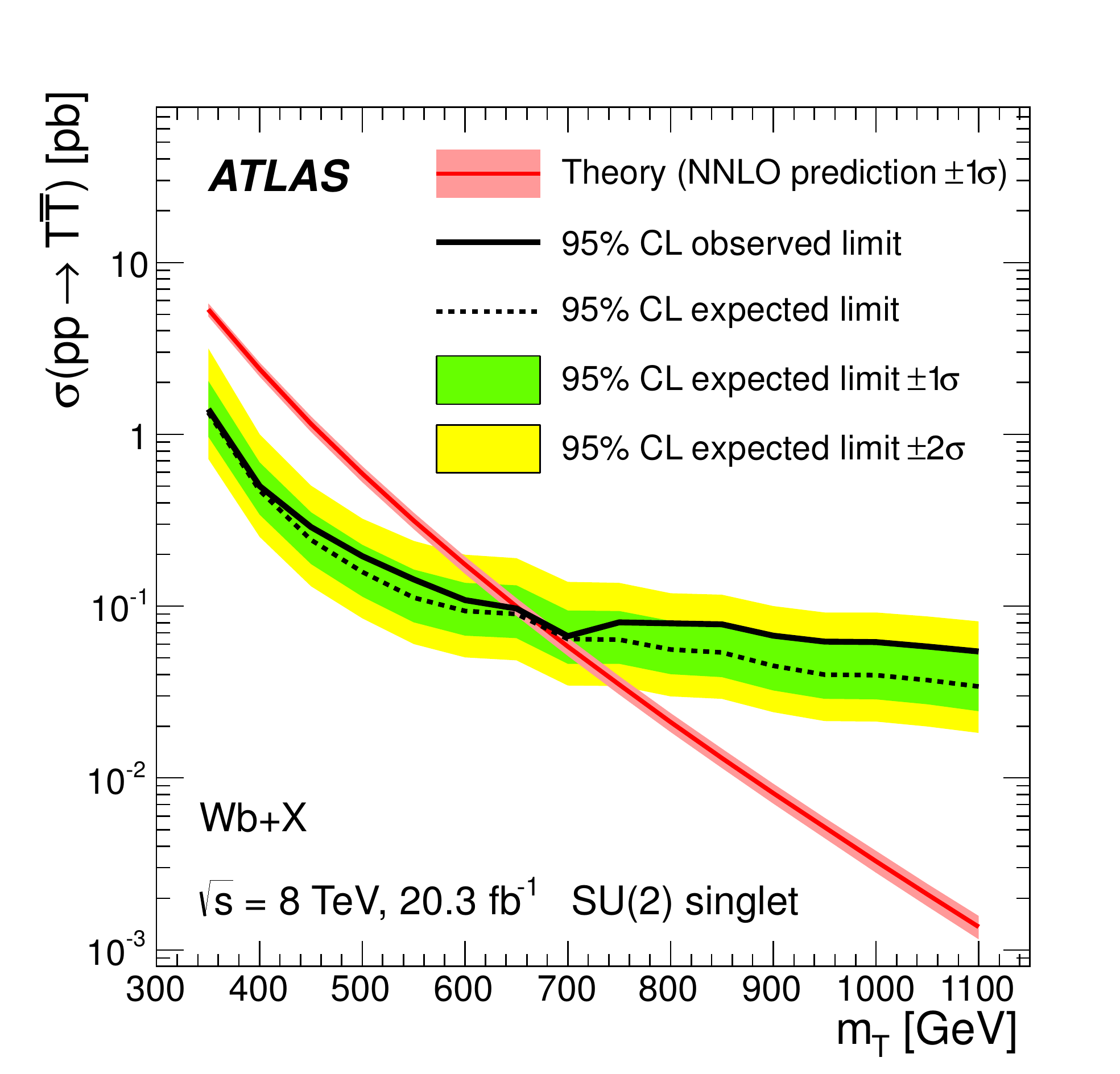}}
\subfloat[]{\includegraphics[width=0.45\textwidth]{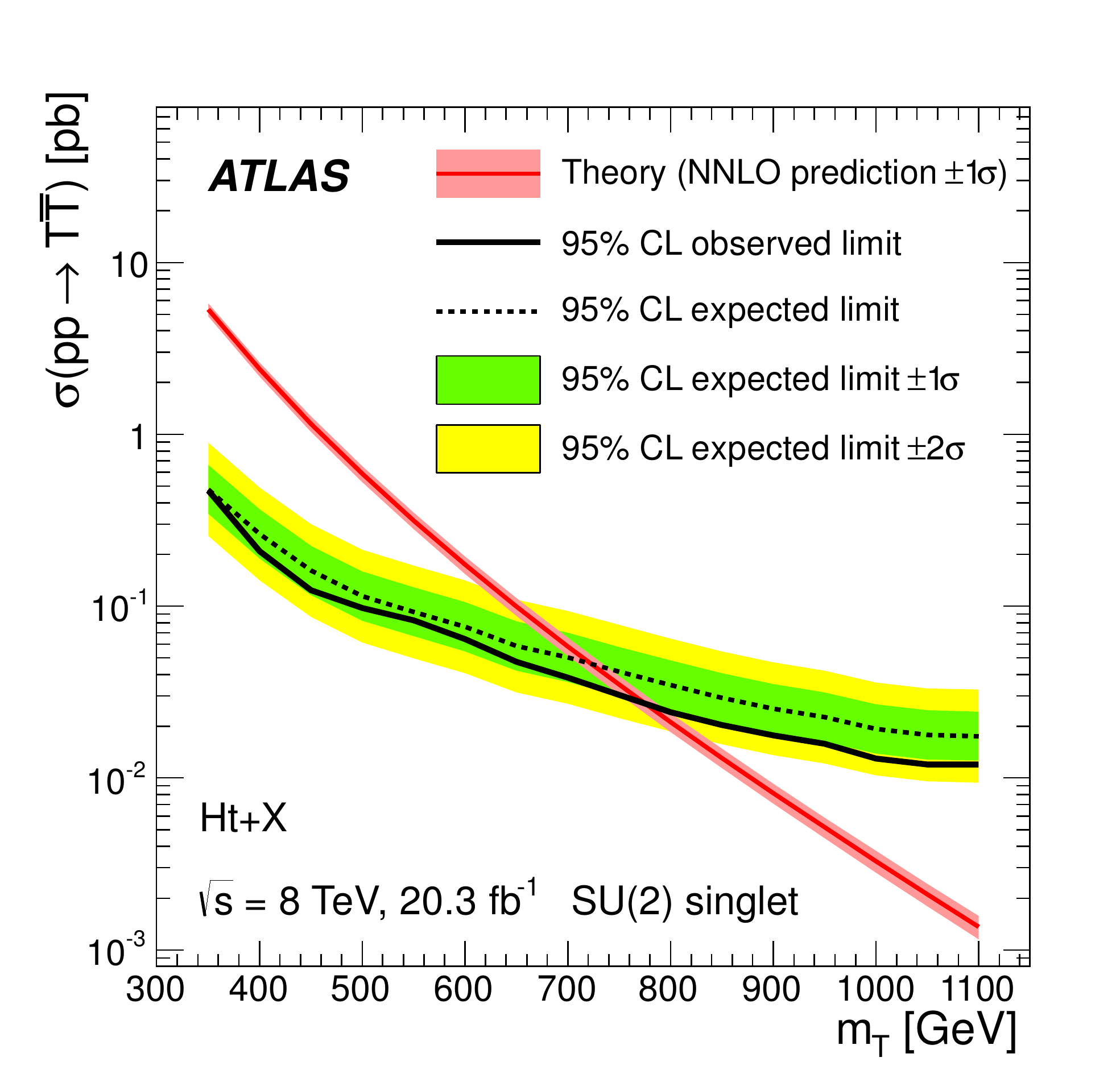}}
\caption{Observed (solid line) and expected (dashed line) 95\% CL upper limits on the $T\bar{T}$ cross section 
for a vector-like singlet $T$ quark as a function of the $T$ quark mass from (a) the $T\bar{T} \to Wb$+X search and (b) $T\bar{T} \to Ht$+X search. 
The surrounding shaded bands correspond to $\pm1$ and $\pm2$ standard deviations around the expected limit. 
The thin red line and band show the theoretical prediction and its $\pm1$ standard deviation uncertainty.}
\label{fig:limits1D_singlet}
\end{figure*}

\begin{figure*}[tbp]
\centering
\includegraphics[width=0.9\textwidth]{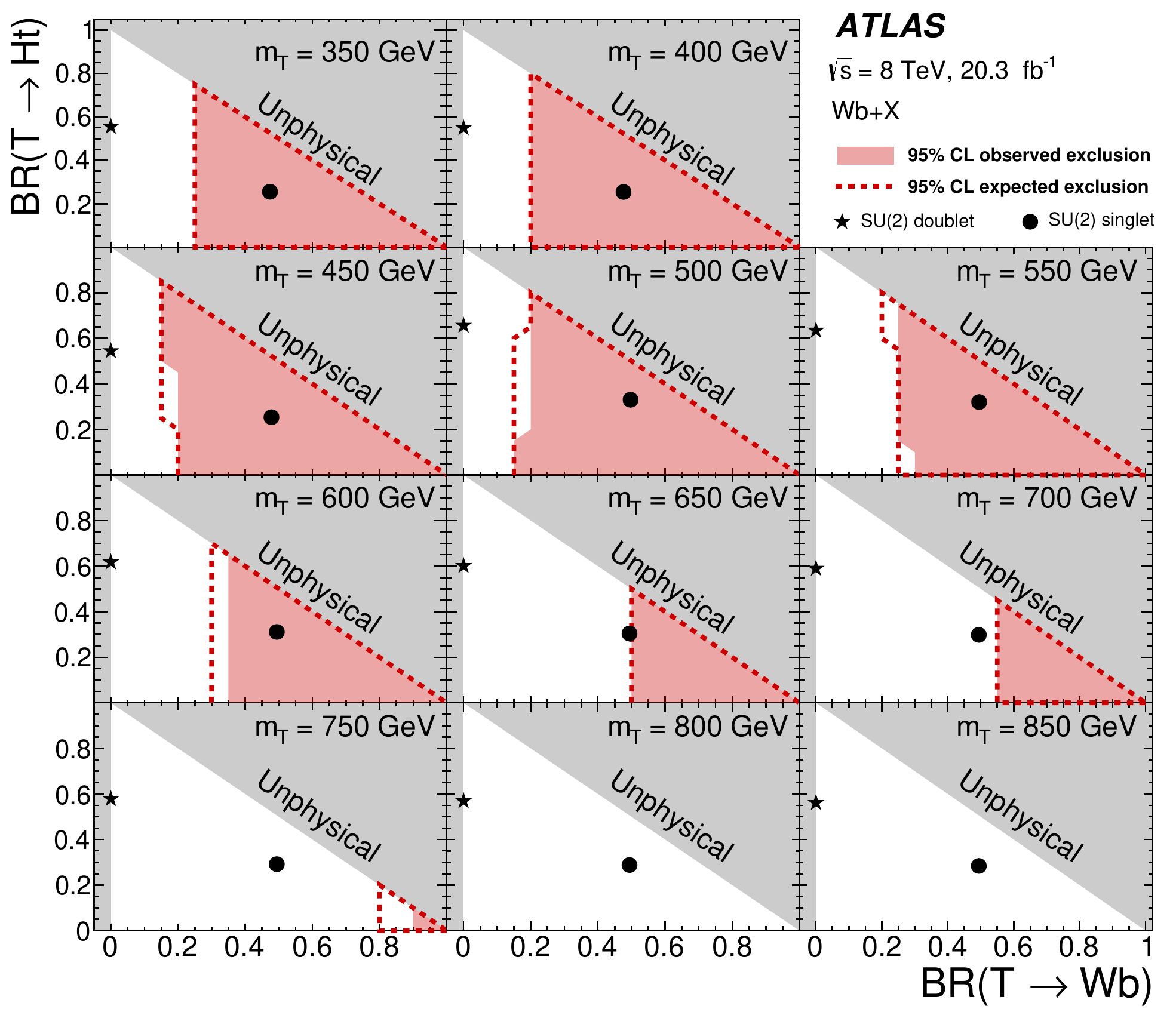}
\caption{
Observed (red filled area) and expected (red dashed line) 95\% CL exclusion in the plane of
$\BR(T \to Wb)$ versus $\BR(T \to Ht)$ for the $T\bar{T} \to Wb$+X search, 
for different values of the vector-like $T$ quark mass.
The grey (dark shaded) area corresponds to the unphysical region where the sum of branching ratios exceeds unity. 
The default branching ratio values from the {\sc Protos} event generator for the weak-isospin singlet and doublet cases 
are shown as plain circle and star symbols respectively. 
\label{fig:limits2D_WbX}}
\end{figure*}

\begin{figure*}[tbp]
\centering
\subfloat[]{\includegraphics[width=0.48\textwidth]{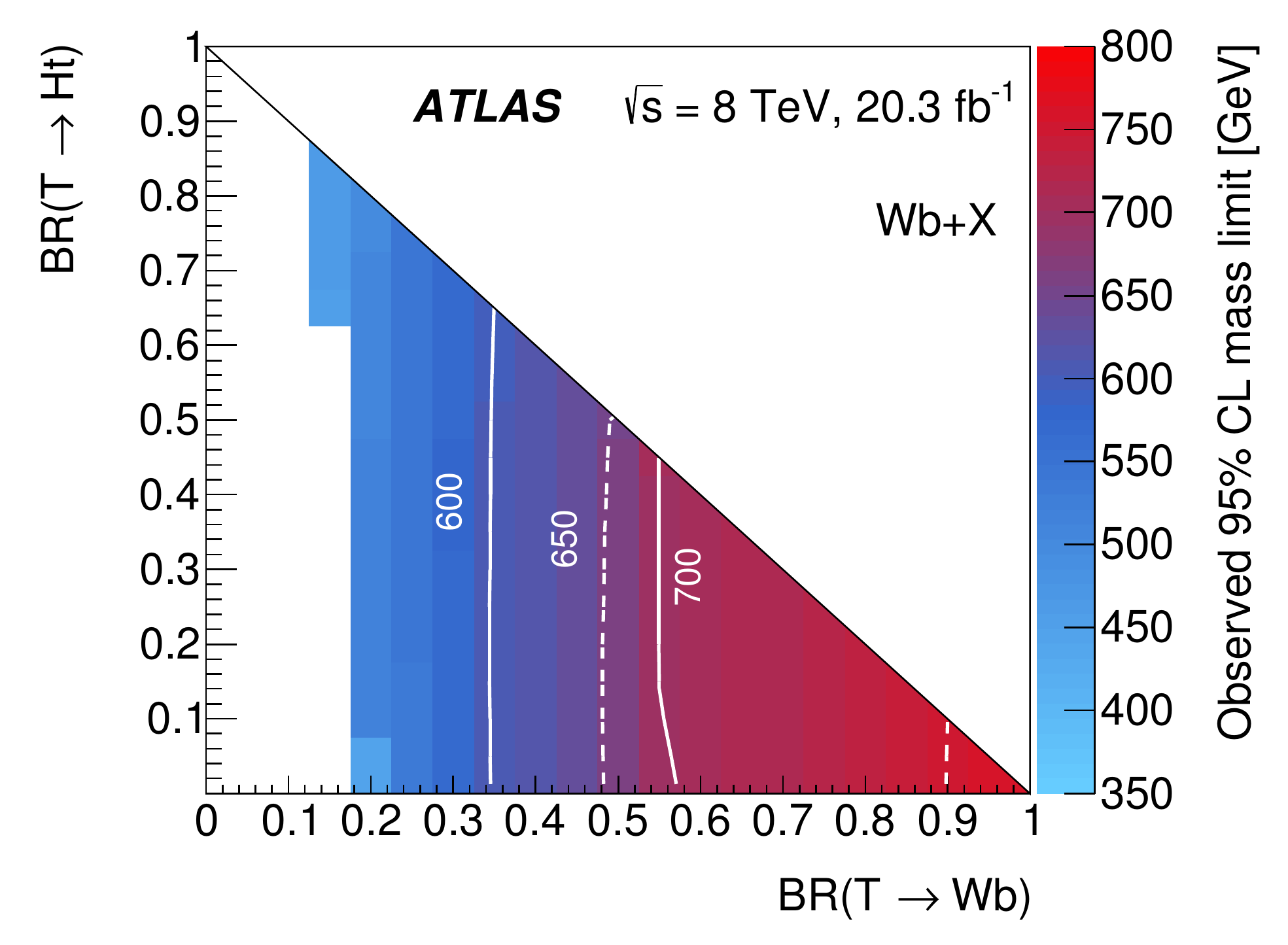}}
\subfloat[]{\includegraphics[width=0.48\textwidth]{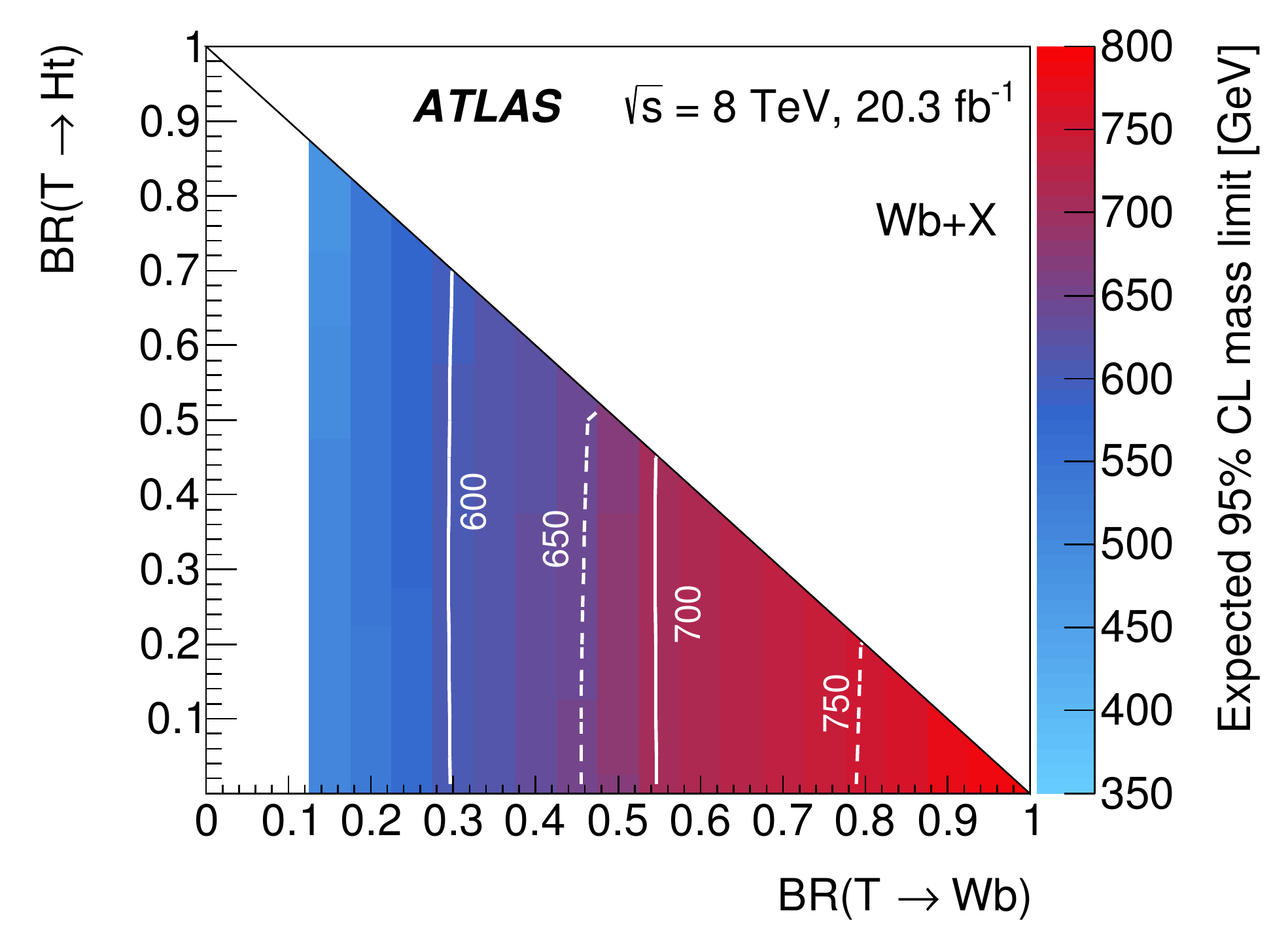}}
\caption{(a) Observed and (b) expected limit (95\% CL) on the mass of the $T$ quark in the plane 
of $\BR(T \to Ht)$ versus $\BR(T \to Wb)$ for the $T\bar{T} \to Wb$+X search. 
Contour lines are provided to guide the eye.
The region shown in white is not excluded for any values of the $T$ quark mass probed.}
\label{fig:limits2D_WbX_temp}
\end{figure*}

\begin{figure}[h!]
\centering
\includegraphics[width=0.9\textwidth]{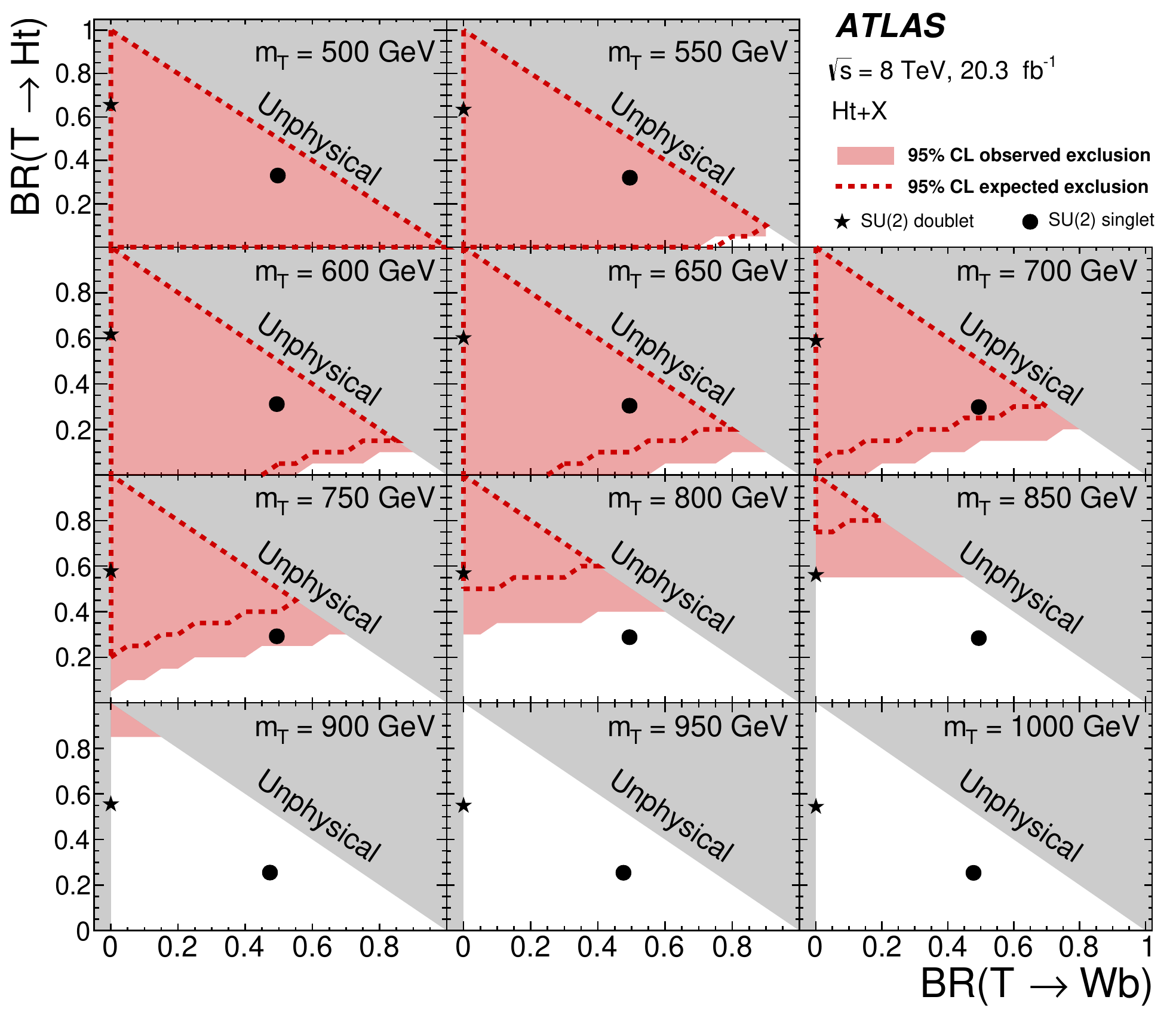}
\caption{
Observed (red filled area) and expected (red dashed line) 95\% CL exclusion in the plane of
$\BR(T \to Wb)$ versus $\BR(T \to Ht)$ for the $T\bar{T} \to Ht$+X search, 
for different values of the vector-like $T$ quark mass.
The grey (dark shaded) area corresponds to the unphysical region where the sum of branching ratios exceeds unity. 
The default branching ratio values from the {\sc Protos} event generator for the weak-isospin singlet and doublet cases 
are shown as plain circle and star symbols respectively. 
\label{fig:limits2D_HtX}}
\end{figure}

\begin{figure*}[tbp]
\centering
\subfloat[]{\includegraphics[width=0.48\textwidth]{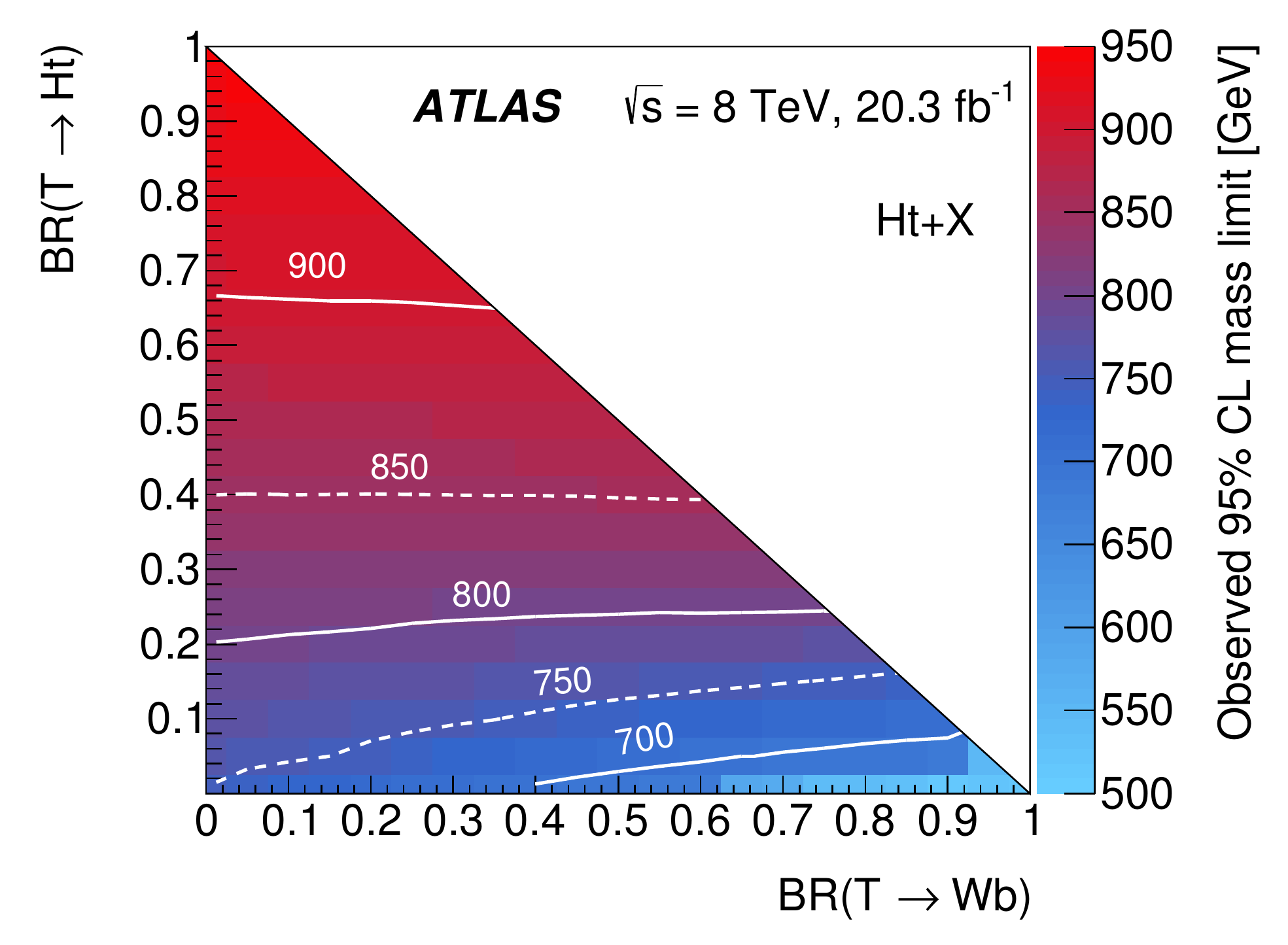}}
\subfloat[]{\includegraphics[width=0.48\textwidth]{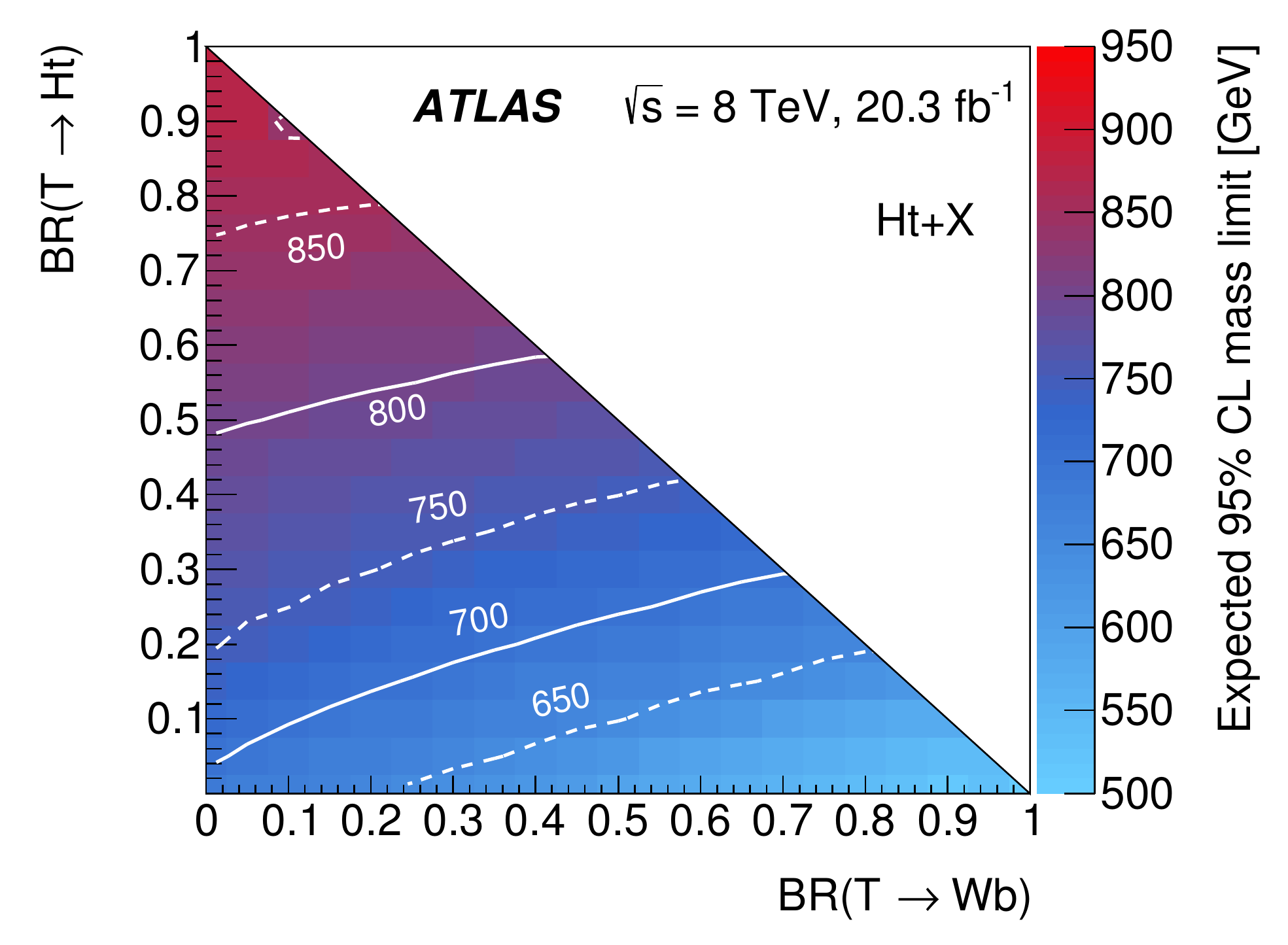}}
\caption{(a) Observed and (b) expected limit (95\% CL) on the mass of the $T$ quark in the plane 
of $\BR(T \to Ht)$ versus $\BR(T \to Wb)$ for the $T\bar{T} \to Ht$+X search.
Contour lines are provided to guide the eye.}
\label{fig:limits2D_HtX_temp}
\end{figure*}


\FloatBarrier

\printbibliography

%
%
%

\newpage 
\begin{flushleft}
{\Large The ATLAS Collaboration}

\bigskip

G.~Aad$^{\rm 85}$,
B.~Abbott$^{\rm 113}$,
J.~Abdallah$^{\rm 151}$,
O.~Abdinov$^{\rm 11}$,
R.~Aben$^{\rm 107}$,
M.~Abolins$^{\rm 90}$,
O.S.~AbouZeid$^{\rm 158}$,
H.~Abramowicz$^{\rm 153}$,
H.~Abreu$^{\rm 152}$,
R.~Abreu$^{\rm 30}$
% ...
\bigskip
\\
% ...
$^{11}$ Institute of Physics, Azerbaijan Academy of Sciences, Baku, Azerbaijan\\
% ...
$^{30}$ CERN, Geneva, Switzerland\\
% ...
$^{85}$ CPPM, Aix-Marseille Universit{\'e} and CNRS/IN2P3, Marseille, France\\
% ...
$^{90}$ Department of Physics and Astronomy, Michigan State University, East Lansing MI, United States of America\\
% ...
$^{107}$ Nikhef National Institute for Subatomic Physics and University of Amsterdam, Amsterdam, Netherlands\\
% ...
$^{113}$ Homer L. Dodge Department of Physics and Astronomy, University of Oklahoma, Norman OK, United States of America\\
% ...
$^{151}$ Institute of Physics, Academia Sinica, Taipei, Taiwan\\
$^{152}$ Department of Physics, Technion: Israel Institute of Technology, Haifa, Israel\\
$^{153}$ Raymond and Beverly Sackler School of Physics and Astronomy, Tel Aviv University, Tel Aviv, Israel\\
% ...
$^{158}$ Department of Physics, University of Toronto, Toronto ON, Canada\\
% ...
$^{*}$ Deceased
\end{flushleft}


\end{document}